\documentclass[12pt]{article}
\pdfoutput=1

\usepackage[usenames,dvipsnames,svgnames,table]{xcolor} 
\usepackage[obeyspaces,hyphens,spaces]{url}
\usepackage{jcapmod}
\usepackage{verbatim}

\usepackage{booktabs}
\usepackage[english]{babel}
\usepackage{amsmath, amssymb, amsbsy, amstext, amsthm, simplewick}
\usepackage{hyperref}
\usepackage{graphicx}
\usepackage{amsfonts}
\usepackage{upgreek}
\usepackage{framed}
\usepackage{tensor}
\usepackage{pifont}
\usepackage{latexsym, mathrsfs}
\usepackage{array}
\usepackage{hyperref}
\usepackage{xspace}
\usepackage{longtable}
\usepackage{multirow}
\usepackage{cases}
\usepackage{empheq}

\usepackage{bm}
\usepackage{bbm}
\usepackage{float}
\usepackage{afterpage}
\usepackage{setspace}

\usepackage[load-configurations=astronomy, range-units=brackets, range-phrase=-, per-mode=reciprocal, mode=math]{siunitx}
\usepackage{threeparttable}
\usepackage{subfig}

\usepackage{tcolorbox}

\allowdisplaybreaks 

\usepackage{braket}

\usepackage{cleveref} 
\Crefname{equation}{Eq.}{Eqs.}
\Crefname{section}{Sec.}{Secs.}
\Crefname{figure}{Fig.}{Figs.}

\usepackage{ragged2e}

\usepackage{hhline}
\usepackage{array}
\newcolumntype{P}[1]{>{\centering\arraybackslash}p{#1}}
\usepackage{booktabs}
\usepackage{tikz}
\usetikzlibrary{calc,shadings,patterns,tikzmark,fadings}

\definecolor{Blue}{rgb}{0.25, 0.41, 0.88}
\definecolor{Red}{rgb}{0.92,0.,0.}
\definecolor{darkorange}{rgb}{1.0,0.549,0.}
\definecolor{cobalt}{RGB}{44, 98, 120}
\definecolor{Mathematica1}{rgb}{0.368417, 0.506779, 0.709798}
\definecolor{Mathematica2}{rgb}{0.880722, 0.611041, 0.142051}
\definecolor{Mathematica3}{rgb}{0.560181, 0.691569, 0.194885}
\definecolor{Mathematica4}{rgb}{0.922526, 0.385626, 0.209179}
\definecolor{Mathematica5}{rgb}{0.528488, 0.470624, 0.701351}
\definecolor{Mathematica6}{rgb}{0.772079, 0.431554, 0.102387}
\definecolor{Mathematica7}{rgb}{0.363898, 0.618501, 0.782349}
\definecolor{Mathematica8}{rgb}{1, 0.75, 0}
\definecolor{Mathematica9}{rgb}{0.647624, 0.37816, 0.614037}
\definecolor{plotBlue}{RGB}{94, 130, 181}
\definecolor{plotRed}{RGB}{233, 85, 54}
\definecolor{plotGreen}{RGB}{142, 176, 50}
\definecolor{plotPurple}{RGB}{135, 120, 178}

\newcolumntype{C}[1]{>{\centering\let\newline\\\arraybackslash\hspace{0pt}}m{#1}}

\def\x{{\bf x}}

\makeatletter
\newlength{\apb@width}
\newcommand{\autoparbox}[2][c]{\settowidth{\apb@width}{#2}\parbox[#1]{\apb@width}{#2}}

\makeatother

\makeatletter
\newsavebox\myboxA
\newsavebox\myboxB
\newlength\mylenA

\newcommand*\xoverline[2][0.75]{
    \sbox{\myboxA}{$\m@th#2$}%
    \setbox\myboxB\null
    \ht\myboxB=\ht\myboxA%
    \dp\myboxB=\dp\myboxA%
    \wd\myboxB=#1\wd\myboxA
    \sbox\myboxB{$\m@th\overline{\copy\myboxB}$}
    \setlength\mylenA{\the\wd\myboxA}
    \addtolength\mylenA{-\the\wd\myboxB}%
    \ifdim\wd\myboxB<\wd\myboxA%
       \rlap{\hskip 0.5\mylenA\usebox\myboxB}{\usebox\myboxA}%
    \else
        \hskip -0.5\mylenA\rlap{\usebox\myboxA}{\hskip 0.5\mylenA\usebox\myboxB}%
    \fi}
\makeatother

\numberwithin{equation}{section}

\def\beq{\begin{equation}}
\def\eeq{\end{equation}}

\def\bea{\begin{eqnarray}}
\def\eea{\end{eqnarray}}

\def\Tr{{\rm Tr}}

\def\beq{\begin{equation}}
\def\eeq{\end{equation}}
\def\bea{\begin{eqnarray}}
\def\eea{\end{eqnarray}}

\def\Tr{{\rm Tr}}

\DeclareRobustCommand{\SkipTocEntry}[4]{}

\usepackage{colortbl}
\definecolor{blue2}{cmyk}{1, 0.1, 0.1, 0.1}

\definecolor{pyBlue}{RGB}{31, 119, 180}
\definecolor{pyRed}{RGB}{214, 39, 40}
\definecolor{pyGreen}{RGB}{44, 160, 44}
\definecolor{pyBlue2}{RGB}{0, 111, 237}
\definecolor{pyRed2}{RGB}{224, 52, 36}

\begin{document}

\tcbset{colframe=black,arc=0mm,box align=center,halign=left,valign=center,top=-10pt}

\renewcommand{\thefootnote}{\fnsymbol{footnote}}

\pagenumbering{roman}
\begin{titlepage}
\baselineskip=5.5pt \thispagestyle{empty}

\bigskip\

\vspace{0.2cm}
\begin{center}
{\Huge \textcolor{Sepia}{\bf \sffamily T}}{\large \textcolor{Sepia}{\bf\sffamily HE}} {\Huge \textcolor{Sepia}{\bf\sffamily G}}{\large \textcolor{Sepia}{\bf\sffamily ENERALIZED}} {\Huge \textcolor{Sepia}{\bf\sffamily OTOC}} \\ \vspace{0.3cm} {\Huge \textcolor{Sepia}{\bf\sffamily F}}{\large \textcolor{Sepia}{\bf\sffamily ROM}} \\ \vspace{0.25cm} {\Huge \textcolor{Sepia}{\bf\sffamily S}}{\large \textcolor{Sepia}{\bf\sffamily UPERSYMMETRIC}} {\Huge \textcolor{Sepia}{\bf\sffamily Q}}{\large \textcolor{Sepia}{\bf\sffamily UANTUM}} {\Huge \textcolor{Sepia}{\bf\sffamily M}}{\large \textcolor{Sepia}{\bf\sffamily ECHANICS}}\\ \vspace{0.25cm}
	{\footnotesize \textcolor{Sepia}{\textbf{\sffamily \selectfont{Study of Random Fluctuations from Eigenstate Representation of Correlation Functions}}}}
\end{center}

\vspace{0.05cm}

\begin{center}
{\fontsize{12}{18}\selectfont Kaushik Y. Bhagat${}^{\textcolor{Sepia}{1}}$},
{\fontsize{12}{18}\selectfont Baibhab Bose${}^{\textcolor{Sepia}{2}}$},
{\fontsize{12}{18}\selectfont Sayantan Choudhury${}^{\textcolor{Sepia}{3,4,5}}$\footnote{{\sffamily \textit{ Corresponding author, E-mail}} : {\ttfamily sayantan.choudhury@aei.mpg.de, sayanphysicsisi@gmail.com}}}${{}^{,}}$
\footnote{{\sffamily \textit{ NOTE: This project is the part of the non-profit virtual international research consortium ``Quantum Aspects of Space-Time \& Matter" (QASTM)} }. }, 
			{\fontsize{12}{18}\selectfont Satyaki Chowdhury${}^{\textcolor{Sepia}{4,5}}$},
			{\fontsize{12}{18}\selectfont Rathindra N. Das${}^{\textcolor{Sepia}{6}}$},
			{\fontsize{12}{18}\selectfont Saptarshhi G. Dastider${}^{\textcolor{Sepia}{7}}$},
			{\fontsize{12}{18}\selectfont Nitin Gupta${}^{\textcolor{Sepia}{8}}$},
			{\fontsize{12}{18}\selectfont Archana Maji${}^{\textcolor{Sepia}{6}}$},\\
		{\fontsize{12}{18}\selectfont Gabriel D. Pasquino${}^{\textcolor{Sepia}{9}}$},
		{\fontsize{12}{18}\selectfont Swaraj Paul${}^{\textcolor{Sepia}{10}}$}
\end{center}


\begin{center}
\vskip4pt
\scriptsize{ \sffamily
\textit{${}^{1}$Indian Institute of Science, Bengaluru, Karnataka-560012, India}\\
\textit{${}^{2}$Department of Physics \& Astrophysics, University of Delhi, Delhi-11007, India}\\
\textit{${}^{3}$Quantum Gravity and Unified Theory and Theoretical Cosmology Group, \\Max Planck Institute for Gravitational Physics (Albert Einstein Institute),\\
	Am M$\ddot{u}$hlenberg 1,
	14476 Potsdam-Golm, Germany.}
	\\
\textit{${}^{4}$National Institute of Science Education and Research, Bhubaneswar, Odisha - 752050, India}\\
\textit{${}^{5}$Homi Bhabha National Institute, Training School Complex, Anushakti Nagar, Mumbai - 400085, India}\\
\textit{${}^{6}$Department of Physics, Indian Institute of Technology Bombay, Powai, Mumbai - 400076, India}\\
\textit{${}^{7}$Sree Chaitanya College, Prafullanagar, Habra, West Bengal - 743268}\\
\textit{${}^{8}$Department of Physical Sciences, Indian Institute of Science Education \& Research Mohali, Punjab - 140306, India}\\
\textit{${}^{9}$University  of Waterloo, 200 University Ave W, Waterloo, ON, Canada, N2L 3G1}\\
\textit{${}^{10}$Discipline of Mathematics, Indian Institute of Technology Indore, Indore 453 552, India}
}

\end{center}

\vspace{0.2cm}
\hrule 
\begin{center}
\textbf{Abstract}
\end{center}

\noindent
The concept of the out-of-time-ordered correlation (OTOC) function is treated as a very strong theoretical probe of quantum randomness, using which one can study both chaotic and non-chaotic phenomena in the context of quantum statistical mechanics. In this paper, we define a general class of OTOC, which can perfectly capture quantum randomness phenomena in a better way. Further, we demonstrate an equivalent formalism of computation using a general time-independent Hamiltonian having well-defined eigenstate representation for integrable supersymmetric quantum systems. We found that one needs to consider two new correlators apart from the usual one to have a complete quantum description. To visualize the impact of the given formalism we consider the two well-known models viz. Harmonic Oscillator and one dimensional potential well within the framework of supersymmetry. For the Harmonic Oscillator case, we obtain similar periodic time dependence but dissimilar parameter dependences compared to the results obtained from both micro-canonical and canonical ensembles in quantum mechanics without supersymmetry. On the other hand, for one dimensional potential well problem, we found significantly different time scale and the other parameter dependence compared to the results obtained from non-supersymmetric quantum mechanics. Finally, to establish the consistency of the prescribed formalism in the classical limit, we demonstrate the phase space averaged version of the classical version of OTOCs from a model-independent Hamiltonian along with the previously mentioned these well-cited models.

\vskip10pt
\hrule
\vskip10pt

\noindent
\text{Keywords:  OTOC, Supersymmetry, Out-of-equilibrium quantum statistical mechanics.} 

\end{titlepage}
\begin{center}
$~~~~~~~~~~~~~~~~~~~~~~~~~~~~~~~~~~~~~~~~~~~~~~~~~~~~~~~~~~~~~~~~~~~~~~~~~~~~~~~~~~~$\\
$~~~~~~~~~~~~~~~~~~~~~~~~~~~~~~~~~~~~~~~~~~~~~~~~~~~~~~~~~~~~~~~~~~~~~~~~~~~~~~~~~~~$\\
$~~~~~~~~~~~~~~~~~~~~~~~~~~~~~~~~~~~~~~~~~~~~~~~~~~~~~~~~~~~~~~~~~~~~~~~~~~~~~~~~~~~$\\
$~~~~~~~~~~~~~~~~~~~~~~~~~~~~~~~~~~~~~~~~~~~~~~~~~~~~~~~~~~~~~~~~~~~~~~~~~~~~~~~~~~~$\\
$~~~~~~~~~~~~~~~~~~~~~~~~~~~~~~~~~~~~~~~~~~~~~~~~~~~~~~~~~~~~~~~~~~~~~~~~~~~~~~~~~~~$\\
$~~~~~~~~~~~~~~~~~~~~~~~~~~~~~~~~~~~~~~~~~~~~~~~~~~~~~~~~~~~~~~~~~~~~~~~~~~~~~~~~~~~$\\
$~~~~~~~~~~~~~~~~~~~~~~~~~~~~~~~~~~~~~~~~~~~~~~~~~~~~~~~~~~~~~~~~~~~~~~~~~~~~~~~~~~~$\\
$~~~~~~~~~~~~~~~~~~~~~~~~~~~~~~~~~~~~~~~~~~~~~~~~~~~~~~~~~~~~~~~~~~~~~~~~~~~~~~~~~~~$\\
$~~~~~~~~~~~~~~~~~~~~~~~~~~~~~~~~~~~~~~~~~~~~~~~~~~~~~~~~~~~~~~~~~~~~~~~~~~~~~~~~~~~$\\
$~~~~~~~~~~~~~~~~~~~~~~~~~~~~~~~~~~~~~~~~~~~~~~~~~~~~~~~~~~~~~~~~~~~~~~~~~~~~~~~~~~~$\\
$~~~~~~~~~~~~~~~~~~~~~~~~~~~~~~~~~~~~~~~~~~~~~~~~~~~~~~~~~~~~~~~~~~~~~~~~~~~~~~~~~~~$\\
$~~~~~~~~~~~~~~~~~~~~~~~~~~~~~~~~~~~~~~~~~~~~~~~~~~~~~~~~~~~~~~~~~~~~~~~~~~~~~~~~~~~$\\
$~~~~~~~~~~~~~~~~~~~~~~~~~~~~~~~~~~~~~~~~~~~~~~~~~~~~~~~~~~~~~~~~~~~~~~~~~~~~~~~~~~~$\\
$~~~~~~~~~~~~~~~~~~~~~~~~~~~~~~~~~~~~~~~~~~~~~~~~~~~~~~~~~~~~~~~~~~~~~~~~~~~~~~~~~~~$\\
$~~~~~~~~~~~~~~~~~~~~~~~~~~~~~~~~~~~~~~~~~~~~~~~~~~~~~~~~~~~~~~~~~~~~~~~~~~~~~~~~~~~$\\
$~~~~~~~~~~~~~~~~~~~~~~~~~~~~~~~~~~~~~~~~~~~~~~~~~~~~~~~~~~~~~~~~~~~~~~~~~~~~~~~~~~~$\\
$~~~~~~~~~~~~~~~~~~~~~~~~~~~~~~~~~~~~~~~~~~~~~~~~~~~~~~~~~~~~~~~~~~~~~~~~~~~~~~~~~~~$\\
$~~~~~~~~~~~~~~~~~~~~~~~~~~~~~~~~~~~~~~~~~~~~~~~~~~~~~~~~~~~~~~~~~~~~~~~~~~~~~~~~~~~$\\
\textit{ \fontsize{53}{56}\selectfont  \bfseries \textcolor{Sepia}{We would like to dedicate this work for the people those who are helping us to fight against
COVID-19 pandemic across the globe.}}
\end{center}
\newpage
\thispagestyle{empty}
\setcounter{page}{2}
\begin{spacing}{1.03}
	\tableofcontents
\end{spacing}

\clearpage
\pagenumbering{arabic}
\setcounter{page}{1}

\renewcommand{\thefootnote}{\arabic{footnote}}

\textcolor{Sepia}{\section{\sffamily Introduction}\label{sec:introduction}}
The concept of out-of-time-ordered-correlators (OTOC) first introduced by the author duo, {\it Larkin} and {\it Ovchinnikov} to describe the semi-classical correlation in the context of superconductivity \cite{larkin1969quasicl}, which was mostly used in various condensed matter systems to study various out-of-equilibrium phenomena in the quantum regime \cite{Choudhury:2018rjl}. However, recently it has attracted the attention of theoretical physicists from other branches in very different context finding applications in the finite-temperature extension of quantum field theories, bulk gravitational theories, quantum black holes, and many more sensational topics in the list\cite{Haehl:2017eob, Haehl:2017qfl, Chaudhuri:2018ihk, Chaudhuri:2018ymp, Chakrabarty:2018dov, Gharibyan:2019sag}. It is considered to be one of the strongest theoretical probes for quantifying quantum chaos in terms of quantum {\it Lyapunov exponent} \cite{Gharibyan:2018fax}, quantum theories of {\it stochasticity} and  {\it randomness} among the theoretical physics community. Besides playing a key role in investigating the holographic duality \cite{Heemskerk:2009pn, Heemskerk:2010ty, Czech:2015qta, kitaev2015si} between a strongly correlated quantum system and a gravitational dual system, it also characterizes the {\it chaotic behaviour} and {\it information scrambling} \cite{Zhuang:2019jyq, Hartmann:2019cxq, Han:2018bqy, Yoshida:2019qqw, Yan:2020wkt, Anous:2019yku} in the context of many-body quantum systems \cite{Sahu:2020xdn, Swingle:2018ekw, Gharibyan:2018jrp}. The detailed study of OTOCs reveal an intimate relationship between three entirely different physical concepts, namely {\it holographic duality}, {\it quantum chaos} and {\it information scrambling}. The key idea of OTOCs can be best understood as the growth of the non-commutativity of quantum mechanical operators\footnote{Specifically this non-commutative structure of the quantum operators describe the unequal time commutation relations (UETCRs) within the framework of quantum mechanics. However, the mathematical structure as well as the physical consequences of these correlators in the quantum regime is completely different from the concept of formulating advanced and the retarded correlators. In the later part of this paper we will explicitly demonstrate such differences or the correlators within the framework of micro-canonical and canonical quantum statistical systems.} which are defined at different time scales, and hence can be described using the Poisson Brackets for its classical counterpart. Not only that but also the quantum mechanical thermal ensemble average or equivalently the quantum mechanical trace operation can be described by using the phase space average in the classical limit. It is considered as the quantum mechanical analogue of the classical sensitiveness to the initial conditions in the time dynamics of a quantum system. The exponential growth of these correlators indicates the presence of chaos in the quantum system, which has led to discussions of the ``{\it butterfly effect}" in black holes \cite{Shenker:2013pqa, Shenker:2014cwa, Addazi:2015gna} with a saturation bound on the {\it quantum Lyapunov exponent} and for various spin models \cite{Aleiner:2016eni, Roberts:2014ifa}. 

There has been a growing interest to understand the behaviour of OTOC even for systems where quantum chaos is expected to be absent, the most relevant example being the study of OTOC in the quantum Ising spin chain model, where power law growth of OTOC's is observed as opposed to the exponential growth in non-integrable models in support of non-chaoticity \cite{Lin:2018tce, Kukuljan:2017xag, Huang:2016knw}. Another interesting revelation came from a recent study of OTOC's for a  quantum system with discrete energy levels, weakly coupled to a non-adiabatic dissipative thermal environment. This type of system is commonly known as {\it open quantum system} (OQS), where the OTOC was found to saturate exponentially in contrast to the exponential growth for a {\it quantum chaotic system} \cite{Syzranov}.
OTOC's has been of prime theoretical interest for diagnosing also the {\it rate of growth of chaos} with respect to the different time scales involved in the quantum system through the operators and hence for studying the scrambling of quantum information in black holes and strongly correlated quantum mechanical systems. It serves as a strong theoretical probe for investigating various bulk gravitational dual theories in the framework of AdS/CFT correspondence. Among others, the existence of shock waves inside black holes \cite{Roberts:2014isa, Shenker:2013yza, Stanford:2014jda} and the maximum saturation bound of the quantum version of the {\it Lyapunov exponent} are the most famous examples where out of time ordered correlation functions have proven to be useful in the Ads/CFT correspondence. This maximum saturation bound is famously known as the {\it Maldacena-Shenker-Stanford (MSS) bound} \cite{Maldacena:2015waa}. The SYK model \cite{Sachdev:1992fk, Maldacena:2016hyu, Fu:2016vas, Rosenhaus:2018dtp, Gross:2017aos, Polchinski:2016xgd, Choudhury:2017tax, Klebanov:2020kck, Marcus:2018tsr, Kobrin:2020xms, Gu:2019jub, Das:2017eiw, Das:2017wae, Almheiri:2019jqq, Nosaka:2019tcx, Turiaci:2017zwd, Witten:2016iux, Li:2017hdt, Gurau:2019qag, Klebanov:2016xxf, Klebanov:2017nlk, Bulycheva:2017ilt, Giombi:2018qgp, Klebanov:2018fzb, Kim:2019upg, Kitaev:2017awl, Gurau:2017xhf, Gurau:2017qna, Benedetti:2017fmp, Benedetti:2018goh, Gurau:2016lzk} is a well known model which saturates this well known bound and shows the signature of maximal chaos both in its 1D CFT and gravitational dual 2D balck hole counterpart. In the ref.~\cite{Murthy:2019fgs} this bound on {\it quantum Lyapunov exponent} was further generalized for many body systems using the well known {\it Eigenstate Thermalization Hypothesis (ETH)}.  Very recently in the ref.~\cite{Choudhury:2020yaa}, the author used the tools and techniques of computing OTOC in the context of Cosmology by following the underlying slogan {\it Cosmology meets Condensed Matter Physics} to study the quantum mechanical correlation functions from random primordial fluctuations appearing in the context of cosmological perturbation theory of background spatially flat FLRW metric. Specifically, these fluctuations are appearing in the context of stochastic particle production during inflation, during the epoch of reheating and also for the primordial phenomena which is governed by the quantum generalization of {\it Brownian motion} i.e. for the cosmological epochs in the time line of the universe where the physics of out-of-equilibrium phenomena play significant role. 

In ref.~\cite{Hashimoto:2017oit}, the earlier discussed fact regarding the exponential growth of the OTOCs in the associated time scales to describe the chaotic fluctuations for non-integrable systems has been established for various well known quantum mechanical systems. A study of the same for integrable models suggest non-chaotic quantum mechanical fluctuations in the quantum regime. In the present context, the phrase {\it quantum randomness} describes a physical phenomena which describes chaotic or non-chaotic i.e. in principle any random behaviour of a system with respect to the associated time scales of the system. For the physical systems such {\it quantum randomness} can be described by the following two-fold formalisms:
\begin{enumerate}
\item \underline{\textcolor{red}{\bf Formalism~I:}}\\ \\
The first approach is based on the construction and the mathematical from of the solution of the {\it Fokker Planck equation}, using which various stochastic phenomena in the quantum out-of-equilibrium regime can be studied. One of the famous example is the stochastic cosmological particle production phenomena which can be directly mapped to a problem of solving Schr$\ddot{o}$dinger equation with an impurity potential within the framework of quantum mechanics, which is actually describing propagation of electrons inside an electrical wire in presence of an impurity or defect. Within the framework of quantum statistical mechanics instead of solving the Schr$\ddot{o}$dinger problem directly or may be solving the dynamical equation for the quantum mechanical fluctuations during the stochastic particle production one can think about a cumulative probability distribution function of of this stochastic process, ${\cal P}(n,\tau)$ which depend on two crucial quantities, which are - the number density of the produced particle and the associated time scale of the stochastic process. Using a detailed physical arguments and computation one can explicitly show that this probability distribution function, ${\cal P}(n,\tau)$, satisfy {\it Fokker Planck equation}, which is given by:
\bea &&\underline{\textcolor{red}{\bf At~ infinite~temperature~~(\beta\rightarrow 0):}}~~~~~~~\nonumber\\
&&~~~~~~~~~\frac{1}{\mu}\frac{\partial {\cal P}(n,\tau)}{\partial\tau}=\underbrace{n(n+1)\frac{\partial^2 {\cal P}(n,\tau)}{\partial n^2}}_{\textcolor{red}{\bf Diffusion}}+\underbrace{(1+2n)~\frac{\partial {\cal P}(n,\tau)}{\partial n}}_{\textcolor{red}{\bf Drift}}~~,~~~~~~~\\
 && \underline{\textcolor{red}{\bf At~ finite~temperature~~(\beta\neq 0 ):}}~~~~~~~\nonumber\\
 &&~~~~~~~~\frac{1}{\mu}\frac{\partial {\cal P}(n,\tau)}{\partial\tau}=\underbrace{n(n+1)\frac{\partial^2 {\cal P}(n,\tau)}{\partial n^2}}_{\textcolor{red}{\bf Diffusion}}+\underbrace{(1+2n)~\frac{\partial {\cal P}(n,\tau)}{\partial n}}_{\textcolor{red}{\bf Drift}}~~\nonumber\\
 &&~~~~~~~~+\underbrace{\beta\left\{n(n+1)\left[\frac{\partial V(n)}{\partial n}~\frac{\partial  {\cal P}(n,\tau)}{\partial n}+\frac{\partial^2 V(n)}{\partial n^2}~  {\cal P}(n,\tau)\right]+(2n+1)~\frac{\partial V(n)}{\partial n}~  {\cal P}(n,\tau)\right\}}_{\textcolor{red}{\bf Finite~temperature~contribution}}.~~~~~~~\eea
where $\mu$ represents the mean stochastic particle production rate and $V(n)$ associated potential which is only significant at finite temperature. By solving these set of equations in presence of appropriate initial conditions one can get to know about the related profile of the stochastic process semi classically in the present context. Just not only that, but also one can treat these versions of the {\it Fokker Planck equation} as the semi classical statistical moment generating equations because by replacing the profile function ${\cal P}(n,\tau)$ with the appropriate moment generating function $ {\cal F}(n)$ one can compute all the moments. To serve this purpose one need to use the following fundamental equation:
\bea \textcolor{red}{\bf Statistical~ average~ of~ moment~ generator:}~~\langle {\cal F}(n)\rangle:=\int dn~{\cal F}(n)~{\cal P}(n,\tau),~~~~~~\eea
which physically represents the expectation value or the statistical average value of the number density dependent moment generating any arbitrary function ${\cal F}(n)$. Further substituting the appropriate form of this function in the moment dependent {\it Fokker Planck equations} one can explicitly compute the expression for all the physically relevant statistical moments i.e. $\langle n\rangle$, $\langle n^2\rangle$, $\cdots$ explicitly without explicitly knowing about the particular mathematical structure of the profile of the related stochastic dynamical process at infinite or finite temperature for a given structure of number density potential function. These moments are extremely important in the present context of study as all of them semi classically compute the expressions for all the equal time quantum correlation functions required to study the out-of-equilibrium aspects, such as stochastic effects, disorder, random fluctuations etc. both at infinite and finite temperatures. See the references \cite{Choudhury:2018rjl, Choudhury:2018bcf, Amin:2015ftc, Garcia:2019icv, Garcia:2020mwi} where all authors have studied the physical impact of this formalism to describe out-of-equilibrium aspects in various different contexts. 

 Now let us speak about some applicability issues related to this particular formalism. Since this formalism only allows us to know about the effect of the semi-classical correlations at equal time that might be not very interesting when we are actually thinking of doing the computation of the correlations and its rate of change at different time scales associated with the quantum mechanical system of study. For an example, if we are interested to compute the any general $N$-point semi-classical correlation function as defined by the following expression:
\bea \langle \prod^{N}_{i=1}n(\tau_i)\rangle=\langle n(\tau_1)n(\tau_2)\cdots n(\tau_N)\rangle~~~~\forall~~\tau_i~(i=1,2,\cdots,N)~~{\rm \textcolor{red}{\bf contain~disorder}},~~~~~~~~ \eea
then this particular formalism will not work, as using this formalism one cannot capture the effect of disorder effect in the associated time scale of the system. On the other by following the usual tools and techniques one can only compute the above mentioned $N$-point correlators either in time ordered sense, where $\tau_1>\tau_2>\cdots>\tau_N$ or in the anti-time ordered sense, where $\tau_1<\tau_2<\cdots<\tau_N$. So from the technical ground defining this 
$N$-point correlators including the effect of disorder in the time scale at any arbitrary temperature is also a very important topic of research and here comes the crucial role of the next formalism where we can allowed to define and explicitly compute such quantum effects at the level of correlation functions. Another important aspect we want to point out here that, the present formalism don't bother about any {\it Lagrangian} or the {\it Hamiltonian} formulation of the associated quantum mechanical system under consideration. So if we really interested to know about the effect of time disordering in the expressions for the quantum mechanical correlation functions which are defined in terms of the fundamental operators appearing in the quantum version of the {\it Lagrangian} or the {\it Hamiltonian} of the system under consideration and also want to explicitly know about time variation with respect to different time scales associated with the system, which are actually the source of time disordering, then instead of the present formalism it is obviously technically correct and easier to think about the implementation of the second formalism, which gives us the better understanding of time scale disordering. In the next point and in the rest of the paper we will follow the second formalism to compute the quantum correlation functions from the fundamental operators from the quantum mechanical systems under study which can explicitly capture the effect of disordering in the time scale. Not only us, but also the present trend in the research suggesting to make use of the next formalism to get better understanding of time disordering phenomena in quantum mechanical system.

\item  \underline{\textcolor{red}{\bf Formalism~II:}}\\ \\
The second approach is based on finding quantum correlation functions including the time disordering effect and through out the paper we have followed this formalism to study effects of out-of-equilibrium physics in physical systems\cite{Choudhury:2020yaa}. The present computational methodology helps us to know more about the underlying unexplored physical facts regarding the quantum mechanical aspects of various stochastic random process where time ordering or anti time ordering is not at all important and instead of that disorder in the time scale can be captured in the quantum correlations at very early time scale. This method not only helps us to know about the early time behaviour of quantum correlations in the out-of-equilibrium regime of the quantum statistical mechanics, but also give crucial information regarding the late time equilibrium behaviour of the quantum correlations of a specific quantum system. However for all the systems in nature the above interpretation of the quantum mechanical aspects of the randomness phenomena are not same. Based on all these types of quantum systems one can categorize the random time disordering phenomena as, A. Chaotic system which shows exponential growth in the quantum correlators, B. Non-chaotic system which shows periodic or aperiodic or irregular random fluctuations in the quantum correlators. The best possible theoretical measure of all such time disorder averaging  phenomena for various statistical ensembles, micro-canonical and canonical ensembles are described by out-of-time-ordered correlation (OTOC) function within the framework of quantum statistical mechanics. Let us define, a set of operators, ${\cal O}_i(t_j)~\forall~i,j=1,2$ with $i\neq j$ or $i=j$ possibilities. The time disorder thermal average over statistical ensemble is described by the following expression:
\bea C^{(ij)}_{N}(t_1,t_2):=-\langle \left[{\cal O}_i(t_1),{\cal O}_j(t_2)\right]^{N}\rangle_{\beta}&=&-{\rm Tr}\left[\rho_{\beta}~ \left[{\cal O}_i(t_1),{\cal O}_j(t_2)\right]^{N}\right]~~~~\forall~~i,j=1,2,~~~~~~~~~\eea
where the thermal density matrix $\rho_{\beta}$ is defined as:
\bea \rho_{\beta}:=\frac{1}{Z}~\exp\left(-\beta H\right)~~{\rm with}~~Z={\rm Tr}\left[\exp(-\beta H)\right].~~~~\eea
In the present context, $C^{(ij)}_{N}(t_1,t_2)$, represent three possible types of OTOC out of which only $i\neq j$ possibility which will describe only one OTOC have been explored in earlier works in this area. The other two OTOCs which are appearing from the $i=j$ possibility will be explicitly studied in this paper. The prime objective is to incorporate two more type of OTOCs along with the well known other OTOC is to study the all possible signatures of time disordering average from a quantum mechanical system. Our expectation is all these three types of OTOCs can able to describe the more general structure of stochastic randomness or any simple type of random process in the quantum regime. Though, this idea was revived by {\it Kitaev}, then followed by {\it Maldacena, Shenker} and {\it Stanford} (MSS) and many more to study the quantum mechanical signature of chaos, which is actually the $i\neq j$ case in the above definition, but the mathematical structure of the other two OTOCs represented by the $i=j$ case suggests that any non-chaotic behaviour, such as, periodic or aperiodic time dependent behaviour, any time dependent growth in the correlators which are different from any type of exponential growth, any type of decaying behaviour in the correlators can be explained in a better way compared to just studying the time dependent behaviour from the well known OTOC which are commonly used in the literature. So in short, to give a complete picture of any kind of time disordering phenomena it is better to study all these possible three types of OTOCs to finally comment on the properties of any physical systems in quantum mechanical regime. Few other important things we want to point out here for better understanding the structure of all these OTOCs capturing the underlying physics of disordering averaging phenomena. First of all here we have to mention that in the definition that we have provided in this paper using which we can able to compute three sets of $2N$ point OTOCs. Though in the further computation we have restricted our study in the paper by considering $N=1$ and $N=2$ cases, but to study the general time disorder averaging process one may study the any even multipoint (i.e.~$2N$ point) correlation functions from the present definition. Now here the $N=1$ case is basically representing a non-zero UTCR and can be treated as the building block of the full computation as this particular case is mimicking the computation of the Green's function in presence of time disordering. More technically, one can interpret that this contribution is made up of two disconnected time disorder averaged thermal correlator. These disconnected contributions are extremely significant if we wait for a large time scale, in literature we usually identified this time scale as a dissipation time scale on which one can explicitly factorize any higher $2N$ point correlators in terms of the non-vanishing disconnected contributions. For this specific reason one can treat $N=1$ case result as the building block of the any higher $2N$ point thermal correlators. However, for most of the quantum systems the $N=1$ case shows random but decaying behaviour with respect to the associated time scales which are explicitly appearing in the quantum operators of the theory. For this reason study of any $N>1$ play significant role to give a better understanding of the time disordering phenomena. For this purpose next we have studied the $N=2$ case, which represents the four point thermal correlator in the present context of study and one of the most significant quantity in the present day research of this area which can capture better information regarding the time disorder averaging compared to the $N=1$ case. In a future version of this work we have a plan to extend the present computation to study the physical implications of $N>2$ quantum correlators to better understand the time disorder averaging phenomena. Now we will comment on the technical side of the present formalism using which one can explicitly compute these OTOCs in the present context. First of all we talk about the time independent Hamiltonian of a quantum system which have their own eigenstates with specific energy eigenvalue spectrum. In this case, construction of the OTOCs describing the time disorder thermal averaging over a canonical ensemble is described by two crucial components, the Boltzmann factor on which the general eigenstate dependent spectrum appears and also the temperature independent micro-canonical part of the OTOCs. At the end we need to take the sum over all possible eigenstates, which will finally give a simplified closed expressions for OTOCs in the present context. Due to the appearance of the eigenstates from the time independent Hamiltonian this particular procedure will reduce the job extremely to study the time dependent behaviour of all the previously mentioned OTOCs that we have defined earlier in this paper. In the rest of the paper we have followed this prescription which is only valid for time independent Hamiltonian which have their own well defined eigenstates. For more details see the rest of the computations and related discussion that we have studied in this paper. Most importantly, using this formalism we can compute all of these OTOCs in a very simple model -independent way. The other technique is more complicated than the previously discussed one. In this case one starts with a time dependent Hamiltonian of the theory and using the well known, {\it Schwinger Keldysh formalism}, which is a general path integral framework at finite temperature for the study of the time evolution of a quantum mechanical system which is in the out-of-equilibrium state. At the early time scale once a small perturbation or a response is provided to a quantum system, then it is described by a out-of-equilibrium process within the framework of quantum statistical mechanics and the present formalism provides us the sufficient tools and techniques using which one can compute the expressions for the OTOCs. Not only that, the late time behaviour of such OTOC is described by a saturation behaviour for chaotic systems from which one can compute the various characteristic features of large time equilibrium behaviour from these OTOCs.
\newpage

\item  \underline{\textcolor{red}{\bf Formalism~III:}}\\ 
 The third approach is based on the {\it circuit quantum complexity},\cite{Bhattacharyya:2019kvj, Bhattacharyya:2018bbv, Susskind:2018pmk, Susskind:2018fmx, Brown:2019whu, Cotler:2017jue, Brown:2016wib} which is relatively a very new concept and physically defined as the minimum number of unitary operators, commonly known {\it as quantum gates},  that are specifically required to construct the desired target quantum state from a suitable reference quantum mechanical state. In amore generalized physical prescription, quantum mechanical complexity can serve as one of several strong diagnostics for
probing the time disorder averaging phenomena of a quantum mechanical system or quantum randomness.
The underlying physical concept of circuit complexity can essentially provide essential
information regarding various aspects of quantum mechanical randomness, such as the concept of {\it scrambling time}, {\it Lyapunov exponent}~\footnote{The associated time scale when the quantum circuit complexity starts to
grow is usually identified to be the {\it scrambling time} scale and in the representative plot with respect to the time scale particularly the magnitude of the slope of the linear portion of the curve is physically interpreted as the {\it Lyapunov exponent} for the specific systems where the general quantum randomness or the time disorder averaging phenomena is described by {\it quantum chaos}.}, etc, which are particularly the key features of the study of quantum mechanical chaos.
One can further compare between the physical outcomes of the two strongest measures of quantum randomness, which are appearing from out-of-time-order correlators (OTOCs) and the quantum mechanical circuit complexity and comment on further that for a specific quantum system which one is capturing the more information regarding the description of quantum mechanical randomness.
\end{enumerate}
In this paper we generalize the study of OTOCs for investigating the phenomena of quantum randomness in various supersymmetric integrable quantum mechanical models. 
The main motivation behind introducing the concept of supersymmetry lies in the well established fact that for any Supersymmetric quantum mechanical model, the original Hamiltonian is always associated with a partner Hamiltonian, which in general is widely different from its original counterpart in terms of eigenstates. Though not always but it is possible that the quantum mechanical model under consideration, attains vastly different properties in the context of quantum randomness, due to the introduction of supersymmetry within the framework of quantum mechanics. It is our expectation that this generalization of the study of all the classes of OTOCs would provide an understanding about the significant role that supersymmetry plays in modifying the randomness behaviour of the quantum mechanical models under investigation. Most importantly, from the present study it will be clear that the additional inclusion of symmetries in the form of supersymmetry will at all effect, if so then how much it will effect the time disorder averaging phenomena for a canonical and micro-canonical statistical ensemble studied within the framework of out-of-equilibrium quantum statistical mechanics. In the eigenstate representation of the OTOCs we have actually studied three types of OTOC in this paper, out of that one of them is commonly studied in this literature mostly used to explain the phenomena of quantum mechanical randomness (not only chaotic behaviour, but also a general feature including any types of non-chaotic behaviour), and the other two OTOCs that we have included in this paper might not be completely independent physical information of each other, but extremely important to capture the complete quantum mechanical effect in the quantum mechanical correlation functions to give a more general physical interpretation and a complete and detailed description of time disorder averaging phenomena for various quantum statistical ensembles within the framework of supersymmetric quantum mechanics. Now if we can't able to find out the eigenstate representation of a given time independent Hamiltonians within the framework of supersymmetry then the prescribed methodology for computing the general class of OTOCs in terms of the simplest eigenstate representation will not work. This can happen for the physical systems which is actually described by the time dependent Hamiltonians. In that case to compute all of these previously mentioned general class of OTOCs one need to use the quantum mechanical path integral generalization of the present framework, which is commonly described by the, {\it Schwinger Keldysh formalism}, at finite temperature within the framework of supersymmetric quantum mechanics. The good news is physical outcomes of such a generalization also have not studied yet, which we have future plan to look into in detail and we are also hopeful to get non-trivial and better understanding of various supersymmetric quantum mechanical models out of these computations.
\\
\\
\textbf{Organization of the Paper is as follows- }
\begin{itemize} 

\item In \Cref{sec:revSUSYQM} we provide a brief review of the concept of Supersymmetric Quantum Mechanics. 

\item In \Cref{sec:qrandcorr}, we explain how the phenomenon of Quantum Randomness can be diagnosed through the out of time ordered correlators.

\item  In \Cref{sec:eigenrepcorr}, we provide a model independent eigenstate representation of the 2 and the 4-pt correlators of all the three kinds defined equally well for any QM model with well defined eigenstates.

\item  In \Cref{sec:susyqmho} we explicitly calculate the correlators for the supersymmetric Harmonic Oscillator. 

\item In \Cref{sec:qbox} we provide the numerical calculations of the correlators for the Supersymmetric 1D infinite potnetial well. 

\item In \Cref{sec:classlimit} we discuss the semiclassical analogue results for the two Supersymmetric QM model. 

\item  Finally we conclude with the most important observations from our analysis of the considered Supersymmetric QM models.

\end{itemize}

\begin{figure}[h!]
	\centering
	\includegraphics[width=.86\textwidth,height=.50\textheight,page=1]{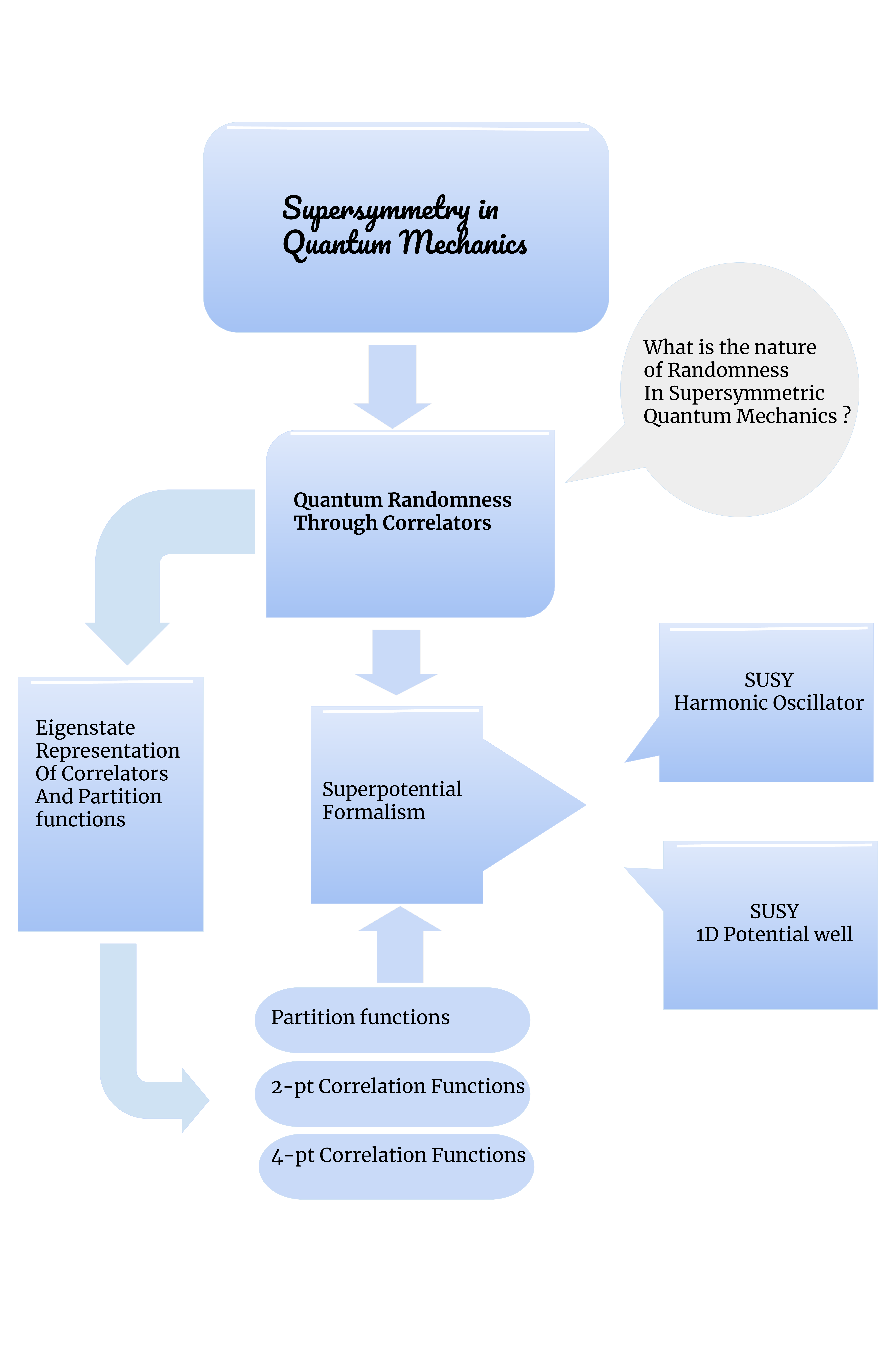}
	\includegraphics[width=.86\textwidth,height=.50\textheight,page=2]{OTOCDesign-2-new}
	\caption{This mnemonic diagram shows the organization of the entire paper.}
	\label{fig:OTOCDesign2}
\end{figure}

\begin{figure}[h]
	\centering
	\includegraphics[width=1.1\textwidth, height=0.9\textheight]{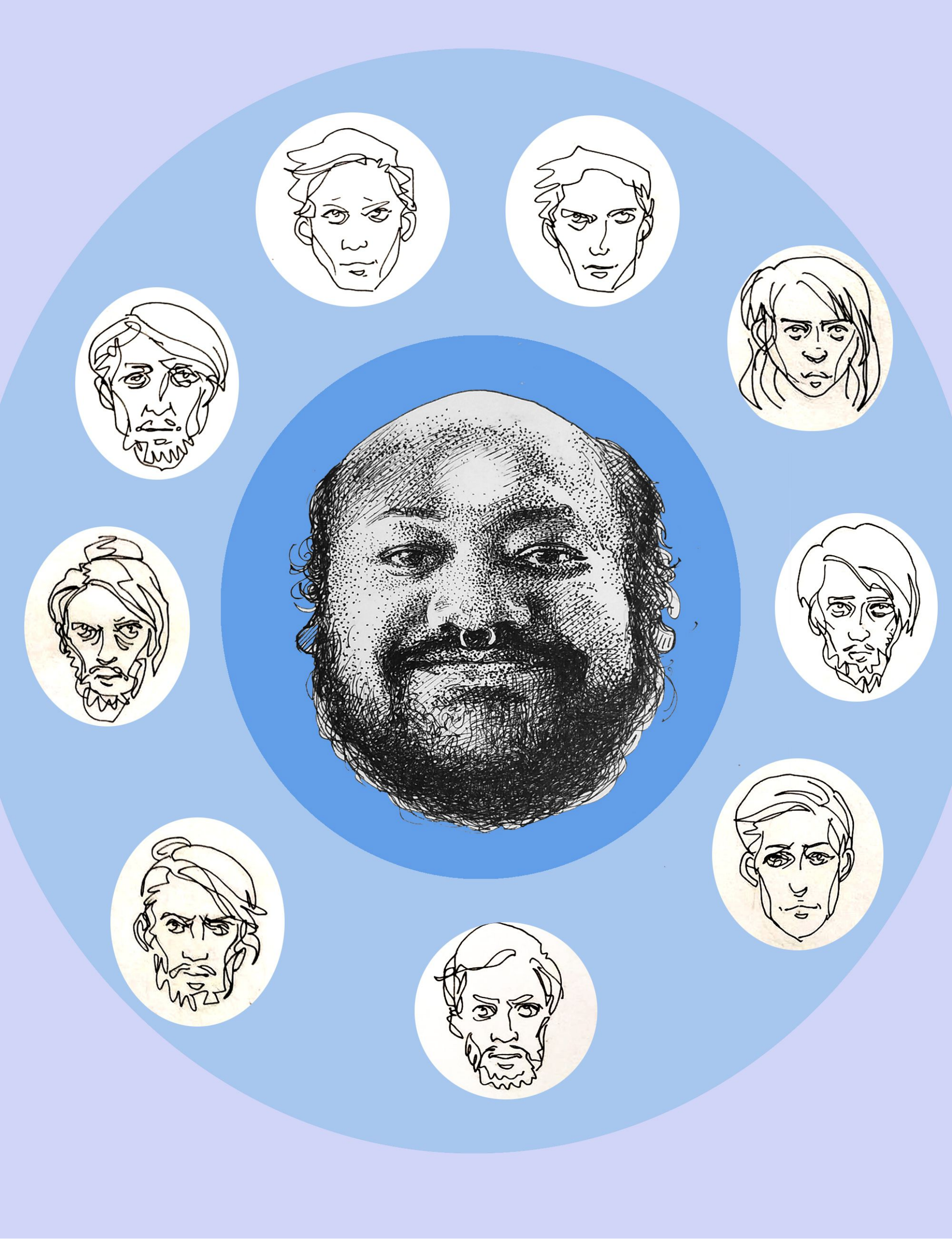}
	\caption{The OTOC Team}
	\label{fig:OTOCDesign1}
\end{figure}

\clearpage

\textcolor{Sepia}{\section{\sffamily Lexicography}\label{sec:lexicography}}
\begin{center}
	\begin{tabular}{||m{6.5cm}|m{8cm} ||}
		\hline
		\textcolor{red}{\textbf{Symbol}} & \textcolor{red}{\textbf{Meaning}}  \\ [0.5ex] 
		\hline\hline
		W(x) & Superpotential \\ 
		\hline
		$H_{SUSY}$ & Hamiltonian of the Supersymmetric Quantum mechanical model. \\
		\hline
		$\ket{\Psi_n}$ & Eigenstate of the Supersymmetric QM model in the direct sum Hilbert space.  \\
		\hline
		$E_{nm}$ = $E_{n} - E_{m}$ & Energy difference between the nth and mth energy eigenstate.  \\
		\hline
		$y_m^{(1)}(t_1,t_2)$  & Microcanonical 2-pt correlator of first kind.  \\ [1ex] 
		\hline
		$y_m^{(2)}(t_1,t_2)$  & Microcanonical 2-pt correlator of second kind. \\
		\hline
		$y_m^{(3)}(t_1,t_2)$  &  Microcanonical 2-pt correlator of third kind. \\
		\hline
		$Y^{(1)}(t_1,t_2)$ = -$\braket{[x(t_1),p(t_2)]}_{\beta}$ & Un-normalized 2-pt correlator of first kind.  \\ [1ex] 
		\hline
		$Y^{(2)}(t_1,t_2)$ = -$\braket{[x(t_1),x(t_2)]}_{\beta}$& Un-normalized 2-pt correlator of second kind. \\
		\hline
		$Y^{(3)}(t_1,t_2)$ = -$\braket{[p(t_1),p(t_2)]}_{\beta}$&  Un-normalized 2-pt correlator of third kind. \\
		\hline
		$c_m^{(1)}(t_1,t_2)$  & Microcanonical 4-pt correlator of first kind.  \\ [1ex] 
		\hline
		$c_m^{(2)}(t_1,t_2)$ & Microcanonical 4-pt correlator of second kind. \\
		\hline
		$c_m^{(3)}(t_1,t_2)$ &  Microcanonical 4-pt correlator of third kind. \\
		\hline
		$C^{(1)}(t_1,t_2)$ = -$\braket{[x(t_1),p(t_2)]^2}_{\beta}$ & Un-normalized 4-pt canonical correlator of first kind. \\
		\hline
		$C^{(2)}(t_1,t_2)$ = -$\braket{[x(t_1),x(t_2)]^2}_{\beta}$ & Un-normalized 4-pt canonical correlator of second kind.\\
		\hline
		$C^{(3)}(t_1,t_2)$ = -$\braket{[p(t_1),p(t_2)]^2}_{\beta}$ & Un-normalized 4-pt canonical correlator of third kind. \\
		\hline
		$\widetilde{Y}^{(1)}(t_1,t_2)$ = -$\frac{\braket{[x(t_1),p(t_2)]}_{\beta}}{\braket{x(t_1)x(t_1)}_{\beta} \braket{p(t_2)p(t_2)}_{\beta}}$ & Normalized 2-pt correlator of first kind. \\
		\hline
		$\widetilde{Y}^{(2)}(t_1,t_2)$ = -$\frac{\braket{[x(t_1),x(t_2)]}_{\beta}}{\braket{x(t_1)x(t_1)}_{\beta} \braket{x(t_2)x(t_2)}_{\beta}}$ & Normalized 2-pt correlator of second kind. \\
		\hline
		$\widetilde{Y}^{(3)}(t_1,t_2)$ = -$\frac{\braket{[p(t_1),p(t_2)]}_{\beta}}{\braket{p(t_1)p(t_1)}_{\beta} \braket{p(t_2)p(t_2)}_{\beta}}$ & Normalized 2-pt correlator of third kind. \\
		\hline
		$\widetilde{C}^{(1)}(t_1,t_2)$ = -$\frac{\braket{[x(t_1),p(t_2)]^2}_{\beta}}{\braket{x(t_1)x(t_1)}_{\beta} \braket{p(t_2)p(t_2)}_{\beta}}$ & Normalized 4-pt correlator of first kind. \\
		\hline
		$\widetilde{C}^{(2)}(t_1,t_2)$ = -$\frac{\braket{[x(t_1),x(t_2)]^2}_{\beta}}{\braket{x(t_1)x(t_1)}_{\beta} \braket{x(t_2)x(t_2)}_{\beta}}$ & Normalized 4-pt correlator of second kind. \\
		\hline
		$\widetilde{C}^{(3)}(t_1,t_2)$ = -$\frac{\braket{[p(t_1),p(t_2)]^2}_{\beta}}{\braket{p(t_1)p(t_1)}_{\beta} \braket{p(t_2)p(t_2)}_{\beta}}$ & Normalized 4-pt correlator of third kind. \\
		\hline
		
	\end{tabular}
\end{center}

\textcolor{Sepia}{\section{\sffamily A short review of Supersymmetric Quantum Mechanics}\label{sec:revSUSYQM}}

The theory of Supersymmetric Quantum Mechanics \cite{Bagchi:2001dx, Cooper:1994eh, Khare:2004kn, wipf2000, ramadevi, wellman} relates quantum eigenfunctions and the corresponding eigenvalues between two partner Hamiltonians through an intertwining relationship using the so called charge operators. It generally uses the technique of factorizing the Hamiltonian in terms of the intertwining operators and hence to determine the superpotential using the well known \textit{Riccati equation}. The idea of factorizing generally allows one to express the superpotential in terms of the ground state wave-function of the original Hamiltonian of the quantum mechanical model under consideration. The process of factorization is mathematically described by the following equation:
\begin{align}
H_1=A^{\dagger}A ,
\end{align}
where A and $A^{\dagger}$ are the intertwining operators which are defined from the superpotential as:
\begin{align}
A=\frac{1}{\sqrt{2}}\ \frac{d}{dx}+W(x) \quad ; \quad
A^{\dagger}= -\frac{1}{\sqrt{2}}\ \frac{d}{dx}+W(x) .
\end{align} 
Within the framework of supersymmetry, the ground state energy is usually taken to be zero, which is well justified because it is only the relative energy difference that matters. For a zero energy ground state, the {\it Schr$\ddot{o}$dinger equation} can be written as:
\begin{align}
H_{1} \ket{\Psi_{0}} = 0.
\end{align}
Substituting the expressions for $A$ and $A^{\dagger}$ in the above equation it is not very hard to derive the \textit{Riccati equation}, which further gives a way of writing the potential in terms of the superpotential given by the following expression:
\begin{align}
V_1(x)= W^2(x)-\frac{1}{\sqrt{2}}W'(x),
\end{align}
where $'$ corresponds to $d/dx$ in the above equation.
It is often an {\it overrated fact} that one needs to know the form of the potential guiding the Hamiltonian to have an idea about the wave functions of the quantum mechanical system and the fact that the knowledge of the ground state wave function allows one to exactly know the potential associated with the system is overlooked. However, in Supersymmetric quantum mechanics one generally utilizes this unappreciated fact to construct the potential from the known ground state wave function with zero modes:
\begin{align}
\label{superpotential}
	W(x)=-\frac{1}{\sqrt{2m}}\frac{\psi_0'(x)}{\psi_0(x)}.
\end{align} 
The knowledge of superpotential allows one to determine the supersymmetric partner potential via the following equation:
\begin{align}
\label{partnerpotential}
	V_2(x)= W^2(x)+\frac{1}{\sqrt{2m}}\ W'(x).
\end{align} 
The partner Hamiltonian is constructed by reversing the order of the intertwining operators used in the factorization of the original Hamiltonian. The energy eigenvalues and the eigenstates of the original and the partner Hamiltonian are not independent of each other and that's where the beauty of supersymmetry lies. Knowing the original Hamiltonian and its ground state one can easily determine the energy spectrum of the partner Hamiltonian. The eigenvalues of $H_1$ (original Hamiltonian) and $H_2$ (partner Hamiltonian) are related via the following equation:
\begin{align}
	E_n^{(2)}= E_{n+1}^{(1)} \quad ; \quad E_0^{(1)}=0 \quad \forall~ n = 0,1,2,\dots
\end{align}
It is easy to show that the knowledge of the eigen functions of $H_1$ can be used to derive the eigen functions of $H_2$ using the A operator and the eigen functions of $H_1$ from that of $H_2$ using the $A^{\dagger}$ operator. The role of the operators A and $A^{\dagger}$, apart from the conversion of an eigenfunction of the original Hamiltonian to that of its partner Hamiltonian with the same energy, it also destroys or creates one node in the eigenfunction. This justifies the absence of the zero energy or the ground energy state of the partner Hamiltonian. One can put this argument simply by stating that the operator A converts an energy state of the original Hamiltonian into a lower energy state of the partner Hamiltonian keeping the energy value of the state constant. A$^{\dagger}$ on the other hand does the opposite conversion i.e takes an energy state of the partner Hamiltonian and converts it into an higher energy state of the original Hamiltonian keeping the energy eigenvalue fixed.

A supersymmetric quantum mechanical model is generally described by a Hamiltonian having the following form:
\begin{align}
\label{eq:hamil}
H_{\textnormal{SUSY}}=\begin{pmatrix}
H_1 & &  0 \\
0 & & H_2
\end{pmatrix}.
\end{align}
In general, for such a Hamiltonian a quantum state is represented by: 
\begin{align} \label{eq:totwavefn-wrong}
\ket{\Psi_{n}}^{\textnormal{T}} = \left(\ \ket{\psi^{(1)}} \ \ket{\psi^{(2)}}\ \right),
\end{align}
where, $\ket{\psi^{(1)}}$ and $\ket{\psi^{(2)}}$ are the wave functions of the original and the partner Hamiltonian respectively.

The prime objective of this paper is to provide an eigenstate representation of the desired OTOCs, that we have already defined in the introduction, using which we study the various well known quantum mechanical models in the context of supersymmetry to study the general aspects of time disorder averaging phenomena. With this aim, one would generally look for an eigenstate of the Hamiltonian under inspection. Remembering the relation between the energy eigenvalues and eigenstates of the original and the partner Hamiltonian it can be easily verified that the wave function given by \Cref{eq:totwavefn-wrong} is not an eigenstate of the Hamiltonian given in \Cref{eq:hamil}. We therefore take the wave function of the Hamiltonian to be of the form
\begin{align} \label{eq:wavefn1}
	\ket{\Psi_{n}}^{\textnormal{T}} = \frac{1}{\sqrt{2}}\ \left(\ \ket{\psi^{(1)}} \ \ket{\psi^{(2)}}\ \right),
\end{align}
which indeed represents a normalized eigenfunction of the Hamiltonian of the supersymmetric quantum mechanical systems considered in this work. 
\begin{align*}
H_{SUSY}\ket{\Psi_n} &=\frac{1}{\sqrt{2}} \begin{pmatrix}
H_1 & & 0 \\
0 & & H_2
\end{pmatrix}\begin{pmatrix} \ket{\psi_n^{(1)}} \\
\ket{\psi_{n-1}^{(2)}}
\end{pmatrix} = \frac{1}{\sqrt{2}}\begin{pmatrix} E_n^{(1)}\ket{\psi_n^{(1)}} \\
E_{n-1}^{(2)}|\psi_{n-1}^{(2)}\rangle
\end{pmatrix}=\frac{E_n^{(1)}}{\sqrt{2}}\begin{pmatrix} \ket{\psi_n^{(1)}} \\
\ket{\psi_{n-1}^{(2)}}
\end{pmatrix},
\end{align*}
where we have used the relation $E_{n-1}^{(2)}=E_{n}^{(1)}$. If we use \Cref{eq:totwavefn-wrong} then ${\ket{\Psi_{n}}}$ does not remain an eigenstate of $H_{\textnormal{SUSY}}$ as we show below.
\begin{align*}
H_{\textnormal{SUSY}}\ket{\Psi_n} &= \begin{pmatrix}
H_1 & & 0 \\
0 & &  H_2
\end{pmatrix}\begin{pmatrix}\ket{\psi_n^{(1)}} \\
\ket{\psi_{n}^{(2)}}
\end{pmatrix} = \begin{pmatrix} E_n^{(1)}\ket{\psi_n^{(1)}} \\
E_{n}^{(2)}\ket{\psi_n^{(2)}}
\end{pmatrix}.
\end{align*}
Since the energy associated with the ${n^{\textnormal{th}}}$ energy eigenstate of the original Hamiltonian ${H_{1}}$ is not equal to that of its associated partner Hamiltonian ${H_{2}}$, i.e $E_{n}^{(1)} \neq E_{n}^{(2)}$, ${\ket{\Psi_{n}}}$ fails to be an eigenstate of ${H_{\textnormal{SUSY}}}$.

\noindent

\noindent

\textcolor{Sepia}{\section{\sffamily General remarks on time disorder averaging and thermal OTOCs}\label{sec:qrandcorr}}

{\it Quantum randomness}, using which we have prime objective to technically demonstrate the time disorder averaging phenomena is actually a very broad topic of research in theoretical physics and there are many ways and possibilities using which one can explicitly quantify this phenomena in the quantum regime. Quantum correlators of different orders are one of them. When a quantum state evolves to reach equilibrium at the late time scales in that case the overall amplitude of the correlators also evolves with the evolutionary time scales, which are usually described in terms of the fundamental quantum operators and the time evolution of these correlators can show the presence of time disorder averaging in the form of chaotic or non-chaotic, periodic or aperiodic random behaviour in the quantum mechanical system under study. Thermal average over the canonical statistical ensemble of a quantum operator is a very powerful technique using which one can explicitly study the time dependent exponential growth (chaotic) or some other time dependent non-chaotic random behaviour of an operator for a quantum system that is in out of equilibrium after giving an external response.For a very longer time it was not very clear that how actually one can quantify these quantum correlation functions within the framework of out-of-equilibrium quantum statistical physics. Following the previous set of works the present work helps us to quantify as well as to physically understand the impact of them in the present context. In this paper we are actually interested in three specific kinds of OTOCs which are described by six set of correlators which are given by:
\begin{enumerate}
\item \textcolor{red}{\bf 2-point~OTOCs:}~~~~$\langle\left[x(t_1),x(t_2)\right]\rangle$, $\langle\left[p(t_1),p(t_2)\right]\rangle$, $\langle\left[x(t_1),p(t_2)\right]\rangle$,

\item  \textcolor{red}{\bf 4-point~OTOCs:}~~~~$\langle\left[x(t_1),x(t_2)\right]^2\rangle$, $\langle\left[p(t_1),p(t_2)\right]^2\rangle$, $\langle\left[x(t_1),p(t_2)\right]^2\rangle$.
\end{enumerate}  
where $x$ is the quantum position operator and $p$ is the associated canonically conjugate momentum and most importantly both the quantum operators are defined at different time scales, which is one of the prime requirement to study the effect of time disorder averaging phenomena through the above set of OTOCs. Also, it is important to note that, the symbol $\langle\cdots\rangle$ actually represents the thermal average of a time dependent quantum operator over a canonical ensemble within the framework of quantum mechanics, which is technically defined as:
\bea \langle {\cal O}(t)\rangle:=\frac{1}{Z(t)}~{\rm Tr}\left[\exp(-\beta H){\cal O}(t)\right]={\rm Tr}\left[\rho_{\beta}{\cal O}(t)\right],\eea
where the partition function $Z$ and thermal density matrix operator $\rho_{\beta}$ in terms of a quantum system Hamiltonian, $H$, are already defined in the introduction of this paper. Since we are dealing with quantum mechanical operators using which we are trying to understand the impact of random features in the quantum regime, it is quite expected to start with the fact that thermal one point function of the position operator $x$ and momentum operator $p$ defined at a specific time scale are zero, which can be technically demonstrated as, \bea \langle x(t)\rangle=0, ~~\langle p(t)\rangle=0,\eea where $t$ is the associated time scale on which both of the quantum operators are evolving. For this specific reason, the explicit study and the computation of these one point functions are not at all important in the present context of discussion. On the other hand, due to the time translational symmetry in these thermal correlators, which is actually described by the well known, {\it Kubo Martin Schwinger} condition, all the odd point OTOCs appearing in the present context will be trivially zero and for that reason not the object of interest in the present context of study. This can be further technically demonstrated as, \bea &&\underline{\textcolor{red}{\bf Odd~x~correlator:}}\nonumber\\
&&~~~~~~~~~~~~~~~~\langle x(t_1)x(t_2)x(t_3)\cdots \textcolor{blue}{\rm odd~number~of~terms}\rangle=0,\\
&&\underline{\textcolor{red}{\bf Odd~p~correlator:}}\nonumber\\
&&~~~~~~~~~~~~~~~~  \langle p(t_1)p(t_2)p(t_3)\cdots \textcolor{blue}{\rm odd~number~of~terms}\rangle=0, \\
&&\underline{\textcolor{red}{\bf One~p~\& ~even~x~correlator:}}\nonumber\\
&&~~~~~~~~~~~~~~~~ \langle p(t_1)x(t_2)x(t_3)\cdots \textcolor{blue}{\rm odd~number~of~terms}\rangle=0,\\
&& \underline{\textcolor{red}{\bf One~x~\& ~even~p~correlator:}}\nonumber\\
&&~~~~~~~~~~~~~~~~ \langle x(t_1)p(t_2)p(t_3)\cdots \textcolor{blue}{\rm odd~number~of~terms}\rangle=0.\eea This implies that we are only left with all even order OTOCs out of which in this paper we are explicitly computing the physical outcomes from the sex set of time dependent correlators, describe by the previously mentioned two point and the four point all possible OTOCs.  

Among these correlators two are made of different operators which will show the perturbation of one operator measured at one time scale to the other measured at a different time and vice versa. The other four operators are the new sets of two and four point correlators which are defined in terms of the same quantum operators, which are  basically capturing the quantum effect of the self-correlation of one operator on itself having any arbitrary time dependent profile in general to describe the phenomena of time disordering within the framework of quantum mechanics. In specific cases it may happen that these newly defined operators show exponential growth or some other kind of growth which is periodic or aperiodic in the corresponding associated time scales of the quantum mechanical operator on which we are interested in the present context. In a more general context one can see by studying different kinds of physical systems available in the literature where the two-point self-correlation will decay exponentially with an associated time scale of $t_d$, widely known as {\it ``dissipation time scale''}. This particular time scale is playing the role of {\it transition scale} in the present context after reaching that the four-point correlators can be factorised into the product of two two-point correlation functions representing disconnected diagrams within the framework of quantum field theory. Also, it is important to note that, here at the {\it ``dissipation time scale''} all other terms are exponentially suppressed by the factor $e^{-t/t_d}$, which will completely disappear from the factorized version of the four point correlators on which we are interested in this paper in the large time limit given by, $t\gg t_d$ with $t\rightarrow \infty$. In usual prescription the {\it ``dissipation time scale''}, $t_d$ is identified with the inverse temperature $\beta$ i.e. $t_d\sim \beta=1/T$, where we use Boltzmann constant $k_{\rm B}=1$ and $T$ physically represents the equilibrium temperature of the quantum statistical ensemble on which we are interested in. So one can translate the large time limit in terms of the associated equilibrium temperature as, $tT\gg 1$, which is not obviously true for zero temperature case, but can be justifiable in any (small or large) temperature of the system under consideration. For example, we now look into a specific four point thermal correlation function at the vicinity of the previously mentioned {\it dissipation time scale}, around which one can factorize it in the following specific form:
\begin{align}
\underbrace{\braket{ x(t_1)x(t_1)p(t_2)p(t_2)}}_{\textcolor{red}{\bf Unequal~time~4-point~correlator}} \approx \underbrace{\braket{x(t_1)x(t_1)}}_{\textcolor{red}{\bf Equal~time~2-point~correlator}}~~~  \underbrace{\braket{p(t_2)p(t_2)}}_{\textcolor{red}{\bf Equal~time~2-point~correlator}} \nonumber\\ + \underbrace{\mathcal{O}(e^{-t/t_d})}_{\textcolor{red}{\bf Subdominant~decaying~contribution}},
\end{align}
\begin{figure}[!ht]  
	\subfloat[Correlator : ${ -\braket{[x(t_{1}),p(t_{2})]}}$]{%
		\includegraphics[width=6cm, height=7cm]{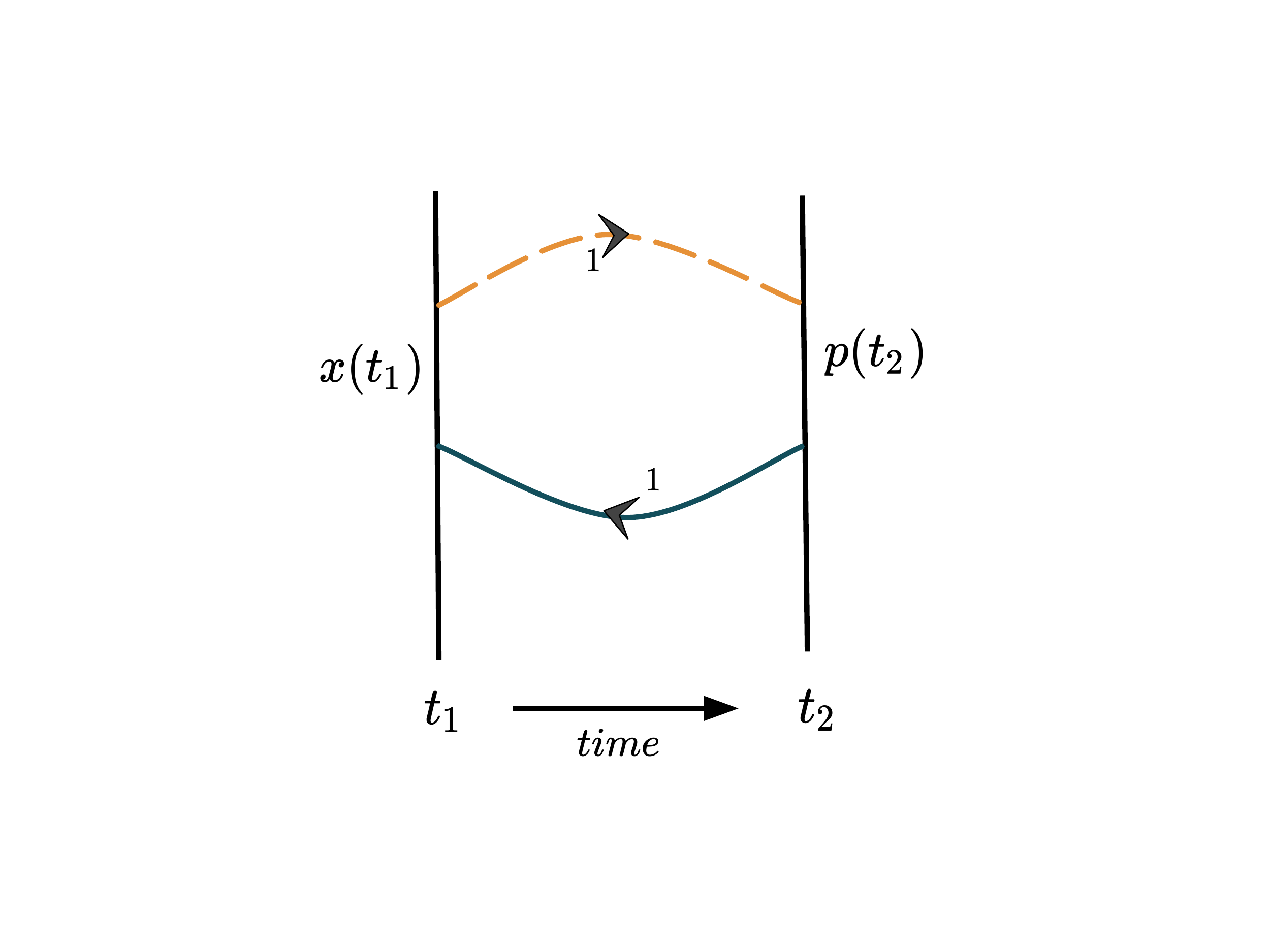} 
	} 
	\hfill 
	\subfloat[Correlator : ${ -\braket{[x(t_{1}),x(t_{2})]}}$]{%
		\includegraphics[width=6cm, height=7cm]{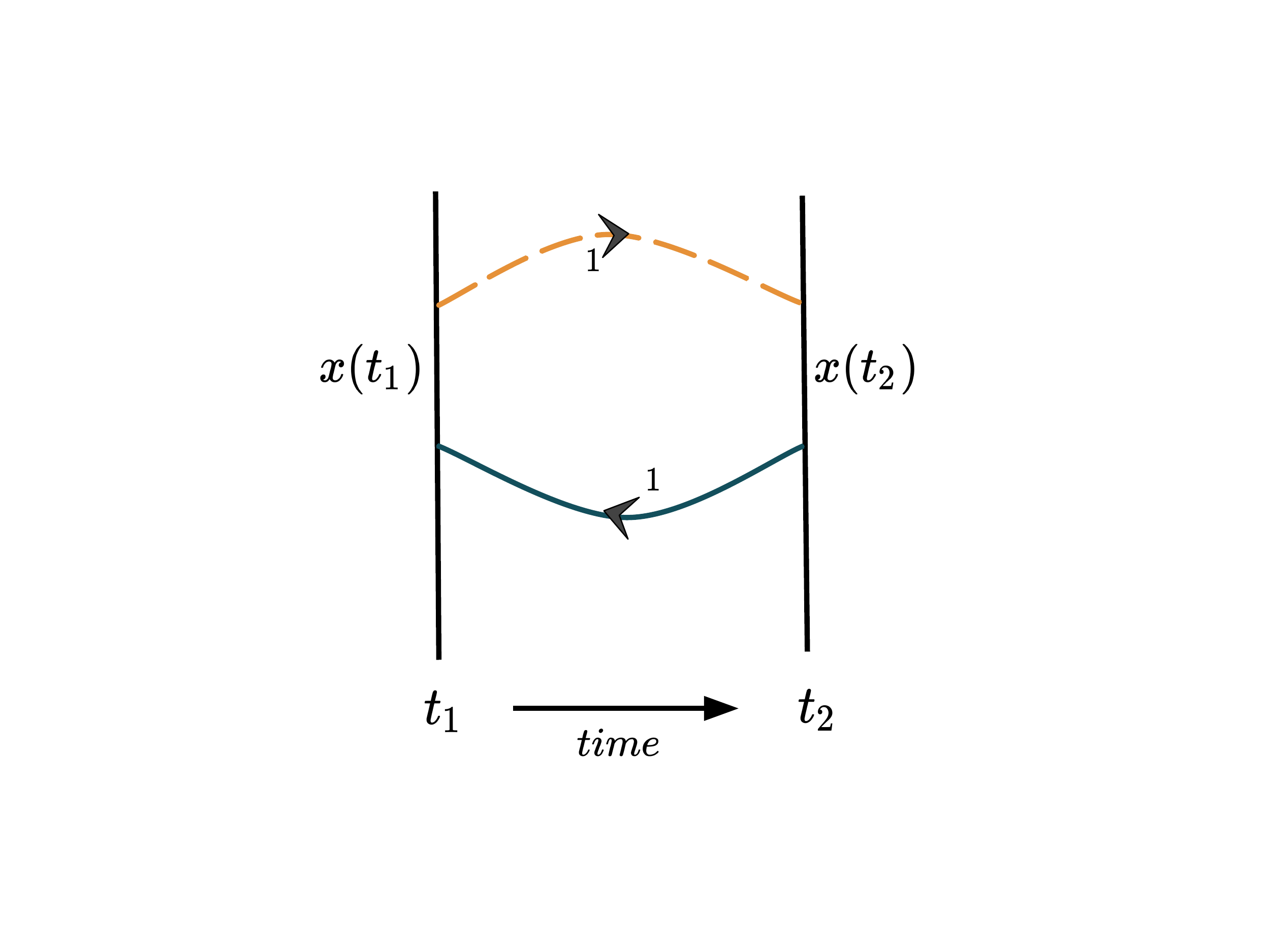} 
	}
	\subfloat[Correlator : ${ -\braket{[p(t_{1}),p(t_{2})]}}$]{%
		\includegraphics[width=6cm, height=7cm]{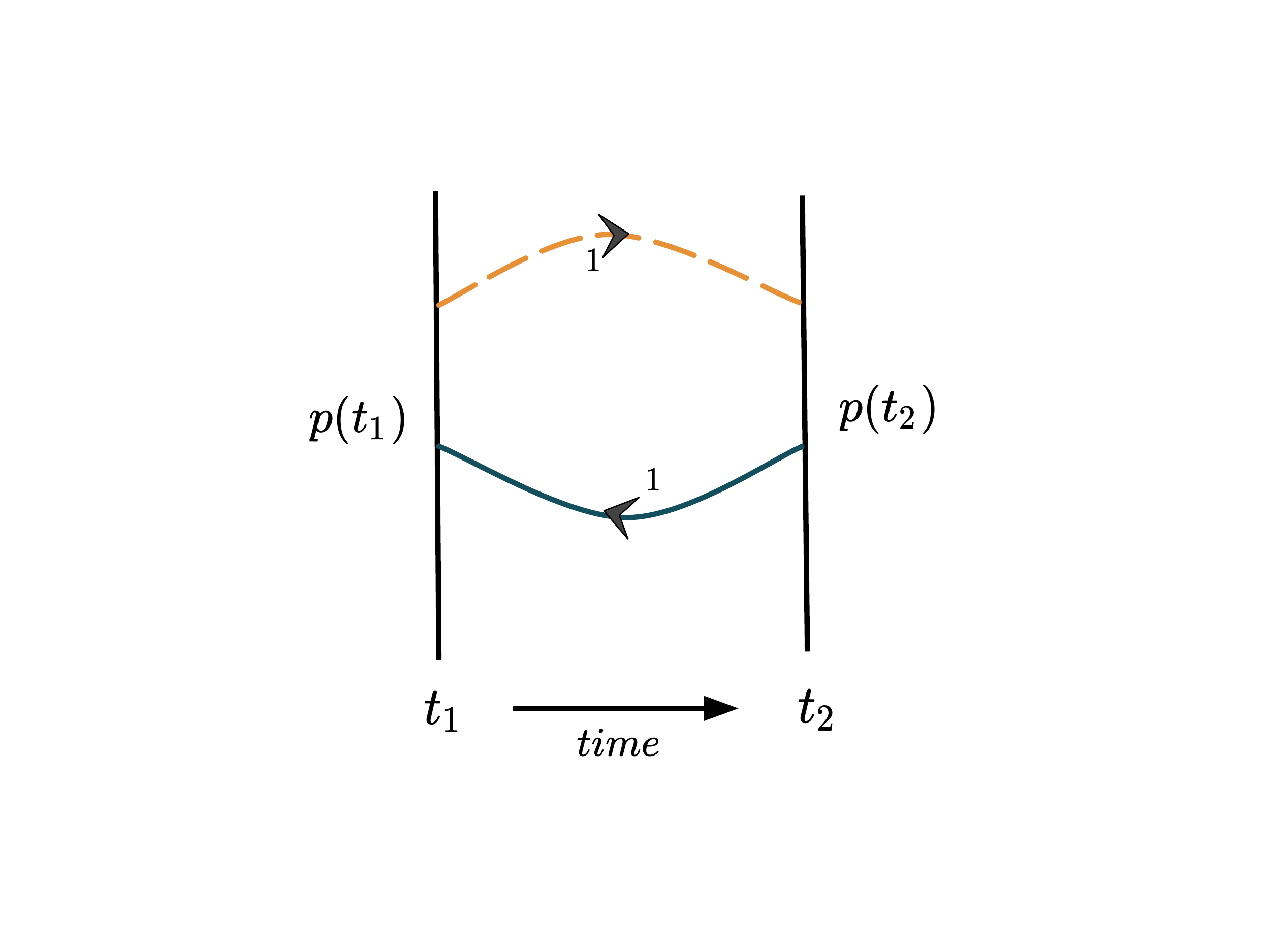} 
	}\\
	\subfloat[Correlator : ${ -\braket{[x(t_{1}),p(t_{2})]}}$]{%
		\includegraphics[width=6cm, height=7cm]{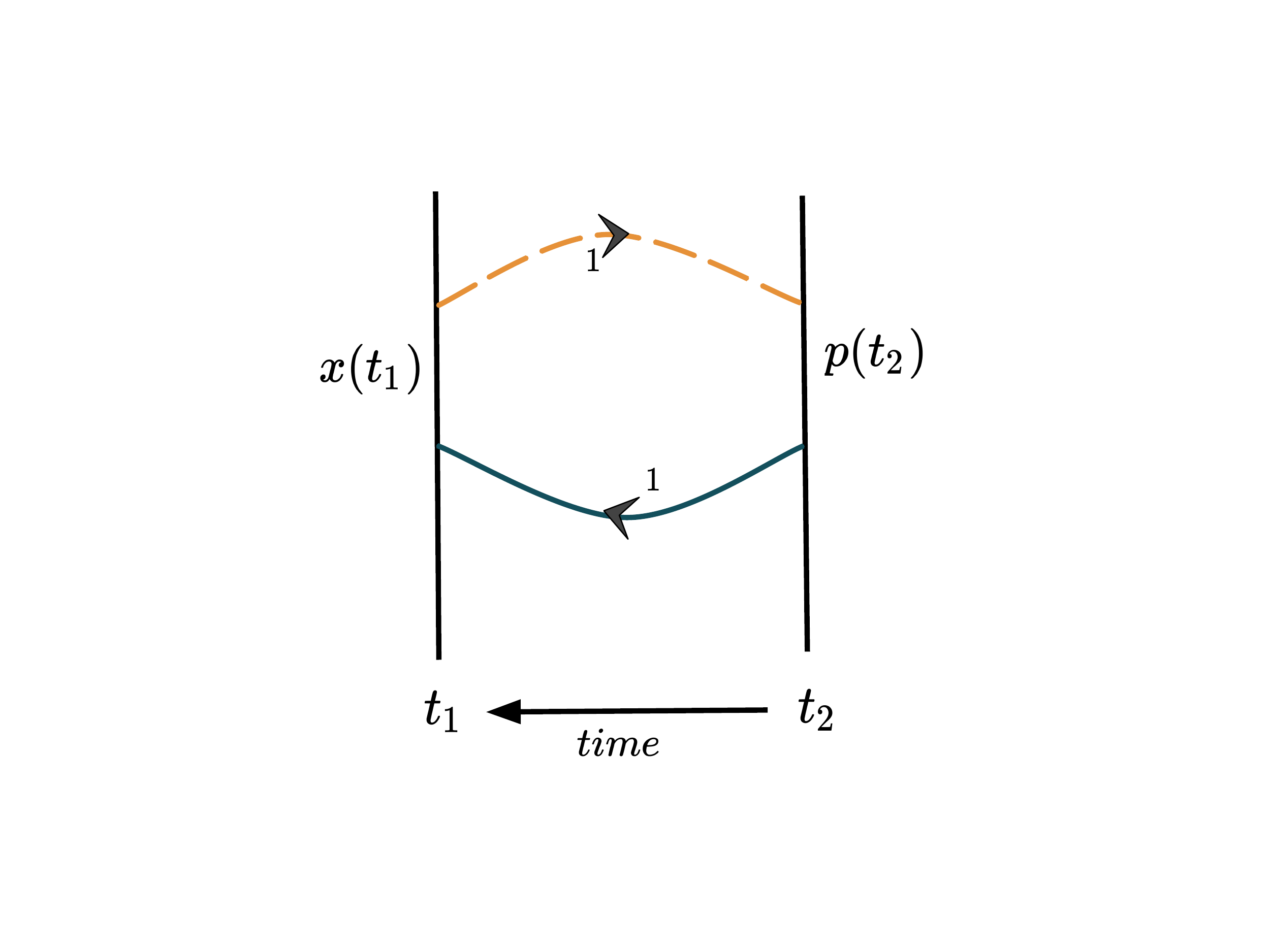} 
	} 
	\hfill 
	\subfloat[Correlator : ${ -\braket{[x(t_{1}),x(t_{2})]}}$]{%
		\includegraphics[width=6cm, height=7cm]{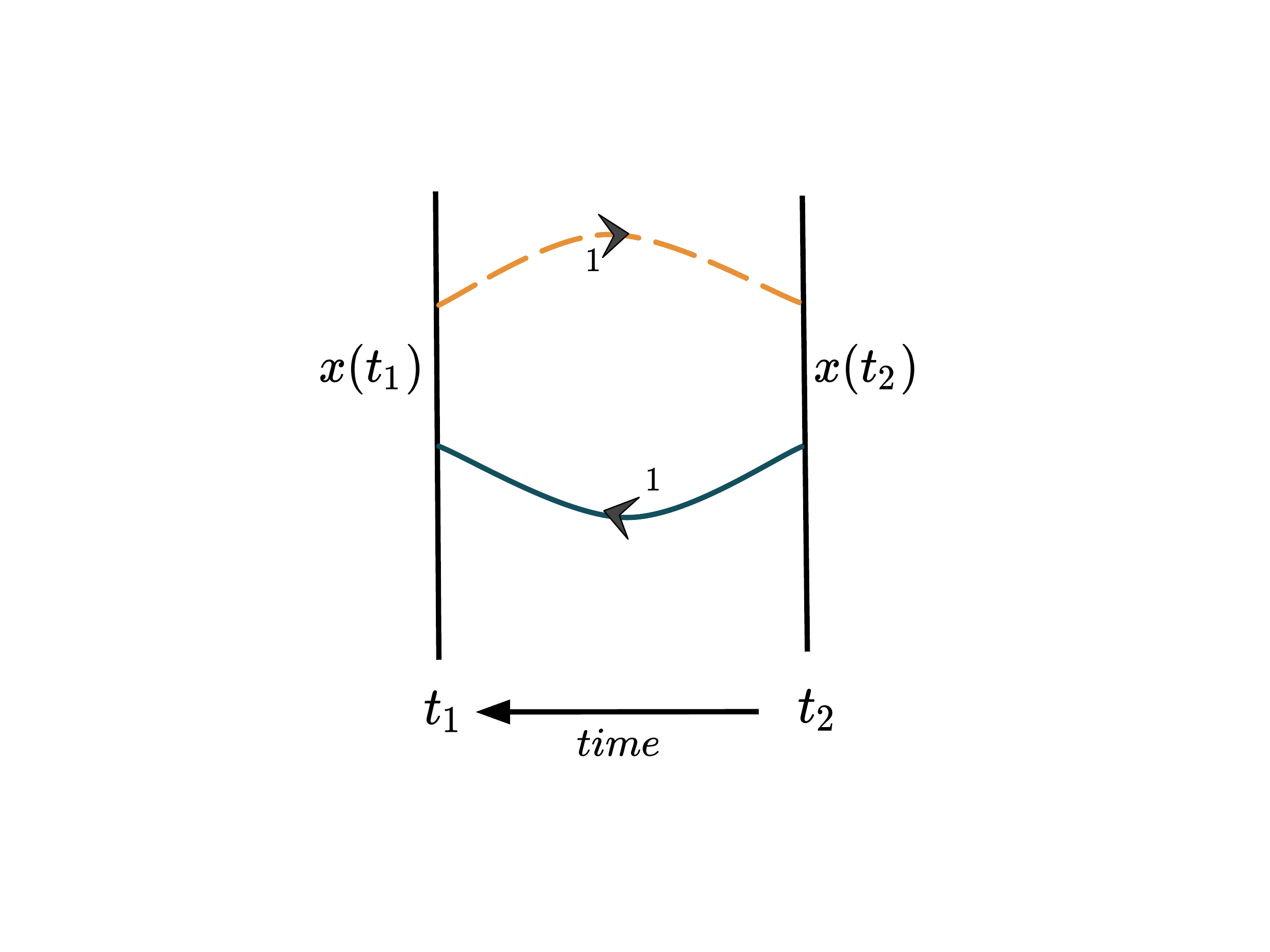} 
	}
	\subfloat[Correlator : ${ -\braket{[p(t_{1}),p(t_{2})]}}$]{%
		\includegraphics[width=6cm, height=7cm]{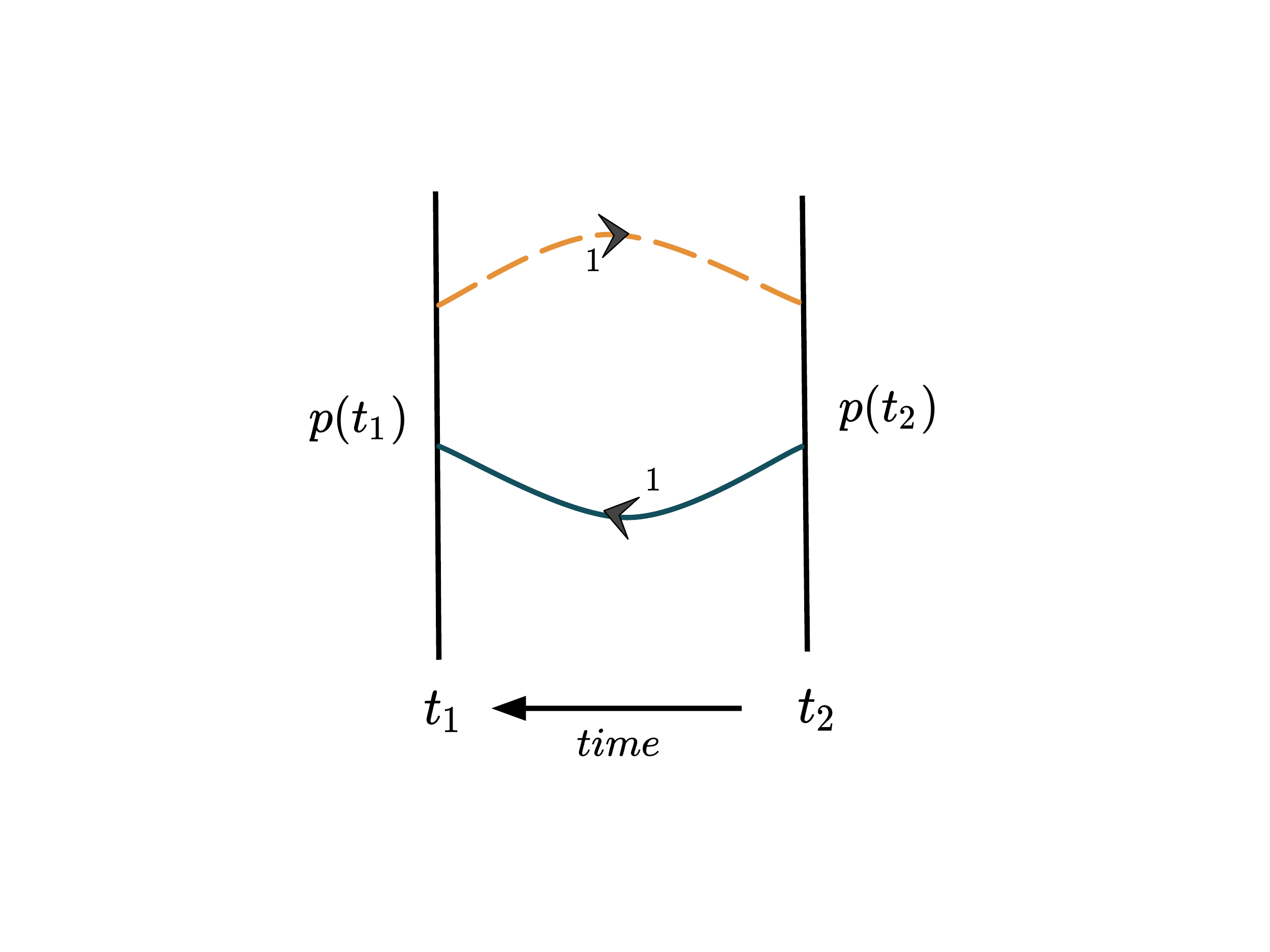} 
	}
	\caption{Diagrammatic representation of all possible two-point OTOCs.}
	\label{vennin1}
\end{figure}

\begin{figure}[!ht]
	\subfloat[Correlator : ${ -\braket{[x(t_{1}),p(t_{2})]^2}}$]{%
		\includegraphics[width=6cm, height=7cm]{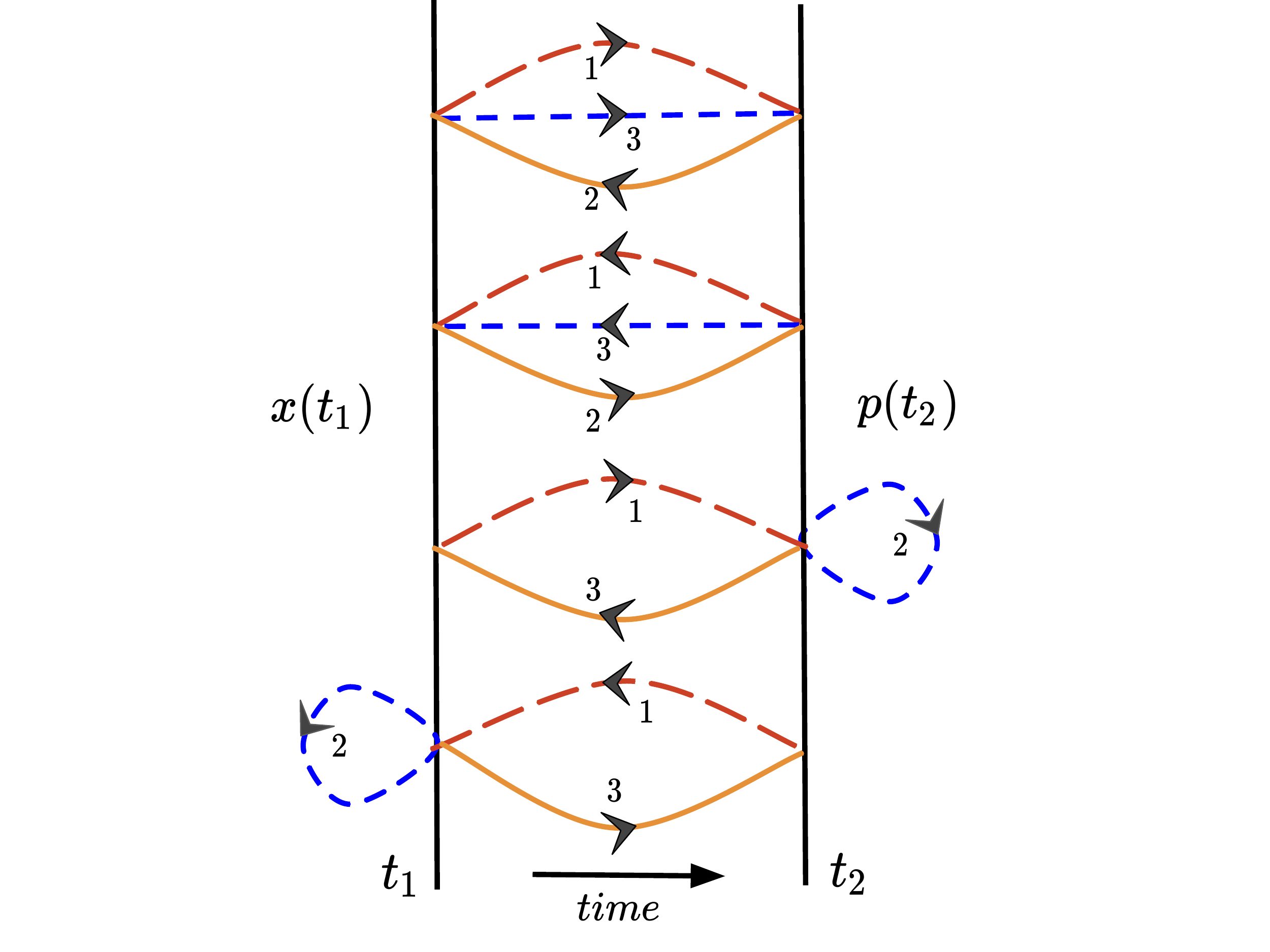} 
	}
	\hfill 
	\subfloat[Correlator : ${ -\braket{[x(t_{1}),x(t_{2})]^2}}$]{%
		\includegraphics[width=6cm, height=7cm]{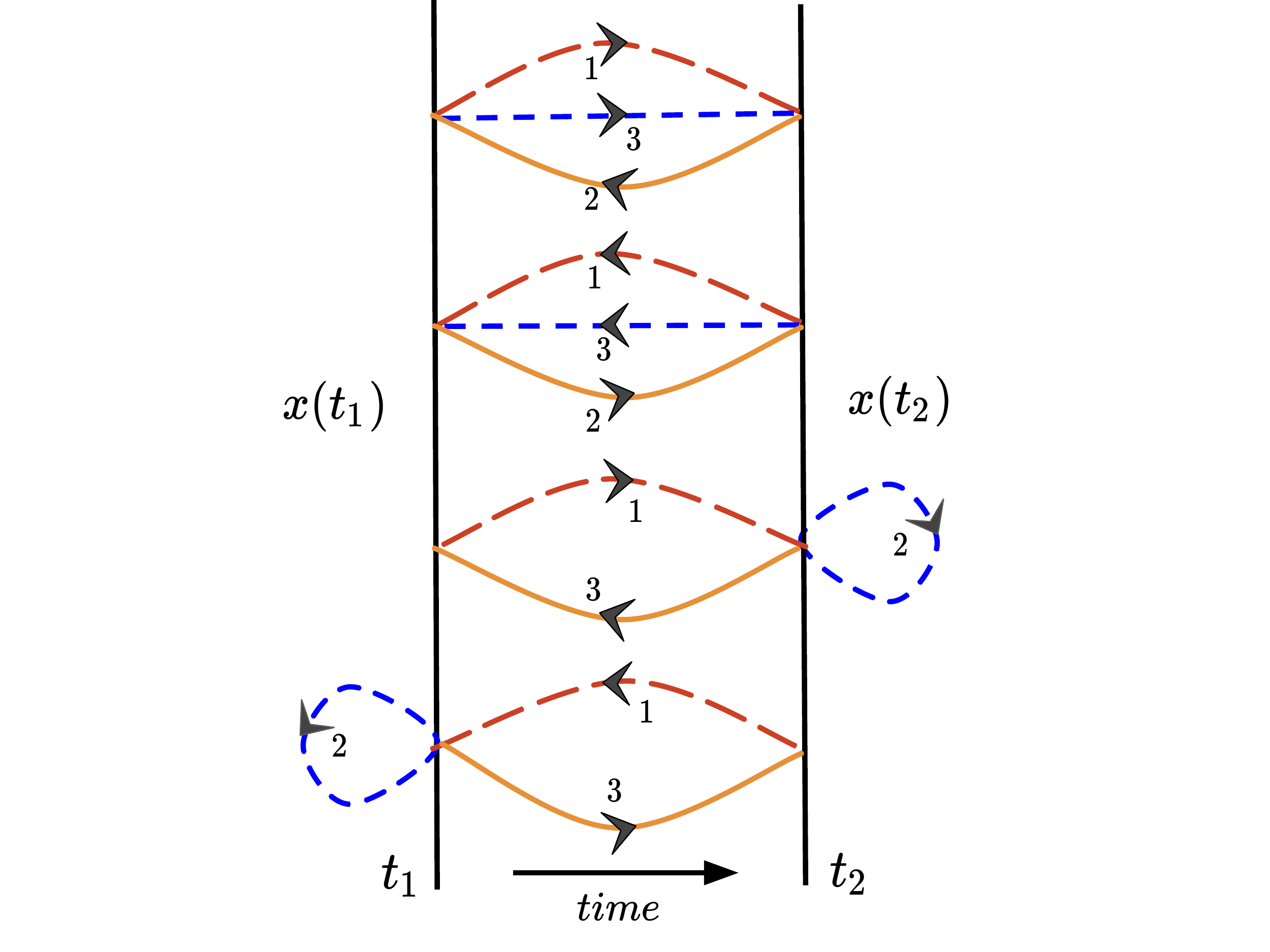} 
	} 
	\subfloat[Correlator : ${ -\braket{[p(t_{1}),p(t_{2})]^2}}$]{%
		\includegraphics[width=6cm, height=7cm]{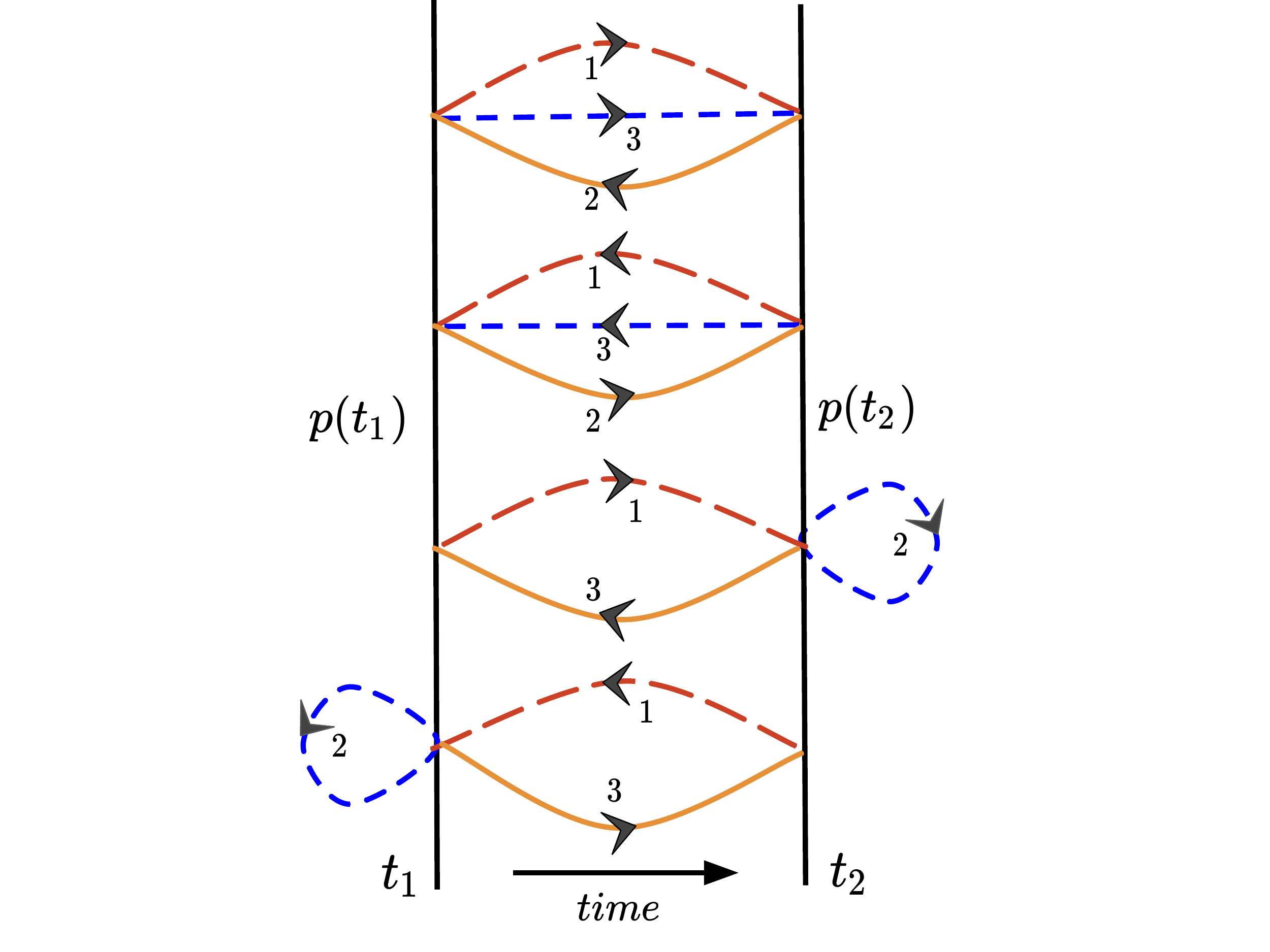} 
	} \\
	\subfloat[Correlator : ${ -\braket{[x(t_{1}),p(t_{2})]^2}}$]{%
		\includegraphics[width=6cm, height=7cm]{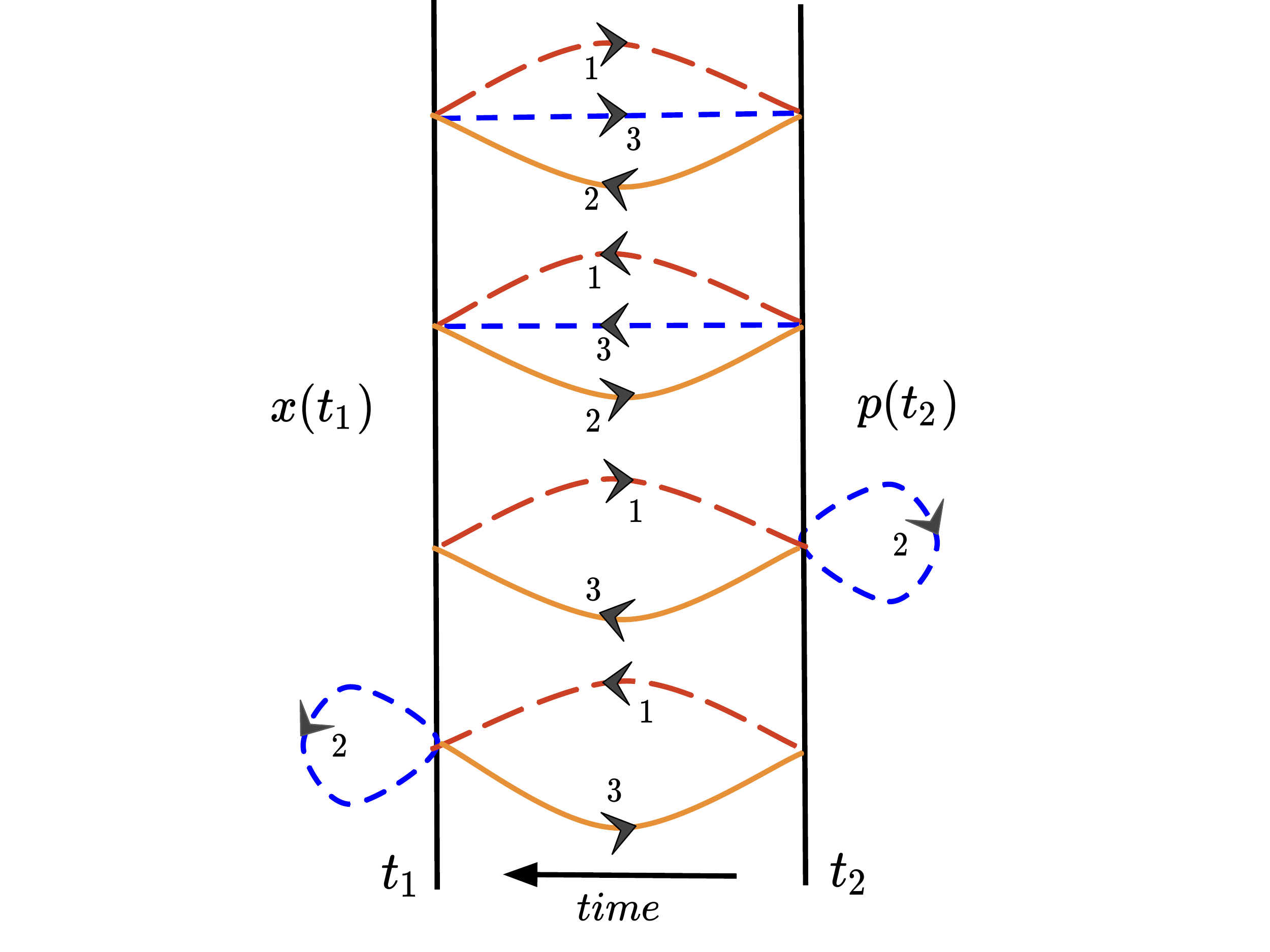} 
	}
	\hfill 
	\subfloat[Correlator : ${ -\braket{[x(t_{1}),x(t_{2})]^2}}$]{%
		\includegraphics[width=6cm, height=7cm]{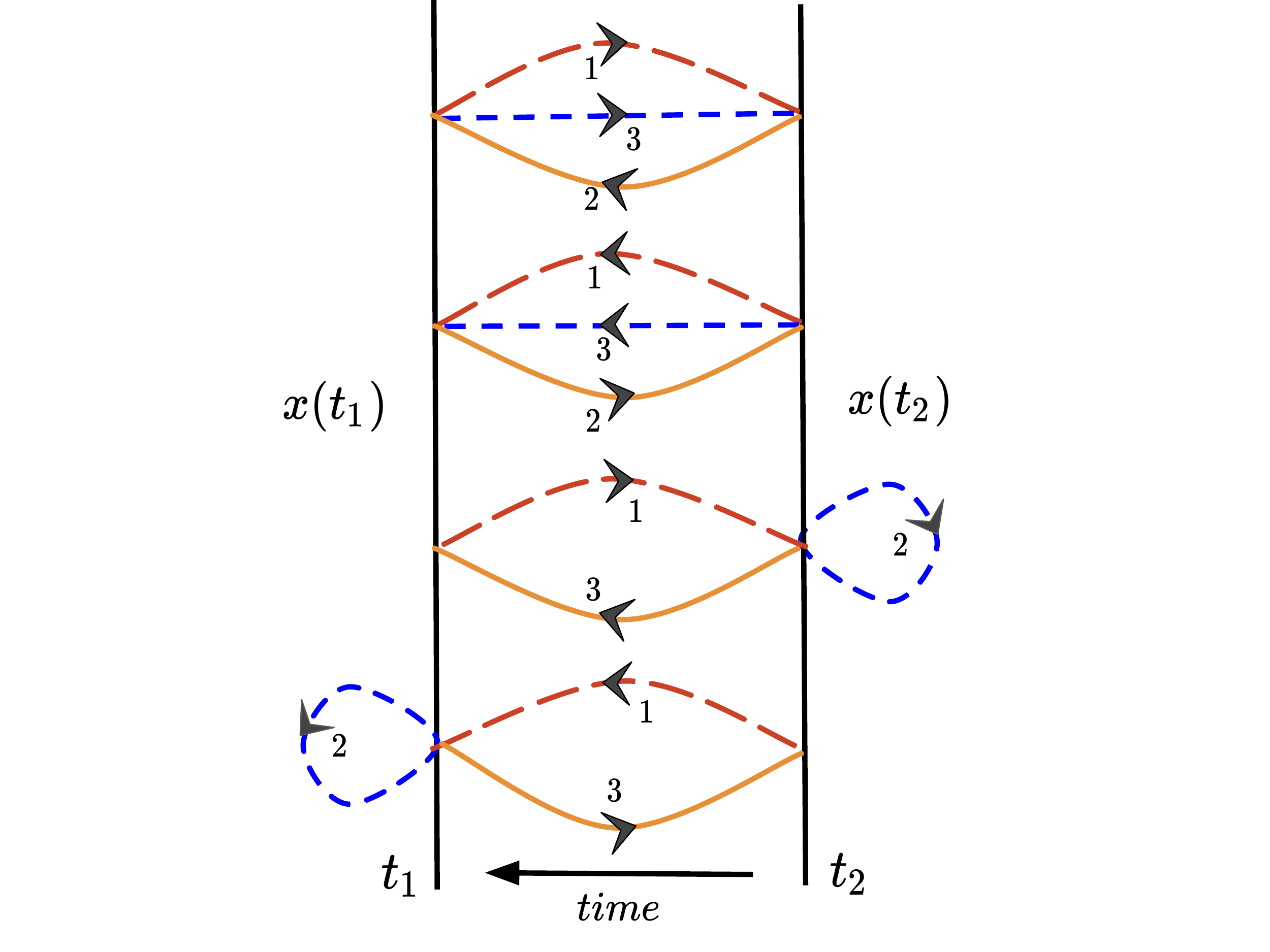} 
	} 
	\subfloat[Correlator : ${ -\braket{[p(t_{1}),p(t_{2})]^2}}$]{%
		\includegraphics[width=6cm, height=7cm]{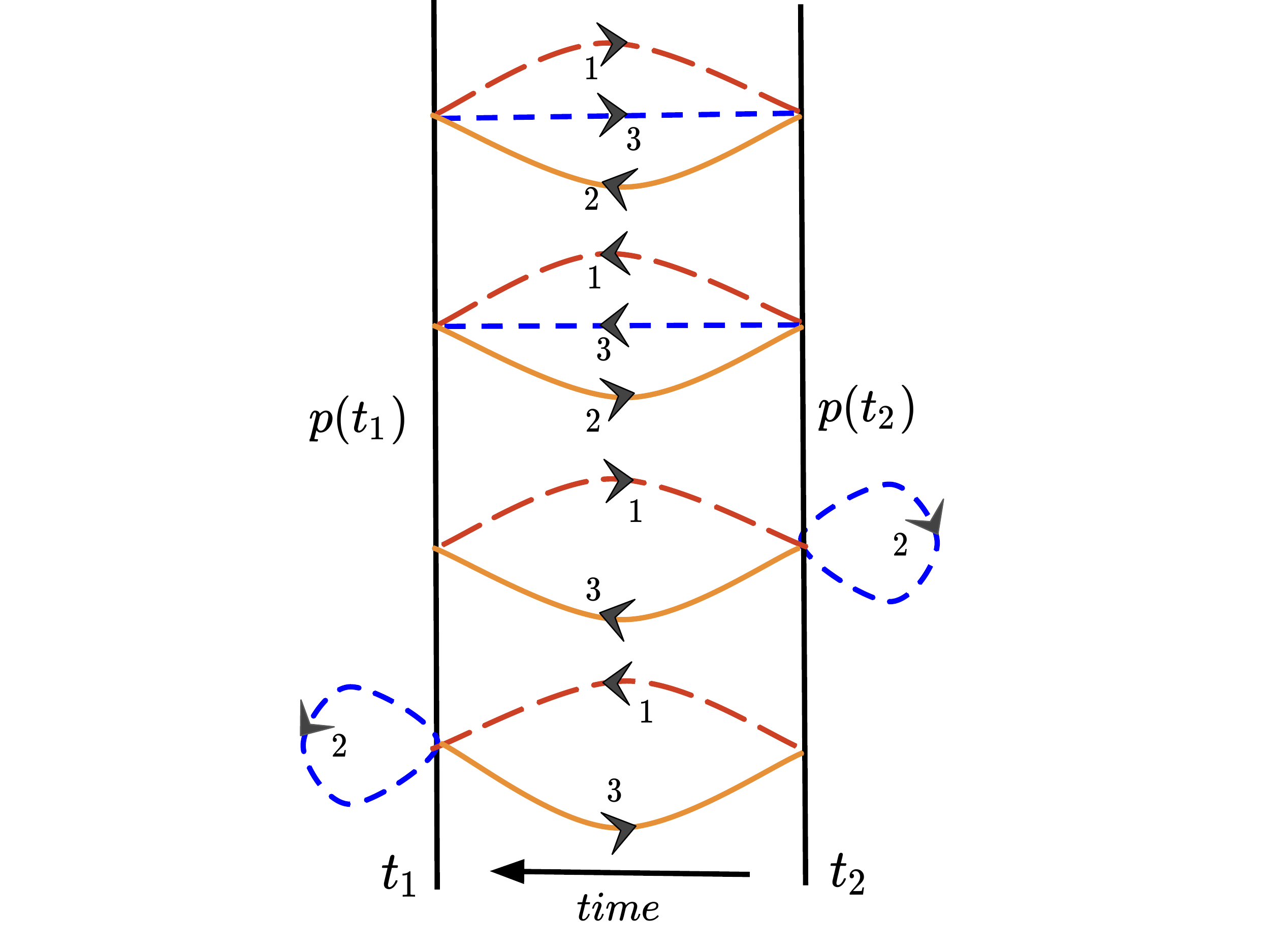} 
	} 
		\caption{Diagrammatic representation of all possible four-point OTOCs.}
		\label{vennin2}
\end{figure}
where in this above mentioned factorization $\mathcal{O}(e^{-t/t_d})$ are the higher order correction terms which are actually sub-dominant at the vicinity of the {\it dissipation time scale}. One more thing we can observe from the above mentioned factorization is that, the individual contributions of the two point contributions don't mix up the time scales and for this reason they can be written in terms of the product of two equal time two point correlators. Also for this particular example when we are thinking of doing the computation with two different operators, after doing the factorization one can easily observe that the two different operators don't mix with each other at the level of two-point correlators. So, we are mainly interested in the terms that can not be written in time ordered form and those terms provide more insight to the randomness present in a system. Now we would like to go one step further and normalize the OTOC, which is basically related to this factorization process of the four-point correlators in terms disconnected equal time two point contributions. The process of normalization actually helps us to reduce the unwanted fluctuations from the computed OTOCs which further allows us to give a clearer picture of the time dependent behaviour of the desired OTOCs in which we are interested in this paper. We do not normalize the two-point correlation functions as they are the main building blocks of our computation of OTOCs and hence we only normalize the four-point OTOC as given by the following expression. We can write $-\braket{\left[x(t_1),p(t_2)\right]^2}$ in the following simplified mathematical form:
\begin{align}
- \braket{\left[x(t_1),p(t_2)\right]^2}=2\left\{\left[\braket{x(t_1)x(t_1)p(t_2)p(t_2)}\right] - \textnormal{Re}\left[\braket{x(t_1)p(t_2)x(t_1)p(t_2)}\right]\right\}.
\end{align}
After imposiing the previously mentioned constrained obtained at the vicinity of the {\it dissipation time scale}, one can further simplify the expression for the mentioned correlator as given by the following expression:
 \begin{align}
- \braket{\left[x(t_1),p(t_2)\right]^2}=2\left\{\braket{x(t_1)x(t_1)}  \braket{ p(t_2)p(t_2)} - \textnormal{Re}\left[\braket{x(t_1)p(t_2)x(t_1)p(t_2)} \right]\right\} + \mathcal{O}(e^{-t/t_d}) .
\end{align}
Now we normalize this above mentioned quantity using two different equal time two-point correlators, which we have obtained from the factorization in terms of the disconnected pieces. Consequently, we get the following simplified form of the normalized OTOC:
\begin{align}
\label{general}
\mathcal{C}^{(1)}(t_{1},t_{2}) = \frac{ - \braket{\left[x(t_1),p(t_2)\right]^2}}{\braket{x(t_1)x(t_1)}  \braket{p(t_2)p(t_2)}}
= 2\left[1- \frac{\textnormal{Re}\left[\braket{x(t_1)p(t_2)x(t_1)p(t_2)} \right]}{\braket{x(t_1)x(t_1)} \braket{p(t_2)p(t_2)}}\right] .
\end{align}
The first expression in \Cref{general} is a universal contribution, which will always appear for the quantum systems where the previously mentioned factorization process works at the vicinity of the {\it dissipation time scale}. On the other hand, the second term of \Cref{general} is basically representing a normalized version of the previously mentioned four-point function which can only appear after dissipation time. The other two non-trivial OTOCs in which we are also interested in this paper are of the form $-\braket{\left[x(t_1),x(t_2)\right]^2}$ and $-\braket{\left[p(t_1),p(t_2)\right]^2}$, and by following the same logical argument at the vicinity of the {\it dissipation time scale} we can normalize them as well and write them in the following simplified mathematical forms:
\begin{align}
C^{(2)}(t_{1},t_{2}) = \frac{-\braket{\left[x(t_1),x(t_2)\right]^2}}{\braket{x(t_1)x(t_1)}  \braket{x(t_2)x(t_2)}} = 2\left[1- \frac{\textnormal{Re}\left[\braket{x(t_1)x(t_2)x(t_1)x(t_2)} \right]}{\braket{x(t_1)x(t_1)} \braket{x(t_2)x(t_2)}}\right],
\end{align}
and
\begin{align}
C^{(3)}(t_{1},t_{2}) = \frac{-\braket{\left[p(t_1),p(t_2)\right]^2}}{\braket{p(t_1)p(t_1)} \braket{p(t_2)p(t_2)}}= 2\left[1- \frac{\textnormal{Re}\left[\braket{p(t_1)p(t_2)p(t_1)p(t_2)} \right]}{\braket{p(t_1)p(t_1)}  \braket{p(t_2)p(t_2)}}\right].
\end{align}

Now, since these OTOCs acts as a theoretical probe to  know about the generic chaotic or non-chaotic time dependent behaviour of quantum system, it has to satisfy the following constraints in terms of the four point correlations, which survived in the vicinity of the previously introduced {\it dissipation time scale}, as given by:
\begin{eqnarray}
&& \frac{\textnormal{Re}\left[\braket{x(t_1)p(t_2)x(t_1)p(t_2)} \right]}{\braket{x(t_1)x(t_1)} \braket{p(t_2)p(t_2)}}= \large \left\{
     \begin{array}{lr}
  \displaystyle  1-{\cal A}^{(1)}_{12}~e^{\lambda_1 \left(\frac{t_1+t_2}{2}\right)} ~~{\rm with}~~\lambda_1\leq \frac{2\pi}{\beta_1},&~\text{\textcolor{red}{\bf Chaotic}}\\ \\
 \displaystyle     1-{\cal B}^{(1)}_{12}~f_1(t_1,t_2).& \text{\textcolor{red}{\bf Non-Chaotic}}  \end{array}
   \right.~~~~~~~\\
   && \frac{\textnormal{Re}\left[\braket{x(t_1)p(t_2)x(t_1)p(t_2)} \right]}{\braket{x(t_1)x(t_1)} \braket{p(t_2)p(t_2)}}= \large \left\{
     \begin{array}{lr}
  \displaystyle  1-{\cal A}^{(2)}_{12}~e^{\lambda_2 \left(\frac{t_1+t_2}{2}\right)} ~~{\rm with}~~\lambda_2\leq \frac{2\pi}{\beta_2},&~\text{\textcolor{red}{\bf Chaotic}}\\ \\
 \displaystyle     1-{\cal B}^{(2)}_{12}~f_2(t_1,t_2).& \text{\textcolor{red}{\bf Non-Chaotic}}  \end{array}
   \right.~~~~~~~\\
   && \frac{\textnormal{Re}\left[\braket{p(t_1)p(t_2)p(t_1)p(t_2)} \right]}{\braket{p(t_1)p(t_1)} \braket{p(t_2)p(t_2)}}= \large \left\{
     \begin{array}{lr}
  \displaystyle  1-{\cal A}^{(3)}_{12}~e^{\lambda_3 \left(\frac{t_1+t_2}{2}\right)} ~~{\rm with}~~\lambda_3\leq \frac{2\pi}{\beta_3},&~\text{\textcolor{red}{\bf Chaotic}}\\ \\
 \displaystyle     1-{\cal B}^{(3)}_{12}~f_3(t_1,t_2).& \text{\textcolor{red}{\bf Non-Chaotic}}  \end{array}
   \right.~~~~~~~
\end{eqnarray}
The above mentioned expression captures all the possibilities which one can observe in different quantum mechanical systems available in nature. Here ${\cal A}^{(1)}_{12}$, ${\cal A}^{(2)}_{12}$ and ${\cal A}^{(3)}_{12}$ are the quantum mechanical model dependent pre-factors which shows exponential growth (chaotic behaviour) in the four point correlator with respect to the time scales associated with the system under study. On the other hand, ${\cal B}^{(1)}_{12}$, ${\cal B}^{(2)}_{12}$ and ${\cal B}^{(3)}_{12}$ are the quantum mechanical model dependent pre-factors which shows any type of time dependent fluctuations (non-chaotic behaviour). Also, for the general prescription the quantum {\it Lyapunov exponents}, $\lambda_1$, $\lambda_2$ and $\lambda_3$ are not same for which the MSS bound on quantum chaos from these three cases are also not the same. Consequently, the equilibrium saturation temperatures at the late time scales from these three OTOCs also differ from each other i.e. $\beta_1\neq \beta_2\neq\beta_3$. Also it is important to point out that, the mathematical structure of the time dependent functions $f_1(t_1,t_2)$, $f_2(t_1,t_2)$ and $f_3(t_1,t_2)$ are also different for general quantum mechanical set ups. When we are considering the quantum mechanical models are described by time dependent Hamiltonian in that case these expectations and all sorts of predictions work very well. However, when we are thinking about particularly quantum mechanical models which are described by the time independent Hamiltonian and the eigenstate representation of OTOCs, in that situations one might have further simplifications. There might be possibility to have underlying connection between the two functions $f_2(t_1,t_2)$ and $f_3(t_1,t_2)$ in the eigenstate representation of the OTOCs in the non-chaotic case and for this reason they might not be capturing completely independent information of the time disorder averaging. On the other hand, in the chaotic case if three of the OTOCs independently show exponential growth in the time scale the {\it quantum Lyapunov exponents} and the related equilibrium saturation temperatures are not at all same even in the eigenstate representation. But if the first OTOC is showing the chaotic behaviour and other two are not, in that case the previous connection between the two functions $f_2(t_1,t_2)$ and $f_3(t_1,t_2)$ holds good in the eigenstate representation.

In figure, \ref{vennin1} and \ref{vennin2}, we have presented the {\it diagrammatic} representations of all types of OTOCs in which we are interested in this paper. Particularly in figure \ref{vennin1}, we have explicitly depicted the possible two point OTOCs. Here we have three possibilities, which are given by, ${ -\braket{[x(t_{1}),p(t_{2})]}}$, ${ -\braket{[x(t_{1}),x(t_{2})]}}$ and ${ -\braket{[p(t_{1}),p(t_{2})]}}$. Since each OTOC are made up of commutator bracket in the quantum mechanical description, for each case we have two different contributions having overall opposite signatures. Further to draw the representative diagrams we need to consider the flow of time scale from $t_1$ to $t_2$ or $t_2$ to $t_1$. In all the representative diagrams, the two vertical solid thick line corresponds to the specific time slice having time, $t=t_1$ and $t=t_2$ respectively. It is understandable from the mathematical structure of the mentioned two point OTOCs that since the correlator involve only two time scales , that is why drawing two vertical parallel lines are physically justifiable in the present context. Because of the previously mentioned flow of time scale from $t_1$ to $t_2$ or $t_2$ to $t_1$ for each two point OTOCs we have two possible diagrams. So as a whole for the two-point OTOCs one can draw six possibilities. Also by studying each of the diagrammatic representations we can also observe that each of the contributions of the two point functions are represented by separate lines with representative arrows on that which is completely depend on the structure of the individual two point correlators. Two differentiate between these two contributions we have used red dotted line and blue solid line in the representative diagrams. Further, in figure \ref{vennin2}, we have explicitly shown the possible four point OTOCs in the present context of discussion.  Here we can draw three possible representative diagrams, which are coming from the three possible OTOCs, as given by, ${ -\braket{[x(t_{1}),p(t_{2})]^2}}$, ${ -\braket{[x(t_{1}),x(t_{2})]^2}}$ and ${ -\braket{[p(t_{1}),p(t_{2})]^2}}$.  Here as we can see that the each of the OTOCs are made up of commutator bracket squared contributions in the quantum description, for each case we have four contributions if we expand them. Out of these four four point thermal correlators two of them having positive signature and other two of them overall negative signature in the front of each contributions. Just like the two point OTOCs here also we need to consider the flow of time scale from $t_1$ to $t_2$ or from $t_2$ to $t_1$. It is understandable from the mathematical structure of the mentioned four point OTOCs that since the correlator involve only two time scales instead of four different time scales , that is why drawing two vertical parallel lines to represent the time slice at $t=t_1$ and $t=t_2$ are physically justifiable in the present context. Now, because of the time scale flow each of the four point OTOCs have two contributions in the diagrammatic representation. Also for a given time flow we have four possible diagrams which we shown in a single diagram for the sake of simplicity. So as whole consider both the possibilities of the time flow we have cumulatively $24$ diagrams from the four point OTOCs. Like the previous case, here also to differentiate between each of the individual terms for a given OTOC with a specified time flow we have used red dotted line, blue dotted line and red thick line respectively in the representative diagrammatic representations. Last but not the least, here it is important to point that these set of diagrams are the simplest version of the well known Feynman diagrams as appearing in the context of quantum field theory literature. However, within the present framework we don't have exactly similar Feynman diagrams, but to understand the structure of the previously mentioned OTOCs the present version of the diagrammatic  representations play a very crucial role. All of the arrows appearing in all the diagrams representing an underlying time disordering upon which we have to take finally average in the present context.

\textcolor{Sepia}{\section{\sffamily Eigenstate Representation of thermal OTOCs in Supersymmetric Quantum Mechanics}\label{sec:eigenrepcorr}}

In this section our prime objective is to study various thermal correlators, OTOCs and see how they can be expressed in a model independent manner in the framework of quantum mechanics. We also demarcate clearly the effect of Supersymmetry and how it modifies the functional form of the correlators at the end.

To perform this explicit computation we will follow the prescription outlined in the ref.~\cite{Hashimoto:2017oit} to compute all the thermal OTOCs we have mentioned in the previous section of this paper\footnote{Particularly the ref.~\cite{Hashimoto:2017oit} is extremely important for our computation because here the authors first have performed the computation of the first OTOC, ${ -\braket{[x(t_{1}),p(0)]^2}}$ (though in our computation we have generalized this to ${ -\braket{[x(t_{1}),p(t_{2})]^2}}$) in the eigenstate formalism for a time independent quantum mechanical model independent Hamiltonian.}. According to this prescription for any time-independent Hamiltonian we define the expectation value of quantum mechanical operators as thermal expectation value in the present context, which one can easily apply for a canonical quantum statistical ensemble. Let's say, a quantum mechanical time dependent operator ${A(t)}$ is defined on a specific Hilbert Space ${\mathscr{H}}$ with an associated Hamiltonian ${H}$ (which is time independent obviously) and the eigenvalues (eigen energy spectrum) of ${H}$ corresponding to a infinite tower of eigenkets ${\ket{\Psi_{n}}}~~\forall~~n=0,1,\cdots,\infty$, where the corresponding energy levels are characterized by the index $n$. This energy eigen spectrum is represented by ${E_{n}}~~\forall~~n=0,1,\cdots,\infty$. Then for a canonical quantum statistical ensemble the thermal expectation value of the quantum mechanical time dependent operator ${A}(t)$ at inverse temperature ${\beta}$, considering the Boltzmann constant ${k_{B} = 1}$, is defined as: 
\begin{align*}
\braket{A(t)}_{\beta} &= \frac{1}{Z}\ \Tr\bigl[e^{-\beta H}A(t)\bigr] ,
\end{align*}
\noindent
where the thermal partition function is represented by ${Z}$ such that ${Z = \Tr \left[ e^{-\beta H} \right]}$. Here our job is to represent this mathematical trace operation in terms of the sum over all possible eigenstates starting from the ground state ($n=0$).  Once we can able to express this operation clearly then the rest of the story is very usual and following this one can easily compute all of the mentioned OTOCs in the eigen state representation. It pays to use the thermal representation for expectation values of quantum mechanical operators because there are many physical models and physically relevant toy models which have well-studied structure of time independent Hamiltonian and hence we can utilize this property to give an eigenstate representation to the correlators as presented elaborately in this work. Consequently, the analysis presented in this paper is physically justifiable, applicable and believable, as such, to quantum mechanical models with well defined Hamiltonian. In the eigenstate representation the thermal expectation value of a quantum mechanical time dependent operator ${A}(t)$ can be written as : 

\begin{align*}
\braket{A(t)}_{\beta} = \frac{1}{Z}\ \Tr\biggl[e^{-\beta H} A(t) \underbrace{ \sum_{n}^{}\ket{\Psi_{n}}\bra{\Psi_{n}}}_{\textnormal{\textcolor{red}{\bf Identity~operator}}}\biggr] = \frac{\displaystyle \sum_{n}^{} e^{-\beta E_{n}} \braket{\Psi_{n} | A(t) | \Psi_{n}} }{\displaystyle \sum_{n}^{} e^{-\beta E_{n}} }
.\end{align*}
Here we work in the {\it Heisenberg Picture} where the operators evolve with time as given by, ${\mathcal{O}_{H}(t) = e^{iHt}\ \mathcal{O}_{S}(0)\ e^{-iHt}}$, where ${\mathcal{O}_{H}(t)}$ represents operator ${\mathcal{O}}_{H}$ at time scale ${t}$ in {\it Heisenberg picture} and ${\mathcal{O}_{S}(0)}$ represents its {\it Schr$\ddot{o}$dinger picture} representation at all time scales since operators do not evolve with time in {\it Schr$\ddot{o}$dinger picture}. The later one is denoted as ${\mathcal{O}_{S}}$ to simplify the notation and make it easier to read. 
\\
\\
\textcolor{red}{\large \bf What we learn here?}
\\
\\
\begin{tcolorbox}
	\begin{align}
	&\textbf{\sffamily{\textcolor{MidnightBlue}{Thermal Expectation Value : }}} \braket{A(t)}_{\beta} = \frac{1}{Z}\ \Tr\bigl[e^{-\beta H}A(t)\bigr] \\
	&\textbf{\sffamily{\textcolor{MidnightBlue}{Heisenberg Repsentation of Operators : }}} \mathcal{O}_{H}(t) = e^{iHt}\ \mathcal{O}_{S}\ e^{-iHt} \label{eq:heisenbergrep} \\
	&\textbf{\sffamily{\textcolor{MidnightBlue}{Eigenstate Rep. of Thermal Exp. : }}} \braket{A(t)}_{\beta} = \frac{\displaystyle \sum_{n}^{} e^{-\beta E_{n}} \braket{\Psi_{n} | A(t) | \Psi_{n}} }{\displaystyle \sum_{n}^{} e^{-\beta E_{n}} }
	\end{align}
\end{tcolorbox}
\vspace{0.5cm}

To demonstrate the computation further let us consider a general time independent Hamiltonian of the following form:
\begin{align}
H(q_i,p_i)= \sum_{i=1}^{N} p_{i}^{2}+ V(q_i),	
\end{align}
where we have used the fact that the mass of all $N$ number of particles are same and given by, $m_i=m=1/2$ to make the further computation simpler. Now here it is very easy to prove the following relation (see \cite{Hashimoto:2017oit}), which relates the quantum mechanical momentum operator matrix elements with that of the position and the energy operator matrix elements by the following expression: 
\begin{align} \label{eq:pnmxnmrel}
p_{km} = \frac{i}{2} E_{km} x_{km}
.\end{align}

\noindent
Furthermore, the role Supersymmetry plays in modification of non-supersymmetric quantum mechanical observables can be seen vis-a-vis the matrix elements of observables. Let's consider a simple matrix element for the position operator ${x}$. In the following chart we show exactly show how the matrix elements in non-supersymmetric quantum mechanical theories and supersymmetric quantum mechanical theories are connected:
\begin{align*}
	x_{mk} &= \braket{\Psi_{m} | x | \Psi_{k}} = \underbrace{\frac{1}{2} \left\{ \ \underbrace{\braket{\psi_{m}^{\textnormal{B}} | x | \psi_{k}^{\textnormal{B}}}}_{ \substack{ \textbf{ \sffamily \textcolor{MidnightBlue}{bosonic part} } \\ \textbf{ \sffamily \textcolor{MidnightBlue}{non-SUSY \& SUSY}}}} + \underbrace{\braket{\psi_{m-1}^{\textnormal{F}} | x | \psi_{k-1}^{\textnormal{F}}}}_{ \substack{ \textbf{ \sffamily \textcolor{MidnightBlue}{fermionic part} } \\ \textbf{ \sffamily \textcolor{MidnightBlue}{SUSY-only}} }  } \right\} }_{\textbf{ \sffamily \textcolor{MidnightBlue}{neither $\bm {m/k}$ is ground state}}}
\begin{matrix}
\\
&\xrightarrow[ \textbf{\sffamily \textcolor{red}{SUSY}}]{ \textbf{\sffamily \textcolor{red}{switch off}}} & \braket{\psi_{m}^{\textnormal{B}} | x | \psi_{k}^{\textnormal{B}}}, \\
\\
&\xrightarrow[\substack{\textnormal{\textbf{\sffamily \textcolor{red}{if one of $\bm{m/k}$ is}}} \\ \textbf{\sffamily \textcolor{red}{ground state}}}]{\textbf{\sffamily \textcolor{red}{SUSY}}} & \frac{1}{\sqrt{2}} \braket{ \psi_{m}^{\textnormal{B}} | x | \psi_{k}^{\textnormal{B}}}, \\
\\
&\xrightarrow[ \substack{\textbf{\sffamily \textcolor{red}{both} } \textbf{\sffamily \textcolor{red}{$\bm{m\ \&\ k}$} } \textbf{\sffamily\textcolor{red}{are}}\\ \textbf{\sffamily \textcolor{red}{ground state}}} ]{\textbf{\sffamily \textcolor{red}{SUSY}}} &\braket{\psi_{m}^{\textnormal{B}} | x | \psi_{k}^{\textnormal{B}}}.
\end{matrix}
\end{align*}
\noindent 
The modifications due to supersymmetry can clearly be traced back to the fact that we could formulate a wave function of total Hamiltonian as given in \Cref{eq:wavefn1} from the two partner Hamiltonians ${H_{1}}$ and ${H_{2}}$, as appearing from the bosonic and the fermion sectors in supersymmetry. In fact, its the factorization property of ${H_{1}}$ and ${H_{2}}$ which generalizes to a more powerful setting in terms of supersymmetric generalized description of the theory under consideration. One might consider the power of the supersymmetric description in two fold ways:
\begin{enumerate}
\item First, one can merely consider it as a tool to solve non-trivial potentials by means of solutions of their partner ones, provided that the partner ones are easily solvable.

\item Secondly, one can consider it as a more unifying description of a more beautiful theory based on the principles of symmetries of nature.
 It is this second philosophy which has been considered in this work.
\end{enumerate}

\noindent
We will consider the following six correlators in this work as stated below : \\

\begin{tcolorbox}[fonttitle=\sffamily\Large,title=Correlation Functions]
	
	\begin{align}
	&\nonumber\\
	&	\textbf{\sffamily{\textcolor{MidnightBlue}{2-pt Correlator : }}} Y^{(1)}(t_{1},t_{2}) = -\braket{[x(t_{1}),p(t_{2})]}_{\beta} \\
	&	\textbf{\sffamily{\textcolor{MidnightBlue}{2-pt Correlator : }}} Y^{(2)}(t_{1},t_{2}) = -\braket{[x(t_{1}),x(t_{2})]}_{\beta} \\
	&	\textbf{\sffamily{\textcolor{MidnightBlue}{2-pt Correlator : }}} Y^{(3)}(t_{1},t_{2}) = -\braket{[p(t_{1}),p(t_{2})]}_{\beta} \\ \nonumber\\
	&	\textbf{\sffamily{\textcolor{MidnightBlue}{4-pt Correlator : }}} C^{(1)}(t_{1},t_{2}) = -\braket{[x(t_{1}),p(t_{2})]^{2}}_{\beta} \\
	&	\textbf{\sffamily{\textcolor{MidnightBlue}{4-pt Correlator : }}} C^{(2)}(t_{1},t_{2}) = -\braket{[x(t_{1}),x(t_{2})]^{2}}_{\beta} \\
	&	\textbf{\sffamily{\textcolor{MidnightBlue}{4-pt Correlator : }}} C^{(3)}(t_{1},t_{2}) = -\braket{[p(t_{1}),p(t_{2})]^{2}}_{\beta} \\
	&\nonumber
	\end{align}
	
\end{tcolorbox} 
\vspace{0.5cm}
 
\noindent
We will also consider normalized 4-pt Correlation Functions and the discussion pertaining to those is provided in the previous section:
\\
\begin{tcolorbox}[fonttitle=\sffamily\Large,title=Normalized 4-pt Correlation Functions]
	
	\begin{align}
	&\nonumber\\
	&\textbf{\sffamily{\textcolor{MidnightBlue}{ Normalized 4-pt Correlator : }}} \widetilde{C}^{(1)}(t_{1},t_{2}) &= \frac{C^{(1)}(t_{1},t_{2})}{\braket{x(t_{1})x(t_{1})}_{\beta} \braket{p(t_{2}) p(t_{2})}_{\beta}} \nonumber \\
	&&= \frac{-\braket{[x(t_{1}),p(t_{2})]^{2}}_{\beta}}{\braket{x(t_{1})x(t_{1})}_{\beta} \braket{p(t_{2}) p(t_{2})}_{\beta}} \label{eq:normcan4ptcorrxp} \\ \nonumber\\
	&\textbf{\sffamily{\textcolor{MidnightBlue}{Normalized 4-pt Correlator : }}} \widetilde{C}^{(2)}(t_{1},t_{2}) &= \frac{C^{(2)}(t_{1},t_{2})}{ \braket{x(t_{1}) x(t_{1})}_{\beta} \braket{x(t_{2}) x(t_{2})}_{\beta} } \nonumber \\
	&&= \frac{-\braket{[x(t_{1}),x(t_{2})]^{2}}_{\beta}}{ \braket{x(t_{1}) x(t_{1})}_{\beta} \braket{x(t_{2}) x(t_{2})}_{\beta} } \label{eq:normcan4ptcorrxx} \\ \nonumber\\
	&\textbf{\sffamily{\textcolor{MidnightBlue}{Normalized 4-pt Correlator : }}} \widetilde{C}^{(3)}(t_{1},t_{2}) &= \frac{C^{(3)}(t_{1},t_{2})}{ \braket{p(t_{1}) p(t_{1})}_{\beta} \braket{p(t_{2}) p(t_{2})}_{\beta} } \nonumber \\
	&&= \frac{-\braket{[p(t_{1}),p(t_{2})]^{2}}_{\beta}}{ \braket{p(t_{1}) p(t_{1})}_{\beta} \braket{p(t_{2}) p(t_{2})}_{\beta}} \label{eq:normcan4ptcorrpp}
	\end{align}
	
\end{tcolorbox}
\vspace{0.5cm}

\textcolor{Sepia}{\subsection{\sffamily Partition Function from Supersymmetric Quantum Mechanics}\label{ssec:eigensusyqmpart}}

\noindent
In the context of Supersymmetric quantum mechanics the partition function, ${Z}$, can be can be expressed in terms of the eigenstate by the following expression:
\[
\begin{aligned}
Z &= \text{Tr}[e^{-\beta H_{\textnormal{SUSY}}}] = \sum_{m}^{} \braket{\Psi_{m} | e^{-\beta H_{\textnormal{SUSY}}} | \Psi_{m}}. \\
\end{aligned}
\]

\noindent
Here the underlying relation between supersymmetric total system and the component (or partner) systems, as discussed in \Cref{sec:revSUSYQM}, is established through the following relations : 
\[
\begin{aligned}
& H_{\textnormal{SUSY}}= \begin{pmatrix}
H_{1} & 0 \\
0 & H_{2}
\end{pmatrix} \implies e^{-\beta H_{\textnormal{SUSY}} } = \begin{pmatrix}
e^{- \beta H_{1}} & 0 \\
0 & e^{- \beta H_{2}}
\end{pmatrix}, \\
&\ket{\Psi_{m}}^{\textnormal{T}} = \frac{1}{\sqrt{2}} \left( \ \ket{\psi_{m}^{(1)}}\ \ket{\psi_{m-1}^{(2)}} \ \right)  \quad ; \quad \ket{\Psi_{0}}^{\textnormal{T}} = \left(\ \ket{\psi_{m}^{(1)}}\ 0\ \right).
\end{aligned}
\]
So, the supersymmetric quantum partition function in the eigenstate representation can be explicitly written as : 
\begin{align*}
	Z &= \braket{\psi_{m}^{(1)} | e^{-\beta H_{1}} | \psi_{m}^{(1)}} \bigg|_{m=0} +\ \frac{1}{2} \sum_{m>0}^{} \left\{ \braket{\psi_{m}^{(1)} | e^{-\beta H_{1}} | \psi_{m}^{(1)}} + \braket{\psi_{m-1}^{(2)} | e^{-\beta H_{2}} | \psi_{m-1}^{(2)}} \right\}, \\
	  &= \underbrace{ e^{-\beta E_{m}^{(1)}}\braket{\psi_{m}^{(1)}  | \psi_{m}^{(1)}} \bigg|_{m=0}}_{\substack{\textnormal{\sffamily \bf \textcolor{red}{Ground State} }\\ \textnormal{\sffamily \bf \textcolor{red}{Contribution}}}} +\ \frac{1}{2} \sum_{m>0}^{} \biggl\{ e^{-\beta E_{m}^{(1)}} \braket{\psi_{m}^{(1)} | \psi_{m}^{(1)}} + e^{-\beta E_{m-1}^{(2)}} \braket{\psi_{m-1}^{(2)} | \psi_{m-1}^{(2)}} \biggr\}.
\end{align*}

As shown in \Cref{sec:revSUSYQM} in supersymmetric quantum mechanical theories we have the following constraint:
 \bea {E_{m} = E_{m}^{(1)} = E_{m-1}^{(2)}},\eea
  with the requirement that the ground state is eigenstate of ${H_{1}}$ only and all other states are doubly degenerate. It is also important to note that the ground state always has zero energy eigenvalue.  It is to be noted that ${E_{m}}$ refers to the energy eigenvalue of the total Hamiltonian ${H_{\textnormal{SUSY}}}$. Using this sets of requirements, the above mentioned expression for the quantum partition function within the framework of supersymmetry can be further simplified as:\\ \\
\begin{tcolorbox}[colframe=black,arc=0mm]
\begin{align}
	Z &= 1 + \frac{1}{2} \sum_{m>0}^{} \left\{ e^{-\beta E_{m}} \left( \braket{\psi_{m}^{(1)} | \psi_{m}^{(1)}} + \braket{\psi_{m-1}^{(2)} | \psi_{m-1}^{(2)}} \right) \right\} ^{}= 1 + \sum_{m>0}^{} e^{-\beta E_{m}},
\end{align}
\end{tcolorbox}
which is different compared to usual quantum mechanics result without supersymmetry as in the present context ground state energy eigen value $E_0=0$ which is not in general zero for the other case. So this implies that if we separately write down the contribution from the ground state and from all other excited states, instead of writing sum over all eigenstates together, then one can clearly visualize the difference between the results obtained for quantum partition function in the eigenstate representation in both the cases. In the next section, we will provide the summary of all the obtained general model independent results in which we will implement the above mentioned fact to explicitly show that our obtained results for all types of OTOCs in the framework of supersymmetric quantum mechanics are different compared to the results obtained from usual quantum mechanics without supersymmetry.

\textcolor{Sepia}{\subsection{\sffamily Representation of two point OTOC : ${Y^{(1)}(t_{1},t_{2})}$}\label{ssec:2pty1}}

\noindent
The first two point OTOC is given by the thermal average of the operator $-{[x(t_{1}),p(t_{2})]}$, which is described in the eigenstate representation as:
\[
\begin{aligned}
Y^{(1)}(t_{1},t_{2}): &=  - \braket{[x(t_{1}),p(t_{2})]}_{\beta} = - \frac{\displaystyle \sum_{m}^{} e^{-\beta E_{m}} \braket{\Psi_{m} | [x(t_{1}),p(t_{2})] | \Psi_{m}}}{\displaystyle \left(1 + \sum_{m>0}^{} e^{-\beta E_{m}}\right)} .
.\end{aligned}
\]
\noindent
Using \Cref{eq:heisenbergrep} for {\it Heisenberg representation} for ${x(t_{1})}$ and ${p(t_{2})}$ and inserting the identities between the operators this two-point correlator can be written in terms of the micro-canonical correlator, which shows the temperature independent behaviour of the system:
\begin{align} \label{eq:can2ptcorrxp}
Y^{(1)}(t_{1},t_{2}) &= \frac{\displaystyle \sum_{m}^{} e^{-\beta E_{m}} y^{(1)}_{m}(t_{1},t_{2})}{\displaystyle\left(1 + \sum_{m>0}^{} e^{-\beta E_{m}}\right) },
\end{align}
where the micro-canonical two point OTOC is defined as:
\bea y^{(1)}_{m}(t_{1},t_{2}) = -\braket{\Psi_{m} | [x(t_{1}),p(t_{2})] | \Psi_{m}}.\eea
Expanding the commutator in the definition of the micro-canonical correlator, and using the above relation with appropriate insertion of identities between the operators, it can be shown that the eigenstate representation of the micro-canonical correlator ${y_{m}^{(1)}(t_{1},t_{2})}$ takes the following form. 

\begin{tcolorbox}[colframe=black,arc=0mm] 
	\begin{align} \label{eq:miccan2ptcorrxp}
	y_{m}^{(1)}(t_{1},t_{2}) = -i\ \sum_{k}^{}\ E_{km}\ x_{mk}\ x_{km}\ \cos E_{km} \left( t_{1} - t_{2} \right),
	\end{align}
\end{tcolorbox}
\noindent
where we define, ${E_{mk / m,k} =  E_{m} - E_{k}}$ and ${\ x_{mk / m,k} =  \braket{\Psi_{m} | x | \Psi_{k}}\ }$. \\

\noindent
Substituting the above mentioned expression for the micro-canonical OTOC in the definition of the canonical correlator,  ${Y^{(1)}(t_{1},t_{2})}$  we get the following expression : \\

\begin{tcolorbox}[colframe=black,arc=0mm]
	\begin{align} \label{eq:can2ptcorrxpfinal}
	Y^{(1)}(t_{1},t_{2}) = - i~\frac{\displaystyle \sum_{m,k}^{}\ e^{-\beta E_{m}}\ E_{km}\ x_{mk}\ x_{km} \cos E_{km} \ (t_{1}-t_{2}) }{\displaystyle \left(1 + \sum_{m>0}^{} e^{-\beta E_{m}}\right)}\ 
	.\end{align}
\end{tcolorbox}
\vspace{0.5cm}

\noindent

\textcolor{Sepia}{\subsection{\sffamily Representation of two point OTOC : ${Y^{(2)}(t_{1},t_{2})}$}\label{ssec:2pty2}}

\noindent
The second two point OTOC is given by the thermal average of the operator $-{[x(t_{1}),x(t_{2})]}$, which is described in the eigenstate representation as:
\[
\begin{aligned}
Y^{(2)}(t_{1},t_{2}) &=  - \braket{[x(t_{1}),x(t_{2})]}_{\beta} = - \frac{\displaystyle \sum_{m}^{} e^{-\beta E_{m}} \braket{\Psi_{m} | [x(t_{1}),x(t_{2})] | \Psi_{m}}}{\displaystyle \left(1 + \sum_{m>0}^{} e^{-\beta E_{m}}\right) } 
.\end{aligned}
\]
\noindent
Using \Cref{eq:heisenbergrep} for {\it Heisenberg representation} for ${x(t_{1})}$ and ${x(t_{2})}$ the two point correlator in terms of the temperature independent micro-canonical correlator can be written as:
\begin{align} \label{eq:can2ptcorrxx}
Y^{(2)}(t_{1},t_{2}) &= \frac{\displaystyle \sum_{m}^{} e^{-\beta E_{m}}\ y^{(2)}_{m}(t_{1},t_{2})}{\displaystyle \left(1 + \sum_{m>0}^{} e^{-\beta E_{m}}\right) },
\end{align}
where the micro-canonical two point OTOC is defined as:
\bea y^{(2)}_{m}(t_{1},t_{2})=-\braket{\Psi_{m} | [x(t_{1}),x(t_{2})] | \Psi_{m}}.\eea
Expanding the commutator in the definition of the micro-canonical correlator and inserting identity in between the operators, it can be shown that the eigenstate representation for micro-canonical correlator ${y_{m}^{(2)}(t_{1},t_{2})}$ takes the following form:\\

\noindent
\begin{tcolorbox}[colframe=black,arc=0mm]
	\begin{align} \label{eq:miccan2ptcorrxx}
	y_{m}^{(2)}(t_{1},t_{2}) = -2i\ \sum_{k}^{}\ x_{mk}\ x_{km}\  \text{sin} ~E_{mk} (t_{1} - t_{2}) 
	.\end{align}
\end{tcolorbox}
\vspace{0.5cm}

\noindent
where we define, ${E_{mk / m,k} =  E_{m} - E_{k}}$ and ${\ x_{mk / m,k} =  \braket{\Psi_{m} | x | \Psi_{k}}\ }$. \\

\noindent
Substituting \Cref{eq:miccan2ptcorrxx} in \Cref{eq:can2ptcorrxx}, the eigenstate representation for the canonical correlator ${Y^{(2)}(t_{1},t_{2})}$ takes the form : \\

\begin{tcolorbox}[colframe=black,arc=0mm]
	\begin{align} \label{eq:can2ptcorrxxfinal}
	Y^{(2)}(t_{1},t_{2}) = - 2i~\frac{\displaystyle  \sum_{m,k}^{}\ e^{-\beta E_{m}}\ x_{mk}\ x_{km}\ \text{sin} ~E_{mk} (t_{1} - t_{2}) }{\displaystyle\left(1 + \sum_{m>0}^{} e^{-\beta E_{m}}\right)}\ .
	\end{align}
\end{tcolorbox}
\vspace{0.5cm}

\textcolor{Sepia}{\subsection{\sffamily Representation of two point OTOC : ${Y^{(3)}(t_{1},t_{2})}$}\label{ssec:2pty3}}

\noindent
The third two point OTOC is given by the thermal average of the operator $-{[p(t_{1}),p(t_{2})]}$, which is described in the eigenstate representation as:
\[
\begin{aligned}
Y^{(3)}(t_{1},t_{2}) &=  - \braket{[p(t_{1}),p(t_{2})]}_{\beta} = - \frac{\displaystyle \sum_{m}^{} e^{-\beta E_{m}} \braket{\Psi_{m} | [p(t_{1}),p(t_{2})] | \Psi_{m}}}{\displaystyle \left(1 + \sum_{m>0}^{} e^{-\beta E_{m}}\right)} 
.\end{aligned}
\]
\noindent
Proceeding as previously mentioned we obtain the following expression for the OTOC which can be expressed in terms of the temperature independent micro-canonical correlator as: 
\begin{align} \label{eq:can2ptcorrpp}
Y^{(3)}(t_{1},t_{2})
&= \frac{\displaystyle \sum_{m}^{} e^{-\beta E_{m}}\ y^{(3)}_{m}(t_{1},t_{2})}{\displaystyle  \left(1 + \sum_{m>0}^{} e^{-\beta E_{m}}\right)} ,
\end{align}
where the micro-canonical two point OTOC is defined as:
\bea y^{(3)}_{m}(t_{1},t_{2})=-\braket{\Psi_{m} | [p(t_{1}),p(t_{2})] | \Psi_{m}}.\eea
\noindent
Expanding the commutator in the definition of the micro-canonical correlator and inserting identity in between the operators, it can be shown that the eigenstate representation for micro-canonical correlator ${y_{m}^{(3)}(t_{1},t_{2})}$ is given by the following expression:\\

\begin{tcolorbox}[colframe=black,arc=0mm]
	\begin{align} \label{eq:miccan2ptcorrpp}
	y^{(3)}_{m}(t_{1},t_{2}) = -\frac{i}{2} \sum_{k}^{} E_{mk}\ x_{mk}\ E_{km}\ x_{km}\ \text{sin}~  E_{km} \left( t_{1} - t_{2} \right) .
	\end{align}
\end{tcolorbox}
\vspace{0.5cm}

\noindent
Substituting the expression for the micro-canonical correlator obtained from  \Cref{eq:miccan2ptcorrpp} in \Cref{eq:can2ptcorrpp}, the eigenstate representation for the canonical correlator ${Y^{(3)}(t_{1},t_{2})}$ can be written as : \\

\begin{tcolorbox}[colframe=black,arc=0mm]
	\begin{align} \label{eq:can2ptcorrppfinal}
	Y^{(3)}(t_{1},t_{2}) = -\frac{i}{2}~ \frac{\displaystyle \sum_{m,k}^{} e^{-\beta E_{m}} E_{mk}\ x_{mk}\ E_{km}\ x_{km} \sin ~ E_{km} \left( t_{1} - t_{2} \right)}{\displaystyle \left(1 + \sum_{m>0}^{} e^{-\beta E_{m}}\right)}
	.\end{align}
\end{tcolorbox}
\vspace{0.5cm}

\textcolor{Sepia}{\subsection{\sffamily Representation of four point OTOC : ${C^{(1)}(t_{1},t_{2})}$}\label{ssec:4ptc1}}

In this section, we provide the eigenstate representation of the four-point OTOCs that are mainly used for studying the quantum mechanical analogue of phenomena of time disorder averaging , related randomness in quantum regime. Similar to the two-point correlators we define three kinds of four-point correlators. Generally in the literature people study the four-point correlator  $C^{(1)}(t_{1},t_{2})$ (as per our notation) which is the thermal expectation value of the square of the commutator made up of the dynamical operators of different kinds at different time scales. The four-point correlator, ${C^{(1)}(t_{1},t_{2})}$ is proposed as a quantifier of quantum chaos but chaos is a specific kind of randomness. Hence, it is important to consider the correlators constructed from similar operators at different times for having a complete understanding of the underlying phenomenon of randomness. To understand this time disordering phenomena explicitly we have also introduced two more new correlators, ${C^{(2)}(t_{1},t_{2})}$ and ${C^{(3)}(t_{1},t_{2})}$, described in detail in the later half of this paper.

\textcolor{Sepia}{\subsubsection{\sffamily Un-normalized : $C^{(1)}(t_1,t_2)$}\label{sssec:4ptc1un}}

\noindent
The first four point OTOC is given by the thermal average of the operator $-{[x(t_{1}),p(t_{2})]^2}$, which is described in the eigenstate representation as:

\begin{align} \label{eq:can4ptcorrxp}
C^{(1)}(t_{1},t_{2})  &=  - \braket{[x(t_{1}),p(t_{2})]^{2}}_{\beta} 
=  \frac{\displaystyle \sum_{m}^{} e^{- \beta E_{m}}\ c_{m}^{(1)}(t_{1},t_{2})}{\displaystyle \left(1 + \sum_{m>0}^{} e^{-\beta E_{m}}\right)} 
,\end{align}
where $c_{m}^{(1)}(t_{1},t_{2})$ is the micro-canonical correlator and is responsible for the temperature independent behaviour of the correlator. The temperature dependence in the canonical correlator is actually appearing due to the exponential thermal Botzmann factor in the eigenstate representation of the OTOC. Once we will take the sum over all possible eigen states (finite or infinite in number) we get a cumulative dependence on time, temperature and energy eigenstates.
In the present context 
the micro-canonical four-point correlator for \Cref{eq:can4ptcorrxp} is given by the following simplified expression:

\begin{align}
\label{c1mes}
c^{(1)}_{m}(t_{1},t_{2}) = - \braket{\Psi_{m} | [x(t_{1}),p(t_{2})]^{2} | \Psi_{m}} = \sum_{k}^{} b_{mk}^{(1)}(t_{1},t_{2})\ b_{mk}^{(1)*}(t_{1},t_{2}) 
,\end{align}
where we define a time dependent matrix element, $b_{mk}^{(1)}(t_{1},t_{2})$, which is given by: 
\[
\begin{aligned}
b_{mk}^{(1)}(t_{1},t_{2}) &= -i \braket{\Psi_{m} | [x(t_{1}),p(t_{2})]| \Psi_{k}}. \\
.\end{aligned}
\]
Using the {\it Heisenberg picture} evolution equation for an operator and inserting identity between the operators after expanding the commutator, it is not hard to verify that $b_{mk}^{(1)}$ can be written as:
\[
\begin{aligned}
b_{mk}^{(1)*}(t_{1},t_{2}) = \frac{1}{2} \sum_{r}^{} \left\{ E_{rk} e^{it_{1} E_{rm}} e^{i t_{2} E_{kr}} - E_{mr} e^{i t_{2} E_{rm}} e^{i t_{1} E_{kr}} \right\} x_{rm} x_{kr}
.\end{aligned}
\]
Substituting the above expression of $	b_{mk}^{(1)*}(t_{1},t_{2})$ in \Cref{c1mes}, it can be shown that after simplification the eigenstate representation of the micro-canonical correlator for \Cref{eq:can4ptcorrxp} is given by the following expression: \\

\begin{tcolorbox}[colframe=black,arc=0mm]
	\begin{align} \label{eq:miccan4ptcorrxp}
	c_{m}^{(1)}(t_{1},t_{2}) &= \frac{1}{4} \sum_{k,l,r}^{} x_{ml}\ x_{lk}\ x_{rm}\ x_{kr}\ \times \biggl[ E_{rk} E_{lk} e^{i E_{rl} \left( t_{1} - t_{2} \right)}  +  E_{mr} E_{ml} e^{- i E_{rl} \left( t_{1} - t_{2} \right)}  \nonumber \\
	& \hspace{1cm} -  E_{rk} E_{ml} e^{i \left( E_{rm} + E_{lk} \right) \left( t_{1} - t_{2} \right)}  - E_{mr} E_{lk} e^{-i \left( E_{rm} + E_{lk} \right) \left( t_{1} - t_{2} \right)}  \biggr] 
	.\end{align}
\end{tcolorbox}
\vspace{0.5cm}

\noindent
So, the eigenstate representation of canonical correlator from \Cref{eq:can4ptcorrxp} using \Cref{eq:miccan4ptcorrxp} is given as : \\

\begin{tcolorbox}[colframe=black,arc=0mm]
	\begin{align} \label{eq:can4ptcorrxpfinal}
	C^{(1)}(t_{1},t_{2}) &= \frac{1}{4~\displaystyle  \left(1 + \sum_{m>0}^{} e^{-\beta E_{m}}\right)} \ \sum_{m,k,l,r}^{} e^{-\beta E_{m}}~x_{ml}\ x_{lk}\ x_{rm}\ x_{kr} \nonumber \\
	& \times \biggl[ E_{rk} E_{lk} e^{i E_{rl} \left( t_{1} - t_{2} \right)}  +  E_{mr} E_{ml} e^{- i E_{rl} \left( t_{1} - t_{2} \right)} \nonumber \\
	& -  E_{rk} E_{ml} e^{i \left( E_{rm} + E_{lk} \right) \left( t_{1} - t_{2} \right)}  -  E_{mr} E_{lk} e^{-i \left( E_{rm} + E_{lk} \right) \left( t_{1} - t_{2} \right)}  \biggr] 
	.\end{align}
\end{tcolorbox}
\vspace{0.5cm}

\textcolor{Sepia}{\subsubsection{\sffamily Normalized~~~~ : $\widetilde{C}^{(1)}(t_1,t_2)$}\label{sssec:4ptc1n}}

The normalized first four-point OTOC is from the un-normalized one expressed in  \Cref{eq:normcan4ptcorrxp} can be expressed as: 
\begin{align*}
\widetilde{C}^{(1)}(t_{1},t_{2}) = \frac{C^{(1)}(t_{1},t_{2})}{\braket{x(t_{1})x(t_{1})}_{\beta} \braket{p(t_{2})p(t_{2})}_{\beta}}.
\end{align*}
We have already calculated the numerator ${C^{(1)}(t_{1},t_{2})}$ as given in \Cref{eq:can4ptcorrxpfinal}. So here our only job is compute the two sets of equal time thermal correlators in the eigenstate representation. Since we know the formalism very well that we have developed in this paper, computing these disconnected pieces of equal time correlators which are extremely significant in the large time dissipation limit are not at all complicated. Now we are going to show that how one can compute these contributions.

To serve this purpose we need to compute the following expressions in the eigenstate representation:
\bea && \langle j|x(t_1)x(t_1)|j\rangle=\sum_{l}x_{jl}x_{lj},\\ 
&& \langle j|p(t_2)p(t_2)|j\rangle=\sum_{l}p_{jl}p_{lj}=-\frac{1}{4}\sum_{l}E_{jl}E_{lj}x_{jl}x_{lj}.\eea
Consequently, the desired canonical equal time thermal correlators can be computed as:
\bea &&\braket{x(t_{1})x(t_{1})}_{\beta} =\frac{\displaystyle \sum_{j}e^{-\beta E_j}~\langle j|x(t_1)x(t_1)|j\rangle}{\displaystyle \left(1 + \sum_{m>0}^{} e^{-\beta E_{m}}\right)}=\frac{\displaystyle \sum_{j,l}e^{-\beta E_j}~x_{jl}x_{lj}}{\displaystyle \left(1 + \sum_{m>0}^{} e^{-\beta E_{m}}\right)},\\
 &&\braket{p(t_{2})p(t_{2})}_{\beta} =\frac{\displaystyle \sum_{j}e^{-\beta E_j}~\langle j|p(t_2)p(t_2)|j\rangle}{\displaystyle \left(1 + \sum_{m>0}^{} e^{-\beta E_{m}}\right)}=\frac{\displaystyle \sum_{j,l}e^{-\beta E_j}~p_{jl}p_{lj}}{\displaystyle \left(1 + \sum_{m>0}^{} e^{-\beta E_{m}}\right)}=-\frac{1}{4}~\frac{\displaystyle \sum_{j,l}e^{-\beta E_j}~E_{jl}E_{lj}x_{jl}x_{lj}}{\displaystyle \left(1 + \sum_{m>0}^{} e^{-\beta E_{m}}\right)}.~~~~~~~~~~~\eea
Finally, the normalized first four-point OTOC can be expressed by the following simplified expression in the eigenstate representation, as given by:\\
\\
\begin{tcolorbox}[colframe=black,arc=0mm]
	\begin{align} \label{eq:dcan4ptcorrxpfinal}
	\widetilde{C}^{(1)}(t_{1},t_{2}) &= \frac{C^{(1)}(t_{1},t_{2})}{\braket{x(t_{1})x(t_{1})}_{\beta} \braket{p(t_{2})p(t_{2})}_{\beta}}\nonumber\\
	 &= -\frac{\displaystyle\left(1 + \sum_{j>0}^{} e^{-\beta E_{j}}\right)}{~\displaystyle  \sum_{j,m,l}^{}\ e^{-\beta (E_j+E_{m})}~E_{jl}E_{lj}x_{jl}x_{lj}} \ \nonumber\\
	 &~~~\times\sum_{m,k,l,r}^{} e^{-\beta E_{m}}~x_{ml}\ x_{lk}\ x_{rm}\ x_{kr} \nonumber \\
	& ~~~~~~~\times \biggl[ E_{rk} E_{lk} e^{i E_{rl} \left( t_{1} - t_{2} \right)}  +  E_{mr} E_{ml} e^{- i E_{rl} \left( t_{1} - t_{2} \right)} \nonumber \\
	& ~~~~~~~ -  E_{rk} E_{ml} e^{i \left( E_{rm} + E_{lk} \right) \left( t_{1} - t_{2} \right)}  -  E_{mr} E_{lk} e^{-i \left( E_{rm} + E_{lk} \right) \left( t_{1} - t_{2} \right)}  \biggr] 
	.\end{align}
\end{tcolorbox}
\vspace{0.5cm}

\textcolor{Sepia}{\subsection{\sffamily Representation of four point OTOC : ${C^{(2)}(t_{1},t_{2})}$}\label{ssec:4ptc2}}

\textcolor{Sepia}{\subsubsection{\sffamily Un-normalized : $C^{(2)}(t_1,t_2)$}\label{sssec:4ptc2un}}

\noindent
The second four point OTOC is given by the thermal average of the operator $-{[x(t_{1}),x(t_{2})]^2}$, which is described in the eigenstate representation as: 
\begin{align} \label{eq:can4ptcorrxx}
C^{(2)}(t_{1},t_{2}) &=  - \braket{[x(t_{1}),x(t_{2})]^{2}}_{\beta} 
=  \frac{\displaystyle \sum_{m}^{} e^{- \beta E_{m}} c_{m}^{(2)}(t_{1},t_{2})}{\displaystyle \left(1 + \sum_{m>0}^{} e^{-\beta E_{m}}\right) } 
,\end{align}
where $c_{m}^{(2)}(t_{1},t_{2})$ is the micro-canonical correlator and is responsible for the temperature independent behaviour of the correlator. The temperature dependence in the canonical correlator is actually appearing due to the exponential thermal Botzmann factor in the eigenstate representation of the OTOC. Once we will take the sum over all possible eigen states (finite or infinite in number) we get a cumulative dependence on time, temperature and energy eigenstates.
In the present context 
the micro-canonical four-point correlator for \Cref{eq:can4ptcorrxx} is given by the following simplified expression:

\begin{align}
\label{cm2}
c_{m}^{(2)}(t_{1},t_{2}) = - \braket{\Psi_{m} | [x(t_{1}),x(t_{2})]^{2} | \Psi_{m}} = - \sum_{k}^{} b_{mk}^{(2)}(t_{1},t_{2}) b_{km}^{(2)}(t_{1},t_{2}) ,
\end{align}
\noindent
where we define a time dependent matrix element, $b_{mk}^{(2)}(t_{1},t_{2})$, which is given by: 
\begin{align*}
b_{mk}^{(2)}(t_{1},t_{2}) &= \braket{\Psi_{m} | [x(t_{1}),x(t_{2})]| \Psi_{k}} .\\
\end{align*}
Using the {\it Heisenberg picture} for the evolution of operators and inserting identity between the operators, it can be shown that $b_{mk}^{(2)}(t_{1},t_{2})$ can be written as:
\begin{align}
\label{bmk2}
b_{mk}^{(2)}(t_{1},t_{2})= \sum_{l}^{} \left\{ e^{i t_{1} E_{ml}} e^{i t_{2} E_{lk}} - e^{i t_{2} E_{ml}} e^{i t_{1} E_{lk}} \right\} x_{ml} x_{lk} .
\end{align}
\noindent
Substituting \Cref{bmk2} in \Cref{cm2} and simplifying the eigenstate representation for the  temperature independent micro-canonical correlator can be expressed by the following simplified expression: \\

\noindent
\begin{tcolorbox}[colframe=black,arc=0mm]
		\begin{align} 
		c_{m}^{(2)}(t_{1},t_{2}) &= 4~\sum_{k,l,r}^{}\ x_{ml}\ x_{lk}\ x_{kr}\ x_{rm} \nonumber \\
		& \hspace{1cm} ~~~\times\sin \left[ \left( E_{r} - \frac{E_{m} + E_{k}}{2} \right) \left( t_{1} - t_{2} \right) \right] \sin \left[ \left( E_{l} - \frac{E_{m} + E_{k}}{2} \right) \left( t_{1} - t_{2} \right) \right]. \label{eq:miccan4ptcorrxxb}
		\end{align}
\end{tcolorbox}
\vspace{0.5cm}

\noindent
The canonical or the temperature dependent correlator can be calculated by substituting the expression of \Cref{eq:can4ptcorrxx} in \Cref{eq:miccan4ptcorrxxb} and it after applying some algebraic manipulation it can be shown that the canonical correlator can be expressed by the following two expressions : \\

\begin{tcolorbox}[colframe=black,arc=0mm]
	
		\begin{align} 
		C^{(2)}(t_{1},t_{2}) &= \frac{4}{\displaystyle \left(1 + \sum_{j>0}^{} e^{-\beta E_{j}}\right)}\ \ \sum_{m, k,l,r}^{}\ e^{-\beta E_{m}}~x_{ml} x_{lk} x_{kr} x_{rm} \nonumber \\
		& \hspace{1cm} \times  \sin \left[ \left( E_{r} - \frac{E_{m} + E_{k}}{2} \right) \left( t_{1} - t_{2} \right) \right] \sin \left[ \left( E_{l} - \frac{E_{m} + E_{k}}{2} \right) \left( t_{1} - t_{2} \right) \right]. \label{eq:can4ptcorrxxfinalb}
		\end{align}
	
\end{tcolorbox}
\vspace{0.5cm}

\textcolor{Sepia}{\subsubsection{\sffamily Normalized~~~~ : $\widetilde{C}^{(2)}(t_1,t_2)$}\label{sssec:4ptc2n}}

The normalized first four-point OTOC is from the un-normalized one expressed in  \Cref{eq:normcan4ptcorrxx} can be expressed as:
\begin{align*}
\widetilde{C}^{(2)}(t_{1},t_{2}) = \frac{C^{(2)}(t_{1},t_{2})}{\braket{x(t_{1})x(t_{1})}_{\beta} \braket{x(t_{2})x(t_{2})}_{\beta}}.
\end{align*}
We have already calculated the numerator ${C^{(2)}(t_{1},t_{2})}$ as given in \Cref{eq:can4ptcorrxxfinalb}. So here our only job is compute the two sets of equal time thermal correlators in the eigenstate representation. Since we know the formalism very well that we have developed in this paper, computing these disconnected pieces of equal time correlators which are extremely significant in the large time dissipation limit are not at all complicated. Now we are going to show that how one can compute these contributions.

To serve this purpose we need to compute the following expressions in the eigenstate representation:
\bea && \langle j|x(t_1)x(t_1)|j\rangle=\sum_{l}x_{jl}x_{lj}=\langle j|x(t_2)x(t_2)|j\rangle.\eea
Consequently, the desired canonical equal time thermal correlators can be computed as:
\bea &&\label{ge1} \braket{x(t_{1})x(t_{1})}_{\beta} =\frac{\displaystyle \sum_{j}e^{-\beta E_j}~\langle j|x(t_1)x(t_1)|j\rangle}{\displaystyle \left(1 + \sum_{m>0}^{} e^{-\beta E_{m}}\right)}=\frac{\displaystyle \sum_{j,l}e^{-\beta E_j}~x_{jl}x_{lj}}{\displaystyle\left(1 + \sum_{m>0}^{} e^{-\beta E_{m}}\right)}=\braket{x(t_{2})x(t_{2})}_{\beta}.~~~~~~~~~~~\eea
Finally, the normalized first four-point OTOC can be expressed by the following simplified expression in the eigenstate representation, as given by:\\
\\
\begin{tcolorbox}[colframe=black,arc=0mm]
	\begin{align} \label{eq:ccan4ptcorrxpfinal}
	\widetilde{C}^{(2)}(t_{1},t_{2}) &= \frac{C^{(2)}(t_{1},t_{2})}{\braket{x(t_{1})x(t_{1})}_{\beta} \braket{x(t_{2})x(t_{2})}_{\beta}}\nonumber\\
	 &= \frac{4\displaystyle\left(1 + \sum_{j>0}^{} e^{-\beta E_{j}}\right)}{\displaystyle \sum_{j,m,l,k}^{} e^{-\beta (E_j+E_m)}~x_{jl}x_{lj}x_{mk}x_{km}}\ \ \sum_{m, k,l,r}^{}\ e^{-\beta E_{m}}~x_{ml} x_{lk} x_{kr} x_{rm} \nonumber \\
		& \hspace{1cm} \times  \sin \left[ \left( E_{r} - \frac{E_{m} + E_{k}}{2} \right) \left( t_{1} - t_{2} \right) \right] \sin \left[ \left( E_{l} - \frac{E_{m} + E_{k}}{2} \right) \left( t_{1} - t_{2} \right) \right]
	.\end{align}
\end{tcolorbox}
\vspace{0.5cm}

\textcolor{Sepia}{\subsection{\sffamily Representation of four point OTOC : ${C^{(3)}(t_{1},t_{2})}$}\label{ssec:4ptc3}}

\textcolor{Sepia}{\subsubsection{\sffamily Un-normalized : $C^{(3)}(t_1,t_2)$}\label{sssec:4ptc3un}}

\noindent
The third four point OTOC is given by the thermal average of the operator $-{[p(t_{1}),p(t_{2})]^2}$, which is described in the eigenstate representation as: 
\begin{align} \label{eq:can4ptcorrpp}
C^{(3)}(t_{1},t_{2}) &=  - \braket{[p(t_{1}),p(t_{2})]^{2}}_{\beta} 
= \frac{\displaystyle \sum_{m}^{} e^{- \beta E_{m}} c_{m}^{(3)}(t_{1},t_{2})}{\displaystyle \left(1 + \sum_{m>0}^{} e^{-\beta E_{m}}\right)} 
,\end{align}
where $c_{m}^{(3)}(t_{1},t_{2})$ is the micro-canonical correlator and is responsible for the temperature independent behaviour of the correlator. The temperature dependence in the canonical correlator is actually appearing due to the exponential thermal Botzmann factor in the eigenstate representation of the OTOC. Once we will take the sum over all possible eigen states (finite or infinite in number) we get a cumulative dependence on time, temperature and energy eigenstates.
In the present context 
the micro-canonical four-point correlator for \Cref{eq:can4ptcorrpp} is given by the following simplified expression:

\begin{align}
\label{cm3}
c_{m}^{(3)}(t_{1},t_{2}) = - \braket{\Psi_{m} | [p(t_{1}),p(t_{2})]^{2} | \Psi_{m}} = - \sum_{k}^{} b_{mk}^{(3)}(t_{1},t_{2}) b_{km}^{(3)}(t_{1},t_{2}) 
,\end{align}
\noindent
where we define a time dependent matrix element, $b_{mk}^{(2)}(t_{1},t_{2})$, which is given by: 
\begin{align}
\label{bmk3}
b_{mk}^{(3)}(t_{1},t_{2}) &= \braket{\Psi_{m} | [p(t_{1}),p(t_{2})]| \Psi_{k}} .
\end{align}
Using the {\it Heisenberg picture} for the evolution of operators and inserting identity between the operators, it can be shown that $b_{mk}^{(3)}(t_{1},t_{2})$ can be simplified into the following form:
\begin{align*}
b_{mk}^{(3)}(t_{1},t_{2}) =\sum_{l}^{} \left\{ e^{i t_{1} E_{ml}} e^{i t_{2} E_{lk}} - e^{i t_{2} E_{ml}} e^{i t_{1} E_{lk}} \right\} p_{ml} p_{lk}.
\end{align*}

\noindent
Substituting the expression of $b_{mk}^{(3)}(t_{1},t_{2})$ i.e \Cref{bmk3} in \Cref{cm3}, and simplifying, the eigenstate representation of the  micro-canonical correlator can be expressed as: \\

\begin{tcolorbox}[colframe=black,arc=0mm]
		\begin{align}
		c_{m}^{(3)}(t_{1},t_{2}) &= \frac{1}{4}~\sum_{k,l,r}^{}\ E_{ml}\ E_{lk}\ E_{kr}\ E_{rm}x_{ml}\ x_{lk}\ x_{kr}\ x_{rm} \nonumber \\
		& \hspace{1cm} ~~\times \sin \left[ \left( E_{r} - \frac{E_{m} + E_{k}}{2} \right) \left( t_{1} - t_{2} \right) \right] \sin \left[ \left( E_{l} - \frac{E_{m} + E_{k}}{2} \right) \left( t_{1} - t_{2} \right) \right]. \label{eq:miccan4ptcorrppb}
		\end{align}
\end{tcolorbox}
\vspace{0.5cm}

\noindent
The eigenstate representation of the canonical correlator from \Cref{eq:can4ptcorrpp} using \Cref{eq:miccan4ptcorrppb} after simplification can be expressed by the following two equations :  \\

\begin{tcolorbox}[colframe=black,arc=0mm]
		\begin{align} 
		C^{(3)}(t_{1},t_{2}) &=  \frac{1}{4\displaystyle \left(1 + \sum_{m>0}^{} e^{-\beta E_{m}}\right)}~\sum_{m,k,l,r}^{}\ e^{-\beta E_m}~E_{ml}\ E_{lk}\ E_{kr}\ E_{rm}x_{ml}\ x_{lk}\ x_{kr}\ x_{rm} \nonumber \\
		& \hspace{1cm} ~~\times \sin \left[ \left( E_{r} - \frac{E_{m} + E_{k}}{2} \right) \left( t_{1} - t_{2} \right) \right] \sin \left[ \left( E_{l} - \frac{E_{m} + E_{k}}{2} \right) \left( t_{1} - t_{2} \right) \right]. \label{eq:can4ptcorrfinalb}
		\end{align} 
\end{tcolorbox}
\vspace{0.5cm}

\textcolor{Sepia}{\subsubsection{\sffamily Normalized~~~~ : $\widetilde{C}^{(3)}(t_1,t_2)$}\label{sssec:4ptc3n}}

The normalized first four-point OTOC is from the un-normalized one expressed in  \Cref{eq:normcan4ptcorrpp} can be expressed as:
\begin{align*}
\widetilde{C}^{(3)}(t_{1},t_{2}) = \frac{C^{(3)}(t_{1},t_{2})}{\braket{p(t_{1})p(t_{1})}_{\beta} \braket{p(t_{2})p(t_{2})}_{\beta}}.
\end{align*}
We have already calculated the numerator ${C^{(3)}(t_{1},t_{2})}$ as given in \Cref{eq:can4ptcorrfinalb}. So here our only job is compute the two sets of equal time thermal correlators in the eigenstate representation. Since we know the formalism very well that we have developed in this paper, computing these disconnected pieces of equal time correlators which are extremely significant in the large time dissipation limit are not at all complicated. Now we are going to show that how one can compute these contributions.

To serve this purpose we need to compute the following expressions in the eigenstate representation:
\bea 
&& \langle j|p(t_1)p(t_1)|j\rangle=\sum_{l}p_{jl}p_{lj}=-\frac{1}{4}\sum_{l}E_{jl}E_{lj}x_{jl}x_{lj}= \langle j|p(t_2)p(t_2)|j\rangle.\eea
Consequently, the desired canonical equal time thermal correlators can be computed as:
\bea 
 &&\label{ge2}\braket{p(t_{1})p(t_{1})}_{\beta} =\frac{\displaystyle \sum_{j,l}e^{-\beta E_j}~p_{jl}p_{lj}}{\displaystyle\left(1 + \sum_{j>0}^{} e^{-\beta E_{j}}\right)}=-\frac{1}{4}~\frac{\displaystyle \sum_{j,l}e^{-\beta E_j}~E_{jl}E_{lj}x_{jl}x_{lj}}{\displaystyle \left(1 + \sum_{j>0}^{} e^{-\beta E_{j}}\right)}=\braket{p(t_{2})p(t_{2})}_{\beta} .~~~~~~~~~~~\eea
Finally, the normalized first four-point OTOC can be expressed by the following simplified expression in the eigenstate representation, as given by:\\
\\
\begin{tcolorbox}[colframe=black,arc=0mm]
	\begin{align} \label{eq:acan4ptcorrxpfinal}
	\widetilde{C}^{(3)}(t_{1},t_{2}) &= \frac{C^{(3)}(t_{1},t_{2})}{\braket{p(t_{1})p(t_{1})}_{\beta} \braket{p(t_{2})p(t_{2})}_{\beta}}\nonumber\\
	 &=  \frac{4\displaystyle \left(1 + \sum_{j>0}^{} e^{-\beta E_{j}}\right)}{\displaystyle \sum_{j,m,k,l}e^{-\beta (E_j+E_m)}~E_{jl}E_{lj}E_{mk}E_{km}~x_{jl}x_{lj}x_{mk}x_{km}}~\nonumber\\
	 &~~~~~~~~\times\sum_{m,k,l,r}^{}\ e^{-\beta E_m}~E_{ml}\ E_{lk}\ E_{kr}\ E_{rm}~x_{ml}\ x_{lk}\ x_{kr}\ x_{rm} \nonumber \\
		& \hspace{1cm} ~~\times \sin \left[ \left( E_{r} - \frac{E_{m} + E_{k}}{2} \right) \left( t_{1} - t_{2} \right) \right] \sin \left[ \left( E_{l} - \frac{E_{m} + E_{k}}{2} \right) \left( t_{1} - t_{2} \right) \right].\nonumber\\
		&
	\end{align}
\end{tcolorbox}
\vspace{0.9cm}

\noindent

\newpage
\textcolor{Sepia}{\subsection{\sffamily Summary of Results}\label{ssec:sumeigenrep}}
\begin{tcolorbox}[fonttitle=\sffamily\Large,title=Eigenstate Representation for Micro-Canonical Correlators]
	
	\begin{align}
	y_{m}^{(1)}(t_{1},t_{2}) &= -i\ \sum_{k}^{}\ E_{km}\ x_{mk}\ x_{km}\ \cos \left( E_{km} \left( t_{1} - t_{2} \right) \right) \tag{\ref{eq:miccan2ptcorrxp}} ,\\
	y_{m}^{(2)}(t_{1},t_{2}) &= -2i\ \sum_{k}^{}\ x_{mk}\ x_{km}\  \text{sin} \left( E_{mk} (t_{1} - t_{2}) \right)  \tag{\ref{eq:miccan2ptcorrxx}}, \\
	y^{(3)}_{m}(t_{1},t_{2}) &= -\frac{i}{2} \sum_{k}^{} E_{mk}\ x_{mk}\ E_{km}\ x_{km}\ \text{sin} \left( E_{km} \left( t_{1} - t_{2} \right) \right) \tag{\ref{eq:miccan2ptcorrpp}}, \\
	c_{m}^{(1)}(t_{1},t_{2}) &= \frac{1}{4} \sum_{k,l,r}^{} x_{ml}\ x_{lk}\ x_{rm}\ x_{kr}\ \nonumber\\
	&~~~~~~~~~\times \biggl[  E_{rk} E_{lk} e^{i E_{rl} \left( t_{1} - t_{2} \right)}  + E_{mr} E_{ml} e^{- i E_{rl} \left( t_{1} - t_{2} \right)}  \nonumber \\
	& ~~~~~~~~~~~~~~~~~- E_{rk} E_{ml} e^{i \left( E_{rm} + E_{lk} \right) \left( t_{1} - t_{2} \right)}  -  E_{mr} E_{lk} e^{-i \left( E_{rm} + E_{lk} \right) \left( t_{1} - t_{2} \right)}  \biggr] \tag{\ref{eq:miccan4ptcorrxp}}, \\
	c_{m}^{(2)}(t_{1},t_{2}) 
	&= 4~\sum_{k,l,r}^{}\ x_{ml}\ x_{lk}\ x_{kr}\ x_{rm} \nonumber \\
	& ~~~\times \sin \left[ \left( E_{r} - \frac{E_{m} + E_{k}}{2} \right) \left( t_{1} - t_{2} \right) \right] \sin \left[ \left( E_{l} - \frac{E_{m} + E_{k}}{2} \right) \left( t_{1} - t_{2} \right) \right], \tag{\ref{eq:miccan4ptcorrxxb}}\\
	c_{m}^{(3)}(t_{1},t_{2}) 
	&= \frac{1}{4}~\sum_{k,l,r}^{}\ E_{ml}\ E_{lk}\ E_{kr}\ E_{rm}x_{ml}\ x_{lk}\ x_{kr}\ x_{rm} \nonumber \\
		& \hspace{1cm} ~~\times \sin \left[ \left( E_{r} - \frac{E_{m} + E_{k}}{2} \right) \left( t_{1} - t_{2} \right) \right] \sin \left[ \left( E_{l} - \frac{E_{m} + E_{k}}{2} \right) \left( t_{1} - t_{2} \right) \right] \tag{\ref{eq:miccan4ptcorrppb}}
	.\end{align}
	
\end{tcolorbox}

\begin{tcolorbox}[fonttitle=\sffamily\large,title=Eigenstate Representation for Canonical Correlators without normalization]
	\begin{align}
	Y^{(1)}(t_{1},t_{2}) &= - i~\frac{\displaystyle \sum_{m,k}^{}\ e^{-\beta E_{m}}\ E_{km}\ x_{mk}\ x_{km} \cos \left( E_{km} \ (t_{1}-t_{2}) \right)}{\displaystyle \left(1+\sum_{m>0}e^{-\beta E_m}\right)}\  \tag{\ref{eq:can2ptcorrxpfinal}}, \\
	Y^{(2)}(t_{1},t_{2}) &= - 2i~\frac{\displaystyle  \sum_{m,k}^{}\ e^{-\beta E_{m}}\ x_{mk}\ x_{km}\  \sin \left( E_{mk} \left( t_{1} - t_{2} \right) \right) }{\displaystyle\left(1+\sum_{m>0}e^{-\beta E_m}\right)}\ \tag{\ref{eq:can2ptcorrxxfinal}} ,\\
	Y^{(3)}(t_{1},t_{2}) &= -\frac{i}{2}~ \frac{\displaystyle \sum_{m,k}^{} e^{-\beta E_{m}} E_{mk}\ x_{mk}\ E_{km}\ x_{km} \sin \left( E_{km} \left( t_{1} - t_{2} \right) \right)}{\displaystyle\left(1+\sum_{m>0}e^{-\beta E_m}\right)} \tag{\ref{eq:can2ptcorrppfinal}} ,\\
	C^{(1)}(t_{1},t_{2}) &= \frac{1}{4~\displaystyle  \left(1+\sum_{m>0}e^{-\beta E_m}\right)} \ \sum_{m,k,l,r}^{} e^{-\beta E_{m}}~x_{ml}\ x_{lk}\ x_{rm}\ x_{kr} \nonumber \\
	& \times \biggl[ E_{rk} E_{lk} e^{i E_{rl} \left( t_{1} - t_{2} \right)}  +  E_{mr} E_{ml} e^{- i E_{rl} \left( t_{1} - t_{2} \right)} \nonumber \\
	& -  E_{rk} E_{ml} e^{i \left( E_{rm} + E_{lk} \right) \left( t_{1} - t_{2} \right)}  -  E_{mr} E_{lk} e^{-i \left( E_{rm} + E_{lk} \right) \left( t_{1} - t_{2} \right)}  \biggr]  \tag{\ref{eq:can4ptcorrxpfinal}},\\
	C^{(2)}(t_{1},t_{2}) &=  \frac{4}{\displaystyle \left(1+\sum_{m>0}e^{-\beta E_m}\right)}\ \ \sum_{m, k,l,r}^{}\ e^{-\beta E_{m}}~x_{ml} x_{lk} x_{kr} x_{rm} \nonumber \\
		& \hspace{1cm} \times  \sin \left[ \left( E_{r} - \frac{E_{m} + E_{k}}{2} \right) \left( t_{1} - t_{2} \right) \right] \sin \left[ \left( E_{l} - \frac{E_{m} + E_{k}}{2} \right) \left( t_{1} - t_{2} \right) \right] \tag{\ref{eq:can4ptcorrxxfinalb}}, \\
	C^{(3)}(t_{1},t_{2}) &= \frac{1}{4\displaystyle \left(1+\sum_{m>0}e^{-\beta E_m}\right)}~\sum_{m,k,l,r}^{}\ e^{-\beta E_m}~E_{ml}\ E_{lk}\ E_{kr}\ E_{rm}x_{ml}\ x_{lk}\ x_{kr}\ x_{rm} \nonumber \\
		& \hspace{1cm} ~~\times \sin \left[ \left( E_{r} - \frac{E_{m} + E_{k}}{2} \right) \left( t_{1} - t_{2} \right) \right] \sin \left[ \left( E_{l} - \frac{E_{m} + E_{k}}{2} \right) \left( t_{1} - t_{2} \right) \right] \tag{\ref{eq:can4ptcorrfinalb}} 
	.\end{align}
\end{tcolorbox}

\begin{tcolorbox}[fonttitle=\sffamily\large,title=Eigenstate Representation for Canonical Correlators with normalization]
	\begin{align}
	\braket{x(t_{1})x(t_{1})}_{\beta} &= \frac{\displaystyle \sum_{j,l}e^{-\beta E_j}~x_{jl}x_{lj}}{\displaystyle \left(1+\sum_{j>0}e^{-\beta E_j}\right)}=\braket{x(t_{2})x(t_{2})}_{\beta} \tag{\ref{ge1}}, \\
	\braket{p(t_{1})p(t_{1})}_{\beta} &=-\frac{1}{4}~\frac{\displaystyle \sum_{j,l}e^{-\beta E_j}~E_{jl}E_{lj}x_{jl}x_{lj}}{\displaystyle \left(1+\sum_{j>0}e^{-\beta E_j}\right)}=\braket{p(t_{2})p(t_{2})}_{\beta} \tag{\ref{ge2}} ,\\
	\widetilde{C}^{(1)}(t_{1},t_{2}) &= \frac{C^{(1)}(t_{1},t_{2})}{\braket{x(t_{1})x(t_{1})}_{\beta} \braket{p(t_{2})p(t_{2})}_{\beta}}\nonumber\\
	 &= -\frac{\displaystyle\sum_{m, k,l,r}^{}\ \left(1+\sum_{j>0}e^{-\beta E_j}\right)e^{-\beta E_{m}}~x_{ml} x_{lk} x_{kr} x_{rm}}{~\displaystyle  \sum_{j,m,l}^{}\ e^{-\beta (E_j+E_{m})}~E_{jl}E_{lj}x_{jl}x_{lj}} \ \nonumber\\
	& ~~~~~~~\times \biggl[ E_{rk} E_{lk} e^{i E_{rl} \left( t_{1} - t_{2} \right)}  +  E_{mr} E_{ml} e^{- i E_{rl} \left( t_{1} - t_{2} \right)} \nonumber \\
	& ~~~~~~~ -  E_{rk} E_{ml} e^{i \left( E_{rm} + E_{lk} \right) \left( t_{1} - t_{2} \right)}  -  E_{mr} E_{lk} e^{-i \left( E_{rm} + E_{lk} \right) \left( t_{1} - t_{2} \right)}  \biggr]   \tag{\ref{eq:dcan4ptcorrxpfinal}},\\
	\widetilde{C}^{(2)}(t_{1},t_{2}) &= \frac{C^{(2)}(t_{1},t_{2})}{\braket{x(t_{1})x(t_{1})}_{\beta} \braket{x(t_{2})x(t_{2})}_{\beta}}\nonumber\\
	 &= \frac{4\displaystyle \sum_{m, k,l,r}^{}\ \left(1+\sum_{j>0}e^{-\beta E_j}\right)e^{-\beta E_{m}}~x_{ml} x_{lk} x_{kr} x_{rm} }{\displaystyle \sum_{j,m,l,k}^{} e^{-\beta (E_j+E_m)}~x_{jl}x_{lj}x_{mk}x_{km}}\ \ \nonumber\\
		& \times  \sin \left[ \left( E_{r} - \frac{E_{m} + E_{k}}{2} \right) \left( t_{1} - t_{2} \right) \right] \sin \left[ \left( E_{l} - \frac{E_{m} + E_{k}}{2} \right) \left( t_{1} - t_{2} \right) \right] \tag{\ref{eq:ccan4ptcorrxpfinal}} ,\\
	\widetilde{C}^{(3)}(t_{1},t_{2}) &= \frac{C^{(3)}(t_{1},t_{2})}{\braket{p(t_{1})p(t_{1})}_{\beta} \braket{p(t_{2})p(t_{2})}_{\beta}}\nonumber\\
	 &=  \frac{4\displaystyle\sum_{m,k,l,r}^{}\ \left(1+\sum_{j>0}e^{-\beta E_j}\right)e^{-\beta E_m}~E_{ml}\ E_{lk}\ E_{kr}\ E_{rm}~x_{ml}\ x_{lk}\ x_{kr}\ x_{rm}}{\displaystyle \sum_{j,m,k,l}e^{-\beta (E_j+E_m)}~E_{jl}E_{lj}E_{mk}E_{km}~x_{jl}x_{lj}x_{mk}x_{km}}~\nonumber\\
		& \times \sin \left[ \left( E_{r} - \frac{E_{m} + E_{k}}{2} \right) \left( t_{1} - t_{2} \right) \right] \sin \left[ \left( E_{l} - \frac{E_{m} + E_{k}}{2} \right) \left( t_{1} - t_{2} \right) \right]. \tag{\ref{eq:acan4ptcorrxpfinal}}  
	\end{align}
\end{tcolorbox}
\newpage
\textcolor{Sepia}{\section{\sffamily Model~I:~Supersymmetric Quantum Mechanical Harmonic Oscillator}\label{sec:susyqmho}}

\textcolor{Sepia}{\subsection{\sffamily Eigen spectrum of the super-partner Hamiltonian}\label{ssec:eigenHO}}

We consider a system described by Hamiltonian ${H_{1}}$ with the harmonic oscillator potential as given by:
\bea V_{1}(x)= \frac{1}{2} \omega^{2} x^{2},\eea
where $\omega$ is the natural frequency of the oscillator and we have assumed that the mass, $m=1$, for the sake of algebraic simplification. Now, we represent $\Psi_{n}^{(1)}(x)$ and $E_{n}^{(1)}$ as the eigenstates and eigenvalues corresponding to Hamiltonian ${H_{1}}$ which can be obtained by solving the {\it Schr$\ddot{o}$dinger equation} as:
\begin{align}
\label{wavefunction1}
\Psi_{n}^{(1)}(x)&= \biggl(\frac{\omega}{\pi}\biggr)^{1/4} \frac{1}{\sqrt{2^{n}n!}} e^{-\frac{\omega}{2}x^{2}} H_{n}(\sqrt{\omega} x) \quad ;\\
\label{wavefunction2}
 \quad E_{n}^{(1)} &=\biggl(n+ \frac{1}{2}\biggr)\omega \quad.
\end{align}
where $H_{m}(y)$ appearing in the eigenfunctions are the well-known {\it Hermite polynomials} of order $m$. As discussed in \Cref{sec:revSUSYQM} the superpotential can be calculated once the ground state wavefunction is known. 
Hence, using the ground state wavefunction obtained by substituting ${n=0}$ in \Cref{wavefunction1}, the superpotential can be calculated using \Cref{superpotential} and for the case of supersymmetric one dimensional harmonic oscillator it is obtained to be the following:
\begin{align}
	W(x)= \frac{x}{\sqrt 2}.
\end{align}
Thus, we can construct the partner Hamiltonian which is of the following form:
\bea H_2= -\frac{1}{2} \frac{d^{2}}{d x^2} + V_2(x),\eea
where $V_2$ is the partner potential and can be computed using \Cref{partnerpotential}. For the supersymmetric one dimensional  harmonic oscillator we get the following expression for the partner potential:
\begin{align}
	V_2(x)= \left( 1+\frac{\omega^2 x^2}{2} \right).
\end{align}
More precisely, in the context of supersymmetric harmonic oscillator, the original Hamiltonian is usually identified as the {\it Bosonic Hamiltonian} and its associated super-partner is identified as the {\it Fermionic Hamiltonian}, which are constructed by subtracting off the ground state energy as:
\begin{align}
& \textcolor{red}{\bf Bosonic~Hamiltonian:}~~~~~~H_{\textnormal{B}} = H_1 - E_{0}^{(1)}, \\
&\textcolor{red}{\bf Fermionic~Hamiltonian:}~~~ H_{\textnormal{F}} = H_2 -E_{0}^{(1)}.
\end{align} 
Now since this is just a constant shift in the energy, the corresponding eigenstates of the partner Hamiltonians remain the same, and can be represented as:
\begin{align}
& \textcolor{red}{\bf Bosonic~eigenstate:}~~~~~~	\Psi_{n}^{(1)}(x)\equiv \Psi_{n}^{(\textnormal{B})}(x),\\
&\textcolor{red}{\bf Fermionic~eigenstate:}~~~\Psi_{n}^{(2)}(x) \equiv \Psi_{n}^{(\textnormal{F})}(x).
\end{align}
However, in this case the energy eigenvalues get shifted by an amount $E_{0}^{(1)}$ and can be described as:
\begin{align}
	& \textcolor{red}{\bf Bosonic~eigenspectrum:}~~~~~~~~~~~~~	E_{n}^{(B)}=n \omega,\\
&\textcolor{red}{\bf Fermionic~eigenspectrum:}~~~~~~~~~~E_{n}^{(F)}=(n+1)\omega,
\end{align} 
where we have used the property of supersymmetric quantum mechanical theories:
\bea E_{n}^{(\textnormal{F})}=E_{n+1}^{(\textnormal{B})}~~~~~~~~ \forall n=0,1,2,\dots. ~.\eea

\noindent
We can compute the energy eigenstates of the partner fermionic system by using the {\it Schr$\ddot{o}$dinger equation} for the Hamiltonian $H_{\textnormal{F}}$ which yields the same wave functions as for ${H_{\textnormal{B}}}$ as given in \Cref{wavefunction1} and \Cref{wavefunction2}.



\textcolor{Sepia}{\subsection{\sffamily Partition Function}\label{ssec:parsp}}

\noindent
The partition function in the eigenstate representation is defined by the following expression:
\begin{align}
Z=\text{Tr}[e^{-\beta H}]= {\sum_{n} \braket{ \Psi_n|e^{-\beta H}|\Psi_n }}.
\end{align}
In the framework of supersymmetric quantum mechanics as already discussed the ground state is only bosonic. The structure of the normalized eigenfunctions are also different for the ground state and the higher energy state. Hence, the partition function can be written in terms of the ground state contribution and in terms of the contribution coming from the excited state as:
\begin{align}
	Z=\underbrace{\braket{\Psi_{0}|\Psi_{0}}}_{\textcolor{red}{\bf Ground~State~Contribution}}+\underbrace{\frac{1}{2}\ {\sum_{n>0}[{\braket{ \Psi_{n}^{\textnormal{B}}|e^{-\beta H_{\textnormal{B}}}|\Psi_{n}^{\textnormal{B}} }} + {\braket{ \Psi_{n-1}^{\textnormal{F}}|e^{-\beta H_{\textnormal{F}}}|\Psi_{n-1}^{\textnormal{F}}}}]}}_{\textcolor{red}{\bf Excited~State~Contribution}}.
\end{align}
Further using the fact that the bosonic and the fermionic Hamiltonians ${n^{\textnormal{th}}}$ energy eigenfunction are exactly identical, the partition function for the supersymmetric one dimensional harmonic oscillator can be computed as:\\ \\
\begin{tcolorbox}[colframe=black,arc=0mm]
\begin{align}
Z= 1+\frac{1}{2}\sum_{n>0}^{} \left[ (e^{-\beta \omega n}+ e^{-\beta \omega n}) \right]
=\frac{1}{1-e^{-\beta\omega}}=\frac{1}{2}\ e^{\frac{\beta \omega}{2}} \textnormal{ cosech} \left(\frac{\beta \omega}{2}\right).
\end{align}
\end{tcolorbox}

\textcolor{Sepia}{\subsection{\sffamily Computation of two point OTOCs}\label{ssec:2pthosp}}

\textcolor{Sepia}{\subsubsection{\sffamily Computation of $Y^{(1)}(t_{1},t_{2})$}\label{ssec:2pthosp1}}
We compute the 2-pt correlators for the Supersymmetric Harmonic oscillator. We begin with the correlator constructed from two different dynamical operators which is defined as: 

\begin{align}
Y^{(1)}(t_{1},t_{2}) &= - \braket{[x(t_{1}),p(t_{2})]}_{\beta} \nonumber\\
&= - \frac{1}{Z} \sum_{n}^{} e^{-\beta E_{n}} \braket{\Psi_{n} | [x(t_{1}),p(t_{2})] | \Psi_{n}} \nonumber\\
		     &= - \frac{1}{Z} \biggl[ \underbrace{\braket{\Psi_{0} | [x(t_{1}),p(t_{2})] | \Psi_{0}}  }_{\textnormal{\sffamily \textcolor{red}{\bf Ground State Contribution}}} + \underbrace{\sum_{n>0}^{} e^{-\beta E_{n}} \braket{\Psi_{n} | [x(t_{1}),p(t_{2})] | \Psi_{n}}}_\text{\sffamily \textcolor{red}{\bf Excited States Contribution}} \biggr] \nonumber\\
		     &=2e^{-\frac{\beta\omega}{2}}\sinh \left(\frac{\beta\omega}{2}\right)~\biggl(y_0^{(1)}(t_{1},t_{2})+\sum_{n>0}e^{-\beta\omega n}~y_n^{(1)}(t_{1},t_{2})\biggr) 
.\end{align}
\noindent
where $y_0^{(1)}(t_{1},t_{2})$ and $y_n^{(1)}(t_{1},t_{2})$ represent the micro-canonical correlators for the ground state and the excited states and the overall temperature dependent normalization function is appearing from the expression for the inverse of the thermal partition function for canonical quantum statistical ensemble. The purpose of taking the ground state and excited states separately is due to the fact that in supersymmetric quantum mechanics the ground state is always bosonic having zero energy eigenvalue, whereas for the excited states there are contributions from the bosonic as well as the fermionic states.
It is easy to see that the ground state contribution can be calculated by expanding the commutator and inserting the completeness relation, $\displaystyle\sum_k \ket{\Psi_k}\bra{\Psi_k}=1$. However, here it is important to note that there is no contribution appearing from the ${k=0}$ term as for the ground state we have, $\braket{\Psi_0|x(t_1)|\Psi_0}=0$. Hence, from the ground state we finally get the following time dependent non-trivial contribution, which is given by:
\begin{align}
\braket{\Psi_{0} | [x(t_{1}),p(t_{2})] | \Psi_{0}}= \frac{i}{2} \cos \omega(t_1-t_2).
\end{align}

Now further to determine the contribution from the excited states we need to expand the commutator and insert the completeness relation between the desired time dependent quantum operators which gives the following expression:
\begin{align}
\braket{\Psi_{n} | [x(t_{1}),p(t_{2})] | \Psi_{n}} &= \frac{i}{2} \cos\omega(t_{1}-t_{2})\ \delta_{n1} \nonumber \\
						   &~~~~~~+ \frac{i}{2}\cos\omega(t_{1}-t_{2})\bigl(1+\sqrt{n(n+1)}-\sqrt{n(n-1)}\bigr).
\end{align}
\noindent
Therefore, the micro-canonical correlators for the ground state and the excited states are given by the following temperature independent simplified expressions:
\begin{align}
\label{gsmicrocorr}
	y_0^{(1)} &= -\frac{i}{2} \cos\omega(t_{1}-t_{2}), \\ 
	\label{hsmicrocorr}
	y_n^{(1)} &= -\frac{i}{2} \cos\omega(t_{1}-t_{2}) \delta_{n1}- \frac{i}{2}\cos\omega(t_{1}-t_{2})\bigl(1+\sqrt{n(n+1)}-\sqrt{n(n-1)}\bigr)\nonumber\\
	&= -\frac{i}{2} \cos\omega(t_{1}-t_{2})\left[1+\delta_{n1}+\sqrt{n(n+1)}-\sqrt{n(n-1)}\right].
\end{align}
Further, using \Cref{gsmicrocorr} and \Cref{hsmicrocorr} the canonical two point thermal correlator can be obtained by the following expression:
\begin{align}
\nonumber
Y^{(1)}(t_1,t_2)&=-ie^{-\frac{\beta\omega}{2}}\sinh \left(\frac{\beta\omega}{2}\right)~\cos\omega(t_{1}-t_{2})~\nonumber\\
&~~~~~~~\times\biggl\{1+e^{-\beta \omega}  +\sum_{n>0}^{}e^{-\beta \omega n} \biggl(1+\sqrt{n(n+1)}-\sqrt{n(n-1)}\biggr) \biggr\}\nonumber\\
&=-\frac{i}{2}\tanh \left(\frac{\beta\omega}{2}\right)~\cos\omega(t_{1}-t_{2})~\nonumber\\
&\displaystyle~~~~~~~\times\biggl\{1+2~\cosh \left(\frac{\beta\omega}{2}\right)\sum_{n>0}^{}e^{-\beta\omega \left(n+\frac{1}{2}\right)} \biggl(1+\sqrt{n(n+1)}-\sqrt{n(n-1)}\biggr) \biggr\} .
\end{align}
\noindent

It can be seen that the obtained OTOC that it is just a periodic function indicating the absence of chaos. However, it is quite interesting that the obtained result has an explicit temperature dependence which mainly comes from the contribution of the ground state. Though the two point OTOC is not of much significance. It stills give us a pretty good idea that at the level two point correlation function is that in the context of supersymmetric one dimensional harmonic oscillator is non-chaotic in nature. 
\\

\begin{tcolorbox}[colframe=black,arc=0mm,top=5mm]
	
	\textcolor{MidnightBlue}{\textbf{Micro-Canonical 2-pt Correlator for Ground State }}:
\begin{align}
	y^{(1)}_{0}(t_{1},t_{2}) &= -\frac{i}{2}\cos\omega(t_1-t_2).
\end{align}

	\textcolor{MidnightBlue}{\textbf{Micro-Canonical 2-pt Correlator for Excited States }}: 
\begin{align}
	y^{(1)}_{n}(t_{1},t_{2}) &=-\frac{i}{2} \cos\omega(t_{1}-t_{2})\left[1+\delta_{n1}+\sqrt{n(n+1)}-\sqrt{n(n-1)}\right].
\end{align}

\end{tcolorbox}

\begin{tcolorbox}[colframe=black,arc=0mm,top=5mm]

\textcolor{MidnightBlue}{\textbf{Canonical 2-pt Correlator }}:
\begin{align}\label{x1}
	Y^{(1)}(t_{1},t_{2}) 
&=-\frac{i}{2}\tanh \left(\frac{\beta\omega}{2}\right)~\cos\omega(t_{1}-t_{2})~\nonumber\\
&\displaystyle~~~~~~~\times\biggl\{1+2~\cosh \left(\frac{\beta\omega}{2}\right)\sum_{n>0}^{}e^{-\beta\omega \left(n+\frac{1}{2}\right)} \biggl(1+\sqrt{n(n+1)}-\sqrt{n(n-1)}\biggr) \biggr\} .
\end{align}

\end{tcolorbox}
\vspace{0.5cm}
As discussed earlier the correlator $Y^{(1)}(t_1,t_2)$ alone does not provide the complete knowledge about the quantum randomness of the system. Hence, we need to calculate two other correlators constructed from similar operators at different time scales. The other two correlators will be : $-\braket{[x(t_1),x(t_2)]}_{\beta}$ and $-\braket{[p(t_1),p(t_2)]}_{\beta}$ which we will compute in the next part of the paper.

\textcolor{Sepia}{\subsubsection{\sffamily Computation of $Y^{(2)}(t_{1},t_{2})$}\label{ssec:2pthosp2}}

The second correlator is constructed from the commutator of the two position operators defined at different time scales and can be represented as:
\begin{align}
Y^{(2)}(t_{1},t_{2}) &= - \braket{[x(t_{1}),x(t_{2})]}_{\beta}\nonumber\\
&= - \frac{1}{Z} \biggl[ \underbrace{\braket{\Psi_{0} | [x(t_{1}),x(t_{2})] | \Psi_{0}}  }_{\textnormal{\sffamily \textcolor{red}{\bf Ground State Contribution}}} + \underbrace{\sum_{n>0}^{} e^{-\beta E_{n}} \braket{\Psi_{n} | [x(t_{1}),x(t_{2})] | \Psi_{n}}}_\text{\sffamily \textcolor{red}{\bf Excited States Contribution}} \biggr] \nonumber\\
&= 2e^{-\frac{\beta\omega}{2}}\sinh \left(\frac{\beta\omega}{2}\right)~ \biggl(y_0^{(2)}(t_{1},t_{2})+\sum_{n>0}e^{-\beta\omega n}~y_n^{(2)}(t_{1},t_{2})\biggr) 
.\end{align}
\noindent
The reason for separately considering the ground state from the other states has already been discussed in the calculation of $Y^{(1)}(t_1,t_2)$. Following a similar procedure we get:
\begin{align}
\label{gsmicrocorr2}
y_0^{(2)}(t_1,t_2)&=-\braket{\Psi_{0} | [x(t_{1}),x(t_{2})] | \Psi_{0}} \nonumber\\
&= \frac{i}{2\omega}\sin\omega(t_{1}-t_{2}) ,\\
\label{hsmicrocorr2}
y_n^{(2)}(t_1,t_2)&=-\braket{\Psi_{n} | [x(t_{1}),x(t_{2})] | \Psi_{n}}\nonumber\\
&=- \frac{i}{2\omega}\sin\omega(t_{1}-t_{2}) \delta_{n1} \nonumber \\
						   &-\frac{i}{2\omega}\sin\omega(t_{1}-t_{2})\biggl[1+\sqrt{n(n+1)}-\sqrt{n(n-1)}\biggr]\nonumber\\
						   &=-\frac{i}{2\omega}\sin\omega(t_{1}-t_{2})\biggl[1+\delta_{n1} +\sqrt{n(n+1)}-\sqrt{n(n-1)}\biggr].
\end{align} 
Using \Cref{gsmicrocorr2} and \Cref{hsmicrocorr2} the canonical correlator can be obtained as:
\begin{align}
\nonumber
Y^{(2)}(t_1,t_2)&=\frac{i}{\omega}e^{-\frac{\beta\omega}{2}}\sinh \left(\frac{\beta\omega}{2}\right)~\sin\omega(t_{1}-t_{2})\nonumber\\
&~~~~~~\times\biggl\{1-e^{-\beta \omega}-\sum_{n>0}^{}e^{-\beta \omega n} \biggl(1+\sqrt{n(n+1)}-\sqrt{n(n-1)}\biggr) \biggr\} \nonumber\\
&=\frac{2i}{\omega}e^{-\beta\omega}\sinh^2 \left(\frac{\beta\omega}{2}\right)~\sin\omega(t_{1}-t_{2})\nonumber\\
&~~~~~~\times\biggl\{1-\frac{1}{2}\textnormal{ cosech} \left(\frac{\beta \omega}{2}\right)\sum_{n>0}^{}e^{-\beta\omega \left(n-\frac{1}{2}\right)} \biggl(1+\sqrt{n(n+1)}-\sqrt{n(n-1)}\biggr) \biggr\}\nonumber\\
&=\frac{i}{\omega}e^{-\beta\omega}\sinh \left(\beta\omega\right)\tanh \left(\frac{\beta\omega}{2}\right)~\sin\omega(t_{1}-t_{2})\nonumber\\
&~~~~~~\times\biggl\{1-\frac{1}{2}\textnormal{ cosech} \left(\frac{\beta \omega}{2}\right)\sum_{n>0}^{}e^{-\beta\omega \left(n-\frac{1}{2}\right)} \biggl(1+\sqrt{n(n+1)}-\sqrt{n(n-1)}\biggr) \biggr\}.
\end{align}
\noindent

\begin{tcolorbox}[colframe=black,arc=0mm,top=5mm]
	\textcolor{MidnightBlue}{\textbf{Micro-Canonical 2-pt Correlator for Ground State }}:
	\begin{align}
	y^{(2)}_{0}(t_{1},t_{2}) &= \frac{i}{2\omega}\sin\omega(t_{1}-t_{2})=-\omega^{-2}~y^{(3)}_{n}(t_{1},t_{2}).
	\end{align}

	\textcolor{MidnightBlue}{\textbf{Micro-Canonical 2-pt Correlator for Excited States }}:
	\begin{align}
	y^{(2)}_{n}(t_{1},t_{2}) &=-\frac{i}{2\omega}\sin\omega(t_{1}-t_{2})\biggl[1+\delta_{n1} +\sqrt{n(n+1)}-\sqrt{n(n-1)}\biggr]=-\frac{y^{(3)}_{n}(t_{1},t_{2})}{\omega^2}.
	\end{align}
\end{tcolorbox}

\begin{tcolorbox}[colframe=black,arc=0mm,top=5mm]
	\textcolor{MidnightBlue}{\textbf{Canonical 2-pt Correlator }}:
	\begin{align}\label{x2}
	Y^{(2)}(t_{1},t_{2}) &=\frac{i}{\omega}e^{-\beta\omega}\sinh \left(\beta\omega\right)\tanh \left(\frac{\beta\omega}{2}\right)~\sin\omega(t_{1}-t_{2})\nonumber\\
&~~~\times\biggl\{1-\frac{1}{2}\textnormal{ cosech} \left(\frac{\beta \omega}{2}\right)\sum_{n>0}^{}e^{-\beta\omega \left(n-\frac{1}{2}\right)} \biggl(1+\sqrt{n(n+1)}-\sqrt{n(n-1)}\biggr) \biggr\}\nonumber\\
&=-\frac{Y^{(3)}(t_{1},t_{2})}{\omega^2}.
\end{align}
\end{tcolorbox} 
\vspace{0.5cm}

\textcolor{Sepia}{\subsubsection{\sffamily Computation of $Y^{(3)}(t_{1},t_{2})$}\label{ssec:2pthosp3}} 
The third correlator is constructed from the commutator of the two momentum operators defined at different time scales and can be represented as:
\begin{align}
Y^{(3)}(t_{1},t_{2}) &= - \braket{[p(t_{1}),p(t_{2})]}_{\beta} \nonumber\\
&= - \frac{1}{Z} \biggl[ \underbrace{\braket{\Psi_{0} | [p(t_{1}),p(t_{2})] | \Psi_{0}}  }_{\textnormal{\sffamily \textcolor{red}{\bf Ground State Contribution}}} + \underbrace{\sum_{n>0}^{} e^{-\beta E_{n}} \braket{\Psi_{n} | [p(t_{1}),p(t_{2})] | \Psi_{n}}}_\text{\sffamily \textcolor{red}{\bf Excited States Contribution}} \biggr] \nonumber\\
&=  2e^{-\frac{\beta\omega}{2}}\sinh \left(\frac{\beta\omega}{2}\right)~  \biggl(y_0^{(3)}(t_{1},t_{2})+\sum_{n>0}e^{-\beta\omega n}~y_n^{(3)}(t_{1},t_{2})\biggr) 
.\end{align}
\noindent
The reason for separately considering the ground state from the other states has already been discussed in the calculation of $Y^{(1)}(t_1,t_2)$. Following a similar procedure we get the following expressions for the micro-canonical correlators for the ground state and excited states as given by:
\begin{align}
\label{gsmicrocorr3}
y_0^{(3)}(t_1,t_2)&=-\braket{\Psi_{0} | [p(t_{1}),p(t_{2})] | \Psi_{0}} \nonumber\\
&=- \frac{i\omega}{2}\sin\omega(t_{1}-t_{2}),\\
\label{hsmicrocorr3}
y_n^{(3)}(t_1,t_2)&=-\braket{\Psi_{n} | [p(t_{1}),p(t_{2})] | \Psi_{n}} \nonumber\\
&= \frac{i\omega}{2}\sin\omega(t_{1}-t_{2})\ \delta_{n1}\nonumber \\
						   &~~~~~~~~+\frac{i\omega}{2}\sin\omega(t_{1}-t_{2})\biggl[1+\sqrt{n(n+1)}-\sqrt{n(n-1)}\biggr]\nonumber\\
						   &= \frac{i\omega}{2}\sin\omega(t_{1}-t_{2})\biggl[1+\delta_{n1}+\sqrt{n(n+1)}-\sqrt{n(n-1)}\biggr].
\end{align}
Using \Cref{gsmicrocorr3} and \Cref{hsmicrocorr3} the canonical correlator can be obtained as:
\begin{align}
\nonumber
Y^{(3)}(t_1,t_2)&=-i\omega~e^{-\frac{\beta\omega}{2}}\sinh \left(\frac{\beta\omega}{2}\right)~\sin\omega(t_{1}-t_{2})\nonumber\\
&~~~~~~~~\times\biggl\{1-e^{-\beta \omega} -\sum_{n>0}^{}e^{-\beta \omega n}\biggl(1+\sqrt{n(n+1)}-\sqrt{n(n-1)}\biggr) \biggr\}\nonumber\\
&=-2i\omega~e^{-\beta\omega}\sinh^2 \left(\frac{\beta\omega}{2}\right)~\sin\omega(t_{1}-t_{2})\nonumber\\
&~~~~~~~~\times\biggl\{1 -\frac{1}{2}\textnormal{ cosech} \left(\frac{\beta \omega}{2}\right)\sum_{n>0}^{}e^{-\beta\omega \left(n-\frac{1}{2}\right)}\biggl(1+\sqrt{n(n+1)}-\sqrt{n(n-1)}\biggr) \biggr\}\nonumber\\
&=-i\omega~e^{-\beta\omega}\sinh \left(\beta\omega\right)\tanh \left(\frac{\beta\omega}{2}\right)~\sin\omega(t_{1}-t_{2})\nonumber\\
&~~~\times\biggl\{1-\frac{1}{2}\textnormal{ cosech} \left(\frac{\beta \omega}{2}\right)\sum_{n>0}^{}e^{-\beta\omega \left(n-\frac{1}{2}\right)} \biggl(1+\sqrt{n(n+1)}-\sqrt{n(n-1)}\biggr) \biggr\}.
\end{align}
\noindent

\begin{tcolorbox}[colframe=black,arc=0mm,top=5mm]
	\textcolor{MidnightBlue}{\textbf{Micro-Canonical 2-pt Correlator for Ground State }}:
	\begin{align}
	y^{(3)}_{0}(t_{1},t_{2}) &= - \frac{i\omega}{2}\sin\omega(t_{1}-t_{2})=-\omega^2~y^{(2)}_{0}(t_{1},t_{2}),
	\end{align}
	\textcolor{MidnightBlue}{\textbf{Micro-Canonical 2-pt Correlator for Excited States }}:
	\begin{align}
	y^{(3)}_{n}(t_{1},t_{2}) &= \frac{i\omega}{2}\sin\omega(t_{1}-t_{2})\biggl[1+\delta_{n1}+\sqrt{n(n+1)}-\sqrt{n(n-1)}\biggr]\nonumber\\
	&=-\omega^2~y^{(2)}_{n}(t_{1},t_{2}).
	\end{align}
\end{tcolorbox}

\begin{tcolorbox}[colframe=black,arc=0mm,top=5mm]
	\textcolor{MidnightBlue}{\textbf{Canonical 2-pt Correlator }}:
	\begin{align}\label{x3}
	Y^{(3)}(t_{1},t_{2}) &= -i\omega~e^{-\beta\omega}\sinh \left(\beta\omega\right)\tanh \left(\frac{\beta\omega}{2}\right)~\sin\omega(t_{1}-t_{2})\nonumber\\
&~~~\times\biggl\{1-\frac{1}{2}\textnormal{ cosech} \left(\frac{\beta \omega}{2}\right)\sum_{n>0}^{}e^{-\beta\omega \left(n-\frac{1}{2}\right)} \biggl(1+\sqrt{n(n+1)}-\sqrt{n(n-1)}\biggr) \biggr\}\nonumber\\
&=-\omega^2~Y^{(2)}(t_{1},t_{2}).
	\end{align}
\end{tcolorbox}
\vspace{0.5cm}
\newpage

\textcolor{Sepia}{\subsection{\sffamily Computation of un-normalized four point OTOCs}\label{sssec:4pthosp}}

Though the building block two point OTOCs gives us some basic insight about the nature of randomness present in the quantum system, it is actually the four point OTOCs that are considered to be the prime observables for determining the degree and the nature of randomness present in the quantum mechanical systems. Similar to the two point case we consider here different types of OTOCs for the four point case from all possible three combinations of the dynamical operators characterising the quantum mechanical system under consideration, here it is one dimensional supersymmetric harmonic oscillator.

\textcolor{Sepia}{\subsubsection{\sffamily Computation of $C^{(1)}(t_{1},t_{2})$}\label{sssec:4pthosp1}} 
The corresponding four point desired OTOC of first kind is defined by the following expression:
\begin{align} 
\label{cor4pt1}
\nonumber
C^{(1)}(t_1,t_2) &=- \braket{ [x(t_1),p(t_2)]^{2}}_{\beta}\nonumber\\
&=- \frac{1}{Z}\sum_{n}\langle\Psi_n|e^{-\beta H}[x(t_1),p(t_2)]^{2} |\Psi_n\rangle \nonumber\\
&=  - \frac{1}{Z}\biggl\{\underbrace{\braket{\Psi_0|[x(t_1),p(t_2)]^{2} |\Psi_0}}_{\textcolor{red}{\bf Ground~State~Contribution}}+ \underbrace{\sum_{n>0}^{}e^{-\beta E_n}\braket{\Psi_n|[x(t_1),p(t_2)]^{2} |\Psi_n}}_{\textcolor{red}{\bf Excited~State~Contribution}}\biggr\} \nonumber\\
&=2e^{-\frac{\beta\omega}{2}}\sinh \left(\frac{\beta\omega}{2}\right)~  \left[c_{0}^{(1)}(t_1,t_2)+\sum_{n>0} e^{-\beta \omega n}c_{n}^{(1)}(t_1,t_2)\right],
\end{align}
where the ground and the excited contribution of the micro-canonical part of the four point temperature independent OTOCs are defined by the following expressions:
\begin{align}
\nonumber
\label{cn1}
c_{0}^{(1)}(t_1,t_2) &=-\braket{\Psi_0|[x(t_1),p(t_2)]^{2} |\Psi_0} \\   
&=\sum_{m}b_{0 m}^{(1)}(t_1,t_2)b_{0 m}^{(1)*}(t_1,t_2)\nonumber\\
&= b_{0 0}^{(1)}(t_1,t_2)b_{0 0}^{(1)*}(t_1,t_2)+\sum_{m>0}b_{0 m}^{(1)}(t_1,t_2)b_{0 m}^{(1)*}(t_1,t_2),
\end{align}
and
\begin{align}
\nonumber
\label{cn1}
c_{n}^{(1)}(t_1,t_2) &= -\braket{\Psi_n|[x(t_1),p(t_2)]^{2} |\Psi_n} \\   
&=\sum_{m}b_{n m}^{(1)}(t_1,t_2)b_{n m}^{(1)*}(t_1,t_2)\nonumber\\
&= b_{n 0}^{(1)}(t_1,t_2)b_{n 0}^{(1)*}(t_1,t_2)+\sum_{m>0}b_{n m}^{(1)}(t_1,t_2)b_{n m}^{(1)*}(t_1,t_2),
\end{align}
where  $b_{nm}^{(1)}(t_1,t_2)$, $	b_{0 m}^{(1)}(t_1,t_2)$, $	b_{m 0}^{(1)}(t_1,t_2)$ and $	b_{0 0}^{(1)}(t_1,t_2)$ used in the above expression are given by: 
\begin{align}
\label{55}
&b_{n m}^{(1)}(t_1,t_2)= -i \braket{\Psi_n|[x(t_1),p(t_2)] |\Psi_m},   \\
&b_{0 m}^{(1)}(t_1,t_2)= -i \braket{\Psi_0|[x(t_1),p(t_2)] |\Psi_m},  \\
&b_{m 0}^{(1)}(t_1,t_2)= -i \braket{\Psi_m|[x(t_1),p(t_2)] |\Psi_0},  \\
&b_{0 0}^{(1)}(t_1,t_2)= -i \braket{\Psi_0|[x(t_1),p(t_2)] |\Psi_0}.
\end{align}
Now an extremely important fact to keep in mind while performing this calculation is that the ground state is always bosonic, this enforces considering the ground state separately each time an identity is inserted. 
On expanding the commutator and inserting the identity operator ($\displaystyle\mathbb{I}:=\sum_{k} \ket{\Psi_k} \bra{\Psi_k}$) and considering the ${k=0}$ term separately from the ${k>0}$ term above sets of equations can be written as:
\begin{align}
\nonumber
	b_{n m}^{(1)}(t_1,t_2) &=-i\biggl(\braket{\Psi_n|x(t_1)\ket{\Psi_{0}}\bra{\Psi_{0}}p(t_2) |\Psi_m}- \braket{\Psi_n|p(t_2)\ket{\Psi_{0}}\bra{\Psi_{0}}x(t_1) |\Psi_m}\\ 
	&~~~+\sum_{k>0}^{} \braket{\Psi_n|x(t_1)\ket{\Psi_{k}}\bra{\Psi_{k}}p(t_2) |\Psi_m}- \braket{\Psi_n|p(t_2)\ket{\Psi_{k}}\bra{\Psi_{k}}x(t_1) |\Psi_m}\biggr),~~~~~~~~
\end{align}
and a similar expression will be obtained for $	b_{0 m}^{(1)}(t_1,t_2)$, $	b_{m 0}^{(1)}(t_1,t_2)$ and $	b_{0 0}^{(1)}(t_1,t_2)$. Keeping all the above discussed facts it is not difficult to show that for supersymmetric one dimensional harmonic oscillator we get:
\begin{align}
\label{mic4pt1}
b_{n m}^{(1)}(t_1,t_2) &=-\frac{i}{2} \cos\omega (t_{1}-t_{2})\biggl[ \delta_{n1}+\left(1+ \sqrt{n(n+1)} - \sqrt{n(n-1)} \right)\delta_{n m}\biggr], \\
\label{mic4pt2ndpart}
b_{0 m}^{(1)}(t_1,t_2)&=-\frac{i}{2} \cos\omega (t_{1}-t_{2}) \delta_{m0}=-\frac{i}{2} \cos\omega (t_{1}-t_{2}) \delta_{0m}=b_{m0}^{(1)}(t_1,t_2),\\
\label{mic4pt2ndpart00}
b_{0 0}^{(1)}(t_1,t_2)&=-\frac{i}{2} \cos\omega (t_{1}-t_{2}).
\end{align}
Here one can explicitly show that for the supersymmetric one dimensional harmonic oscillator case the following contributions are trivially vanish:
\bea &\displaystyle \sum_{m>0}b_{0 m}^{(1)}(t_1,t_2)b_{0 m}^{(1)*}(t_1,t_2)=0,~~~~\& ~~~~~
 b_{n 0}^{(1)}(t_1,t_2)b_{n 0}^{(1)*}(t_1,t_2)=0~~ \forall~~ n>0.\eea
The above equations can be used to calculate the micro-canonical correlator which is the temperature independent part of the total OTOC. The temperature dependent part comes from the canonical part of the correlator which can be calculated by substituting \Cref{mic4pt1} and \Cref{mic4pt2ndpart} in \Cref{cn1} and for the supersymmetric one dimensional harmonic oscillator we get: 
\begin{align}
\label{cn1resultw1}
c_{0}^{(1)}(t_1,t_2)&=\frac{1}{4}\cos^{2}\omega (t_{1}-t_{2}), \\
\label{cn1resultw2}
c_{n}^{(1)}(t_1,t_2)&= \frac{1}{4}\biggl\{\delta_{n1}\delta_{n1}  + \left(1+ \sqrt{n(n+1)} - \sqrt{n(n-1)} \right)^{2}\biggr\}\cos^{2}\omega (t_{1}-t_{2}).
\end{align}
Hence the temperature dependent canonical part of the four point OTOC of the first kind for a supersymmetric harmonic oscillator is given by substituting \Cref{cn1resultw1} and \Cref{cn1resultw2} in \Cref{cor4pt1} to obtain the following simplified result:
\begin{align}
\label{corr1stkind}
C^{(1)}(t_1,t_2) &= \frac{1}{2}e^{-\frac{\beta\omega}{2}}\sinh \left(\frac{\beta\omega}{2}\right)~\cos^{2}\omega (t_{1}-t_{2}) \nonumber \\
		 & ~~~~~~~~~~~~\times \biggl\{1+ e^{-\beta \omega} +\sum_{n>0}^{} e^{-\beta \omega n} \left(1+\sqrt{n(n+1)}-\sqrt{n(n-1)} \right)^{2}\biggr\}\nonumber\\
&= \frac{1}{2}e^{-\beta\omega}\sinh \left(\beta\omega\right)~\cos^{2}\omega (t_{1}-t_{2}) \nonumber \\
		 & ~~~~~~~~~~~~\times \biggl\{1+\frac{1}{2}{\rm sech}\left(\frac{\beta\omega}{2}\right)\sum_{n>0}^{} e^{-\beta \omega n} \left(1+\sqrt{n(n+1)}-\sqrt{n(n-1)} \right)^{2}\biggr\}.		 
\end{align}

\begin{tcolorbox}[colframe=black,arc=0mm,top=5mm]
	 \textcolor{MidnightBlue}{\textbf{Micro-Canonical 4-pt Correlator for Ground State }}:
	\begin{align}
	c^{(1)}_{0}(t_{1},t_{2}) &=\frac{1}{4}\cos^{2}\omega (t_{1}-t_{2}),
	\end{align}
	 \textcolor{MidnightBlue}{\textbf{Micro-Canonical 4-pt Correlator for Excited States }}:
	 \begin{align}
	c^{(1)}_{n}(t_{1},t_{2}) &= \frac{1}{4}\cos^{2}\omega (t_{1}-t_{2}) \biggl\{\delta_{n1}\delta_{n1}  + \left(1+ \sqrt{n(n+1)} - \sqrt{n(n-1)} \right)^{2}\biggr\}.
	\end{align}
\end{tcolorbox}
\vspace{0.5cm}

\begin{tcolorbox}[colframe=black,arc=0mm,top=5mm]
	\textcolor{MidnightBlue}{\textbf{Canonical 4-pt Correlator }}:
	\begin{align}\label{r1}
		C^{(1)}(t_{1},t_{2}) &= \frac{1}{2}e^{-\beta\omega}\sinh \left(\beta\omega\right)~\cos^{2}\omega (t_{1}-t_{2}) \nonumber \\
		 & ~~~~~~~~~~~~\times \biggl\{1+\frac{1}{2}{\rm sech}\left(\frac{\beta\omega}{2}\right)\sum_{n>0}^{} e^{-\beta \omega n} \left(1+\sqrt{n(n+1)}-\sqrt{n(n-1)} \right)^{2}\biggr\}.
	\end{align}
\end{tcolorbox}
\vspace{0.5cm}
\textcolor{Sepia}{\subsubsection{\sffamily Computation of $C^{(2)}(t_{1},t_{2})$}\label{sssec:4pthosp2}} 

The corresponding four point desired OTOC of second kind is defined by the following expression:
\begin{align} 
\label{cor4pt1v}
\nonumber
C^{(2)}(t_1,t_2) &=- \braket{ [x(t_1),x(t_2)]^{2}}_{\beta}\nonumber\\
&=- \frac{1}{Z}\sum_{n}\langle\Psi_n|e^{-\beta H}[x(t_1),x(t_2)]^{2} |\Psi_n\rangle \nonumber\\
&=  - \frac{1}{Z}\biggl\{\underbrace{\braket{\Psi_0|[x(t_1),x(t_2)]^{2} |\Psi_0}}_{\textcolor{red}{\bf Ground~State~Contribution}}+ \underbrace{\sum_{n>0}^{}e^{-\beta E_n}\braket{\Psi_n|[x(t_1),x(t_2)]^{2} |\Psi_n}}_{\textcolor{red}{\bf Excited~State~Contribution}}\biggr\} \nonumber\\
&=2e^{-\frac{\beta\omega}{2}}\sinh \left(\frac{\beta\omega}{2}\right)~  \left[c_{0}^{(2)}(t_1,t_2)+\sum_{n>0} e^{-\beta \omega n}c_{n}^{(2)}(t_1,t_2)\right],
\end{align}
where the ground and the excited contribution of the micro-canonical part of the four point temperature independent OTOCs are defined by the following expressions:
\begin{align}
\nonumber
\label{cn1}
c_{0}^{(2)}(t_1,t_2) &=-\braket{\Psi_0|[x(t_1),x(t_2)]^{2} |\Psi_0} \\   
&=\sum_{m}b_{0 m}^{(2)}(t_1,t_2)b_{0 m}^{(2)*}(t_1,t_2)\nonumber\\
&= b_{0 0}^{(2)}(t_1,t_2)b_{0 0}^{(2)*}(t_1,t_2)+\sum_{m>0}b_{0 m}^{(2)}(t_1,t_2)b_{0 m}^{(2)*}(t_1,t_2),
\end{align}
and
\begin{align}
\nonumber
\label{cn1}
c_{n}^{(2)}(t_1,t_2) &= -\braket{\Psi_n|[x(t_1),x(t_2)]^{2} |\Psi_n} \\   
&=\sum_{m}b_{n m}^{(2)}(t_1,t_2)b_{n m}^{(2)*}(t_1,t_2)\nonumber\\
&= b_{n 0}^{(2)}(t_1,t_2)b_{n 0}^{(2)*}(t_1,t_2)+\sum_{m>0}b_{n m}^{(2)}(t_1,t_2)b_{n m}^{(2)*}(t_1,t_2),
\end{align}
where  $b_{nm}^{(2)}(t_1,t_2)$, $	b_{0 m}^{(2)}(t_1,t_2)$, $	b_{m 0}^{(2)}(t_1,t_2)$ and $	b_{0 0}^{(2)}(t_1,t_2)$ used in the above expression are given by: 
\begin{align}
\label{55}
&b_{n m}^{(2)}(t_1,t_2)= -i \braket{\Psi_n|[x(t_1),x(t_2)] |\Psi_m},   \\
&b_{0 m}^{(2)}(t_1,t_2)= -i \braket{\Psi_0|[x(t_1),x(t_2)] |\Psi_m},  \\
&b_{m 0}^{(2)}(t_1,t_2)= -i \braket{\Psi_m|[x(t_1),x(t_2)] |\Psi_0},  \\
&b_{0 0}^{(2)}(t_1,t_2)= -i \braket{\Psi_0|[x(t_1),x(t_2)] |\Psi_0}.
\end{align}
Now an extremely important fact to keep in mind while performing this calculation is that the ground state is always bosonic, this enforces considering the ground state separately each time an identity is inserted. 
On expanding the commutator and inserting the identity operator ($\displaystyle\mathbb{I}:=\sum_{k} \ket{\Psi_k} \bra{\Psi_k}$) and considering the ${k=0}$ term separately from the ${k>0}$ term above sets of equations can be written as:
\begin{align}
\nonumber
	b_{n m}^{(2)}(t_1,t_2) &=-i\biggl(\braket{\Psi_n|x(t_1)\ket{\Psi_{0}}\bra{\Psi_{0}}x(t_2) |\Psi_m}- \braket{\Psi_n|x(t_2)\ket{\Psi_{0}}\bra{\Psi_{0}}x(t_1) |\Psi_m}\\ 
	&~~~+\sum_{k>0}^{} \braket{\Psi_n|x(t_1)\ket{\Psi_{k}}\bra{\Psi_{k}}x(t_2) |\Psi_m}- \braket{\Psi_n|x(t_2)\ket{\Psi_{k}}\bra{\Psi_{k}}x(t_1) |\Psi_m}\biggr),~~~~~~~~
\end{align}
and a similar expression will be obtained for $	b_{0 m}^{(2)}(t_1,t_2)$, $	b_{m 0}^{(2)}(t_1,t_2)$ and $	b_{0 0}^{(2)}(t_1,t_2)$. Keeping all the above discussed facts it is not difficult to show that for supersymmetric one dimensional harmonic oscillator we get:
\begin{align}
\label{mic4pt2}
b_{n m}^{(2)}(t_1,t_2) &=\frac{i}{2\omega} \sin\omega (t_{1}-t_{2})\biggl[ \delta_{n1}- \left(1+ \sqrt{n(n+1)} - \sqrt{n(n-1)} \right)\delta_{n m}\biggr], \\
\label{mic4pt22ndpart}
b_{0 m}^{(2)}(t_1,t_2)&=\frac{i}{2\omega} \sin\omega (t_{1}-t_{2}) \delta_{m0}=\frac{i}{2\omega} \sin\omega (t_{1}-t_{2}) \delta_{0m}=b_{m0}^{(2)}(t_1,t_2),\\
\label{mic4pt22ndpart00}
b_{0 0}^{(2)}(t_1,t_2)&=\frac{i}{2\omega} \sin\omega (t_{1}-t_{2}).
\end{align}
Here one can explicitly show that for the supersymmetric one dimensional harmonic oscillator case the following contributions are trivially vanish:
\bea &\displaystyle \sum_{m>0}b_{0 m}^{(2)}(t_1,t_2)b_{0 m}^{(2)*}(t_1,t_2)=0,~~~~\& ~~~~~
 b_{n 0}^{(2)}(t_1,t_2)b_{n 0}^{(2)*}(t_1,t_2)=0~~ \forall~~ n>0.\eea
The above equations can be used to calculate the micro-canonical correlator which is the temperature independent part of the total OTOC and the corresponding ground and excited state contributions are explicitly given by the following expressions:
\begin{align}
\label{cn2result1}
c_{0}^{(2)}(t_1,t_2)&=\frac{1}{4\omega^2}\sin^{2}\omega (t_{1}-t_{2}), \\
\label{cn2result2}c_{n}^{(2)}(t_1,t_2)&= \frac{1}{4\omega^2}\sin^{2}\omega (t_{1}-t_{2}) \biggl\{\delta_{n1}\delta_{n1} - \left(1+ \sqrt{n(n+1)} - \sqrt{n(n-1)} \right)^{2}\biggr\}.
\end{align}
The canonical part of the four point thermal OTOC of the second kind for a supersymmetric Harmonic Oscillator is given by substituting \Cref{cn2result1} and \Cref{cn2result2} in \Cref{cor4pt1v} to obtain the following simplified result:
\newpage
\begin{align}
\label{corr2ndkind}
C^{(2)}(t_1,t_2) &= \frac{1}{2\omega^2}e^{-\frac{\beta\omega}{2}}\sinh \left(\frac{\beta\omega}{2}\right)~\sin^{2}\omega (t_{1}-t_{2}) \nonumber \\
& ~~~~~~~\times \biggl\{1+ e^{-\beta \omega} -\sum_{n>0}^{} e^{-\beta \omega n} \left(1+\sqrt{n(n+1)}-\sqrt{n(n-1)} \right)^{2}\biggr\}\nonumber\\
&=\frac{1}{2\omega^2}e^{-\beta\omega}\sinh \left(\beta\omega\right)~\sin^{2}\omega (t_{1}-t_{2}) \nonumber \\
& ~~~~~~~\times \biggl\{1 -\frac{1}{2}{\rm sech}\left(\frac{\beta\omega}{2}\right)\sum_{n>0}^{} e^{-\beta \omega n} \left(1+\sqrt{n(n+1)}-\sqrt{n(n-1)} \right)^{2}\biggr\}.
\end{align}

\begin{tcolorbox}[colframe=black,arc=0mm,top=5mm]
	\textcolor{MidnightBlue}{\textbf{Micro-Canonical 4-pt Correlator for Ground State }}:
	\begin{align}
	c^{(2)}_{0}(t_{1},t_{2}) &=\frac{1}{4\omega^2}\sin^{2}\omega (t_{1}-t_{2}),
	\end{align}
	\textcolor{MidnightBlue}{\textbf{Micro-Canonical 4-pt Correlator for Excited States }}:
	\begin{align}
	c^{(2)}_{n}(t_{1},t_{2}) &= \frac{1}{4\omega^2} \sin^{2}\omega (t_{1}-t_{2}) \biggl\{\delta_{n1}\delta_{n1} - \bigl(1+ \sqrt{n(n+1)} - \sqrt{n(n-1)} \bigr)^{2}\biggr\}. 	
	\end{align}
\end{tcolorbox}
\vspace{0.5cm}

\begin{tcolorbox}[colframe=black,arc=0mm,top=5mm]
	 \textcolor{MidnightBlue}{\textbf{Canonical 4-pt Correlator }}:
	\begin{align}\label{r2}
	C^{(2)}(t_{1},t_{2}) &=\frac{1}{2\omega^2}e^{-\beta\omega}\sinh \left(\beta\omega\right)~\sin^{2}\omega (t_{1}-t_{2}) \nonumber \\
& ~~~~~~~\times \biggl\{1 -\frac{1}{2}{\rm sech}\left(\frac{\beta\omega}{2}\right)\sum_{n>0}^{} e^{-\beta \omega n} \left(1+\sqrt{n(n+1)}-\sqrt{n(n-1)} \right)^{2}\biggr\}.
	\end{align}
\end{tcolorbox}
\vspace{0.5cm}
\textcolor{Sepia}{\subsubsection{\sffamily Computation of $C^{(3)}(t_{1},t_{2})$}\label{sssec:4pthosp2}} 

The corresponding four point desired OTOC of third kind is defined by the following expression:
\begin{align} 
\label{cor4pt1v1}
\nonumber
C^{(3)}(t_1,t_2) &=- \braket{ [p(t_1),p(t_2)]^{2}}_{\beta}\nonumber\\
&=- \frac{1}{Z}\sum_{n}\langle\Psi_n|e^{-\beta H}[p(t_1),p(t_2)]^{2} |\Psi_n\rangle \nonumber\\
&=  - \frac{1}{Z}\biggl\{\underbrace{\braket{\Psi_0|[p(t_1),p(t_2)]^{2} |\Psi_0}}_{\textcolor{red}{\bf Ground~State~Contribution}}+ \underbrace{\sum_{n>0}^{}e^{-\beta E_n}\braket{\Psi_n|[p(t_1),p(t_2)]^{2} |\Psi_n}}_{\textcolor{red}{\bf Excited~State~Contribution}}\biggr\} \nonumber\\
&=2e^{-\frac{\beta\omega}{2}}\sinh \left(\frac{\beta\omega}{2}\right)~  \left[c_{0}^{(3)}(t_1,t_2)+\sum_{n>0} e^{-\beta \omega n}c_{n}^{(3)}(t_1,t_2)\right],
\end{align}
where the ground and the excited contribution of the micro-canonical part of the four point temperature independent OTOCs are defined by the following expressions:
\begin{align}
\nonumber
\label{cn1}
c_{0}^{(3)}(t_1,t_2) &=-\braket{\Psi_0|[x(t_1),x(t_2)]^{2} |\Psi_0} \\   
&=\sum_{m}b_{0 m}^{(3)}(t_1,t_2)b_{0 m}^{(3)*}(t_1,t_2)\nonumber\\
&= b_{0 0}^{(3)}(t_1,t_2)b_{0 0}^{(3)*}(t_1,t_2)+\sum_{m>0}b_{0 m}^{(3)}(t_1,t_2)b_{0 m}^{(3)*}(t_1,t_2), 
\end{align}
and
\begin{align}
\nonumber
\label{cn1}
c_{n}^{(3)}(t_1,t_2) &= -\braket{\Psi_n|[x(t_1),x(t_2)]^{2} |\Psi_n} \\   
&=\sum_{m}b_{n m}^{(3)}(t_1,t_2)b_{n m}^{(3)*}(t_1,t_2)\nonumber\\
&= b_{n 0}^{(3)}(t_1,t_2)b_{n 0}^{(3)*}(t_1,t_2)+\sum_{m>0}b_{n m}^{(3)}(t_1,t_2)b_{n m}^{(3)*}(t_1,t_2), 
\end{align}
where $b_{nm}^{(3)}(t_1,t_2)$, $	b_{0 m}^{(3)}(t_1,t_2)$, $	b_{m 0}^{(3)}(t_1,t_2)$ and $	b_{0 0}^{(3)}(t_1,t_2)$ used in the above expression are given by: 
\begin{align}
\label{55}
&b_{n m}^{(3)}(t_1,t_2)= -i \braket{\Psi_n|[p(t_1),p(t_2)] |\Psi_m},   \\
&b_{0 m}^{(3)}(t_1,t_2)= -i \braket{\Psi_0|[p(t_1),p(t_2)] |\Psi_m},  \\
&b_{m 0}^{(3)}(t_1,t_2)= -i \braket{\Psi_m|[p(t_1),p(t_2)] |\Psi_0},  \\
&b_{0 0}^{(3)}(t_1,t_2)= -i \braket{\Psi_0|[p(t_1),p(t_2)] |\Psi_0}.
\end{align}
Now an extremely important fact to keep in mind while performing this calculation is that the ground state is always bosonic, this enforces considering the ground state separately each time an identity is inserted. 
On expanding the commutator and inserting the identity operator ($\displaystyle\mathbb{I}:=\sum_{k} \ket{\Psi_k} \bra{\Psi_k}$) and considering the ${k=0}$ term separately from the ${k>0}$ term above sets of equations can be written as:
\begin{align}
\nonumber
	b_{n m}^{(3)}(t_1,t_2) &=-i\biggl(\braket{\Psi_n|p(t_1)\ket{\Psi_{0}}\bra{\Psi_{0}}p(t_2) |\Psi_m}- \braket{\Psi_n|p(t_2)\ket{\Psi_{0}}\bra{\Psi_{0}}p(t_1) |\Psi_m}\\ 
	&~~~+\sum_{k>0}^{} \braket{\Psi_n|p(t_1)\ket{\Psi_{k}}\bra{\Psi_{k}}p(t_2) |\Psi_m}- \braket{\Psi_n|p(t_2)\ket{\Psi_{k}}\bra{\Psi_{k}}p(t_1) |\Psi_m}\biggr)~~~~~~~~
\end{align}
and a similar expression will be obtained for $	b_{0 m}^{(3)}(t_1,t_2)$, $	b_{m 0}^{(3)}(t_1,t_2)$ and $	b_{0 0}^{(3)}(t_1,t_2)$. Keeping all the above discussed facts it is not difficult to show that for supersymmetric one dimensional harmonic oscillator we get:
\begin{align}
\label{mic4pt3}
b_{n m}^{(3)}(t_1,t_2) &= \frac{i\omega}{2} \sin\omega (t_{1}-t_{2})\biggl[\delta_{n1}-\left(1+\sqrt{n(n+1)}- \sqrt{n(n-1)}  \right)\delta_{n m}\biggr], \\
\label{mic4pt23rdpart}
b_{0 m}^{(3)}(t_1,t_2)&= \frac{i\omega}{2} \sin\omega (t_{1}-t_{2})\ \delta_{n0}=\frac{i\omega}{2} \sin[\omega (t_{1}-t_{2})]\ \delta_{0n}=b_{m0}^{(3)}(t_1,t_2),\\
\label{sd00}
b_{0 0}^{(3)}(t_1,t_2)&= \frac{i\omega}{2} \sin\omega (t_{1}-t_{2}).
\end{align}
Using the above equations the ground state and the excited state contributions to the micro-canonical OTOC can be calculated as follows:
\begin{align}
\label{cn3result}
c_{0}^{(3)}(t_1,t_2)&=\frac{\omega^2}{4}\sin^{2}\omega (t_{1}-t_{2}), \\
\label{cn3result1}
c_{n}^{(3)}(t_1,t_2)&= \frac{\omega^2}{4} \sin^{2}\omega (t_{1}-t_{2}) \biggl\{\delta_{n1}\delta_{n1}  - \biggl(1+ \sqrt{n(n+1)} - \sqrt{n(n-1)} \biggr)^{2}\biggr\}.\end{align}
Then the canonical part of the thermal four point OTOC of the third kind for a supersymmetric harmonic oscillator is given by substituting \Cref{cn3result} and \Cref{cn3result1} in \Cref{cor4pt1v1} to obtain the following simplified result:
\begin{align}
\label{corr3rdkind}
C^{(3)}(t_1,t_2) &= \frac{\omega^2}{2}e^{-\frac{\beta\omega}{2}}\sinh \left(\frac{\beta\omega}{2}\right)~\sin^{2}\omega (t_{1}-t_{2}) \nonumber \\
& ~~~~~~~\times \biggl\{1+ e^{-\beta \omega} -\sum_{n>0}^{} e^{-\beta \omega n} \left(1+\sqrt{n(n+1)}-\sqrt{n(n-1)} \right)^{2}\biggr\}\nonumber\\
&=\frac{\omega^2}{2}e^{-\beta\omega}\sinh \left(\beta\omega\right)~\sin^{2}\omega (t_{1}-t_{2}) \nonumber \\
& ~~~~~~~\times \biggl\{1 -\frac{1}{2}{\rm sech}\left(\frac{\beta\omega}{2}\right)\sum_{n>0}^{} e^{-\beta \omega n} \left(1+\sqrt{n(n+1)}-\sqrt{n(n-1)} \right)^{2}\biggr\}.\end{align}

\begin{tcolorbox}[colframe=black,arc=0mm,top=5mm]
	\textcolor{MidnightBlue}{\textbf{Micro-Canonical 4-pt Correlator for Ground State }}:
	\begin{align}
	c^{(3)}_{0}(t_{1},t_{2}) &=\frac{\omega^2}{4}\sin^{2}\omega (t_{1}-t_{2}),
	\end{align}
	\textcolor{MidnightBlue}{\textbf{Micro-Canonical 4-pt Correlator for Excited States }}:
	\begin{align}
	c^{(3)}_{n}(t_{1},t_{2}) &= \frac{\omega^2}{4} \sin^{2}\omega (t_{1}-t_{2}) \biggl\{\delta_{n1}\delta_{n1} - \biggl(1+ \sqrt{n(n+1)} - \sqrt{n(n-1)} \biggr)^{2}\biggr\}.	
\end{align}
\end{tcolorbox}
\vspace{0.5cm}

\begin{tcolorbox}[colframe=black,arc=0mm,top=5mm]
	\textcolor{MidnightBlue}{\textbf{Canonical 4-pt Correlator }}:
	\begin{align}\label{r3}
	C^{(3)}(t_{1},t_{2}) &= \frac{\omega^2}{2}e^{-\beta\omega}\sinh \left(\beta\omega\right)~\sin^{2}\omega (t_{1}-t_{2}) \nonumber \\
& ~~~~~~~\times \biggl\{1 -\frac{1}{2}{\rm sech}\left(\frac{\beta\omega}{2}\right)\sum_{n>0}^{} e^{-\beta \omega n} \left(1+\sqrt{n(n+1)}-\sqrt{n(n-1)} \right)^{2}\biggr\}.
	\end{align}
\end{tcolorbox}
\vspace{0.5cm}

\textcolor{Sepia}{\subsection{\sffamily Computation of normalized four point OTOCs}}
In this section our prime objective is to normalize all of the derived un-normalized three types of OTOCs with appropriate normalization factors, which we have already introduced in the earlier half of the paper for model independent eigenstate representation. In this section we will explicitly derive the expressions for the normalization factors for supersymmetric ine dimensional harmonic oscillator model in its eigenstate representation and derive all of the possible three types of OTOCs after normalization.

\textcolor{Sepia}{\subsubsection{\sffamily Computation of $\widetilde{C}^{(1)}(t_{1},t_{2})$}\label{sssec:4pthosp1ccv1}} 
To normalize the obtained OTOC ${C}^{(1)}(t_{1},t_{2})$ we need to compute the appropriate factors which we are going compute in this subsection.

First of all, we need to evaluate the following two point equal time thermal correlator, which is given by:
\begin{align}
\label{normspxx}
\braket{ x(t_1)x(t_1)}_{\beta}&= \frac{1}{Z} \text{Tr}(e^{-\beta H} x(t_1) x(t_1))\nonumber \\ \nonumber
&= \frac{1}{Z}\sum_{n} \braket{ \Psi_n|e^{-\beta H} x(t_1) x(t_1)|\Psi_n } 
\nonumber\\
&= \frac{1}{Z}\biggl[\underbrace{\braket{\Psi_0|x(t_1)x(t_1)|\Psi_0}}_{\textnormal{\textcolor{red}{\bf Ground State Contribution}}}+\underbrace{\sum_{n>0}^{}e^{-\beta E_n}~\braket{\Psi_n|x(t_1)x(t_1)|\Psi_n}}_{\textnormal{\textcolor{red}{\bf Excited State contribution}}}\biggr].
\end{align}

Inserting the completeness relation between the operators and using the Heisenberg picture equation for the evolution of an operator, the normalization factor involving the position operator can be written as:
\\

\begin{tcolorbox}
	\begin{align}\label{nora}
	\braket{x(t_1)x(t_1)}_{\beta}&=\frac{1}{2\omega}e^{-\frac{\beta\omega}{2}}\sinh \left(\frac{\beta\omega}{2}\right)~\nonumber\\
	&~~~~~~~~~~~\times\biggl(1+ e^{-\beta \omega}+\sum_{n>0} e^{-\beta \omega n} [2n + \sqrt{n(n+1)}+ \sqrt{n(n-1)} ] \biggr)\nonumber\\
	&=\frac{1}{2\omega}e^{-\beta\omega}\sinh \left(\beta\omega\right)~\nonumber\\
	&~~~~~~~~~~~\times\biggl(1+ \frac{1}{2}{\rm sech}\left(\frac{\beta\omega}{2}\right)\sum_{n>0} e^{-\beta \omega n} [2n + \sqrt{n(n+1)}+ \sqrt{n(n-1)} ] \biggr).
	\end{align}
	\end{tcolorbox}

\vspace{0.5cm}

A similar calculation can carried out for computing the two point equal time thermal correlator involving the momentum operators, which is given by:

\begin{align}
\label{normsppp}
\braket{p(t_2)p(t_2)}_{\beta}&= \frac{1}{Z} \text{Tr}(e^{-\beta H} p(t_2) p(t_2)) \nonumber\\
&= \frac{1}{Z}\sum_{n} \langle \Psi_n|e^{\beta H} p(t_2) p(t_2)|\Psi_n \rangle \nonumber\\
&= \frac{1}{Z}\biggl[\underbrace{\braket{\psi_0|p(t_2)p(t_2)|\psi_0}}_{\textnormal{\textcolor{red}{\bf Ground state contribution}} }+\underbrace{\sum_{n>0}^{}\braket{\psi_n|p(t_2)p(t_2)|\psi_n}}_{\textnormal{\textcolor{red}{\bf Excited state contribution}}}\biggr].
\end{align}

One generally needs to consider the ground state separately from the other higher energy states due to the fact that in Supersymmetric QM, the ground state has contributions only from the original Hamiltonian. There is no contribution of the associated partner Hamiltonian in the ground state. Keeping this interesting fact in mind and adopting a similar approach as taken in the previous case, the normalization factor involving the thermal expectation value of the momentum operators are given by:
\\

\begin{tcolorbox}
	\begin{align}
	\braket{p(t_2)p(t_2)}_{\beta}&=\frac{\omega}{2}e^{-\frac{\beta\omega}{2}}\sinh \left(\frac{\beta\omega}{2}\right)~\nonumber\\
	&~~~~~~~~~~~\times\biggl(1+ e^{-\beta \omega}+\sum_{n>0} e^{-\beta \omega n} [2n + \sqrt{n(n+1)}+ \sqrt{n(n-1)} ] \biggr)\nonumber\\
	&=\frac{\omega}{2}e^{-\beta\omega}\sinh \left(\beta\omega\right)~\nonumber\\
	&~~~~~~~~~~~\times\biggl(1+ \frac{1}{2}{\rm sech}\left(\frac{\beta\omega}{2}\right)\sum_{n>0} e^{-\beta \omega n} [2n + \sqrt{n(n+1)}+ \sqrt{n(n-1)} ] \biggr).
	\end{align}
\end{tcolorbox}
\vspace{0.5cm}

Therefore, the normalization factor ${N_{1}}$ for the 4-point correlator $C^{(1)}(t_1,t_2)$ is given by
\\

\begin{tcolorbox}[fonttitle=\sffamily\Large,title=Normalization factor {$N_{1}$} of {${C^{(1)}(t_{1},t_{2})}$}]
	\begin{align}
	\nonumber
	N_{1} &=\braket{x(t_1) x(t_1)}_{\beta}~\braket{p(t_2) p(t_2)}_{\beta} \\
	&=\frac{1}{4}e^{-\beta\omega}\sinh^2 \left(\frac{\beta\omega}{2}\right)~\nonumber\\
	&~~~~~~~~~~~\times\biggl(1+ e^{-\beta \omega}+\sum_{n>0} e^{-\beta \omega n} [2n + \sqrt{n(n+1)}+ \sqrt{n(n-1)} ] \biggr)^2\nonumber\\
	&=\frac{1}{4}e^{-2\beta\omega}\sinh^2 \left(\beta\omega\right)~\nonumber\\
	&~~~~~~~~~~~\times\biggl(1+ \frac{1}{2}{\rm sech}\left(\frac{\beta\omega}{2}\right)\sum_{n>0} e^{-\beta \omega n} [2n + \sqrt{n(n+1)}+ \sqrt{n(n-1)} ] \biggr)^2\nonumber\\
	&=\omega^2~N_2=\omega^{-2}~N_3.
	\end{align}
\end{tcolorbox}
\vspace{0.5cm}
Consequently, the normalized OTOC of the first kind can be expressed as:\\ \\
\begin{tcolorbox}[colframe=black,arc=0mm,top=5mm]
	\textcolor{MidnightBlue}{\textbf{Normalized ${C^{(1)}(t_{1},t_{2})}$ }}:
	\begin{align}
		\widetilde{C}^{(1)}(t_{1},t_{2}) &=2e^{\beta\omega}{\rm cosech} \left(\beta\omega\right)~\cos^{2}\omega (t_{1}-t_{2}) \nonumber \\
		 & ~~~~~~~~~~~~\times \frac{\displaystyle\biggl\{1+\frac{1}{2}{\rm sech}\left(\frac{\beta\omega}{2}\right)\sum_{n>0}^{} e^{-\beta \omega n} \left(1+\sqrt{n(n+1)}-\sqrt{n(n-1)} \right)^{2}\biggr\}}{\displaystyle\biggl\{1+ \frac{1}{2}{\rm sech}\left(\frac{\beta\omega}{2}\right)\sum_{n>0} e^{-\beta \omega n} [2n + \sqrt{n(n+1)}+ \sqrt{n(n-1)} ] \biggr\}^2}.\label{g1}
	\end{align}
\end{tcolorbox}

\textcolor{Sepia}{\subsubsection{\sffamily Computation of $\widetilde{C}^{(2)}(t_{1},t_{2})$}\label{sssec:4pthosp1ccv2}} 
To normalize the obtained OTOC ${C}^{(2)}(t_{1},t_{2})$ we need to compute the appropriate factors which we are going compute in this subsection.

Similarly the normalization factor ${N_{2}}$ for the four point OTOC $C^{(2)}(t_1,t_2)$ can be computed explicitly, which is given by the following simplified expression:
\\

\begin{tcolorbox}[fonttitle=\sffamily\Large,title=Normalization factor {$N_{2}$} of {${C^{(2)}(t_{1},t_{2})}$}]
	\begin{align}
	\nonumber
	N_{2} &=\braket{x(t_1) x(t_1)}_{\beta}~\braket{x(t_2) x(t_2)}_{\beta} \\
	&=\frac{1}{4\omega^2}e^{-\beta\omega}\sinh^2 \left(\frac{\beta\omega}{2}\right)~\nonumber\\
	&~~~~~~~~~~~\times\biggl(1+ e^{-\beta \omega}+\sum_{n>0} e^{-\beta \omega n} [2n + \sqrt{n(n+1)}+ \sqrt{n(n-1)} ] \biggr)^2\nonumber\\
	&=\frac{1}{4\omega^2}e^{-2\beta\omega}\sinh^2 \left(\beta\omega\right)~\nonumber\\
	&~~~~~~~~~~~\times\biggl(1+ \frac{1}{2}{\rm sech}\left(\frac{\beta\omega}{2}\right)\sum_{n>0} e^{-\beta \omega n} [2n + \sqrt{n(n+1)}+ \sqrt{n(n-1)} ] \biggr)^2\nonumber\\
	&=\frac{1}{\omega^4}~N_3\nonumber\\
	&=\frac{1}{\omega^2}~N_1.
	\end{align}
\end{tcolorbox}
\vspace{0.5cm}
Consequently, the normalized OTOC of the second kind can be expressed as:\\ \\
\begin{tcolorbox}[colframe=black,arc=0mm,top=5mm]
	\textcolor{MidnightBlue}{\textbf{Normalized ${C^{(2)}(t_{1},t_{2})}$ }}:
	\begin{align}
		\widetilde{C}^{(2)}(t_{1},t_{2}) &=2e^{\beta\omega}{\rm cosech} \left(\beta\omega\right)~\sin^{2}\omega (t_{1}-t_{2}) \nonumber \\
& ~~~~~~~\times \frac{\displaystyle \biggl\{1 -\frac{1}{2}{\rm sech}\left(\frac{\beta\omega}{2}\right)\sum_{n>0}^{} e^{-\beta \omega n} \left(1+\sqrt{n(n+1)}-\sqrt{n(n-1)} \right)^{2}\biggr\}}{\displaystyle \biggl(1+ \frac{1}{2}{\rm sech}\left(\frac{\beta\omega}{2}\right)\sum_{n>0} e^{-\beta \omega n} [2n + \sqrt{n(n+1)}+ \sqrt{n(n-1)} ] \biggr)^2}\nonumber\\
&=\widetilde{C}^{(3)}(t_{1},t_{2}).\label{g2}
	\end{align}
\end{tcolorbox}
\newpage
\textcolor{Sepia}{\subsubsection{\sffamily Computation of $\widetilde{C}^{(3)}(t_{1},t_{2})$}\label{sssec:4pthosp1ccv3}} 
To normalize the obtained OTOC ${C}^{(3)}(t_{1},t_{2})$ we need to compute the appropriate factors which we are going compute in this subsection.

Similarly the normalization factor ${N_{3}}$ for the four point OTOC $C^{(3)}(t_1,t_2)$ can be computed explicitly, which is given by the following simplified expression:
\\

\begin{tcolorbox}[fonttitle=\sffamily\Large,title=Normalization factor {$N_{3}$} of {${C^{(3)}(t_{1},t_{2})}$}]
	\begin{align}
	\nonumber
	N_{3} &=\braket{p(t_1) p(t_1)}_{\beta}~\braket{p(t_2) p(t_2)}_{\beta} \\
	&=\frac{\omega^2}{4}e^{-\beta\omega}\sinh^2 \left(\frac{\beta\omega}{2}\right)~\nonumber\\
	&~~~~~~~~~~~\times\biggl(1+ e^{-\beta \omega}+\sum_{n>0} e^{-\beta \omega n} [2n + \sqrt{n(n+1)}+ \sqrt{n(n-1)} ] \biggr)^2\nonumber\\
	&=\frac{\omega^2}{4}e^{-2\beta\omega}\sinh^2 \left(\beta\omega\right)~\nonumber\\
	&~~~~~~~~~~~\times\biggl(1+ \frac{1}{2}{\rm sech}\left(\frac{\beta\omega}{2}\right)\sum_{n>0} e^{-\beta \omega n} [2n + \sqrt{n(n+1)}+ \sqrt{n(n-1)} ] \biggr)^2\nonumber\\
	&=\omega^4~N_2\nonumber\\
	&=\omega^2~N_1.
	\end{align}
\end{tcolorbox}
\vspace{0.5cm}
Consequently, the normalized OTOC of the third kind can be expressed as:\\ \\
\begin{tcolorbox}[colframe=black,arc=0mm,top=5mm]
	\textcolor{MidnightBlue}{\textbf{Normalized ${C^{(3)}(t_{1},t_{2})}$ }}:
	\begin{align}\label{g3}
		\widetilde{C}^{(3)}(t_{1},t_{2}) &=2e^{\beta\omega}{\rm cosech} \left(\beta\omega\right)~\sin^{2}\omega (t_{1}-t_{2}) \nonumber \\
& ~~~~~~~\times \frac{\displaystyle \biggl\{1 -\frac{1}{2}{\rm sech}\left(\frac{\beta\omega}{2}\right)\sum_{n>0}^{} e^{-\beta \omega n} \left(1+\sqrt{n(n+1)}-\sqrt{n(n-1)} \right)^{2}\biggr\}}{\displaystyle \biggl(1+ \frac{1}{2}{\rm sech}\left(\frac{\beta\omega}{2}\right)\sum_{n>0} e^{-\beta \omega n} [2n + \sqrt{n(n+1)}+ \sqrt{n(n-1)} ] \biggr)^2}\nonumber\\
&=\widetilde{C}^{(2)}(t_{1},t_{2}) .
	\end{align}
\end{tcolorbox}

\newpage
\textcolor{Sepia}{\subsection{\sffamily Summary of Results}\label{ssec:sumeigenrepSHO}}
\begin{tcolorbox}[fonttitle=\sffamily\Large,title=Ground State Contributions for Micro-Canonical OTOC for SUSY 1D Harmonic Oscillator]
	
	\begin{align}
	y_{0}^{(1)}(t_{1},t_{2}) &=
	 -\frac{i}{2} \cos\omega(t_{1}-t_{2}) \tag{\ref{gsmicrocorr}} \\
	y_{0}^{(2)}(t_{1},t_{2}) &=  \frac{i}{2\omega}\sin\omega(t_{1}-t_{2})=-\frac{1}{\omega^2}~y^{(3)}_{0}(t_{1},t_{2})  \tag{\ref{gsmicrocorr2}} \\
	y^{(3)}_{0}(t_{1},t_{2}) &= - \frac{i\omega}{2}\sin\omega(t_{1}-t_{2})=-\omega^2~y_{0}^{(2)}(t_{1},t_{2}) \tag{\ref{gsmicrocorr3}} \\
	c_{0}^{(1)}(t_{1},t_{2}) &= \frac{1}{4}\cos^{2}\omega (t_{1}-t_{2}) \tag{\ref{cn1resultw1}} \\
	c_{0}^{(2)}(t_{1},t_{2}) 
	&=\frac{1}{4\omega^2}\sin^{2}\omega (t_{1}-t_{2})=\frac{1}{\omega^4}~c_{0}^{(3)}(t_{1},t_{2}) \tag{\ref{cn2result1}}\\
	c_{0}^{(3)}(t_{1},t_{2}) 
	&= \frac{\omega^2}{4}\sin^{2}\omega (t_{1}-t_{2})=\omega^4~c_{0}^{(2)}(t_{1},t_{2}) \tag{\ref{cn3result}}
	.\end{align}
	
\end{tcolorbox}

\begin{tcolorbox}[fonttitle=\sffamily\Large,title=Excited State Contributions for Micro-Canonical OTOC for SUSY 1D Harmonic Oscillator]
	
	\begin{align}
	y_{n}^{(1)}(t_{1},t_{2}) &= -\frac{i}{2} \cos\omega(t_{1}-t_{2})\left[1+\delta_{n1}+\sqrt{n(n+1)}-\sqrt{n(n-1)}\right] \tag{\ref{hsmicrocorr}} \\
	y_{n}^{(2)}(t_{1},t_{2}) &= -\frac{i}{2\omega}\sin\omega(t_{1}-t_{2})\biggl[1+\delta_{n1} +\sqrt{n(n+1)}-\sqrt{n(n-1)}\biggr]\nonumber\\
	&=-\frac{1}{\omega^2}~y^{(3)}_{n}(t_{1},t_{2}) \tag{\ref{hsmicrocorr2}} \\
	y^{(3)}_{n}(t_{1},t_{2}) &= \frac{i\omega}{2}\sin\omega(t_{1}-t_{2})\biggl[1+\delta_{n1}+\sqrt{n(n+1)}-\sqrt{n(n-1)}\biggr]\nonumber\\
	&=-\omega^2~y_{n}^{(2)}(t_{1},t_{2}) \tag{\ref{hsmicrocorr3}} \\
	c_{n}^{(1)}(t_{1},t_{2}) &= \frac{1}{4}\cos^{2}\omega (t_{1}-t_{2}) ~\biggl\{\delta_{n1}\delta_{n1}  + \left(1+ \sqrt{n(n+1)} - \sqrt{n(n-1)} \right)^{2}\biggr\}\tag{\ref{cn1resultw2}} \\
	c_{n}^{(2)}(t_{1},t_{2}) 
	&=\frac{1}{4\omega^2}\sin^{2}\omega (t_{1}-t_{2}) \biggl\{\delta_{n1}\delta_{n1} - \left(1+ \sqrt{n(n+1)} - \sqrt{n(n-1)} \right)^{2}\biggr\}\nonumber\\
	&=\frac{1}{\omega^4}~c_{n}^{(3)}(t_{1},t_{2}) \tag{\ref{cn2result2}}\\
	c_{n}^{(3)}(t_{1},t_{2}) 
	&=  \frac{\omega^2}{4} \sin^{2}\omega (t_{1}-t_{2}) \biggl\{\delta_{n1}\delta_{n1}  - \biggl(1+ \sqrt{n(n+1)} - \sqrt{n(n-1)} \biggr)^{2}\biggr\} \nonumber\\
	&=\omega^4~c_{n}^{(2)}(t_{1},t_{2})  \tag{\ref{cn3result1}}
	.\end{align}
	
\end{tcolorbox}

\begin{tcolorbox}[fonttitle=\sffamily\large,title=Un-normalized two and four point Canonical OTOC for SUSY 1D Harmonic Oscillator]
	\begin{align}
	Y^{(1)}(t_{1},t_{2}) &=-\frac{i}{2}\tanh \left(\frac{\beta\omega}{2}\right)~\cos\omega(t_{1}-t_{2})~\nonumber\\
&\displaystyle~~~~~~~\times\biggl\{1+2~\cosh \left(\frac{\beta\omega}{2}\right)\sum_{n>0}^{}e^{-\beta\omega \left(n+\frac{1}{2}\right)} \biggl(1+\sqrt{n(n+1)}-\sqrt{n(n-1)}\biggr) \biggr\} \tag{\ref{x1}} \\
	Y^{(2)}(t_{1},t_{2}) &= \frac{i}{\omega}e^{-\beta\omega}\sinh \left(\beta\omega\right)\tanh \left(\frac{\beta\omega}{2}\right)~\sin\omega(t_{1}-t_{2})\nonumber\\
&~~~\times\biggl\{1-\frac{1}{2}\textnormal{ cosech} \left(\frac{\beta \omega}{2}\right)\sum_{n>0}^{}e^{-\beta\omega \left(n-\frac{1}{2}\right)} \biggl(1+\sqrt{n(n+1)}-\sqrt{n(n-1)}\biggr) \biggr\}\nonumber\\
&=-\frac{1}{\omega^2}~Y^{(3)}(t_{1},t_{2}) \tag{\ref{x2}} \\
	Y^{(3)}(t_{1},t_{2}) &= -i\omega~e^{-\beta\omega}\sinh \left(\beta\omega\right)\tanh \left(\frac{\beta\omega}{2}\right)~\sin\omega(t_{1}-t_{2})\nonumber\\
&~~~\times\biggl\{1-\frac{1}{2}\textnormal{ cosech} \left(\frac{\beta \omega}{2}\right)\sum_{n>0}^{}e^{-\beta\omega \left(n-\frac{1}{2}\right)} \biggl(1+\sqrt{n(n+1)}-\sqrt{n(n-1)}\biggr) \biggr\}\nonumber\\
&=-\omega^2~Y^{(2)}(t_{1},t_{2}) \tag{\ref{x3}} \\
	C^{(1)}(t_{1},t_{2}) &=  \frac{1}{2}e^{-\beta\omega}\sinh \left(\beta\omega\right)~\cos^{2}\omega (t_{1}-t_{2}) \nonumber \\
		 & ~~~~~~~~~~~~\times \biggl\{1+\frac{1}{2}{\rm sech}\left(\frac{\beta\omega}{2}\right)\sum_{n>0}^{} e^{-\beta \omega n} \left(1+\sqrt{n(n+1)}-\sqrt{n(n-1)} \right)^{2}\biggr\} \tag{\ref{r1}}\\
	C^{(2)}(t_{1},t_{2}) &= \frac{1}{2\omega^2}e^{-\beta\omega}\sinh \left(\beta\omega\right)~\sin^{2}\omega (t_{1}-t_{2}) \nonumber \\
& ~~~~~~~\times \biggl\{1 -\frac{1}{2}{\rm sech}\left(\frac{\beta\omega}{2}\right)\sum_{n>0}^{} e^{-\beta \omega n} \left(1+\sqrt{n(n+1)}-\sqrt{n(n-1)} \right)^{2}\biggr\} \nonumber\\
&=\frac{1}{\omega^4}~C^{(3)}(t_{1},t_{2})\tag{\ref{r2}} \\
	C^{(3)}(t_{1},t_{2}) &=  \frac{\omega^2}{2}e^{-\beta\omega}\sinh \left(\beta\omega\right)~\sin^{2}\omega (t_{1}-t_{2}) \nonumber \\
& ~~~~~~~\times \biggl\{1 -\frac{1}{2}{\rm sech}\left(\frac{\beta\omega}{2}\right)\sum_{n>0}^{} e^{-\beta \omega n} \left(1+\sqrt{n(n+1)}-\sqrt{n(n-1)} \right)^{2}\biggr\}\nonumber\\
&=\omega^4~C^{(2)}(t_{1},t_{2})  \tag{\ref{r3}} 
	.\end{align}
\end{tcolorbox}

\begin{tcolorbox}[fonttitle=\sffamily\large,title=Normalized four point Canonical OTOC for SUSY 1D Harmonic Oscillator]
	\begin{align}
	\braket{x(t_{1})x(t_{1})}_{\beta} &=\braket{x(t_{2})x(t_{2})}_{\beta}\nonumber\\
&=\frac{1}{2\omega}e^{-\beta\omega}\sinh \left(\beta\omega\right)~\nonumber\\
	&~~~~~~~~~~~\times\biggl(1+ \frac{1}{2}{\rm sech}\left(\frac{\beta\omega}{2}\right)\sum_{n>0} e^{-\beta \omega n} [2n + \sqrt{n(n+1)}+ \sqrt{n(n-1)} ] \biggr) \nonumber \\
	&=\frac{1}{\omega^2}\braket{p(t_{1})p(t_{1})}_{\beta}=\frac{1}{\omega^2}\braket{p(t_{2})p(t_{2})}_{\beta} \tag{\ref{nora}} \\
	\widetilde{C}^{(1)}(t_{1},t_{2}) &= \frac{C^{(1)}(t_{1},t_{2})}{\braket{x(t_{1})x(t_{1})}_{\beta} \braket{p(t_{2})p(t_{2})}_{\beta}}\nonumber\\
	 &=2e^{\beta\omega}{\rm cosech} \left(\beta\omega\right)~\cos^{2}\omega (t_{1}-t_{2}) \nonumber \\
		 & ~~~~~~~~\times \frac{\displaystyle\biggl\{1+\frac{1}{2}{\rm sech}\left(\frac{\beta\omega}{2}\right)\sum_{n>0}^{} e^{-\beta \omega n} \left(1+\sqrt{n(n+1)}-\sqrt{n(n-1)} \right)^{2}\biggr\}}{\displaystyle\biggl(1+ \frac{1}{2}{\rm sech}\left(\frac{\beta\omega}{2}\right)\sum_{n>0} e^{-\beta \omega n} [2n + \sqrt{n(n+1)}+ \sqrt{n(n-1)} ] \biggr)^2}  \tag{\ref{g1}}\\
	\widetilde{C}^{(2)}(t_{1},t_{2}) &= \frac{C^{(2)}(t_{1},t_{2})}{\braket{x(t_{1})x(t_{1})}_{\beta} \braket{x(t_{2})x(t_{2})}_{\beta}}\nonumber\\
	 &=2e^{\beta\omega}{\rm cosech} \left(\beta\omega\right)~\sin^{2}\omega (t_{1}-t_{2}) \nonumber \\
& ~~~~~~~\times \frac{\displaystyle \biggl\{1 -\frac{1}{2}{\rm sech}\left(\frac{\beta\omega}{2}\right)\sum_{n>0}^{} e^{-\beta \omega n} \left(1+\sqrt{n(n+1)}-\sqrt{n(n-1)} \right)^{2}\biggr\}}{\displaystyle \biggl(1+ \frac{1}{2}{\rm sech}\left(\frac{\beta\omega}{2}\right)\sum_{n>0} e^{-\beta \omega n} [2n + \sqrt{n(n+1)}+ \sqrt{n(n-1)} ] \biggr)^2}\nonumber\\
&=\widetilde{C}^{(3)}(t_{1},t_{2}) \tag{\ref{g2}} \\
	\widetilde{C}^{(3)}(t_{1},t_{2}) &= \frac{C^{(3)}(t_{1},t_{2})}{\braket{p(t_{1})p(t_{1})}_{\beta} \braket{p(t_{2})p(t_{2})}_{\beta}}\nonumber\\
	 &= 2e^{\beta\omega}{\rm cosech} \left(\beta\omega\right)~\sin^{2}\omega (t_{1}-t_{2}) \nonumber \\
& ~~~~~~~\times \frac{\displaystyle \biggl\{1 -\frac{1}{2}{\rm sech}\left(\frac{\beta\omega}{2}\right)\sum_{n>0}^{} e^{-\beta \omega n} \left(1+\sqrt{n(n+1)}-\sqrt{n(n-1)} \right)^{2}\biggr\}}{\displaystyle \biggl(1+ \frac{1}{2}{\rm sech}\left(\frac{\beta\omega}{2}\right)\sum_{n>0} e^{-\beta \omega n} [2n + \sqrt{n(n+1)}+ \sqrt{n(n-1)} ] \biggr)^2}\nonumber\\
&=\widetilde{C}^{(2)}(t_{1},t_{2}). \tag{\ref{g3}}  
	\end{align}
\end{tcolorbox}
\newpage

\textcolor{Sepia}{\section{\sffamily Model II: Supersymmetric One Dimensional Potential Well}\label{sec:qbox}}

\noindent
The one dimensional infinite potential well is characterized by the following potential:
\begin{align}
V_{1}(x) = \left\{\begin{array}{ll}
0 \qquad \qquad &\text{for } 0 \leq x \leq L \\
\infty \qquad \qquad &\text{otherwise}  
\end{array}\right\}
.\end{align}

\noindent
The eigenfunctions and the corresponding energy eigenvalues associated with the Hamiltonian ${H_{1}}$ for this potential is a well known result \cite{ramadevi} and is given by the following expressions: 
\begin{align*}
\psi_{n}^{(1)} = \sqrt{\frac{2}{L}}\ \sin \left[ \frac{(n+1)\pi}{L}\ x \right] \quad \textnormal{and} \quad E_{n}^{(1)} = \frac{(n+1)^{2} \pi^{2} \hbar^{2}}{2m L^{2}} \quad \textnormal{for } n \in \{0,1,2,\dots\}
,\end{align*}
where we have replaced the energy level tagging from ${n}$ to ${n+1}$ compared to the regular expressions one encounters in typical textbooks. This does not change anything physically and with this convention ${n}$ can take values from 0 instead of 1. Also, for the sake of simplicity we consider ${\hbar = L = 2m = 1}$. Furthermore, as we have previously mentioned we need the ground state energy to be zero, so that after subtracting off the ground state energy we get the following simplified results:  
\begin{align}
	\psi_{n}^{(1)} = \sqrt{2}\ \sin [(n+1) \pi x] \quad \text{and} \quad E_{n}^{(1)} = n (n+2)\ \pi^{2} \text{ for } n \in \{0,1,2,...\}
.\end{align}
To obtain the partner potential associated with the original potential, the superpotential needs to be calculated which has been done in \cite{ramadevi} and is given by the following expression:
\begin{align}
W(x) &= - \pi\ \text{cot}(\pi x) .
\end{align}
Once the superpotential of a supersymmetric quantum mechanical model is known it can be used to obtain the eigenspectrum and the associated partner potential as discussed in \Cref{sec:revSUSYQM}. For the supersymmetric one dimensional potential well it can be very easily verify that the associated partner potential is given by the following equation:
\begin{align}
\label{v2susy1d}
V_{2}(x) &= 2 \pi^{2}\ \text{cosec}^{2}(\pi x) \nonumber
.\end{align}

\noindent
A look at \Cref{v2susy1d} immediately suggests that the partner potential is remarkably different from the original one unlike the harmonic oscillator case whose partner potential is identical to that of the original one. The energy eigenfunctions and eigenvalues associated with the partner potential can easily be calculated as:
\begin{align}
	\psi_{n}^{(2)}(x) &= \sqrt{\frac{2}{(n+2)^{2} - 1}}\ \biggl\{ (n+2)\ \text{cos} ((n+2) \pi x) - \text{cot}( \pi x)\ \text{sin}((n+2) \pi x) \biggr\} ;  \\
	E_{n}^{(2)} &=  (n^{2} + 4n + 3)\ \pi^{2} \quad \textnormal{for } n \in \{0,1,2,\dots\} .
\end{align}
	
To compute the correlators we need the partition function and the matrix elements of the position operator between any two arbitrary energy states. However while computing the matrix elements if one of the state is the ground state then the expression can be written in a closed form as in the ground state there is no contribution of the partner Hamiltonian. 

\begin{align}
    Z &= 1+\sum_{m>0}e^{-\beta m(m+2)}, \\
	x_{0k}&=\sqrt{2}\int_{0}^{1}dx~\sin\pi x\sin(k+1)\pi x= \frac{\sqrt{2}}{(k+2)}  \frac{\sin\pi k}{\pi k}\\ ,\nonumber
	x_{mk}&=\frac{1}{4\pi}\biggl[\frac{\sin(k-m)\pi}{(k-m)}-\frac{\sin(k+m+2)\pi}{(k+m+2)}\biggr]\\ \nonumber
	&~~~~~~+\frac{2}{\sqrt{((1+k)^2-1)((1+m)^2-1)}} \\ \nonumber
	&~~~~~~~~~~~~~~~~~~~~\int_{0}^{1}dx~x\biggl((1+k)\cos(1+k)\pi x-\cot\pi x\sin(1+k)\pi x\biggr) \\ &~~~~~~~~~~~~~~~~~~~~~~~~~~~~~~~~~ \biggl((1+m)\cos(1+m)\pi x-\cot\pi x\sin(1+m)\pi x\biggr) .
\end{align}
Substituting the above expressions in the eigenstate representation of the correlators obtained in \Cref{sec:eigenrepcorr}, the correlators for 1D SUSY potential well can be calculated. Now it is important to mention here that the OTOCs computed from this particular model only can be presented in terms of integrals which at the end we need to evaluate using numerical computation. So to avoid writing complicated mathematical expressions in terms of huge size integrals and instead of presenting a detailed calculation here we provide and discuss the results in the later \Cref{sec:results} of this paper.

\textcolor{Sepia}{\section{\sffamily General remarks on the classical limiting interpretation of OTOCs}\label{sec:classlimit}}

In this section our prime objective is to study the classical limit of the thermal OTOCs derived explicitly in the previous sections of this paper in the context of supersymmetric quantum mechanical systems. This computation is essential to understand the time and the temperature dependent behaviour of the two and four point thermal OTOCs in the classical limit. By styudying the behaviour in this limit one can check the consistency of the result obtained of the quantum randomness from the computed thermal OTOCs in the previous sections.

The strategy adopted in calculating the classical limit of OTOCs is usually replacing the quantum mechanical commutator bracket with the usual classical Poisson bracket defined by the following equation:
\begin{align}
\label{poisson}
\{f,g\}_{q_i,p_i}= \sum_{i} \left(\frac{\partial f}{\partial q_i}\frac{\partial g}{\partial p_i}-~\frac{\partial f}{\partial p_i}\frac{\partial g}{\partial q_i}\right).
\end{align} 

In the above equation, the $q_i$ and the $p_i$ are the generalized coordinates and momenta and ${f}$ and ${g}$ are functions of these coordinates and momenta i.e.
 \bea f \equiv f(q_1,\dots, q_n\ ;\ p_1,\dots,p_n,t),\\
 g\equiv g(q_1,\dots, q_n\ ;\ p_1,\dots,p_n,t).\eea
It can be readily checked that the commutator bracket of two quantum mechanical operators satisfy the same properties as that of the classical Poisson bracket and it can be viewed as the outcome of the following limit on the commutator bracket:
\begin{align}
\lim_{\hbar\to 0}\frac{[\hat{f},\hat{g}]}{i\hbar}=\{f,g\}.
\end{align}

The thermal average of the correlators is carried out by the trace operation in the quantum case which can be further simplified in the eigenstate representation. For it's classical counterpart the trace operation is replaced by the phase space integral in the classical limit.  
For a quantum mechanical model in the context of supersymmetry usually one would expect that there are two generalised coordinates and momenta, one coming from the original Hamiltonian of the system and the other one from its supersymmetric part. Below we write down the generic expressions for the Poisson Brackets involving the position and the momentum operators and provide a general expression for the classical limit of the two and four point classical versions of the correlators.\\ 
\begin{tcolorbox}[fonttitle=\sffamily\large,title=Classical limit of two point Canonical OTOCs]
	\begin{align}
	& Y^{(1)}(t_{1},t_{2}) = \frac{1}{Z_{cl}} \iint \frac{dx dp}{2 \pi} e^{-\beta H}~ \{x(t_1),p(t_2)\} \\
	& Y^{(2)}(t_{1},t_{2}) = \frac{1}{Z_{cl}} \iint \frac{dx dp}{2 \pi} e^{-\beta H}~ \{x(t_1),x(t_2)\} \\
	& Y^{(3)}(t_{1},t_{2}) = \frac{1}{Z_{cl}} \iint \frac{dx dp}{2 \pi} e^{-\beta H}~ \{p(t_1),p(t_2)\}\\
	& {\rm where}~Z_{cl}=\iint \frac{dx dp}{2 \pi} e^{-\beta H}. 
	\end{align}
\end{tcolorbox}
\vspace{0.1cm}

\begin{tcolorbox}[fonttitle=\sffamily\large,title=Classical limit of four point Canonical OTOCs]
\begin{align}
	& C^{(1)}(t_{1},t_{2}) = \frac{1}{Z_{cl}} \iint \frac{dx\ dp}{2 \pi}\ e^{-\beta H}\ \{x(t_1),p(t_2)\}^2 \\
	& C^{(2)}(t_{1},t_{2}) = \frac{1}{Z_{cl}} \iint \frac{dx\ dp}{2 \pi}\ e^{-\beta H}\ \{x(t_1),x(t_2)\}^2 \\
	& C^{(3)}(t_{1},t_{2}) = \frac{1}{Z_{cl}} \iint \frac{dx\ dp}{2 \pi}\ e^{-\beta H}\ \{p(t_1),p(t_2)\}^2\\
	& {\rm where}~Z_{cl}=\iint \frac{dx dp}{2 \pi} e^{-\beta H}.
\end{align}
\end{tcolorbox}
\vspace{0.5cm}

\textcolor{Sepia}{\section{\sffamily Classical Limit of OTOC for Supersymmetric One Dimensional Harmonic Oscillator}\label{sec:classlimitho}}

As discussed earlier any supersymmetric quantum mechanical Hamiltonian is associated with a partner Hamiltonian. For a Supersymmetric Harmonic oscillator the potential associated with the original and the partner Hamiltonian are exactly equal apart from a constant factor as shown in \Cref{ssec:eigenHO}. Hence, the classical solutions of the dynamical operators takes identical forms and is obtanied by trivially solving the following differential equation:
\begin{align}
\frac{d^2x}{dt^2}=-\omega^2 x.
\end{align}
In solving the above differential equation we take the initial position and momentum to be ${x(0)}$ and ${p(0)}$. The classical solutions of the operators obtained by solving the above differential equation and the mentioned initial conditions are given by:
\begin{align}
&x(t)= x(0)\cos\omega t + \frac{p(0)}{\omega}\sin\omega t;\\
& p(t) = p(0)\cos\omega t - x(0)\ \omega\sin\omega t.
\end{align}


We want to calculate the classical limit of OTOC using the position operators at different times. Using the distributive property of Poisson Bracket ${\{x(t_1),p(t_2)\}}$ can be expanded for our supersymmetric case in the following way :
\begin{align}
\{x(t_1),p(t_2)\}= \underbrace{\{x_B(t_1),p_B(t_2)\}}_{\textcolor{red}{\bf Bosonic~part}}+\underbrace{\{x_F(t_1),p_F(t_2)\}}_{\textcolor{red}{\bf Fermionic~part}}.
\end{align}
Each of the terms of the above equation can be evaluated using the definition of classical Poisson Bracket and has been done below:
\begin{align}
\{x_B(t_1),p_B(t_2)\} &= \left(\frac{\partial x_B(t_1)}{\partial x_B(0)}\frac{\partial p_B(t_2)}{\partial p_B(0)}-~\frac{\partial x_B(t_1)}{\partial p_B(0)}\frac{\partial p_B(t_2)}{\partial x_B(0)}\right) 
= \cos\omega(t_{1}-t_{2}), \\
\{x_F(t_1),p_F(t_2)\} &= \left(\frac{\partial x_F(t_1)}{\partial x_F(0)}\frac{\partial p_F(t_2)}{\partial p_F(0)}-~\frac{\partial x_F(t_1)}{\partial p_F(0)}\frac{\partial p_F(t_2)}{\partial x_F(0)}\right)
= \cos\omega(t_{1}-t_{2}),
\end{align}   
which implies that the contributions from both the bosonic and the fermionic part of the Poisson brackets are exactly identical for supersymmetric one dimensional harmonic oscillator. 
Therefore, the Poisson Bracket of the position and the momentum at different times is given by:
\begin{align}
\{x(t_1),p(t_2)\}=2\cos\omega(t_{1}-t_{2}).
\end{align}

We want to calculate the classical limit of OTOC using the position variables at different times. Using the distributive property of Poisson Bracket ${\{x(t_1),x(t_2)\}}$ can be expanded for our supersymmetric case in the following way:
\begin{align}
\{x(t_1),x(t_2)\}=\underbrace{ \{x_B(t_1),x_B(t_2)\}}_{\textcolor{red}{\bf Bosonic~contribution}}+\underbrace{\{x_F(t_1),x_F(t_2)\}}_{\textcolor{red}{\bf Fermionic~contribution}}.
\end{align}
Each of the terms of the above equation can be evaluated using the definition of the classical Poisson Bracket and has been done below:
\begin{align}
	\{x_B(t_1),x_B(t_2)\} &= \left(\frac{\partial x_B(t_1)}{\partial x_B(0)}\frac{\partial x_B(t_2)}{\partial p_B(0)}-~\frac{\partial x_B(t_1)}{\partial p_B(0)}\frac{\partial x_B(t_2)}{\partial x_B(0)}\right) 
= -\frac{1}{\omega}\sin\omega(t_{1}-t_{2}), \\
	\{x_F(t_1),x_F(t_2)\} &= \left(\frac{\partial x_F(t_1)}{\partial x_F(0)}\frac{\partial x_F(t_2)}{\partial p_F(0)}-~\frac{\partial x_F(t_1)}{\partial p_F(0)}\frac{\partial x_F(t_2)}{\partial x_F(0)}\right)
= -\frac{1}{\omega}\sin\omega(t_{1}-t_{2}).
\end{align}   
which implies that the contributions from both the bosonic and the fermionic part of the Poisson brackets are exactly identical for supersymmetric one dimensional harmonic oscillator. 
Therefore, the classical Poisson Bracket of the position variables at different times is given by:
\begin{align}
\{x(t_1),x(t_2)\}=-\frac{2}{\omega} \sin\omega(t_{1}-t_{2}).
\end{align}


We want to calculate the classical limit of OTOC using the momentum variables at different times. Using the distributive property of Poisson Bracket ${ \left\{ p(t_{1}), p(t_{2}) \right\}}$ can be expanded for our supersymmetric case and the non zero contribution are given by the following equation:
\begin{align}
\{p(t_1),p(t_2)\}&=\underbrace{ \{p_B(t_1),p_B(t_2)\}}_{\textcolor{red}{\bf Bosonic~contribution}}+\underbrace{\{p_F(t_1),p_F(t_2)\}}_{\textcolor{red}{\bf Fermionic~contribution}}.
\end{align}
Each of the terms of the above equation can be evaluated using the definition of Poisson Bracket and has been done below:
\begin{align}
	\{p_B(t_1),p_B(t_2)\} &= \left(\frac{\partial p_B(t_1)}{\partial x_B(0)}\frac{\partial p_B(t_2)}{\partial p_B(0)}-~\frac{\partial p_B(t_1)}{\partial p_B(0)}\frac{\partial p_B(t_2)}{\partial x_B(0)}\right) 
= -\omega\sin\omega(t_{1}-t_{2}) ,\\
	\{p_F(t_1),p_F(t_2)\} &=  \left(\frac{\partial p_F(t_1)}{\partial x_F(0)}\frac{\partial p_F(t_2)}{\partial p_F(0)}-~\frac{\partial p_F(t_1)}{\partial p_F(0)}\frac{\partial p_F(t_2)}{\partial x_F(0)}\right)
= -\omega\sin\omega(t_{1}-t_{2}).
\end{align}   
which implies that the contributions from both the bosonic and the fermionic part of the Poisson brackets are exactly identical for supersymmetric one dimensional harmonic oscillator. 
Therefore, the classical Poisson Bracket of the momentum variables at different times is given by:
\begin{align}
\{p(t_1),p(t_2)\}=-2\omega\sin\omega(t_{1}-t_{2}).
\end{align}

Finally, the classical limit of OTOC of two point thermal classical version of the OTOCs are given by the following expressions: \\

\begin{tcolorbox}[fonttitle=\sffamily\large,title=Classical limit of two point OTOCs for SUSY 1D HO ]
\begin{align*}
	Y^{(1)}(t_{1},t_{2}) &= 2\cos\omega(t_{1}-t_{2}), \\ 	
	Y^{(2)}(t_{1},t_{2}) &= - \frac{2}{\omega}\sin\omega(t_{1}-t_{2})=\frac{1}{\omega^2}~Y^{(3)}(t_{1},t_{2}),\\ 
	Y^{(3)}(t_{1},t_{2}) &= - 2\omega\sin\omega(t_{1}-t_{2})=\omega^2~Y^{(2)}(t_{1},t_{2}).
\end{align*}
\end{tcolorbox}
\vspace{0.5cm}

\begin{tcolorbox}[fonttitle=\sffamily\large,title=Classical limit of four point OTOCs for SUSY 1D HO]
\begin{align*}
	& C^{(1)}(t_{1},t_{2}) = 4\cos^2(\omega(t_1-t_2)), \\ 
	& C^{(2)}(t_{1},t_{2}) = \frac{4}{\omega^2}\sin^2(\omega(t_1-t_2))=\frac{1}{\omega^4}C^{(3)}(t_{1},t_{2}), \\ 
	& C^{(3)}(t_{1},t_{2}) = 4\omega^2\sin^2(\omega(t_1-t_2))=\omega^4C^{(2)}(t_{1},t_{2}).
\end{align*}
\end{tcolorbox}
\vspace{0.5cm}


From the classical limit result of the Supersymmetric Harmonic oscillator it is clear that the classical statistics do not produce the quantum result. Though we get a similar time dependence which is periodic both in the classical and the quantum case, the results are not identical to each other. The prime difference in the result lies in the fact that the quantum result depends on the energy eigenstate. This explicit dependence of the OTOC on the energy eigenstates is what prevents the quantum results to give the classical result in the high temperature limit. The important factor to note here is the appearance of the factor 2 in the two point classical correlators. It was already discussed in \Cref{sec:revSUSYQM} that for every potential in supersymmetry there is an associated partner potential. For the case of Supersymmetric Harmonic oscillator it can be seen that the structure of the partner potential is exactly similar to the original potential differing in only an overall constant factor. So it is very easy to understand that the classical solutions for both the potentials will be exactly identical. Now while calculating the Poisson Bracket in the context of Supersymmetric, the bosonic and the fermionic part can be considered as two degrees of freedom. Due to the similar solutions of the dynamical variables the contribution of the bosonic and the fermionic degrees of freedom for the Supersymmetric harmonic oscillator are exactly identical, which adds up to give twice the result obtained from one degree of freedom. Another important point to note that the classical limit of the correlators do not depend on the initial conditions of the dynamical variables.

\textcolor{Sepia}{\section{\sffamily Classical Limit of OTOC for Supersymmetric 1D box}\label{sec:classlimitbox}}

The case of supersymmetric infinite potential well is not as trivial as that of the Harmonic oiscillator. The associated partner potential is not identical to that of original potential. As has been derived in the paper \cite{ramadevi} the partner potential for a 1D box of unit length is given by
\begin{align}
V_2(x)=2\pi^2\text{cosec}^2(\pi x).
\end{align}
whereas the well known original potential is given by
\begin{align}
V_{1}(x) = \left\{\begin{array}{ll}
0 \qquad \qquad &\text{for } 0 \leq x \leq 1 \\
\infty \qquad \qquad &\text{otherwise}  
\end{array}\right\}
.\end{align}
The classical solutions of the dynamical operators for the original potential are pretty trivial to calculate  and is obtained by trivially solving the following differential equation
\begin{align}
\frac{d^2x}{dt^2}=0.
\end{align}
The classical solutions of the operators in this case can be explicitly written as:
\begin{align}
\label{classolx}
&x_1(t)= x_1(0)+2p_1(0)\ t \quad ,\\
& p_1(t)= p_1(0).
\end{align}
where ${x_{1}(0)}$ and ${p_{1}(0)}$ are the initial position and momentum of the particle moving in the potential ${V_{1}(x)}$. However, the classical solution of the particle moving in the potential ${V_{2}(x)}$ can be obtained by solving the following differential equation:
\begin{align}
\frac{d^2x_2}{dt^2}=4\pi^3 \cot(\pi x_2)\ \text{cosec}^2(\pi x_2).
\end{align}
The above differential equation can be solved explicitly to give the following solutions of the dynamical variables:
\begin{align}
\label{classolxpar}
x_2(t)&= \frac{1}{\pi}\cos^{-1}\biggl[\sqrt{\frac{c}{c+4\pi^4}}\ \sin \left( \sqrt{c+4\pi^4}\ (c_1-t) \right) \biggr], \\
\label{classolppar}
p_2(t)&= \sqrt{c-4\pi^4 \cot^2(\pi x_2)}.
\end{align}
The constants ${c}$ and ${c_{1}}$ can be fixed from the initial conditions by taking the initial position and momentum to be ${x_{2}(0)}$ and ${p_{2}(0)}$ respectively and can be written as:
\begin{align}
c &= p_2(0)^2 +4 \pi^4 \cot^2 \left[ \pi x_2(0) \right], \\
c_1&= \frac{1}{\sqrt{c+4\pi^4}}\ \sin^{-1}\biggl[\frac{\sqrt{c+4\pi^4} \cos(\pi x_2(0))}{\sqrt{c}}\biggr].
\end{align}
It is to be noted that the classical solution written above is valid before the particle bounces at a boundary. After it experienxes a bounce at the boundary the momentum changes its direction i.e $p(t) \rightarrow -p(t)$. We take this fact into account by considering the infinitesimal deviation of the initial position and fixing the momentum as:
 $$(x(0), p(0)) \rightarrow (x(0)+\delta x(0), p(0)).$$
  The bouncing of the particle at the boundary is given by the time evolution i.e.
   $$\delta x(t)= (-1)^{n} \delta x(0) ,$$ after the ${n^{\textnormal{th}}}$ bounce.


We are interested in calculating the Poisson Bracket of the position operator at a certain time with the momentum operator at another time which gives us the classical limit of the correlator of first kind. Similarly the classical limit of the correlator of the second kind is obtained from the Poisson Bracket of the position operators at two different times. The classical limit of the correlator of the third kind is given by the Poisson Bracket of the momentum operators at two different times.
The Poisson Bracket for the classical limit of first kind of correlator is given by:
\begin{align}
\{x(t_1),p(t_2)\}=\underbrace{ \{x_1(t_1),p_1(t_2)\}_{B}}_{\textcolor{red}{\bf Bosonic~contribution}}+\underbrace{\{x_2(t_1),p_2(t_2)\}_{F}}_{\textcolor{red}{\bf Fermionic~contribution}}.
\end{align}
Using the classical solutions obtained in \Cref{classolx} the Poisson Bracket involving the position and momentum of the particle moving in the potential ${V_1}$ can be written as:
\begin{align}
	\{x_1(t_1),p_1(t_2)\} = \left(\frac{\partial x_1(t_1)}{\partial x_1(0)}\frac{\partial p_1(t_2)}{\partial p_1(0)}-~\frac{\partial x_1(t_1)}{\partial p_1(0)}\frac{\partial p_1(t_2)}{\partial x_1(0)}\right)=(-1)^n .
\end{align}
In a similar manner, the Poisson Bracket involving the classical position variables of the particle associated with the original potential  can be calculated as:
\begin{align}
	\{x_1(t_1),x_1(t_2)\}= \left(\frac{\partial x_1(t_1)}{\partial x_1(0)}\frac{\partial x_1(t_2)}{\partial p_1(0)}-~\frac{\partial x_1(t_1)}{\partial p_1(0)}\frac{\partial x_1(t_2)}{\partial x_1(0)}\right)= 2\ (-1)^n\  (t_2-t_1).
\end{align}
The Poisson Bracket involving the classical momentum variables of the particle associated with the original potential can be calculated as
\begin{align}
\{p_1(t_1),p_1(t_2)\} = \left(\frac{\partial p_1(t_1)}{\partial x_1(0)}\frac{\partial p_1(t_2)}{\partial p_1(0)}-~\frac{\partial p_1(t_1)}{\partial p_1(0)}\frac{\partial p_1(t_2)}{\partial x_1(0)}\right)= 0.
\end{align} 

In the same manner the Poisson Bracket relations of the particle  associated with the partner potential can be calculated in the following way:
\begin{align}
	\{x_2(t_1),p_2(t_2)\} = \left(\frac{\partial x_2(t_1)}{\partial x_2(0)}\frac{\partial p_2(t_2)}{\partial p_2(0)}-~\frac{\partial x_2(t_1)}{\partial p_2(0)}\frac{\partial p_2(t_2)}{\partial x_2(0)}\right)
= (-1)^n\ \frac{\mathcal{X}_2}{\mathcal{Q}_2}.
\end{align}
where the expressions of $\mathcal{X}_2$ and $\mathcal{Q}_2$ are given in \Cref{app:PoissonbrackSUSY1Dpot}.


The Poisson Bracket involving the classical position variables of the particle associated with the partner potential can be calculated as:
\begin{align}
\{x_2(t_1),x_2(t_2)\} &= \left(\frac{\partial x_2(t_1)}{\partial x_2(0)}\frac{\partial x_2(t_2)}{\partial p_2(0)} - \frac{\partial x_2(t_1)}{\partial p_2(0)}\frac{\partial x_2(t_2)}{\partial x_2(0)}\right)= \frac{\mathcal{K}_1}{\mathcal{K}_2}. \\
\end{align}
where the symbols $\mathcal{K}_1$ and $\mathcal{K}_2$ represent the following terms
\begin{align*}
	\mathcal{K}_1 &= (-1)^n\ (1+4\pi^4)\ (t_1-t_2)\cos(\sin^{-1}(\alpha)-t_1\ \beta)\ \cos(\sin^{-1}(\alpha) - t_2\ \beta), \\
	\mathcal{K}_2 &= \sqrt{\ \eta(t_1)\ \eta(t_2)}.
\end{align*}
Similarly, the Poisson Bracket involving the classical momentum variables of the particle associated with the partner potential can be calculated as
\begin{align}
	\{p_2(t_1),p_2(t_2)\} = \left(\frac{\partial p_2(t_1)}{\partial x_2(0)}\frac{\partial p_2(t_2)}{\partial p_2(0)}-~\frac{\partial p_2(t_1)}{\partial p_2(0)}\frac{\partial p_2(t_2)}{\partial x_2(0)}\right)
= \frac{\mathcal{P}_1}{\mathcal{P}_2}.
\end{align} 
The explicit expressions for the symbols used in the above equations has been written in the \Cref{app:PoissonbrackSUSY1Dpot}.
Therefore, the classical limit of OTOC of 2-pt correlators are given by 

\vspace{0.5cm}
\begin{tcolorbox}[fonttitle=\sffamily\large,title=Classical limit of two point Canonical Correlators for SUSY 1D infinite potential well]
	\begin{align*}
	Y^{(1)}(t_{1},t_{2}) &=\frac{1}{Z_{cl}}\iint \frac{dx dp}{2\pi}e^{-\beta H}\biggl[(-1)^n\biggl(1+\frac{\mathcal{X}_2}{\mathcal{Q}_2}\biggr)\biggr], \\ 	
	Y^{(2)}(t_{1},t_{2}) &= \frac{1}{Z_{cl}}\iint \frac{dx dp}{2\pi}e^{-\beta H}\biggl[ 2\ (-1)^n\  (t_2-t_1)+\frac{\mathcal{K}_1}{\mathcal{K}_2}\biggr], \\ 	 
	Y^{(3)}(t_{1},t_{2}) &= \frac{1}{Z_{cl}}\iint \frac{dx dp}{2\pi}e^{-\beta H} \biggl(\frac{\mathcal{P}_1}{\mathcal{P}_2}\biggr). 
	\end{align*}
\end{tcolorbox}
\vspace{0.5cm}

\begin{tcolorbox}[fonttitle=\sffamily\large,title=Classical limit of four point Canonical Correlators for SUSY 1D infinite potential well]
	\begin{align*}
	 C^{(1)}(t_{1},t_{2}) &= \frac{1}{Z_{cl}}\iint \frac{dx dp}{2\pi}e^{-\beta H}\biggl[(-1)^n\biggl(1+\frac{\mathcal{X}_2}{\mathcal{Q}_2}\biggr)\biggr]^2, \\ 	
	 C^{(2)}(t_{1},t_{2}) &= \frac{1}{Z_{cl}}\iint \frac{dx dp}{2\pi}e^{-\beta H}\biggl[ 2\ (-1)^n\  (t_2-t_1)+\frac{\mathcal{K}_1}{\mathcal{K}_2}\biggr]^2, \\ 	  
	 C^{(3)}(t_{1},t_{2}) &= \frac{1}{Z_{cl}}\iint \frac{dx dp}{2\pi}e^{-\beta H} \biggl(\frac{\mathcal{P}_1}{\mathcal{P}_2}\biggr)^2.
	\end{align*}
\end{tcolorbox}
\vspace{0.5cm}

The solutions of the classical dynamical variables obtained for the partner potential associated with the 1D infinite well potential is not trivial as obtained in \Cref{classolxpar,classolppar}. The Poisson Bracket involving the partner potential degrees of freedom have complicated terms whose explicit expressions are given in the \Cref{app:PoissonbrackSUSY1Dpot}. An important observation here is that the contribution of the two degrees of freedom,( one from the original potential and the other from the partner potential) are not identical, which is expected as the structure and hence the classical solutions of the dynamical variables are not same for both the cases. On careful observation of the Poisson Bracket relations of the partner potential degrees of freedom, it can be seen that there is an explicit dependence on the initial values of the dynamical variables.


\textcolor{Sepia}{\section{\sffamily Numerical Results}\label{sec:results}}
 
\noindent
In this section we do the following studies to ascertain the randomness properties of micro-canonical and canonical correlators of the two integrable models we have considered : 

\begin{itemize}
\item \textbf{\underline{\textcolor{red}{Study A :}} }\\
The time dependence of 2-pt and 4-pt micro-canonical correlators for four different states : ${m=1,\ m=2,\ m=5,\ m=10}$. This demonstrates the comparative behavior of micro-canonical correlators for different states under time evolution. 

\item \textbf{\underline{\textcolor{red}{Study B :}}}\\ The time dependence of 2-pt and 4-pt canonical correlators for three different temperatures : ${T = 10,\ T = 50,\ T = 100}$ at fixed time ${t=0.5}$. We have chosen units such that the Boltzmann Constant,  ${k_{\textnormal{B}} = 1}$, so that the inverse temperature, ${\beta = 1/T}$. This demonstrates the comparative behaviour of canonical correlators for different temperature under time evolution. 

\item \textbf{\underline{\textcolor{red}{Study C:}}}\\ The temperature dependence of the 2-pt and 4-pt canonical correlators in the temperature range : ${ 10 \leq T \leq 100}$. This demonstrates the explicit temperature dependence of the canonical correlators. 
\end{itemize}

Technically, one has to take the full infinite dimensional Hilbert Space associated with the Supersymmetric 1D Potential Well for computing the correlators but it is not possible to do that in a numerical evaluation and hence one must choose some finite number of states in the Hilbert Space. This choice of a finite value of total number of states will result in an error and this kind of error is known as \textit{Truncation Error}. The terminology refers to the fact that the error is arising because we have truncated the number of states to a finite value. Here, we have chosen the truncation to be such that all states ${\leq N_{\textnormal{trunc}} = 10}$ for numerical evaluation. 

From the eigenstate representations for the correlators  : ${y_{m}^{(1,2,3)}(t_{1},t_{2})}$, ${Y^{(1,2,3)}(t_{1},t_{2})}$, ${c_{m}^{(1,2,3)}(t_{1},t_{2})}$ and ${C^{(1,2,3)}(t_{1},t_{2})}$ given in \Cref{ssec:sumeigenrep} we know that they depend only on ${t_{1} - t_{2}}$ and hence we have defined ${t = t_{1}- t_{2}}$. So, the negative values of ${t}$ refer to the case of ${t_{2}>t_{1}}$ and so on. 
We proceed to the discussion of obtained numerical results. 

To make any bold comments about whether we are observing randomness or not we need to take the commutator brackets as operators ${\mathcal{O}_{i}}$ \& then calculate : $${\Delta \mathcal{O}_{i} = \sqrt{ \braket{\mathcal{O}_{i}^{2}}_{\beta} - \braket{\mathcal{O}_{i}}_{\beta}^{2} }}$$ and check whether ${\Delta \mathcal{O}_{i}}$ depends on time ${t}$ or not. If it doesn't then what we have is just statistical fluctuation which is arising from the inherent quantum nature of the systems and not from randomness. For example : Consider a QHO which as ${\Delta x \ne 0}$ but it is also independent of time, this means that at each instant of time if we take a large number of copies then there will be some statistical variation in the values of ${x}$ across different copies but that variation will be found exactly the same at each instant of time. If ${\Delta \mathcal{O}_{i} = f(t)}$ then it is indeed a true signature of quantum randomness perhaps depending on functional form of ${f(t)}$. Since, we have calculated both ${ \braket{\mathcal{O}_{i}^{2}}_{\beta}}$ and ${ \braket{\mathcal{O}_{i}}_{\beta}}$ it is really easy to check for any signature of randomness. Furthermore, we can use the same method to calculate higher moments than ${\Delta \mathcal{O}_{i}}$ and get more and more sensitive measurements for randomness which is helpful if the ${f(t)}$ above turns out to be some simple periodic function like ${f(t) \sim \cos \left( a t \right)}$ and we are not satisfied with it because we have really sensitive and amazing technology which can probe for even finer randomness signatures.

To elaborate further consider the following. We don't really know what is quantum uncertainty and hence to work with it we replace it by statistical uncertainty using ensembles and so on. Then uncertainty of an operator ${\mathcal{O}_{i}}$ represents its fluctuation across ensembles and it is not really a true measure of randomness. The fluctuation is merely a sampling fluctuation. Say, we take an ensemble consisting of 10000 copies of a system then the uncertainty tells us the variation in the measured value of ${\mathcal{O}_{i}}$ across the copies / ensemble at a particular time. Now, if we obtain the ${\Delta \mathcal{O}_{i}}$ as a function of time then we can easily probe for randomness depending on the sensitivity of our instrument. If our instrument is really sensitive then we can even use higher moments to check for stronger and stronger conditions of randomness. Obtaining ${\Delta \mathcal{O}_{i}}$ as a function of time means that the fluctuation itself is changing and hence this can be a true signature of randomness this ofcourse depends on what kind of function we get.

\textcolor{Sepia}{\subsection{\sffamily Supersymmetric 1D Infinite Potential Well}\label{ssec:qres_infpotwell}}

\begin{figure}[h!]
	\centering
	\includegraphics[width=17cm,height=13cm]{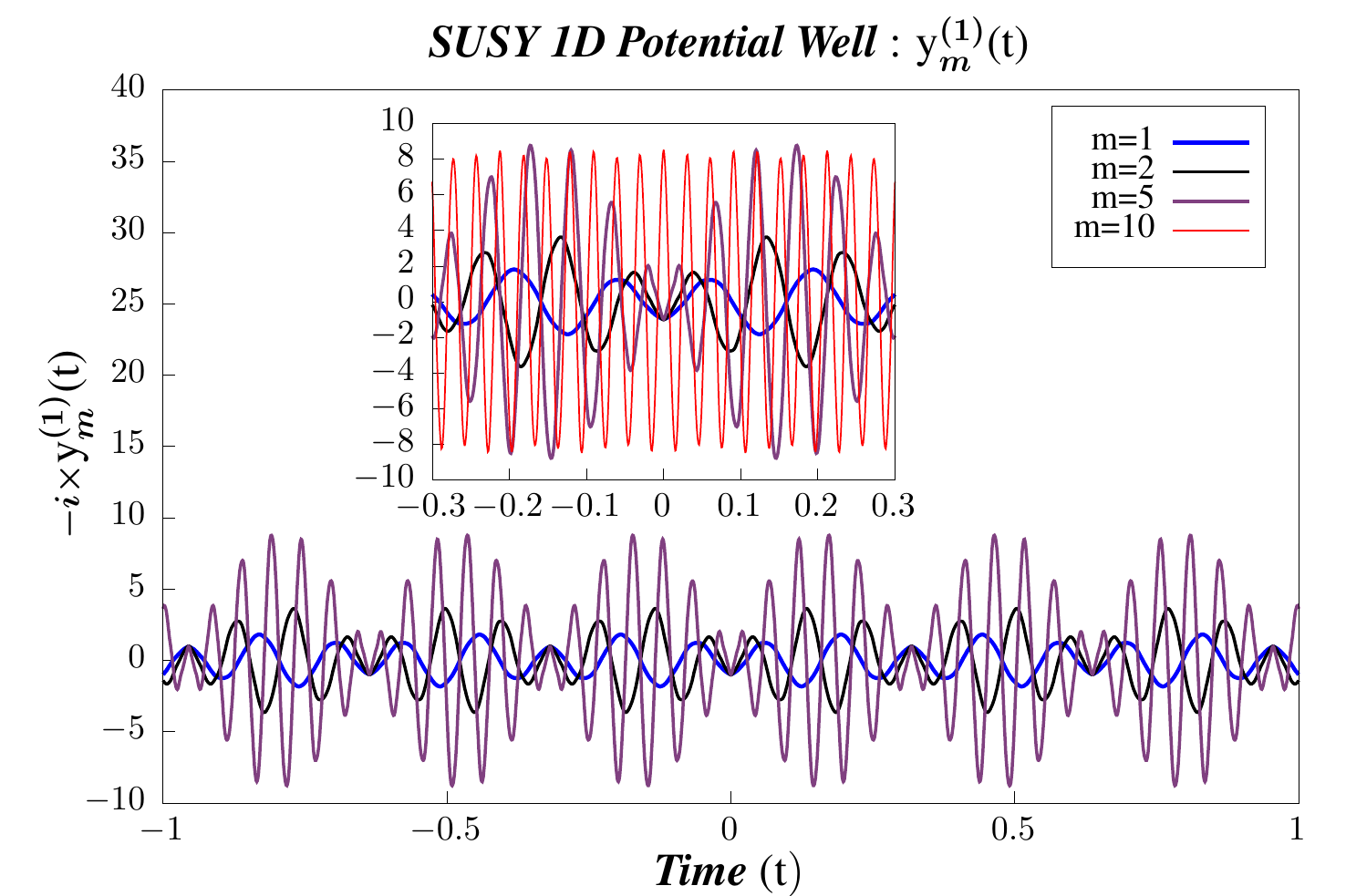}
	\caption{Supersymmetric 1D Infinite Potential Well : Behavior of 2-pt micro-canonical correlator ${y_{m}^{(1)}(t_{1},t_{2}) = - \braket{\Psi_{m} | [x(t_{1}),p({t_{2}})] | \Psi_{m}}}$ with time for different ${m}$. We have chosen ${t_{1} - t_{2} = t}$ as there is only one relevant time parameter.}
	\label{fig:y1IPW}
\end{figure}
\begin{figure}[h!]
	\centering
	\includegraphics[width=17cm,height=13cm]{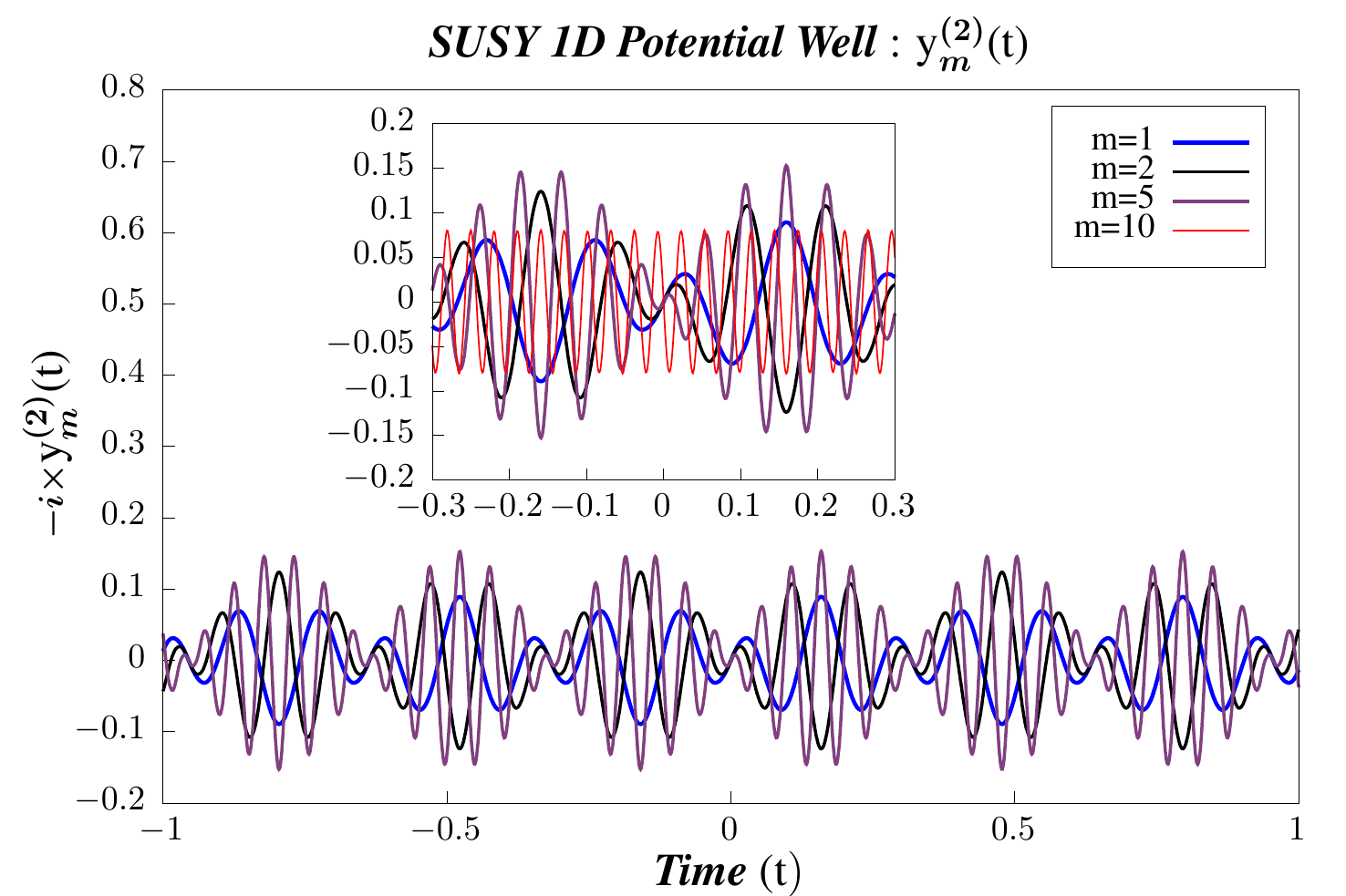}
	\caption{Supersymmetric 1D Infinite Potential Well : Behavior of 2-pt micro-canonical correlator ${y_{m}^{(2)}(t_{1},t_{2}) = - \braket{\Psi_{m} | [x(t_{1}),x({t_{2}})] | \Psi_{m}}}$ with time for different ${m}$. We have chosen ${t_{1} - t_{2} = t}$ as there is only one relevant time parameter.}
	\label{fig:y2IPW}
\end{figure}
\begin{figure}[h!]
	\centering
	\includegraphics[width=17cm,height=13cm]{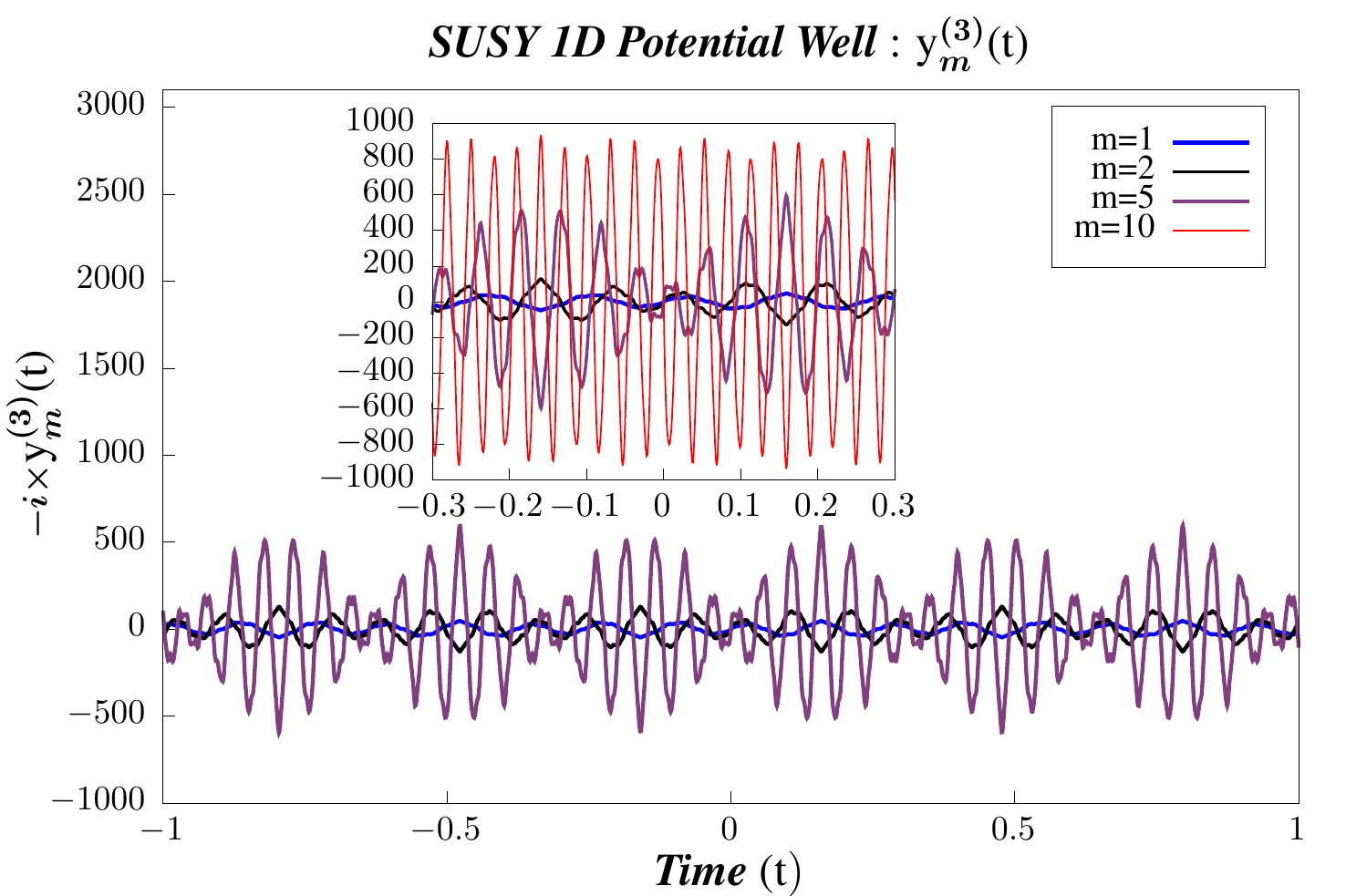}
	\caption{Supersymmetric 1D Infinite Potential Well : Behavior of 2-pt micro-canonical correlator ${y_{m}^{(3)}(t_{1},t_{2}) = - \braket{\Psi_{m} | [p(t_{1}),p({t_{2}})] | \Psi_{m}}}$ with time for different ${m}$. We have chosen ${t_{1} - t_{2} = t}$ as there is only one relevant time parameter.}
	\label{fig:y3IPW}
\end{figure}

\begin{figure}[h!]
	\centering
	\includegraphics[width=17cm,height=13cm]{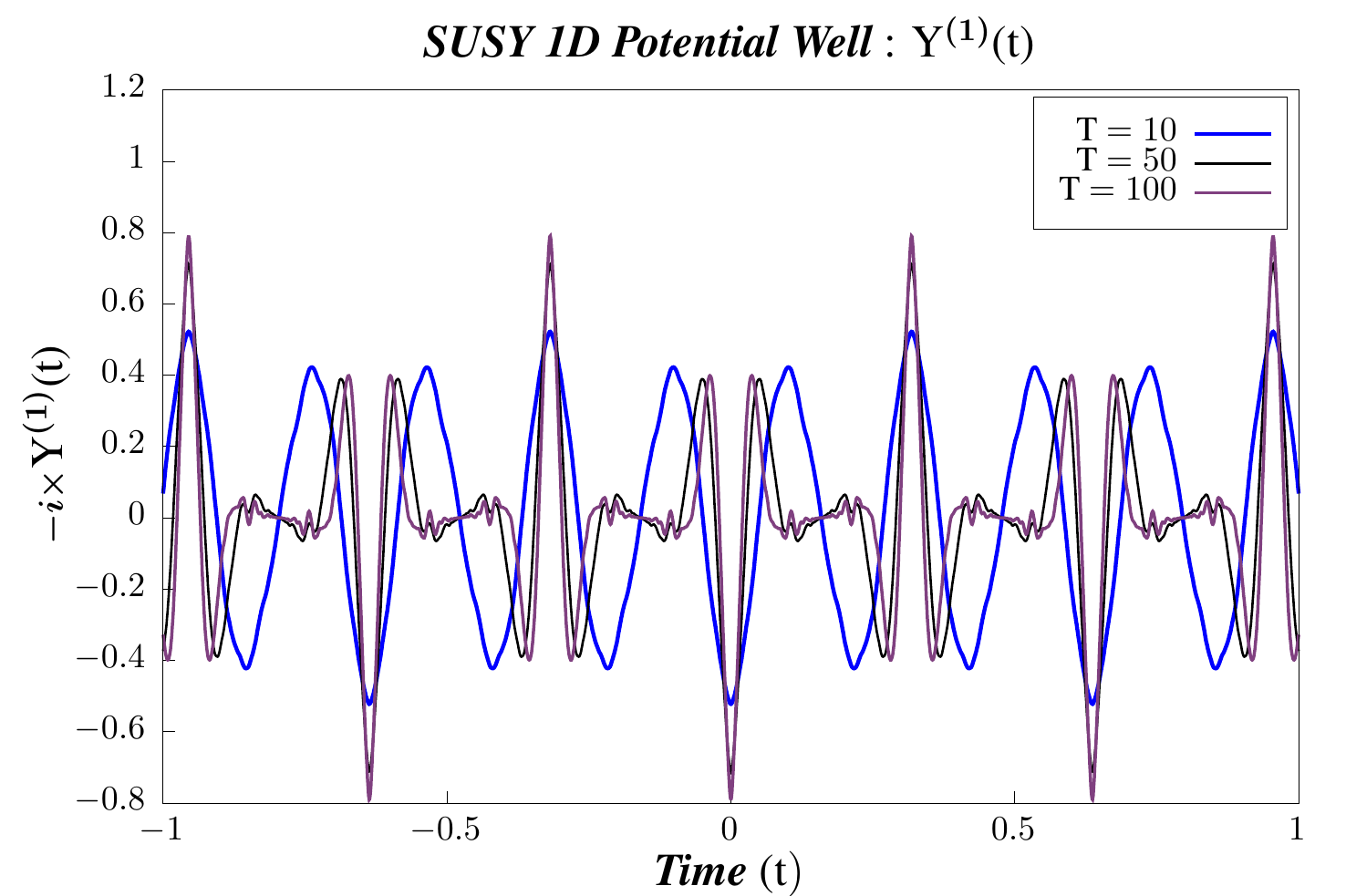}
	\caption{Supersymmetric 1D Infinite Potential Well : Behavior of 2-pt canonical correlator ${Y^{(1)}(t_{1},t_{2}) = - \sum_{m}^{} e^{-\beta E_{m}} \braket{\Psi_{m} | [x(t_{1}),p({t_{2}})] | \Psi_{m}}}$ with time for different temperatures. We have chosen ${t_{1} - t_{2} = t}$ as there is only one relevant time parameter.}
	\label{fig:Y1IPW}
\end{figure}

\begin{figure}[h!]
	\centering
	\includegraphics[width=17cm,height=13cm]{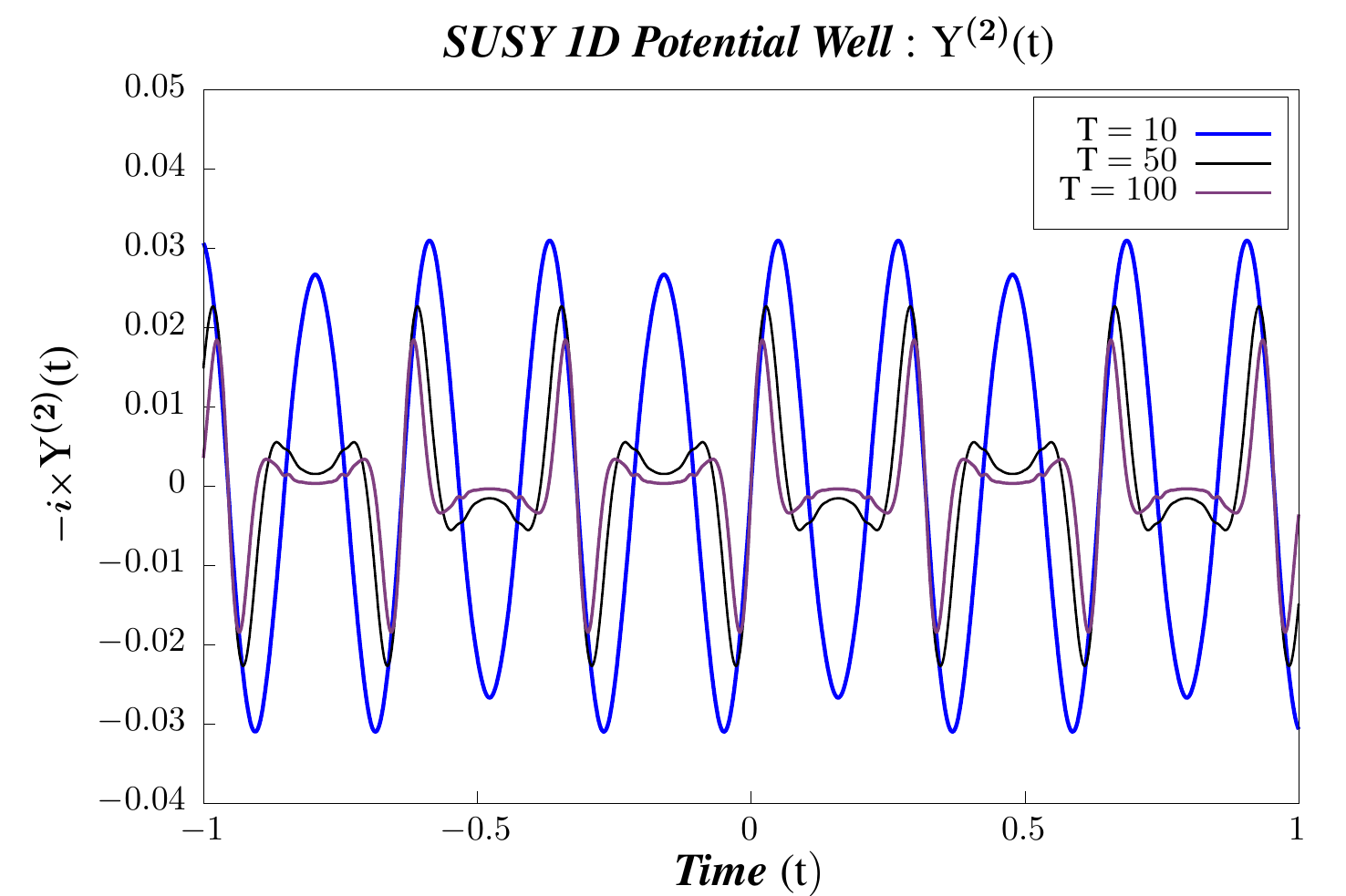}
	\caption{Supersymmetric 1D Infinite Potential Well : Behavior of 2-pt canonical correlator ${Y^{(2)}(t_{1},t_{2}) = - \sum_{m}^{} e^{-\beta E_{m}} \braket{\Psi_{m} | [x(t_{1}),x({t_{2}})] | \Psi_{m}}}$ with time for different temperatures. We have chosen ${t_{1} - t_{2} = t}$ as there is only one relevant time parameter.}
	\label{fig:Y2IPW}
\end{figure}

\begin{figure}[h!]
	\centering
	\includegraphics[width=17cm,height=13cm]{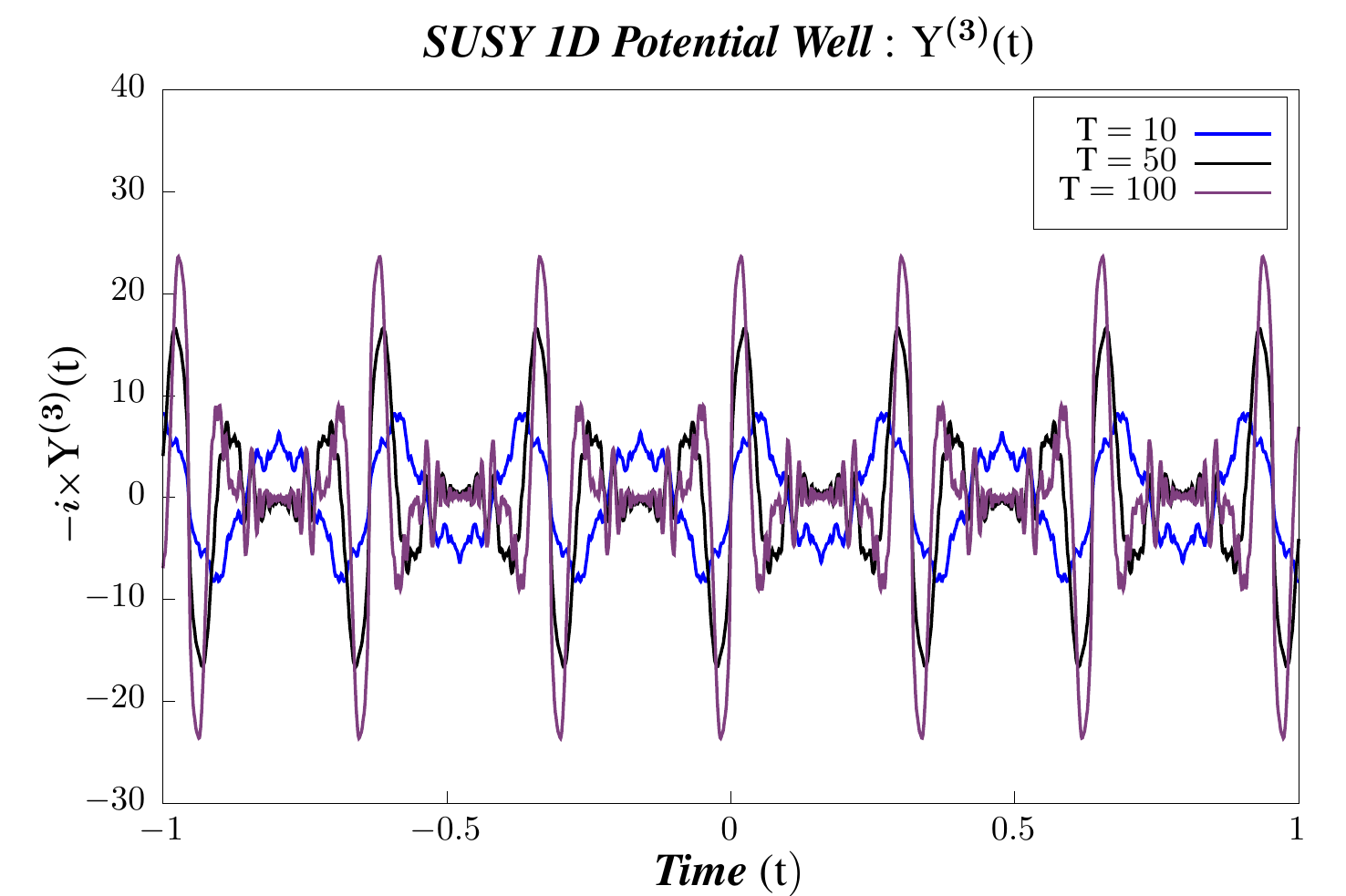}
	\caption{Supersymmetric 1D Infinite Potential Well : Behavior of 2-pt canonical correlator ${Y^{(3)}(t_{1},t_{2}) = - \sum_{m}^{} e^{-\beta E_{m}} \braket{\Psi_{m} | [p(t_{1}),p({t_{2}})] | \Psi_{m}}}$ with time for different temperatures. We have chosen ${t_{1} - t_{2} = t}$ as there is only one relevant time parameter.}
	\label{fig:Y3IPW}
\end{figure}

\begin{figure}[h!]
	\centering
	\includegraphics[width=17cm,height=8.7cm]{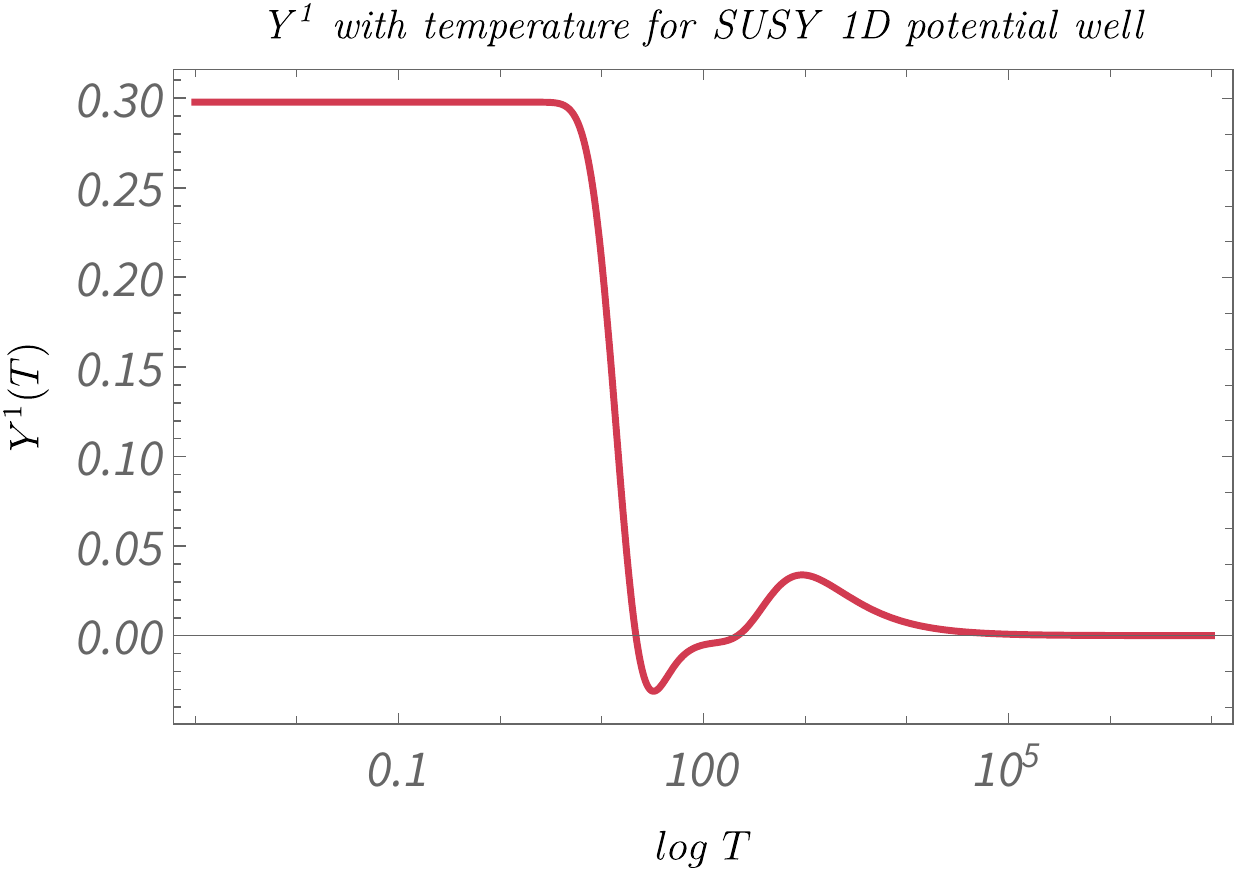}
	\caption{Supersymmetric 1D Infinite Potential Well : Behavior of 2-pt canonical correlators with temperature for a particular value of the time interval . We have chosen ${t_{1} - t_{2} = t}$ as there is only one relevant time parameter.}
	\label{fig:IPWtemp2ptY1}
\end{figure}

\begin{figure}[h!]
	\centering
	\includegraphics[width=17cm,height=8.7cm]{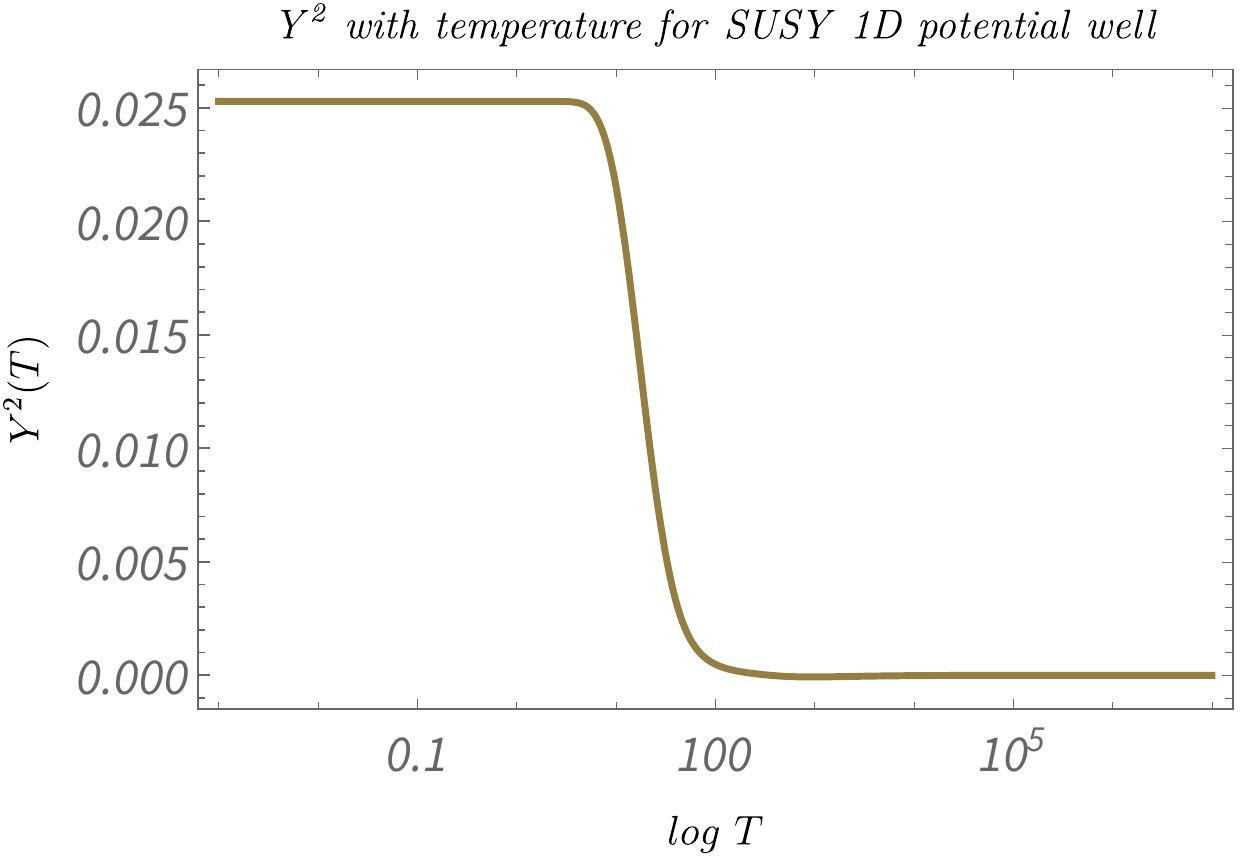}
	\caption{Supersymmetric 1D Infinite Potential Well : Behavior of 2-pt canonical correlators with temperature for a particular value of the time interval . We have chosen ${t_{1} - t_{2} = t}$ as there is only one relevant time parameter.}
	\label{fig:IPWtemp2ptY2}
\end{figure}

\begin{figure}[h!]
	\centering
	\includegraphics[width=17cm,height=8.7cm]{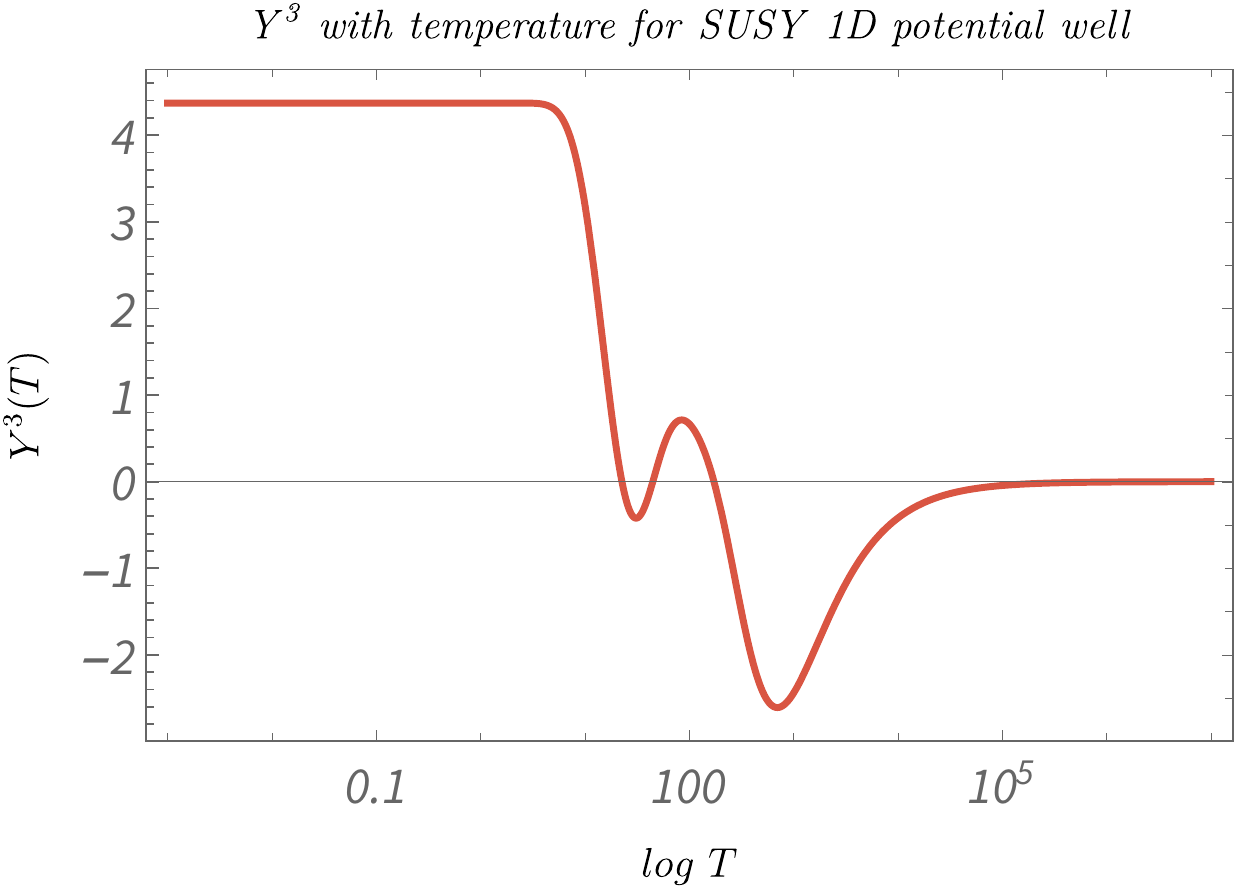}
	\caption{Supersymmetric 1D Infinite Potential Well : Behavior of 2-pt canonical correlators with temperature for a particular value of the time interval . We have chosen ${t_{1} - t_{2} = t}$ as there is only one relevant time parameter.}
	\label{fig:IPWtemp2ptY3}
\end{figure}

\begin{figure}[t]
	\centering
	\includegraphics[width=17cm,height=13cm]{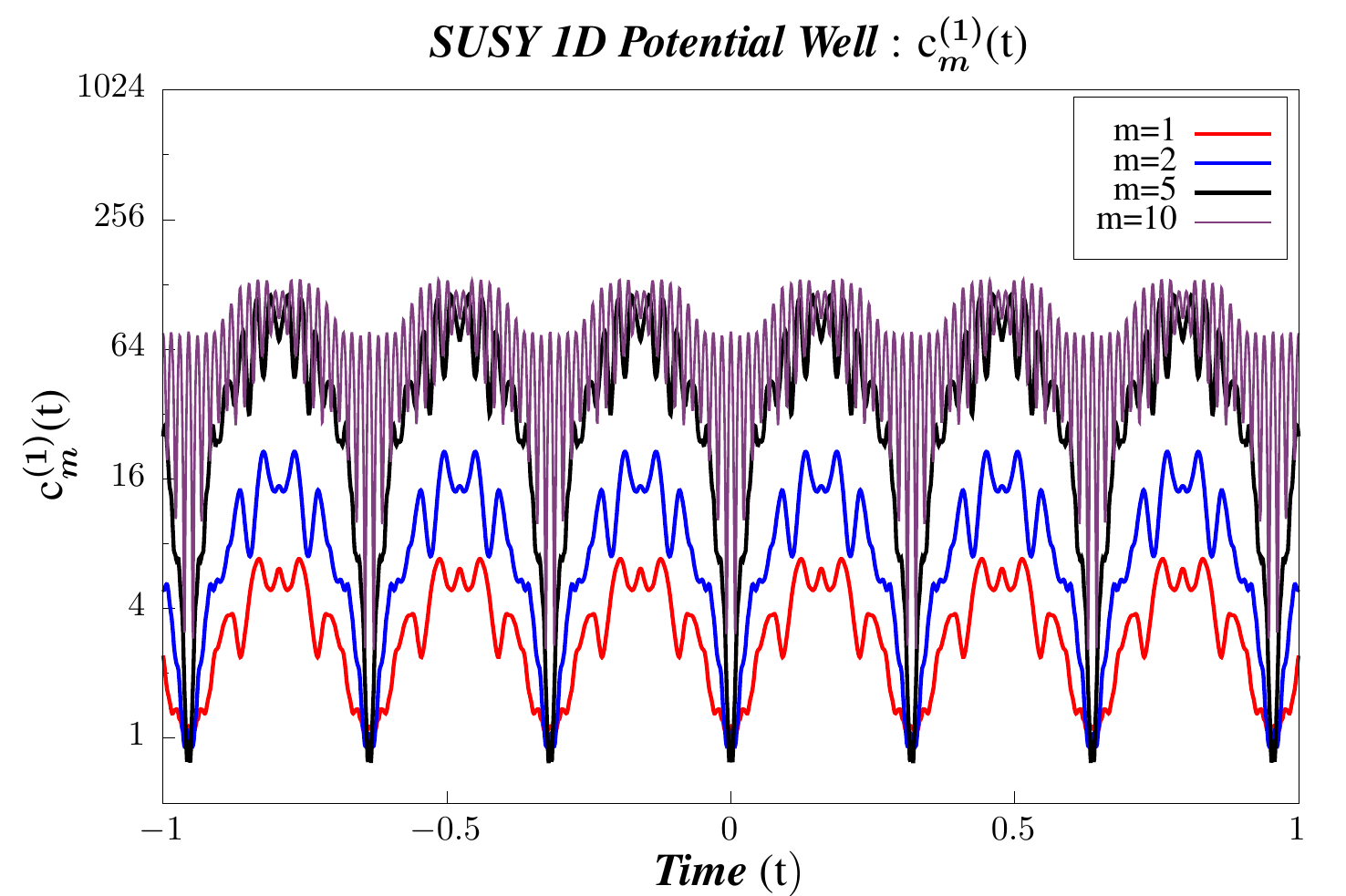}
	\caption{Supersymmetric 1D Infinite Potential Well : Behavior of 4-pt micro-canonical correlator ${c_{m}^{(1)}(t_{1},t_{2}) = - \braket{\Psi_{m} | [x(t_{1}),p({t_{2}})]^{2} | \Psi_{m}}}$ with time for different ${m}$. We have chosen ${t_{1} - t_{2} = t}$ as there is only one relevant time parameter.}
	\label{fig:c1IPW}
\end{figure}
\begin{figure}[!htb]
	\centering
	\includegraphics[width=17cm,height=13cm]{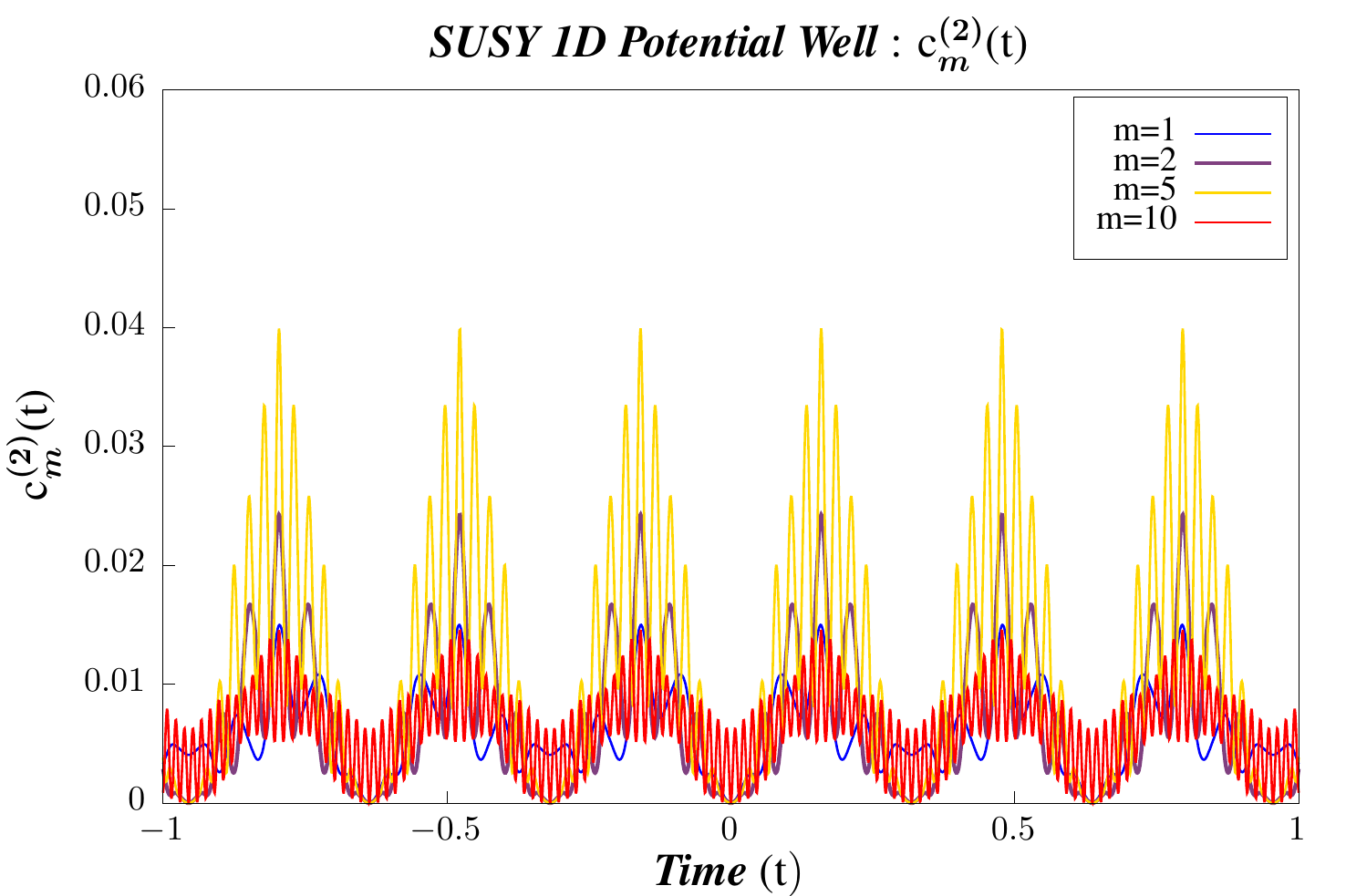}
	\caption{Supersymmetric 1D Infinite Potential Well : Behavior of 4-pt micro-canonical correlator ${c_{m}^{(2)}(t_{1},t_{2}) = - \braket{\Psi_{m} | [x(t_{1}),x({t_{2}})]^{2} | \Psi_{m}}}$ with time for different ${m}$. We have chosen ${t_{1} - t_{2} = t}$ as there is only one relevant time parameter.}
	\label{fig:c2IPW}
\end{figure}
\begin{figure}[!htb]
	\centering
	\includegraphics[width=17cm,height=13cm]{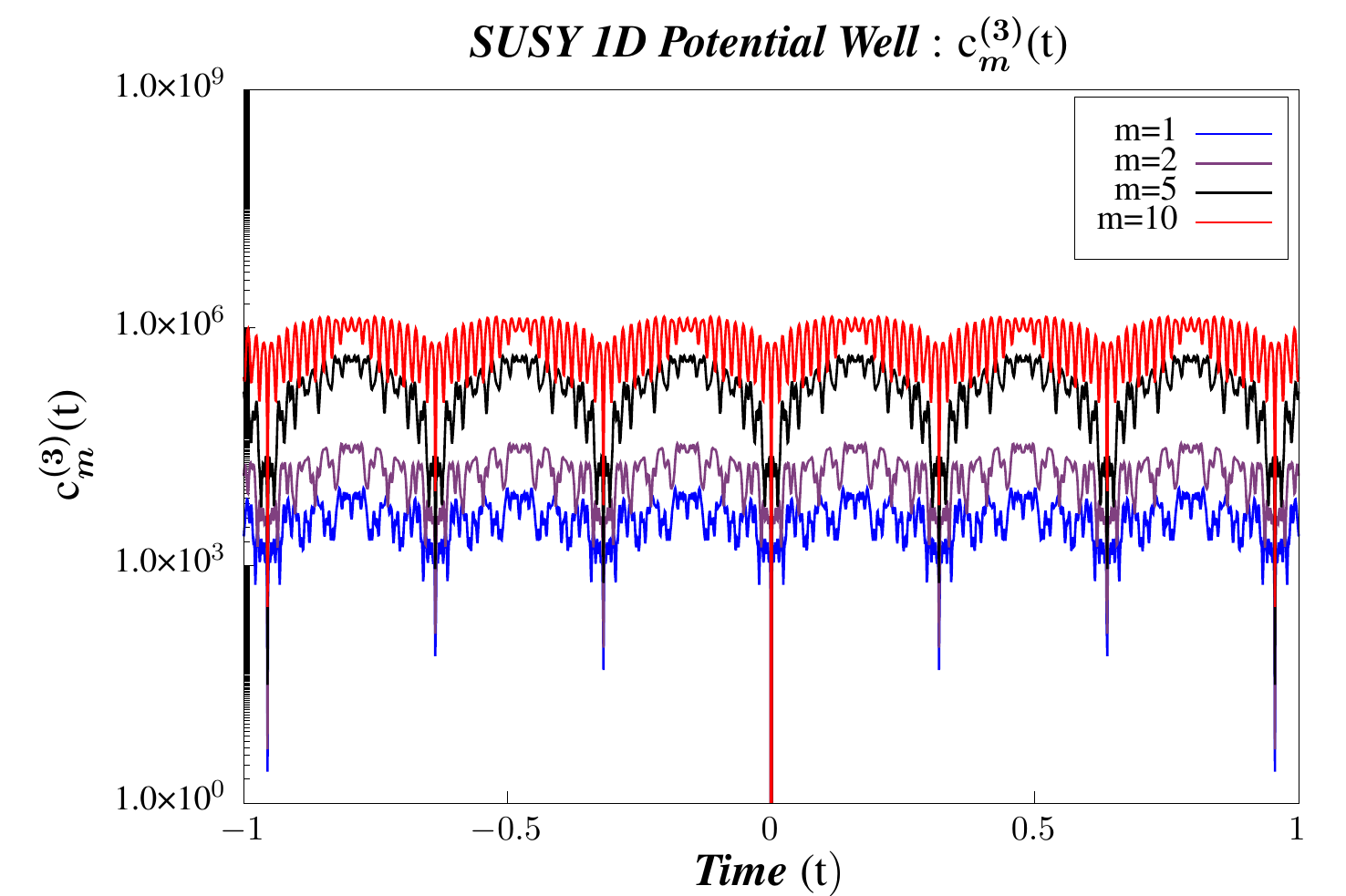}
	\caption{Supersymmetric 1D Infinite Potential Well : Behavior of 4-pt micro-canonical correlator ${c_{m}^{(3)}(t_{1},t_{2}) = - \braket{\Psi_{m} | [p(t_{1}),p({t_{2}})]^{2} | \Psi_{m}}}$ with time for different ${m}$. We have chosen ${t_{1} - t_{2} = t}$ as there is only one relevant time parameter.}
	\label{fig:c3IPW}
	
\end{figure}

\begin{figure}[!htb]
	\centering
	\includegraphics[width=17cm,height=13cm]{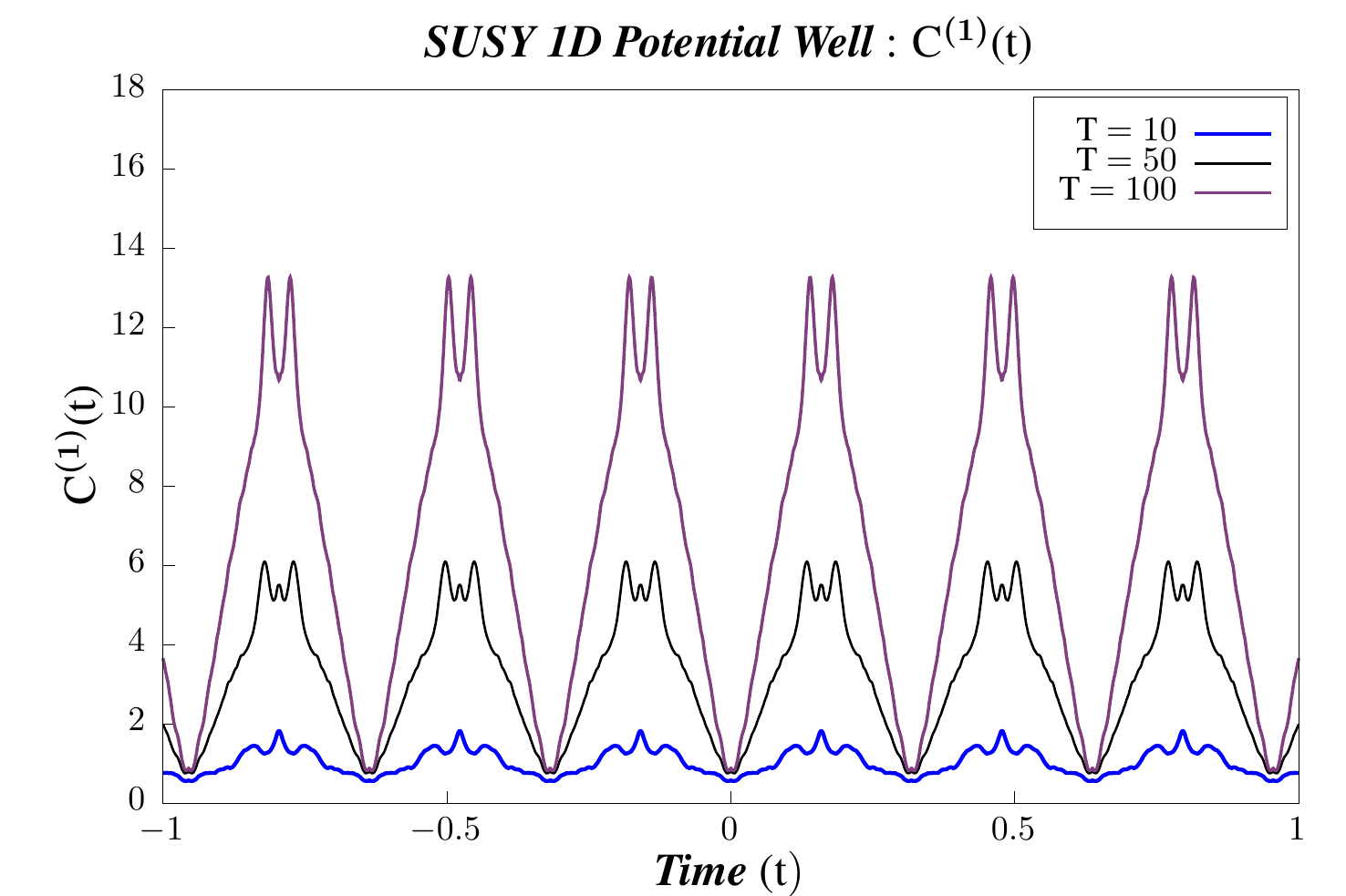}
	\caption{Supersymmetric 1D Infinite Potential Well : Behavior of 4-pt canonical correlator ${C^{(1)}(t_{1},t_{2}) = - \sum_{m}^{} e^{- \beta E_{m}} \braket{\Psi_{m} | [x(t_{1}),p({t_{2}})]^{2} | \Psi_{m}}}$ with time for different temperatures. We have chosen ${t_{1} - t_{2} = t}$ as there is only one relevant time parameter.}
	\label{fig:C1IPW}
\end{figure}

\begin{figure}[!htb]
	\centering
	\includegraphics[width=17cm,height=13cm]{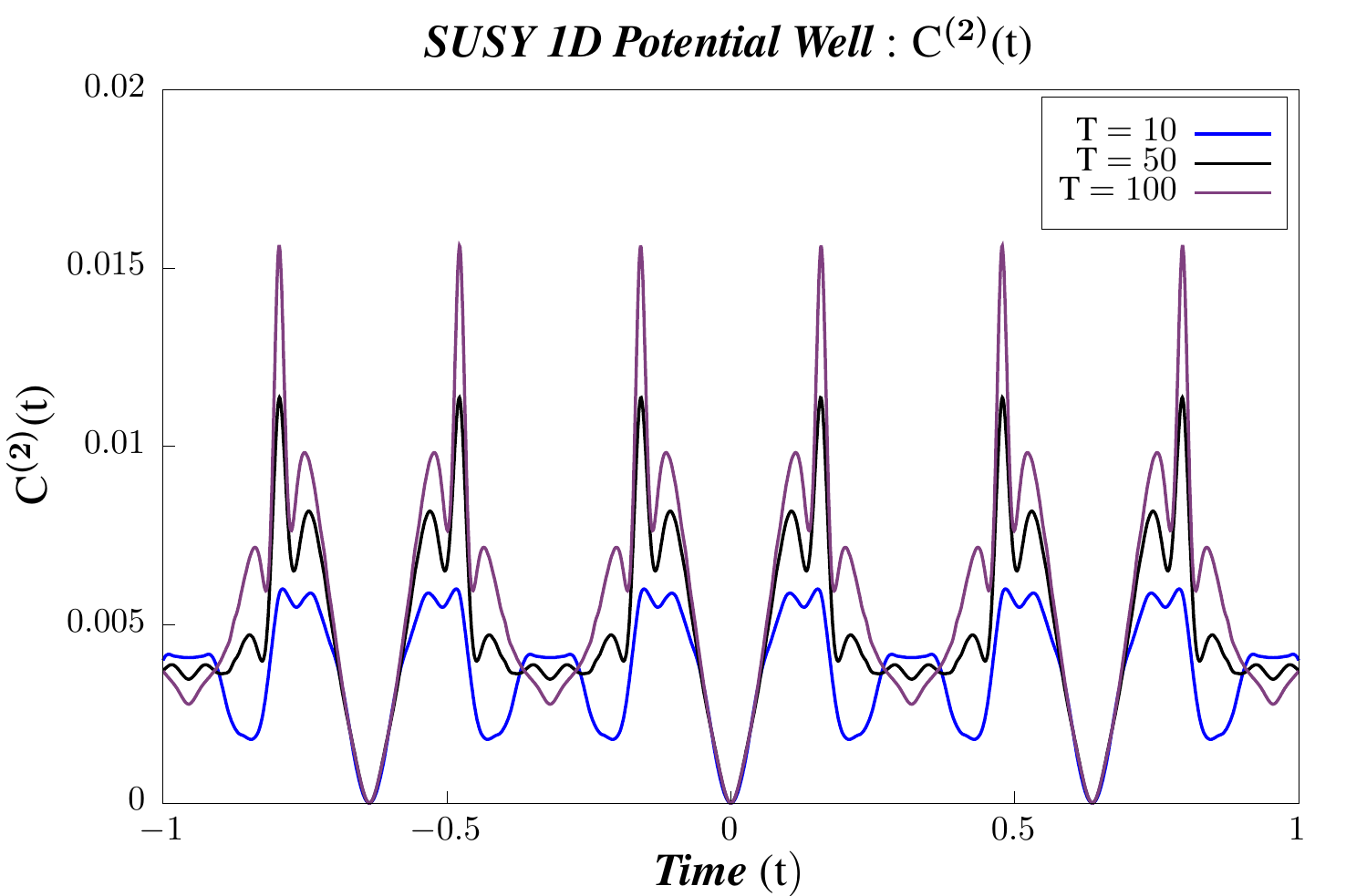}
	\caption{Supersymmetric 1D Infinite Potential Well : Behavior of 4-pt canonical correlator ${C^{(2)}(t_{1},t_{2}) = - \sum_{m}^{} e^{- \beta E_{m}} \braket{\Psi_{m} | [x(t_{1}),x({t_{2}})]^{2} | \Psi_{m}}}$ with time for different temperatures. We have chosen ${t_{1} - t_{2} = t}$ as there is only one relevant time parameter.}
	\label{fig:C2IPW}
\end{figure}

\begin{figure}[h]
	\centering
	\includegraphics[width=17cm,height=13cm]{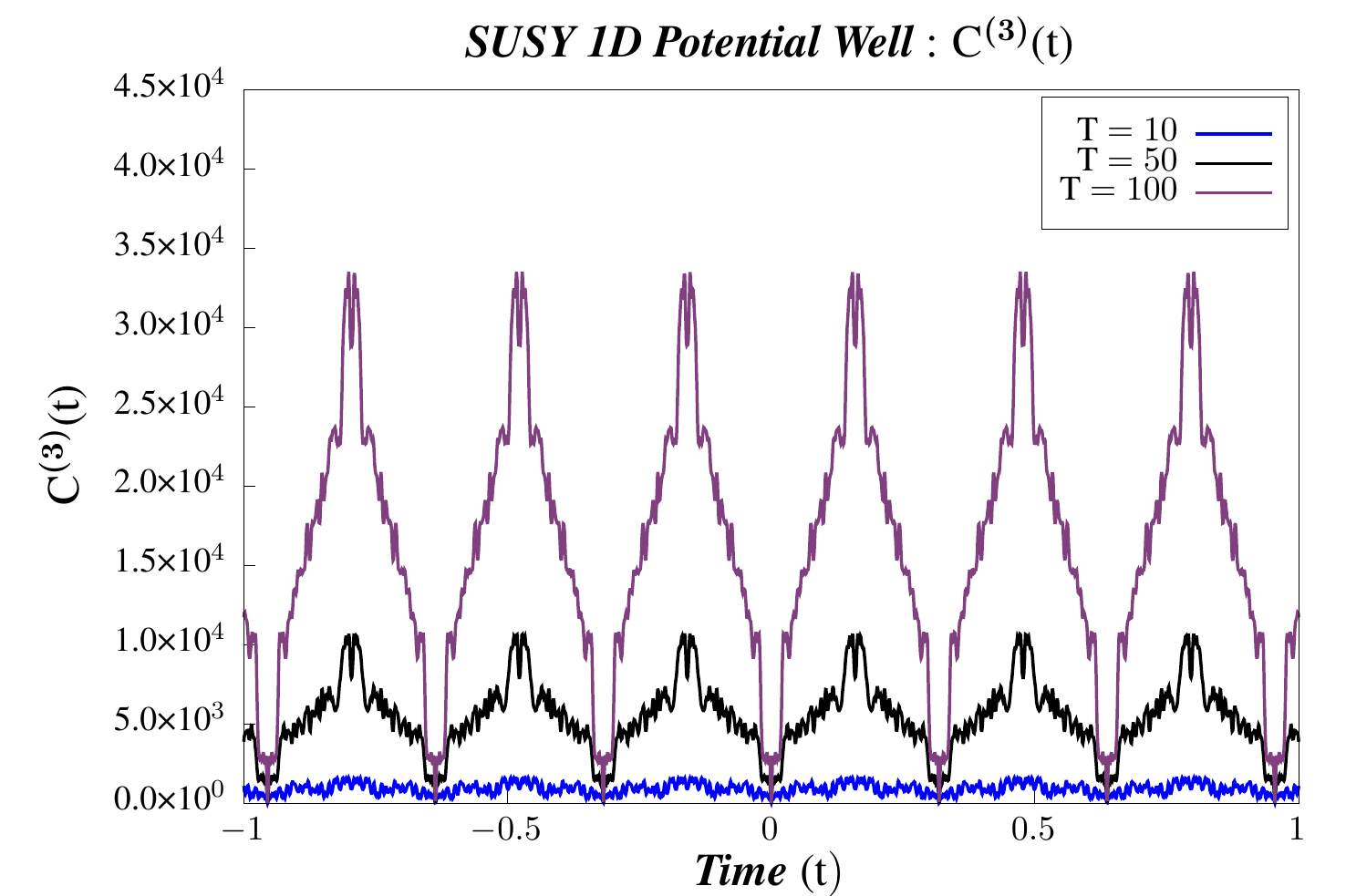}
	\caption{Supersymmetric 1D Infinite Potential Well : Behavior of 4-pt canonical correlator ${C^{(3)}(t_{1},t_{2}) = - \sum_{m}^{} e^{- \beta E_{m}} \braket{\Psi_{m} | [p(t_{1}),p({t_{2}})]^{2} | \Psi_{m}}}$ with time for different temperatures. We have chosen ${t_{1} - t_{2} = t}$ as there is only one relevant time parameter.}
	\label{fig:C3IPW}
\end{figure}

\begin{figure}[htb!]
	\centering
	\includegraphics[width=17cm,height=8.7cm]{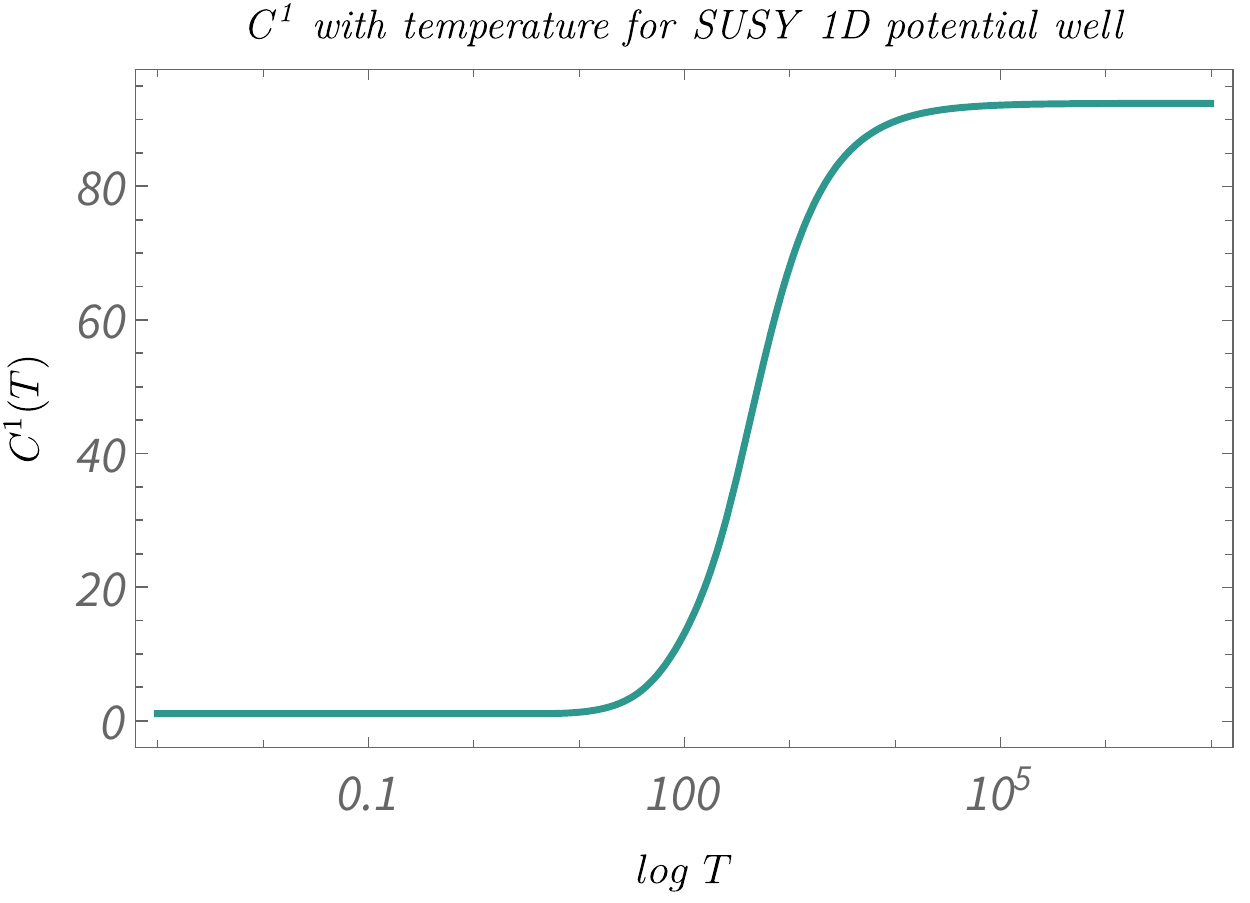}
	\caption{Supersymmetric 1D Infinite Potential Well : Behavior of 4-pt canonical correlators with temperature for a particular value of the time interval . We have chosen ${t_{1} - t_{2} = t}$ as there is only one relevant time parameter.}
	\label{fig:IPWtemp4ptC1}
\end{figure}

\begin{figure}[htb!]
	\centering
	\includegraphics[width=17cm,height=8.7cm]{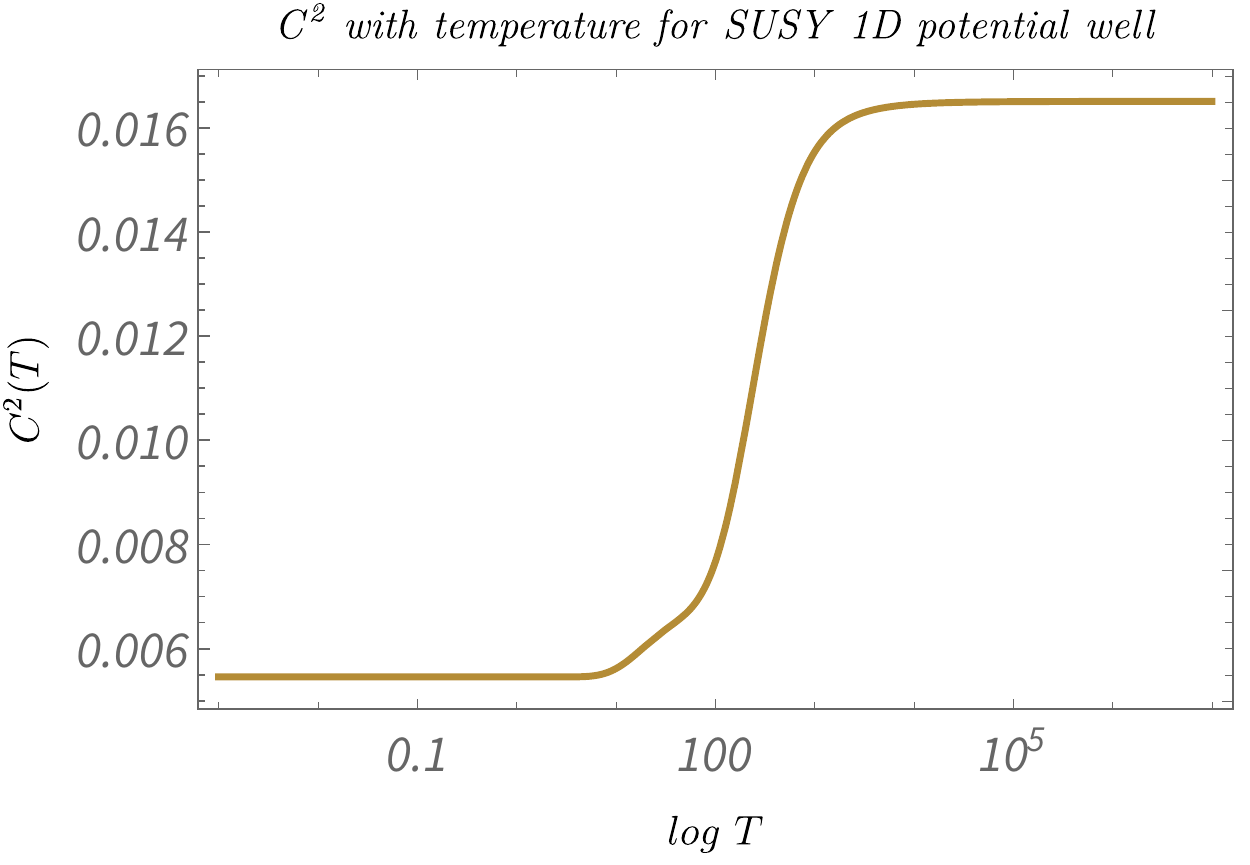}
	\caption{Supersymmetric 1D Infinite Potential Well : Behavior of 4-pt canonical correlators with temperature for a particular value of the time interval . We have chosen ${t_{1} - t_{2} = t}$ as there is only one relevant time parameter.}
	\label{fig:IPWtemp4ptC2}
\end{figure}

\begin{figure}[htb!]
	\centering
	\includegraphics[width=17cm,height=8.7cm]{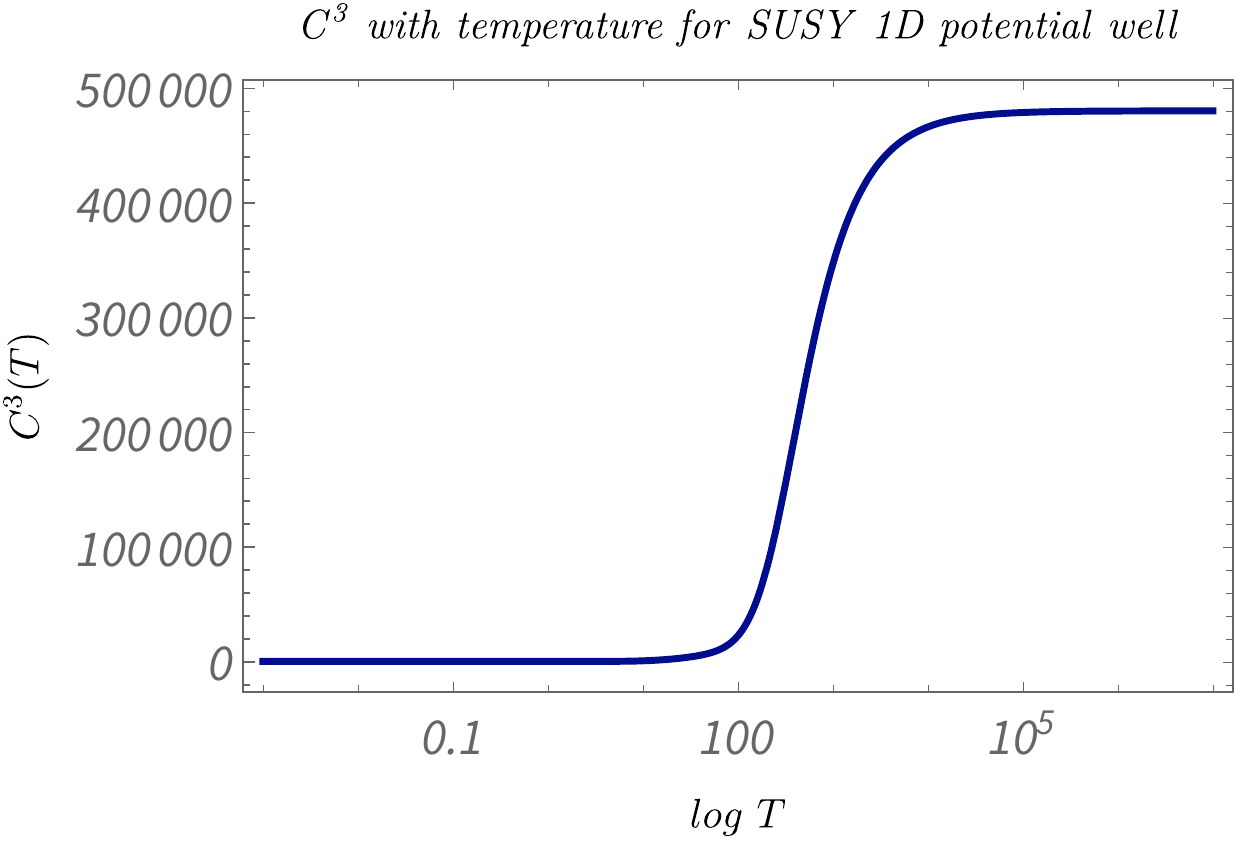}
	\caption{Supersymmetric 1D Infinite Potential Well : Behavior of 4-pt canonical correlators with temperature for a particular value of the time interval . We have chosen ${t_{1} - t_{2} = t}$ as there is only one relevant time parameter.}
	\label{fig:IPWtemp4ptC3}
\end{figure}

\noindent
For numerical evaluation we have chosen : ${L = 1\ \&\ 2m = 1}$, where ${L}$ is the length of the box in which a particle of mass ${m}$ is confined. We also consider ${ \hbar = 1}$. \\

\begin{itemize}

\item In \Cref{fig:y1IPW,,fig:y2IPW,,fig:y3IPW} we perform the \textit{Study A} on the 2-pt micro-canonical correlators ${y_{m}^{(1,2,3)}(t_{1},t_{2})}$ for Supersymmetric 1D Infinite Potential Well.

\begin{itemize}

\item We observe that the correlators ${y_{m}^{(1,2,3)}(t_{1},t_{2})}$ are periodic and that their periodicity does not vary with the state. For the correlator ${y_{m}^{(1)}(t_{1},t_{2})}$ the periodicity is ${\Delta t \simeq 0.65}$. At present, there are no studies for the non-Supersymmetric case for ${y_{m}^{(2,3)}(t_{1},t_{2})}$ and we plan to do the same in a work which is to appear very shortly. 

\item The amplitude of the correlators increases with increasing ${m}$ which primarily comes about because we have : $${-i \times y_{m}^{(1)}(t_{1},t_{2}) \sim E_{mk} = E_{m} - E_{k} = \pi^{2} \left( m^{2} - k^{2} \right)}$$ in the correlator's eigenstate representation which increases with increasing state number.

\item It is also observed that with higher and higher excited state the number of nodes increases and takes on the shape of a wave-packet. This is so because as we go to higher and higher excited states the high frequency modes become more and more prominent. So the comparison of the scaling of the number of nodes here with non-Supersymmetric case is also important job to perform in near future. 

\item In the insets of \Cref{fig:y1IPW,,fig:y2IPW,,fig:y3IPW} we have also plotted ${y_{10}^{(1,2,3)}(t_{1},t_{2})}$ to draw a contrast of the boundary / truncation state with the other states. The ${m=10}$ correlators are lacking in feature, sometimes deceptively so, as compared to the other states which should come as no surprise because we have set our truncation at ${N_{\textnormal{trunc}} = 10}$. Furthermore, this state appears to violate the properties shown by other intermediate states but in fact this is merely an artefact of ${m=10}$ being the truncation state and that contribution of states with ${m>10}$ could not be accommodated in the calculations for ${m=10}$. 

\item The correlators ${y_{m}^{(2,3)}(t_{1},t_{2})}$ largely follow the same patterns and behaviour as shown by ${y_{m}^{(1)}(t_{1},t_{2})}$ with two exceptions. First, the amplitude for ${y_{m}^{(2)}(t_{1},t_{2})}$ correlator is suppressed whereas that of ${y_{m}^{(3)}(t_{1},t_{2})}$ is amplified, both by a factor of ${\mathcal{O}(10^{1})}$, as compared to ${y_{m}^{(1)}(t_{1},t_{2})}$. This comes from the absence of ${E_{mk}}$ factor in ${y_{m}^{(2)}(t_{1},t_{2})}$ and the presence of an additional ${E_{mk}}$ factor in ${y_{m}^{(3)}(t_{1},t_{2})}$ as compared to ${y_{m}^{(1)}(t_{1},t_{2})}$. Second, contrasting behaviour in the symmetry properties in ${t}$. Whereas, ${y_{m}^{(1)}(t_{1},t_{2})}$ is symmetric in ${t}$, ${y_{m}^{(2,3)}(t_{1},t_{2})}$ are anti-symmetric.

\end{itemize}

\item We present the results of performing \textit{Study B} and \textit{Study C} on 2-pt canonical correlators ${Y^{(i)}(t_{1},t_{2})}$ as follows

\begin{itemize}
	\item In \Cref{fig:Y1IPW,,fig:Y2IPW,,fig:Y3IPW} we perform \textit{Study B} on the 2-pt canonical correlators ${Y^{(1,2,3)}(t_1,t_2)}$. We observe that the correlators shows periodic behaviour for the different chosen values of temperature. We observe that for $Y^{1}(t_1,t_2)$, the mid temperature value 50 shows the minimum amplitude, whereas lower value of temperature (10) has greater amplitude. However there is a sudden increase in the amplitude of the correlator for temperatures in the higher value range as can be seen from \Cref{fig:Y1IPW}. In correlators $Y^{2}(t_1,t_2)$ and $Y^{3}(t_1,t_2)$ however we follow a gradual pattern of decreasing amplitude with increasing in temperature as observed from \Cref{fig:Y2IPW} and \Cref{fig:Y3IPW}. To have a better understanding of the temperature dependence of the 2-pt correlators we plot them with varying temperature keeping the time constant.

\item In \Cref{fig:IPWtemp2ptY1,,fig:IPWtemp2ptY2,,fig:IPWtemp2ptY3} we perform \textit{Study C} on the 2-pt canonical correlators ${Y^{(1,2,3)}(t_1,t_2)}$. We plot, respectively, ${Y^{(1,2,3)}(t_{1},t_{2})}$ which are the thermal or canonical correlators corresponding to ${y_{m}^{(1,2,3)}(t_{1},t_{2})}$ respectively with respect to temperature. It is clearly visible that for very low temperatures the canonical correlators are constant. However after a certain value of the temperature the correlators decays rapidly and falls off to zero within a small temperature range 
\end{itemize}

\begin{itemize}
	\item In \Cref{fig:c1IPW,,fig:c2IPW,,fig:c3IPW} we perform the \textit{Study A} on the 4-pt micro-canonical correlators ${c_{m}^{(1,2,3)}(t_{1},t_{2})}$ for Supersymmetric 1D Infinite Potential Well.
	
	\item We observe that the correlators ${c_{m}^{(1,2,3)}(t_{1},t_{2})}$ are periodic and that their periodicity does not vary with the state. For the correlator ${c_{m}^{(1)}(t_{1},t_{2})}$ the periodicity is ${\Delta t \simeq 0.35}$ which is roughly half of the corresponding 2-pt micro-canonical correlator. We note that this is approximately the same periodicity, within numerical error, observed in the case of non-Supersymmetric 1D Infinite Potential Well as obtained by Hashimoto et al. \cite{Hashimoto:2017oit}. Hence, we conclude that introducing Supersymmetry in integrable QMcal models does not affect the periodicity of four-pt micro-canonical correlators. At present, there are no studies for the non-Supersymmetric case for ${c_{m}^{(2,3)}(t_{1},t_{2})}$ and we plan to do the same in a work which is to appear very shortly.

	\item Other properties of ${c^{(i)}_{m}(t_{1},t_{2})}$ are much like ${y^{(i)}_{m}(t_{1},t_{2})}$. We observe similar increase in the amplitude of the correlators with increasing ${m}$. The scaling of amplitudes in the case of 4-pt micro-canonical correlators is with a factor of ${\mathcal{O}(10^{4})}$ instead of a factor ${\mathcal{O}(10^{1})}$ as is the case with ${y_{m}^{(i)}(t_{1},t_{2})}$ such that the relative order of amplitudes can be arranged as : ${c_{m}^{(3)}(t_{1},t_{2}) > c_{m}^{(2)}(t_{1},t_{2}) > c_{m}^{(1)}(t_{1},t_{2})}$.
	
	\item All ${c_{m}^{(i)}(t_{1},t_{2})}$ are symmetric about ${t=0}$ which means that to these 4-pt micro-canonical correlators it does not matter whether ${t_{1} > t_{2}}$ or ${t_{1} < t_{2}}$. 
	
	\item In \Cref{fig:C1IPW,,fig:C2IPW,fig:C3IPW} we perform \textit{Study B} on the 4-pt canonical correlators. We plot the 4-pt canonical correlators for three different temperatures. We observe that all the 4-pt correlators shows periodic behaviour irrespective of the value of the temperature. it is also observed that for each correlator the amplitude increases with the increasing temperature. To have a better understanding of the temperature dependence of the 4-pt correlators we plot them with varying temperature keeping the time constant.
	
	\item In \Cref{fig:IPWtemp4ptC1,,fig:IPWtemp4ptC2,,fig:IPWtemp4ptC3} we perform \textit{Study C} on the 4 point canonical correlators. we plot, respectively, ${C^{(1,2,3)}(t_{1},t_{2})}$ which are the thermal or canonical correlators corresponding to ${c_{m}^{(1,2,3)}(t_{1},t_{2})}$ respectively with respect to temperature. It is observed that for low temperatures the 4-pt correlators show negligible value. However after a certain threshold temperature the 4-pt correlators increases and then saturates to a certain finite value. 
	
\end{itemize}

\item The temperature dependent plots for the 2-pt and the 4-pt canonical correlators suggests that both the 2 and the 4-pt canonical correlators are essential if one wants to have a complete understanding of the phenomenon of Quantum randomness. The plots suggests that 2-pt correlators are a better probe for understanding Quantum randomness at low temperatures whereas at high temperatures it is actually the 4-pt correlators which is more significant. However in the mid-temperature range the 2-pt and the 4-pt correlators have exactly opposite behavior. Hence, to understand the significance of temperature in this range on any Supersymmetric Quantum mechanical model having knowledge of both 2 and 4-pt correlators is of utmost importance.
\end{itemize}

\textcolor{Sepia}{\subsection{\sffamily Supersymmetric 1D Harmonic oscillator}\label{ssec:qres_harmonicoscillator}}

\begin{figure}[!htb]
	\centering
	\includegraphics[width=17cm,height=8.7cm]{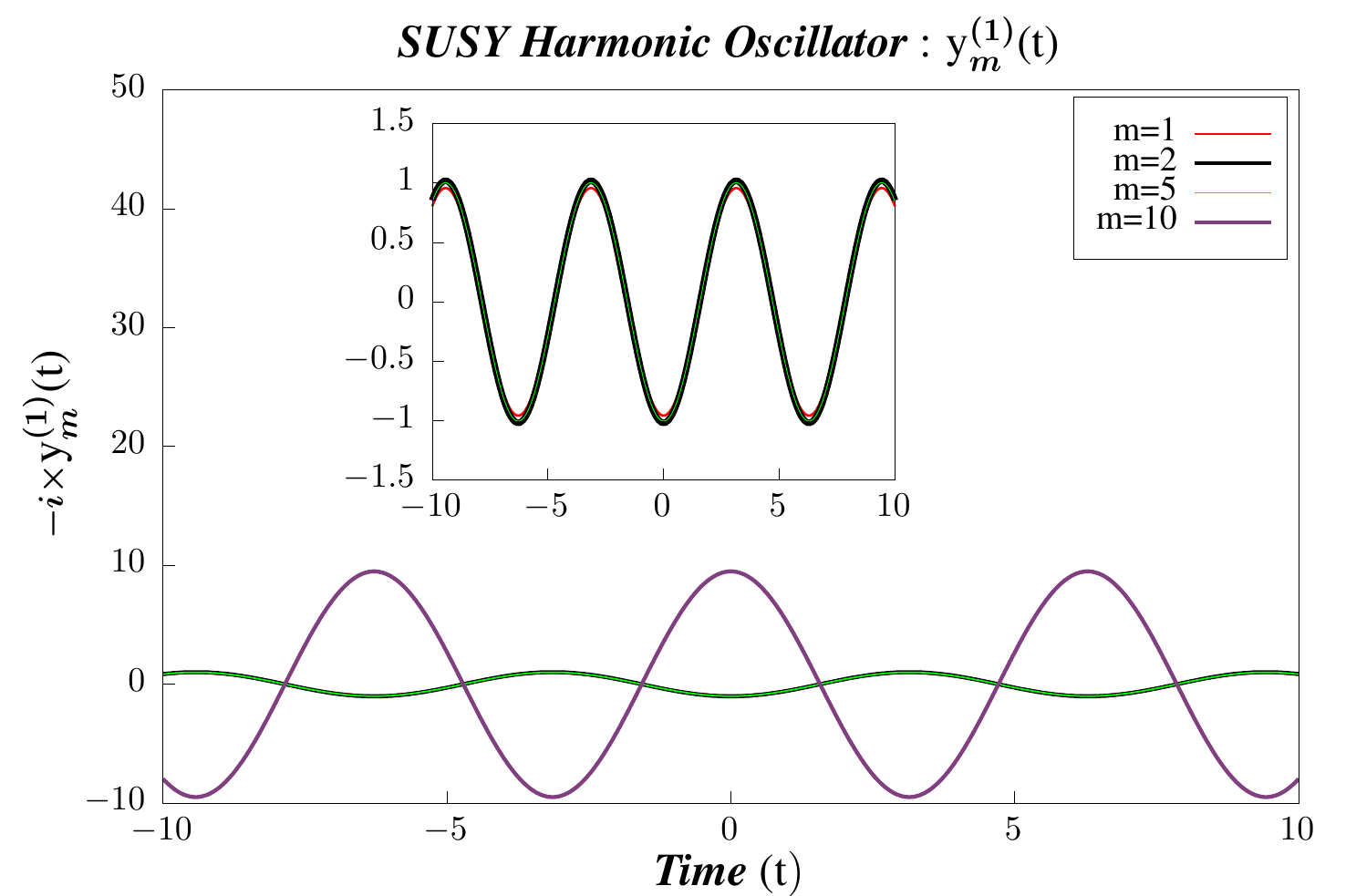}
	\caption{Supersymmetric Harmonic Oscillator : Behavior of 2-pt micro-canonical correlator ${y_{m}^{(1)}(t_{1},t_{2}) = - \braket{\Psi_{m} | [x(t_{1}),p({t_{2}})] | \Psi_{m}}}$ with time for different ${m}$. We have chosen ${t_{1} - t_{2} = t}$ as there is only one relevant time parameter.}
	\label{fig:y1HO}
\end{figure}

\begin{figure}[h]
	\centering
	\includegraphics[width=17cm,height=8.7cm]{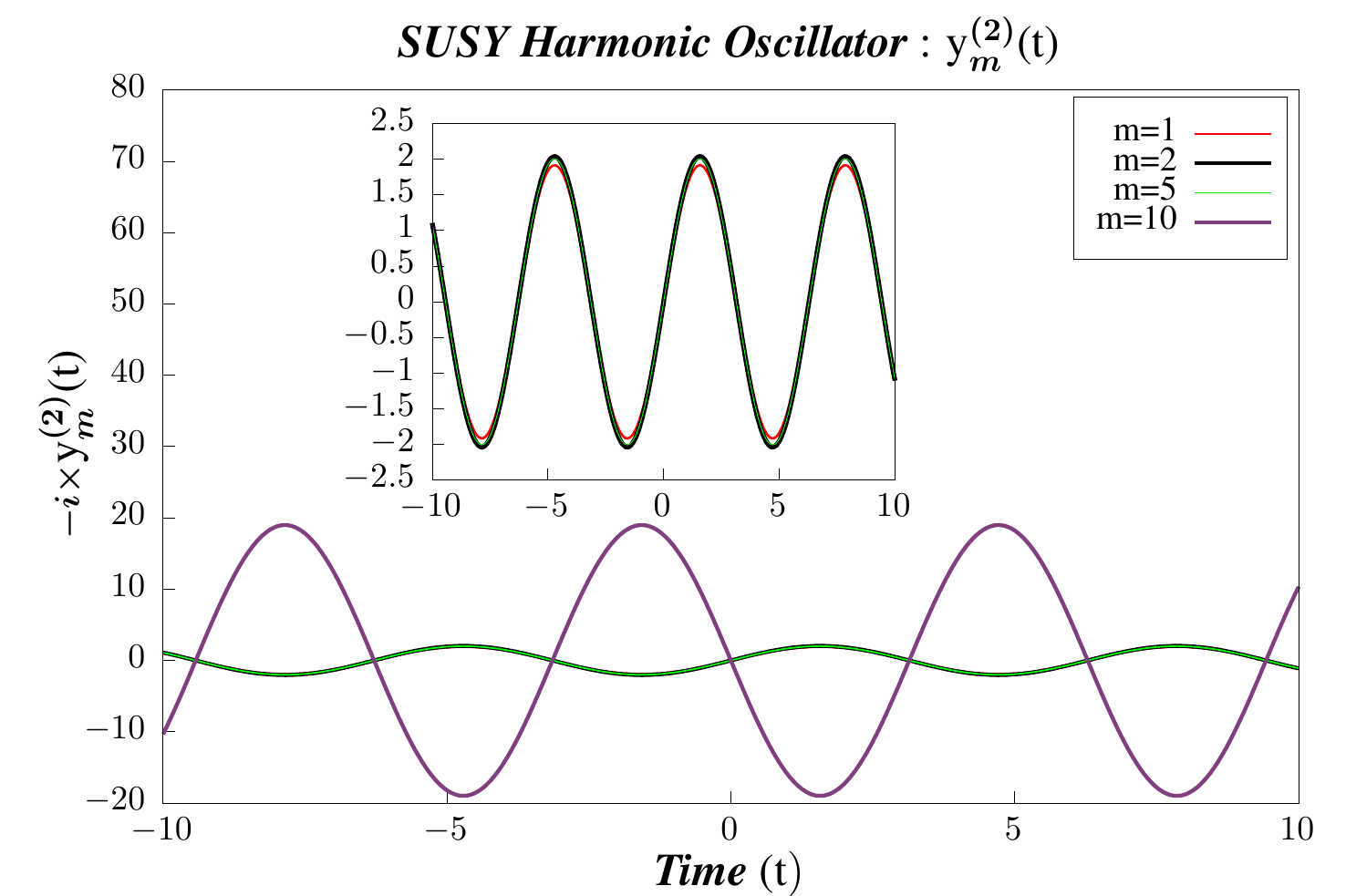}
	\caption{Supersymmetric Harmonic Oscillator : Behavior of 2-pt micro-canonical correlator ${y_{m}^{(2)}(t_{1},t_{2}) = - \braket{\Psi_{m} | [x(t_{1}),x({t_{2}})] | \Psi_{m}}}$ with time for different ${m}$. We have chosen ${t_{1} - t_{2} = t}$ as there is only one relevant time parameter.}
	\label{fig:y2HO}
\end{figure}

\begin{figure}[h]
	\centering
	\includegraphics[width=17cm,height=8.7cm]{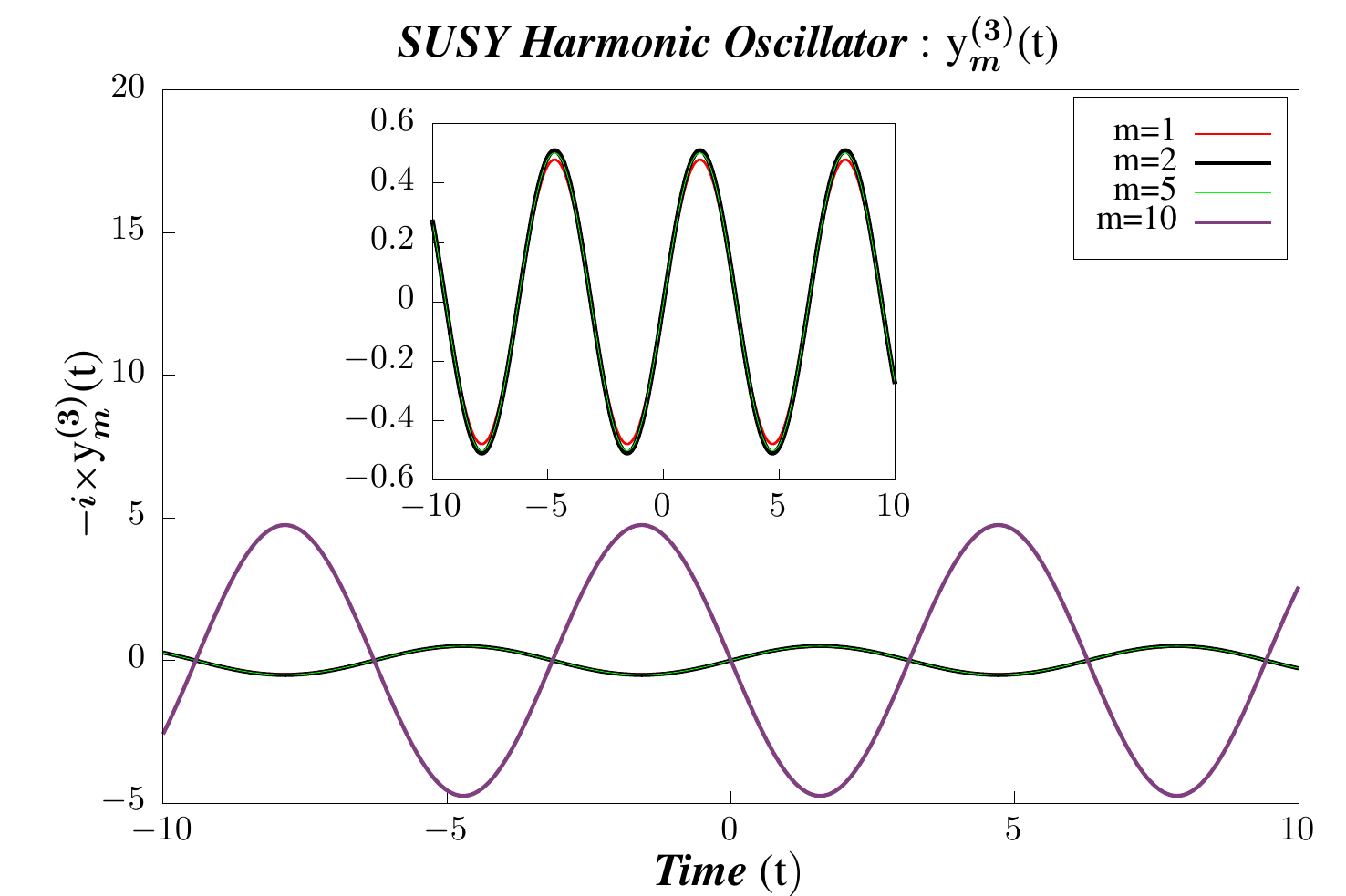}
	\caption{Supersymmetric Harmonic Oscillator : Behavior of 2-pt micro-canonical correlator ${y_{m}^{(3)}(t_{1},t_{2}) = - \braket{\Psi_{m} | [p(t_{1}),p({t_{2}})] | \Psi_{m}}}$ with time for different ${m}$. We have chosen ${t_{1} - t_{2} = t}$ as there is only one relevant time parameter.}
	\label{fig:y3HO}
\end{figure}

\begin{figure}[h]
	\centering
	\includegraphics[width=17cm,height=8.7cm]{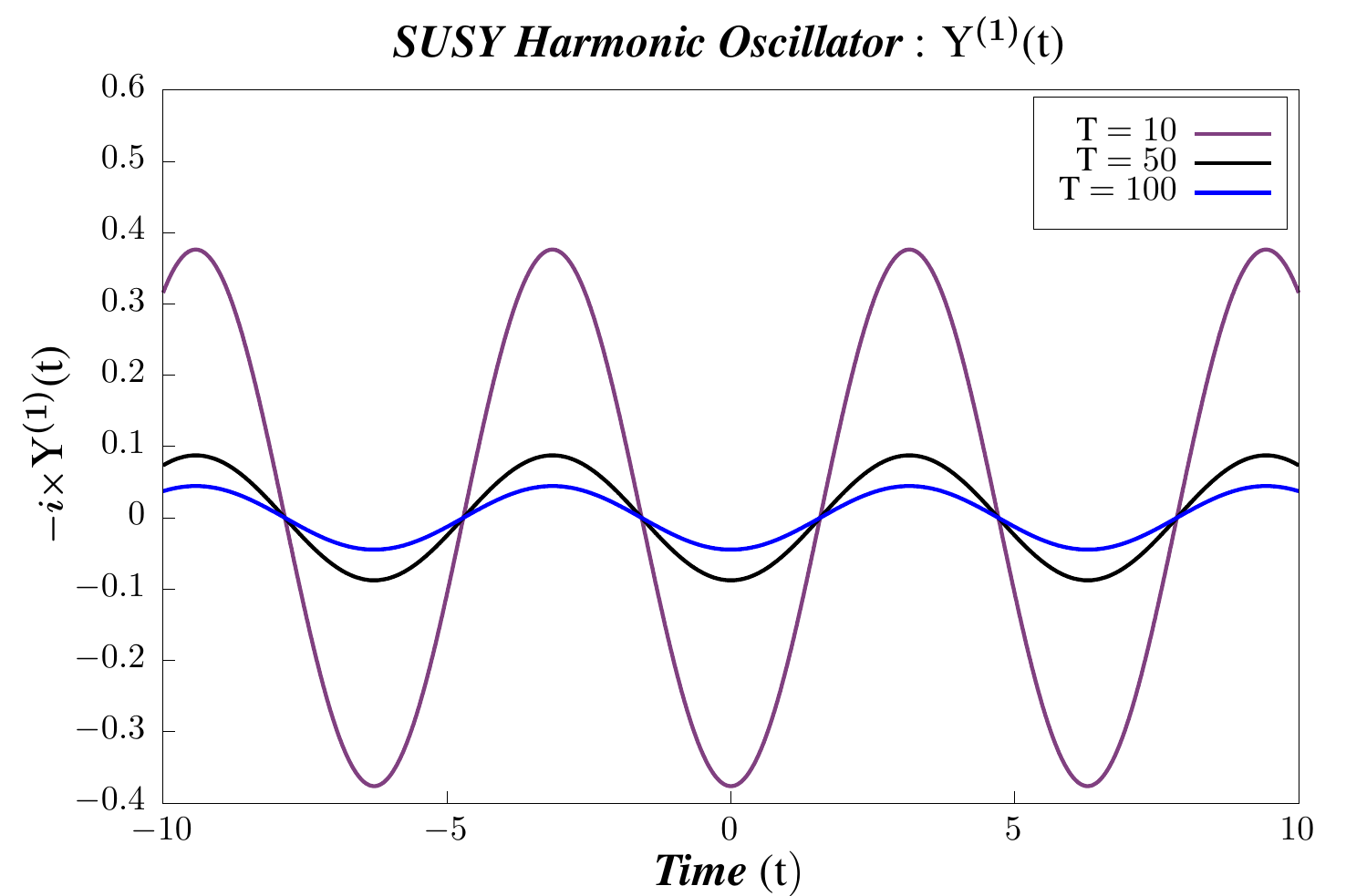}
	\caption{Supersymmetric Harmonic Oscillator : Behavior of 2-pt canonical correlator ${Y^{(1)}(t_{1},t_{2}) = - \sum_{m}^{} e^{-\beta E_{m}} \braket{\Psi_{m} | [x(t_{1}),p({t_{2}})] | \Psi_{m}}}$ with time for different temperatures. We have chosen ${t_{1} - t_{2} = t}$ as there is only one relevant time parameter.}
	\label{fig:Y1HO}
\end{figure}

\begin{figure}[h]
	\centering
	\includegraphics[width=17cm,height=8.7cm]{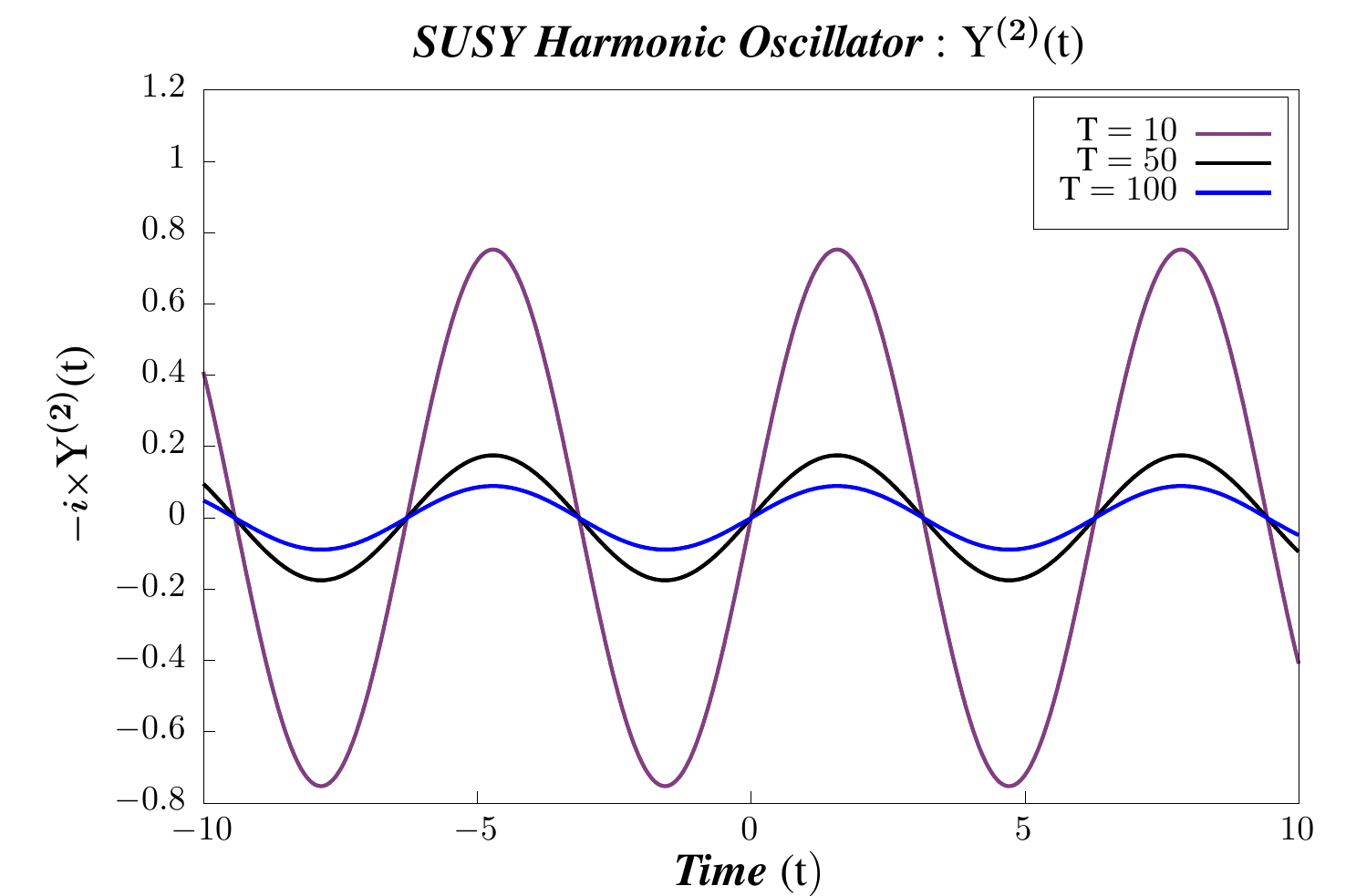}
	\caption{Supersymmetric Harmonic Oscillator : Behavior of 2-pt canonical correlator ${Y^{(2)}(t_{1},t_{2}) = - \sum_{m}^{} e^{-\beta E_{m}} \braket{\Psi_{m} | [x(t_{1}),x({t_{2}})] | \Psi_{m}}}$ with time for different temperatures. We have chosen ${t_{1} - t_{2} = t}$ as there is only one relevant time parameter.}
	\label{fig:Y2HO}
\end{figure}

\begin{figure}[h]
	\centering
	\includegraphics[width=17cm,height=8.7cm]{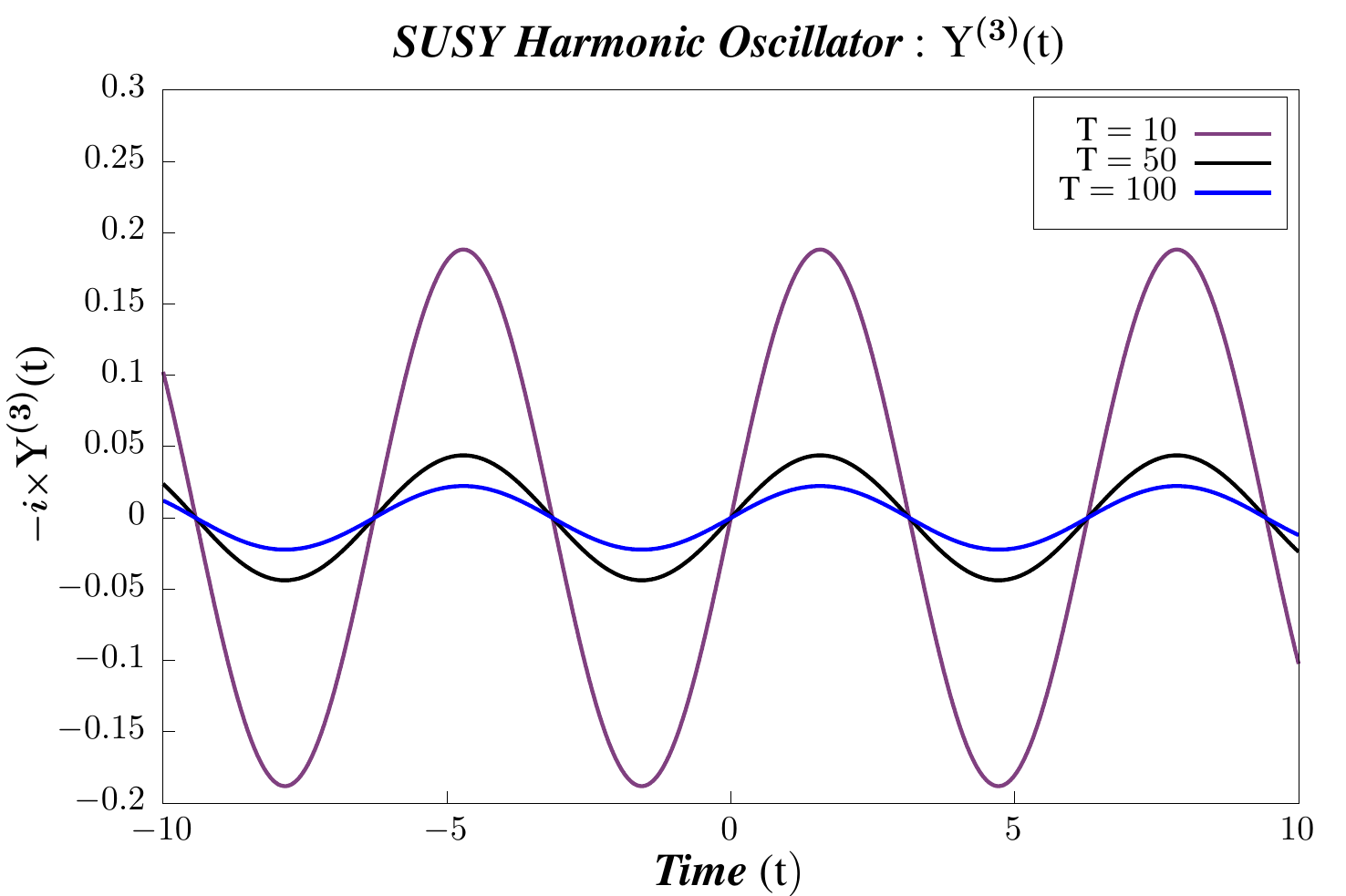}
	\caption{Supersymmetric Harmonic Oscillator : Behavior of 2-pt canonical correlator ${Y^{(3)}(t_{1},t_{2}) = - \sum_{m}^{} e^{-\beta E_{m}} \braket{\Psi_{m} | [p(t_{1}),p({t_{2}})] | \Psi_{m}}}$ with time for different temperatures. We have chosen ${t_{1} - t_{2} = t}$ as there is only one relevant time parameter.}
	\label{fig:Y3HO}
\end{figure}

\begin{figure}[h!]
	\centering
	\includegraphics[width=17cm,height=8.7cm]{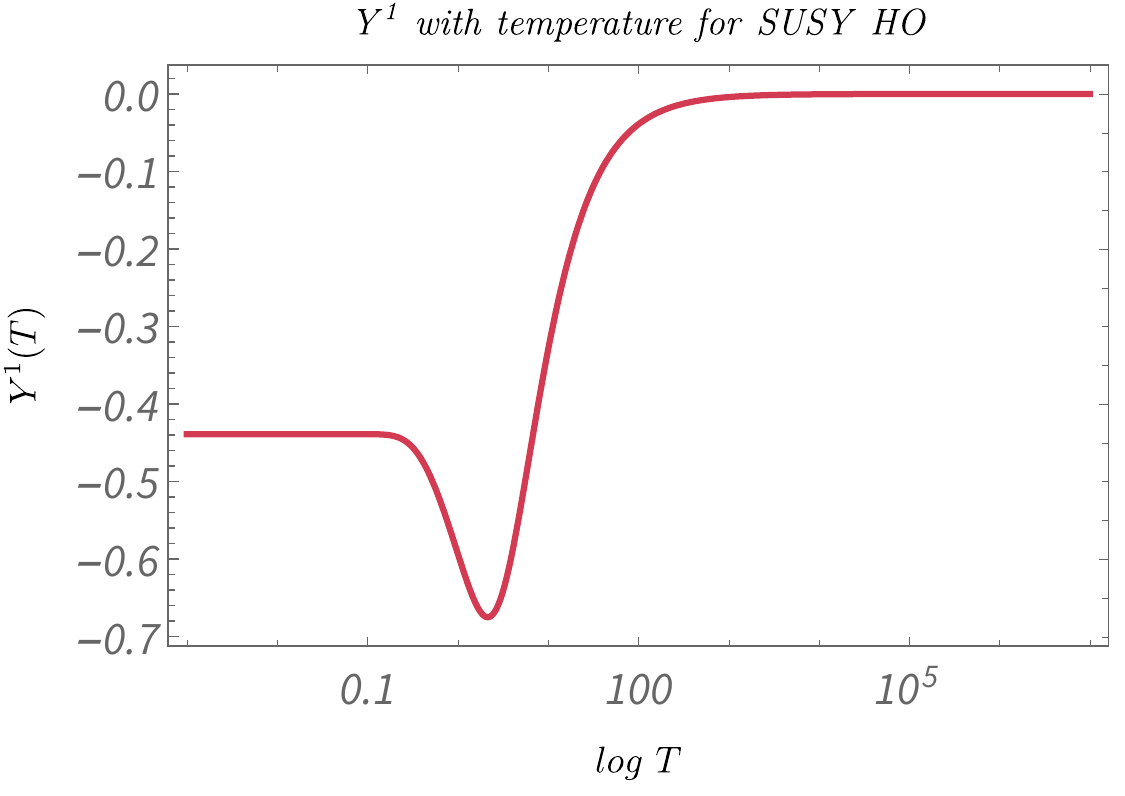}
	\caption{Supersymmetric 1D Harmonic Oscillator: Behavior of 2-pt canonical correlators with temperature for a particular value of the time interval . We have chosen ${t_{1} - t_{2} = t}$ as there is only one relevant time parameter.}
	\label{fig:HOtemp2ptY1}
\end{figure}

\begin{figure}[h!]
	\centering
	\includegraphics[width=17cm,height=8.7cm]{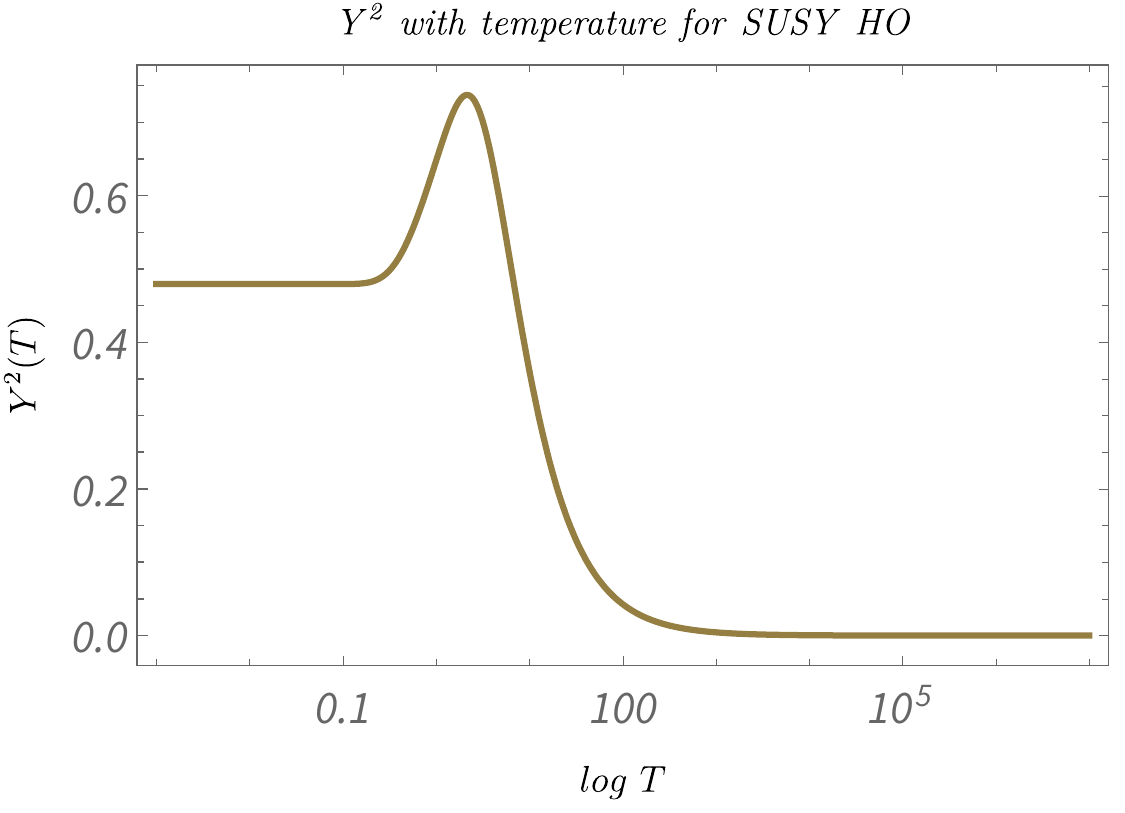}
	\caption{Supersymmetric 1D Harmonic Oscillator: Behavior of 2-pt canonical correlators with temperature for a particular value of the time interval . We have chosen ${t_{1} - t_{2} = t}$ as there is only one relevant time parameter.}
	\label{fig:HOtemp2ptY2}
\end{figure}

\begin{figure}[h!]
	\centering
	\includegraphics[width=17cm,height=8.7cm]{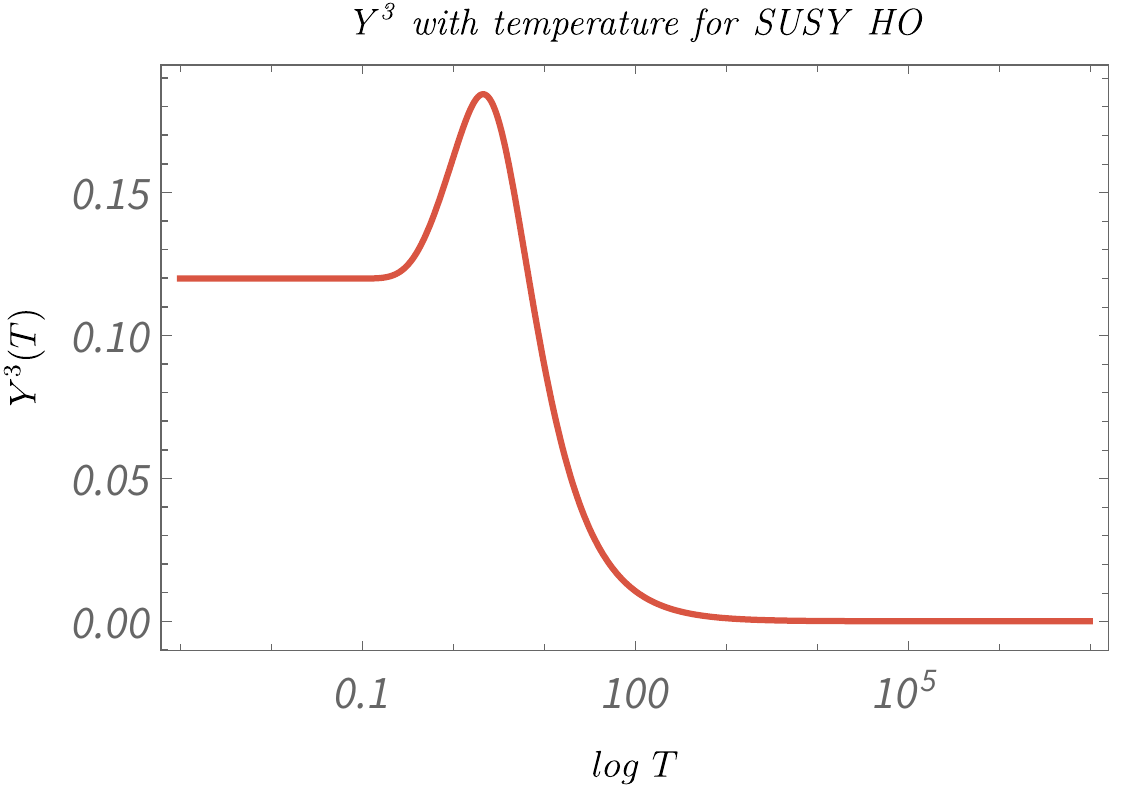}
	\caption{Supersymmetric 1D Harmonic Oscillator: Behavior of 2-pt canonical correlators with temperature for a particular value of the time interval . We have chosen ${t_{1} - t_{2} = t}$ as there is only one relevant time parameter.}
	\label{fig:HOtemp2ptY3}
\end{figure}

\begin{figure}[h]
	\centering
	\includegraphics[width=17cm,height=8.7cm]{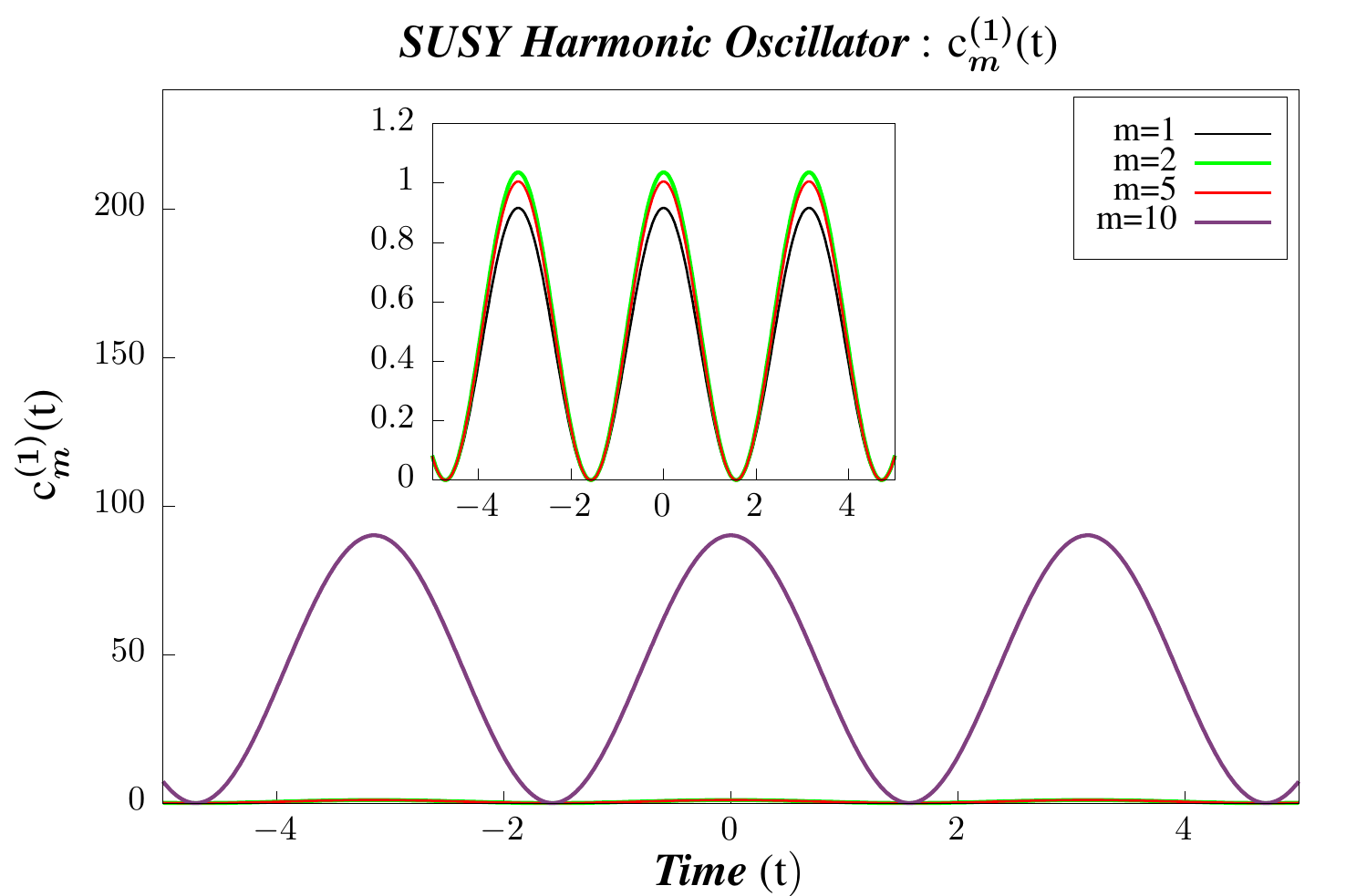}
	\caption{Supersymmetric Harmonic Oscillator : Behavior of 4-pt micro-canonical correlator ${c_{m}^{(1)}(t_{1},t_{2}) = - \braket{\Psi_{m} | [x(t_{1}),p({t_{2}})]^{2} | \Psi_{m}}}$ with time for different ${m}$. We have chosen ${t_{1} - t_{2} = t}$ as there is only one relevant time parameter.}
	\label{fig:c1HO}
\end{figure}

\begin{figure}[h]
	\centering
	\includegraphics[width=17cm,height=8.7cm]{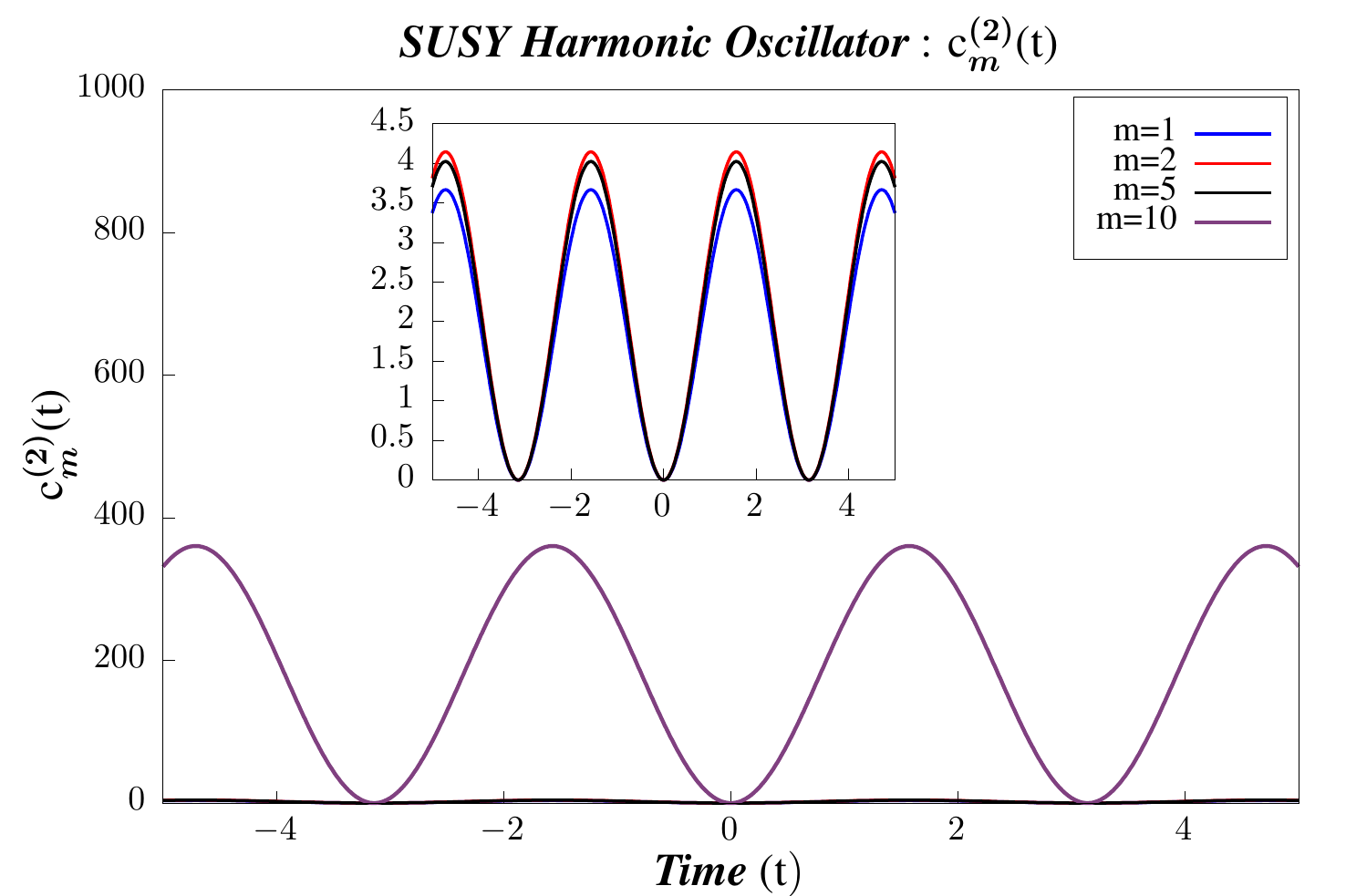}
	\caption{Supersymmetric Harmonic Oscillator : Behavior of 4-pt micro-canonical correlator ${c_{m}^{(2)}(t_{1},t_{2}) = - \braket{\Psi_{m} | [x(t_{1}),x({t_{2}})]^{2} | \Psi_{m}}}$ with time for different ${m}$. We have chosen ${t_{1} - t_{2} = t}$ as there is only one relevant time parameter.}
	\label{fig:c2HO}
\end{figure}

\begin{figure}[h]
	\centering
	\includegraphics[width=17cm,height=8.7cm]{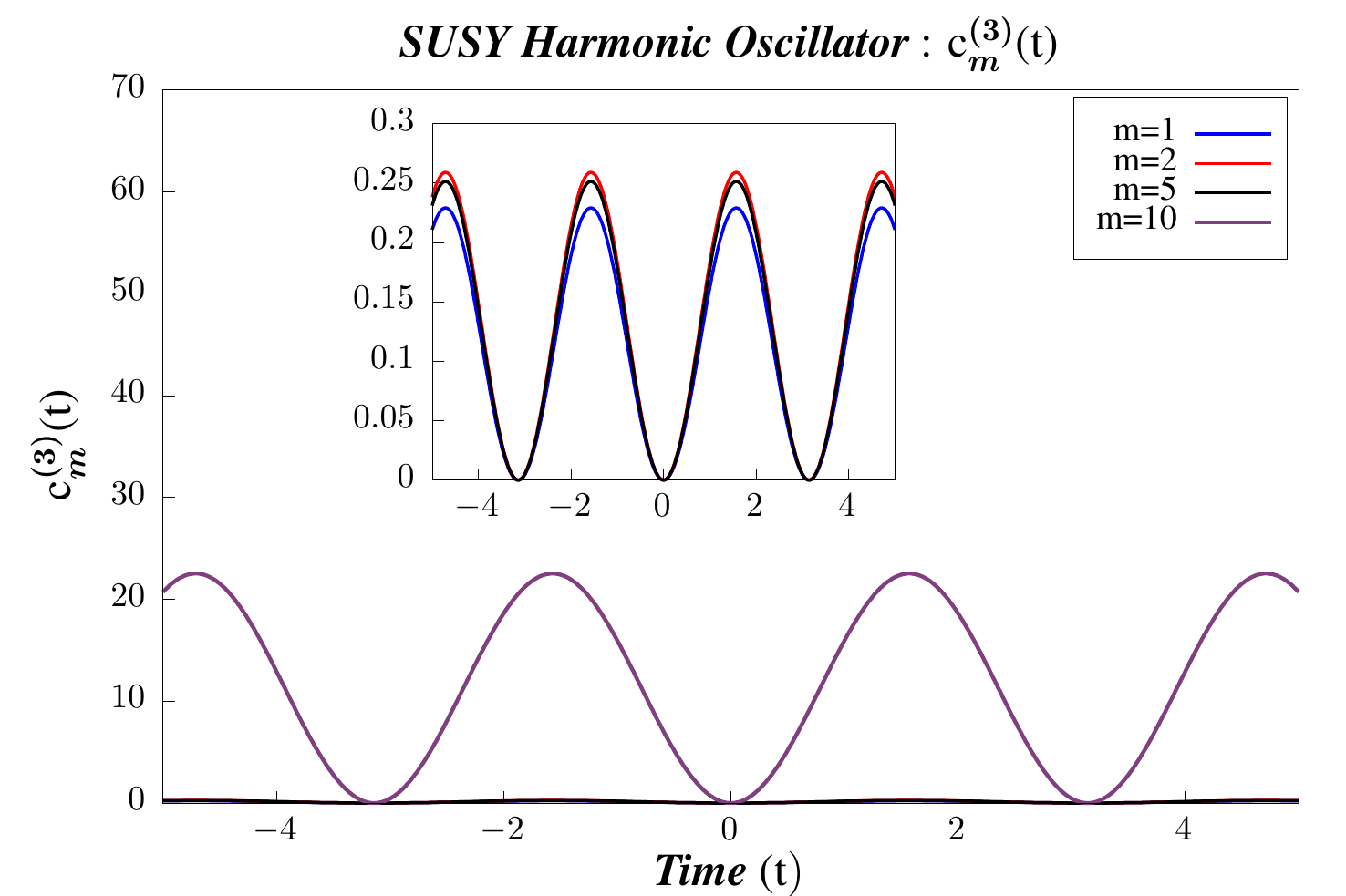}
	\caption{Supersymmetric Harmonic Oscillator : Behavior of 4-pt micro-canonical correlator ${c_{m}^{(3)}(t_{1},t_{2}) = - \braket{\Psi_{m} | [p(t_{1}),p({t_{2}})]^{2} | \Psi_{m}}}$ with time for different ${m}$. We have chosen ${t_{1} - t_{2} = t}$ as there is only one relevant time parameter.}
	\label{fig:c3HO}
\end{figure}

\begin{figure}[h]
	\centering
	\includegraphics[width=17cm,height=8.7cm]{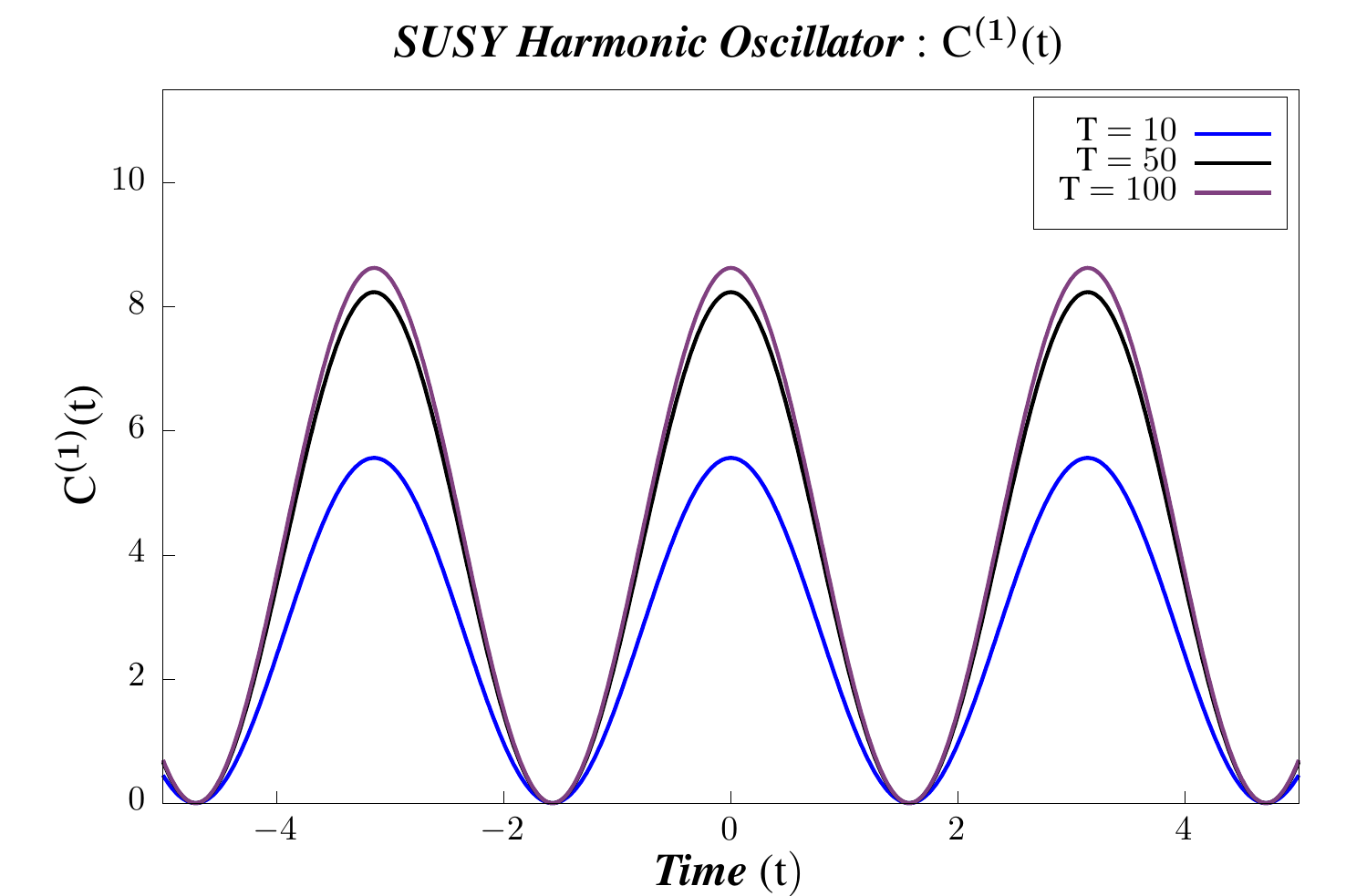}
	\caption{Supersymmetric Harmonic Oscillator : Behavior of 4-pt canonical correlator ${C^{(1)}(t_{1},t_{2}) = - \sum_{m}^{} e^{-\beta E_{m}} \braket{\Psi_{m} | [x(t_{1}),p({t_{2}})]^{2} | \Psi_{m}}}$ with time for different temperatures. We have chosen ${t_{1} - t_{2} = t}$ as there is only one relevant time parameter.}
	\label{fig:C1HO}
\end{figure}

\begin{figure}[h]
	\centering
	\includegraphics[width=17cm,height=8.7cm]{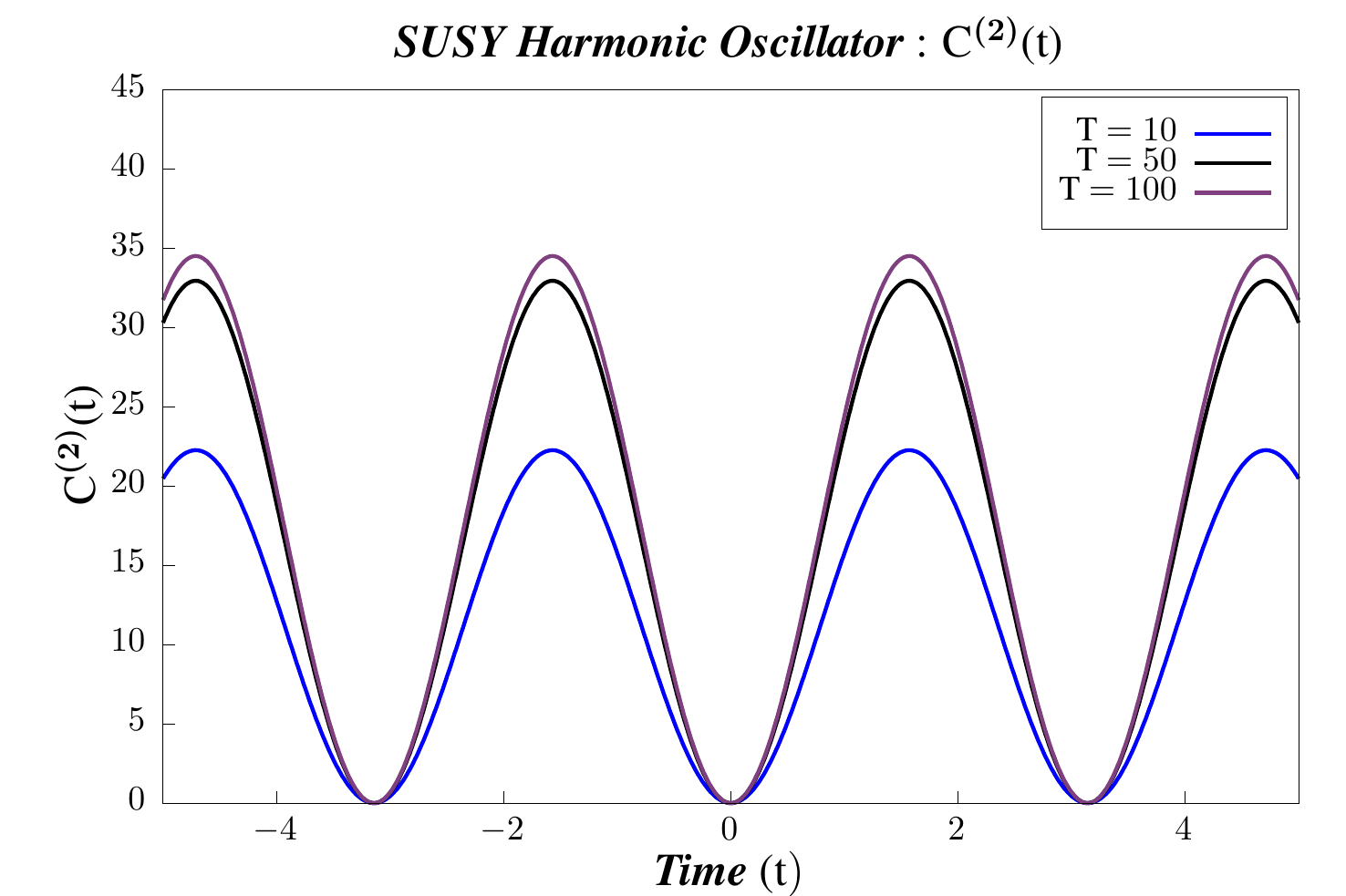}
	\caption{Supersymmetric Harmonic Oscillator : Behavior of 4-pt canonical correlator ${C^{(2)}(t_{1},t_{2}) = - \sum_{m}^{} e^{-\beta E_{m}} \braket{\Psi_{m} | [x(t_{1}),x({t_{2}})]^{2} | \Psi_{m}}}$ with time for different temperatures. We have chosen ${t_{1} - t_{2} = t}$ as there is only one relevant time parameter.}
	\label{fig:C2HO}
\end{figure}

\begin{figure}[h]
	\centering
	\includegraphics[width=17cm,height=8.7cm]{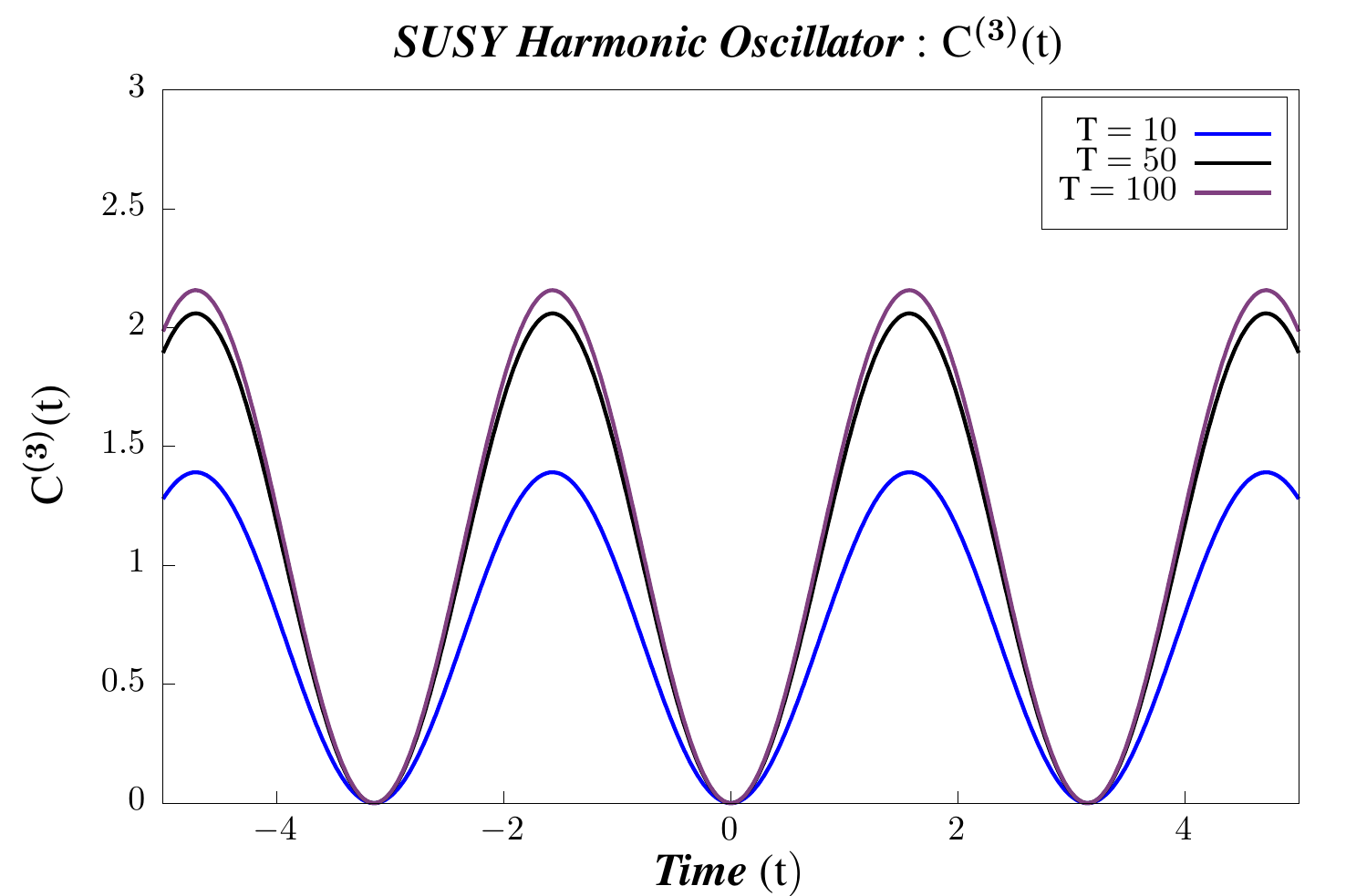}
	\caption{Supersymmetric Harmonic Oscillator : Behavior of 4-pt canonical correlator ${C^{(3)}(t_{1},t_{2}) = - \sum_{m}^{} e^{-\beta E_{m}} \braket{\Psi_{m} | [p(t_{1}),p({t_{2}})]^{2} | \Psi_{m}}}$ with time for different temperatures. We have chosen ${t_{1} - t_{2} = t}$ as there is only one relevant time parameter.}
	\label{fig:C3HO}
\end{figure}

\begin{figure}[htb!]
	\centering
	\includegraphics[width=17cm,height=8.7cm]{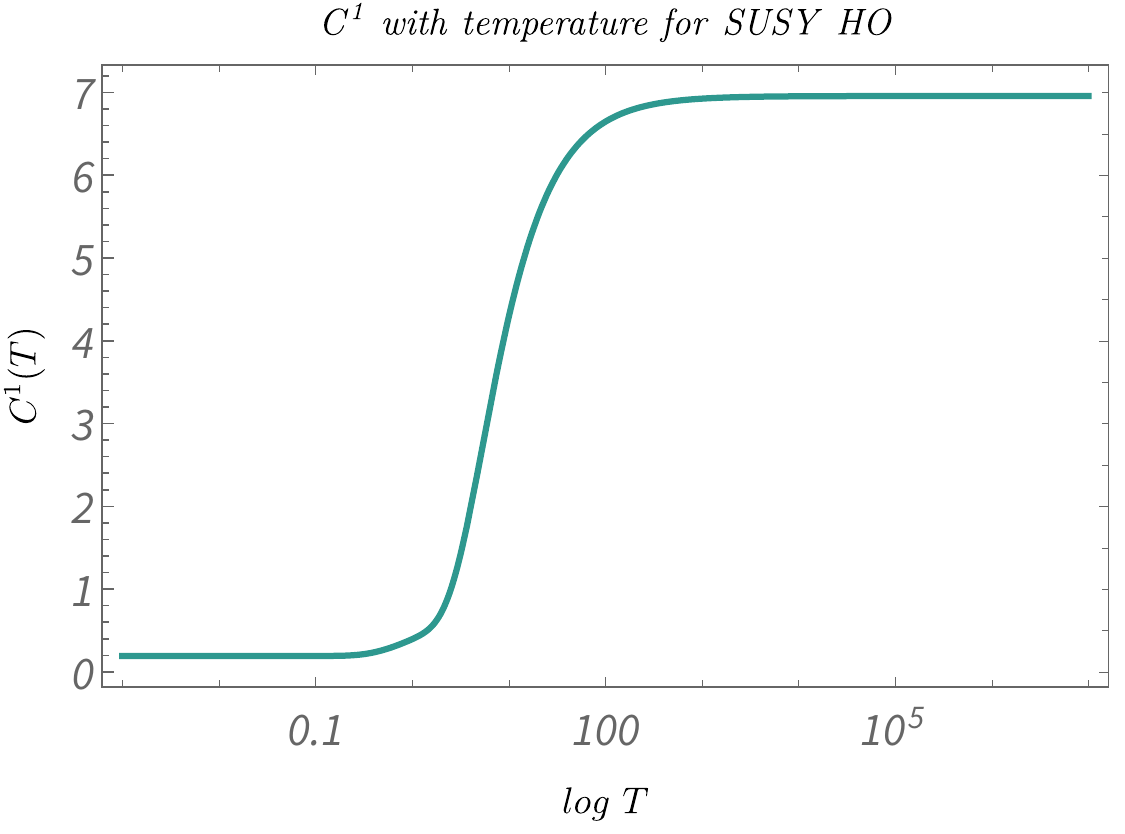}
	\caption{Supersymmetric 1D Harmonic Oscillator : Behavior of 4-pt canonical correlators with temperature for a particular value of the time interval . We have chosen ${t_{1} - t_{2} = t}$ as there is only one relevant time parameter.}
	\label{fig:HOtemp4ptC1}
\end{figure}

\begin{figure}[htb!]
	\centering
	\includegraphics[width=17cm,height=8.7cm]{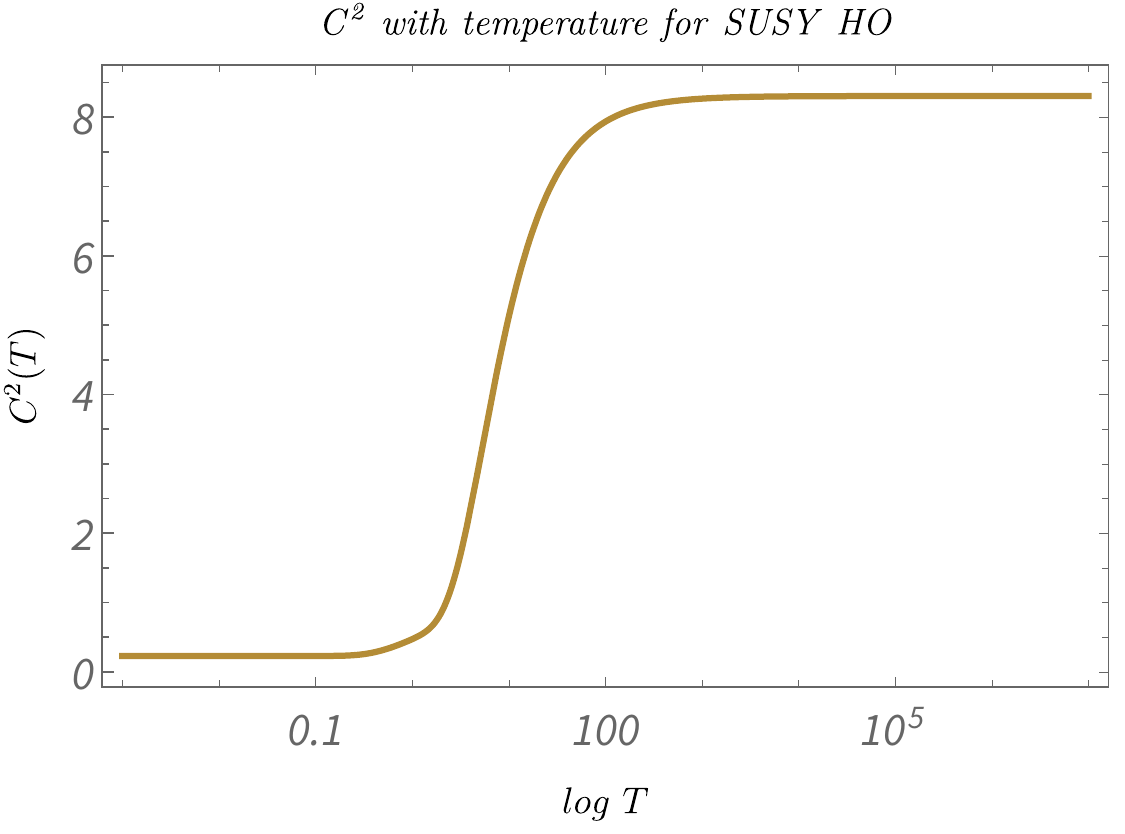}
	\caption{Supersymmetric 1D Harmonic Oscillator : Behavior of 4-pt canonical correlators with temperature for a particular value of the time interval . We have chosen ${t_{1} - t_{2} = t}$ as there is only one relevant time parameter.}
	\label{fig:HOtemp4ptC2}
\end{figure}

\begin{figure}[htb!]
	\centering
	\includegraphics[width=17cm,height=10.7cm]{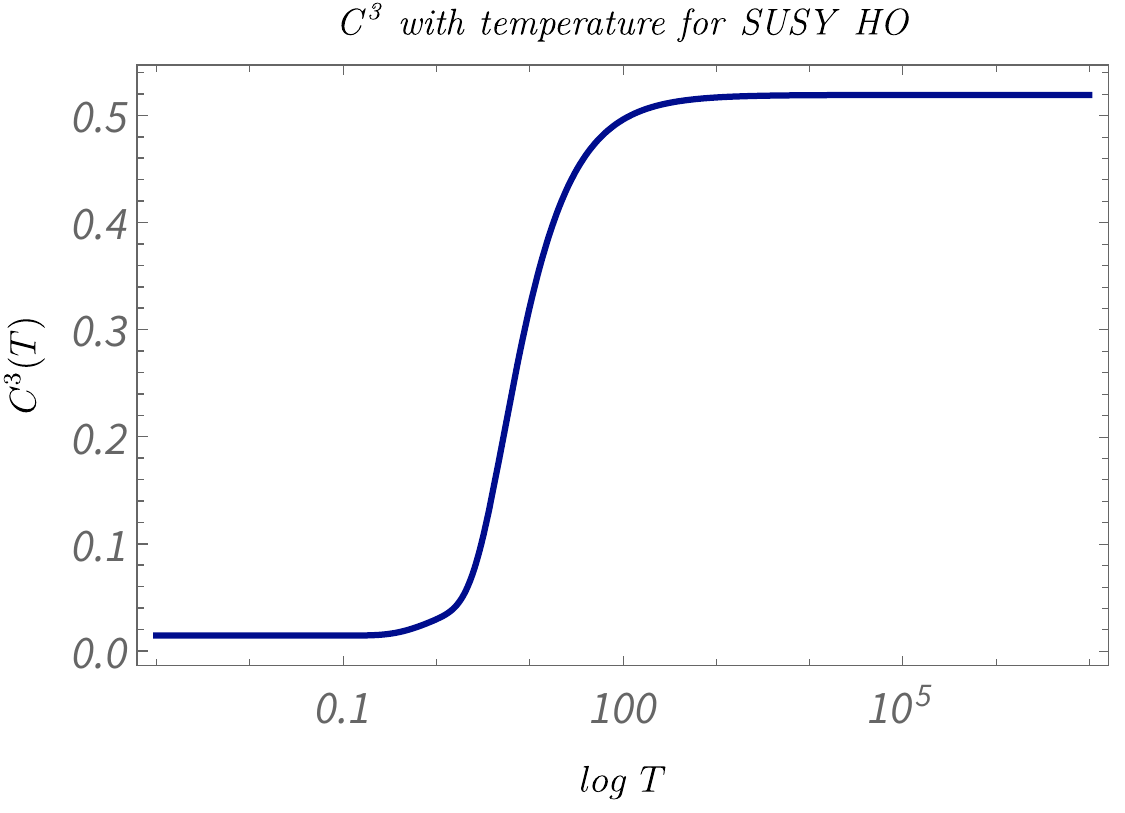}
	\caption{Supersymmetric 1D Harmonic Oscillator : Behavior of 4-pt canonical correlators with temperature for a particular value of the time interval . We have chosen ${t_{1} - t_{2} = t}$ as there is only one relevant time parameter.}
	\label{fig:HOtemp4ptC3}
\end{figure}

\noindent
For numerical evaluation we have chosen : ${\omega = 1\ \&\ 2M = 1}$, where ${\omega}$ is the frequency of the oscillator in which a particle of mass ${M}$ is confined. We also consider ${ \hbar = 1}$. \\

\begin{itemize}
	
	\item In \Cref{fig:y1HO,,fig:y2HO,,fig:y3HO} we perform the \textit{Study A} on the 2-pt micro-canonical correlators ${y_{m}^{(1,2,3)}(t_{1},t_{2})}$ for Supersymmetric 1D Harmonic Oscillator.
	
	\begin{itemize}
		
		\item We observe that the correlators ${y_{m}^{(1,2,3)}(t_{1},t_{2})}$ are periodic and that their periodicity does not vary with the state. 
		
		\item The amplitude of the correlator ${y_{m}^{(1)}(t_{1},t_{2})}$ initially increases with increasing ${m}$ which can be seen from the greater amplitude for m=2  than the amplitude for m=1. However with further increase of m the amplitude of the correlator shows negligible change and the amplitudes of the higher states almost overlaps. This suggests that for the lower energy states the micro-canonical correlators depend on the energy state in which it is calculated. However this state dependency goes away when calculated for the higher energy states. This can also be understood from the analytical expression obtained for the micro-canonical correlators obtained in \Cref{sec:susyqmho} (calculated for Harmonic Oscillator of unit mass i.e ${M=1}$). The micro-canonical correlators have a non trivial state dependence in the form of the factor $\bigl({1+\sqrt{m(m+1)}-\sqrt{m(m-1)}}\bigr)$ which reduces simply to 1 for higher energy states. 
		
		\item In \Cref{fig:y1HO,,fig:y2HO,fig:y3HO} we have also plotted ${y_{10}^{(1,2,3)}(t_{1},t_{2})}$ to draw a contrast of the boundary / truncation state with the other states. The ${m=10}$ correlators are lacking in feature, sometimes deceptively so, as compared to the other states which should come as no surprise because we have set our truncation at ${N_{\textnormal{trunc}} = 10}$. Furthermore, this state appears to violate the properties shown by other intermediate states but in fact this is merely an artefact of ${m=10}$ being the truncation state and that contribution of states with ${m>10}$ could not be accommodated in the calculations for ${m=10}$.
		
		\item The correlators ${y_{m}^{(2,3)}(t_{1},t_{2})}$ largely follow the same patterns and behaviour as shown by ${y_{m}^{(1)}(t_{1},t_{2})}$ with two exceptions. First, the amplitude for ${y_{m}^{(2)}(t_{1},t_{2})}$ correlator is amplified whereas that of ${y_{m}^{(3)}(t_{1},t_{2})}$ is suppressed, as compared to ${y_{m}^{(1)}(t_{1},t_{2})}$. The order of the amplification of the micro-canonical correlator ${y_{m}^{(2)}(t_{1},t_{2})}$ is exactly twice the amplitude of ${y_{m}^{(1)}(t_{1},t_{2})}$ which is merely a reflection of the fact that the mass of the oscillator has been chosen as ${1/2}$. Similarly the suppression of ${y_{m}^{(3)}(t_{1},t_{2})}$ is exactly by the same factor. Second, contrasting behaviour in the symmetry properties in ${t}$. Whereas, ${y_{m}^{(1)}(t_{1},t_{2})}$ is symmetric in ${t}$, ${y_{m}^{(2,3)}(t_{1},t_{2})}$ are anti-symmetric.
		
		\item In \Cref{fig:Y1HO,,fig:Y2HO,,fig:Y3HO} we perform \textit{Study B} on the 2-pt canonical correlators ${Y^{(1,2,3)}(t_1,t_2)}$. We observe that the correlators shows periodic behaviour for the different chosen values of temperature. We observe that each of the 2-pt correlators behave identically with respect to temperature. The amplitude of each of them decreases with increasing temperature. To have a better understanding of the temperature dependence of the 2-pt correlators we plot them with varying temperature keeping the time constant.
		
		\item In \Cref{fig:HOtemp2ptY1,,fig:HOtemp2ptY2,,fig:HOtemp2ptY3} we present the results of performing \textit{Study C} on 2-pt canonical correlators ${Y^{(i)}(t_{1},t_{2})}$. Here we plot ${Y^{(1,2,3)}(t_{1},t_{2})}$ which are the thermal or canonical correlators corresponding to ${y_{m}^{(1,2,3)}(t_{1},t_{2})}$ respectively with respect to temperature. It is clearly visible that for very low temperatures the canonical correlators are constant. After a certain value of the temperature the amplitude of the correlator shows a gradual increase. It reaches a maximum for a particular value of the temperature and then decays exponentially to zero. 
	\end{itemize}
	
	\item In \Cref{fig:c1HO,,fig:c2HO,fig:c3HO} we perform the \textit{Study A} on the 4-pt micro-canonical correlators ${c_{m}^{(1,2,3)}(t_{1},t_{2})}$ for Supersymmetric Harmonic Oscillator. 
	
	\begin{itemize}
		
		\item We observe that the correlators ${c_{m}^{(1,2,3)}(t_{1},t_{2})}$ are periodic and that their periodicity does not vary with the state. For the correlator ${c_{m}^{(1)}(t_{1},t_{2})}$ the periodicity is   roughly half of the corresponding 2-pt micro-canonical correlator. 
		
		\item Other properties of ${c^{(i)}_{m}(t_{1},t_{2})}$ are much like ${y^{(i)}_{m}(t_{1},t_{2})}$. We observe similar change in the amplitude of the correlators with changing ${m}$. The scaling of amplitudes in the case of 4-pt micro-canonical correlators, ${c^{(2)}_{m}(t_{1},t_{2})}$ and ${c^{(3)}_{m}(t_{1},t_{2})}$ is exactly by a factor of 2 than ${c^{(1)}_{m}(t_{1},t_{2})}$ . This is because for the 4-pt correlators there is a mass square dependence unlike the 2-pt correlators which has just mass dependence. The amplitudes of the respective 4-pt correlators can also be found to be exactly half of the amplitudes of its 2-pt counterpart. This is obvious from the time dependent functions appearing in the case of Supersymmetric Harmonic oscillator.
		
		\item All ${c_{m}^{(i)}(t_{1},t_{2})}$ are symmetric about ${t=0}$ which means that to these 4-pt micro-canonical correlators it does not matter whether ${t_{1} > t_{2}}$ or ${t_{1} < t_{2}}$. 
		
		\item In \Cref{fig:C1HO,,fig:C2HO,,fig:C3HO} we perform \textit{Study B} on the 4-pt canonical correlators ${C^{(1,2,3)}(t_1,t_2)}$. We observe that the correlators shows periodic behaviour for the different chosen values of temperature. We observe that each of the 4-pt correlators behave identically with respect to temperature which is exactly opposite in character from the 2-pt canonical correlators. The amplitude of each of them increases with increasing temperature. To have a better understanding of the temperature dependence of the 4-pt correlators we plot them with varying temperature keeping the time constant.
		
		\item In \Cref{fig:HOtemp4ptC1,,fig:HOtemp4ptC2,,fig:HOtemp4ptC3} we present the results of performing \textit{Study C} on 4-pt canonical correlators ${C^{(i)}(t_{1},t_{2})}$. Here we plot ${C^{(1,2,3)}(t_{1},t_{2})}$ which are the thermal or canonical correlators corresponding to ${y_{m}^{(1,2,3)}(t_{1},t_{2})}$ respectively with respect to temperature. It is clearly visible that for very low temperatures the 4-pt canonical correlators have negligible amplitudes. After a certain value of the temperature the amplitude of the correlator starts increasing and finally saturates to a finite value.
	\end{itemize}

\item The temperature dependent plots shows the significance of computing the 4-pt and the 2-pt correlators to study the phenomenon of quantum randomness for a Supersymmetric quantum mechanical model. From the plots it can be seen that for very low temperatures the 2-pt correlators shows a certain finite value whereas the 4-pt correlators are almost negligible. On the other hand at very high temperatures the 2-pt correlators are almost zero whereas the 4-pt correlators shows certain finite value. This suggests that to understand quantum randomness for a Supersymmetric model at very low temperature, the results from the 4-pt correlators can be misleading and similarly at high temperatures the 2-pt correlators might not be an appropriate quantity to study randomness. However to understand quantum randomness at the mid temperature range both 2 and 4-pt correlators play significant role.
We feel this shows the necessity for computing the 2-pt correlators. 
\end{itemize}


\textcolor{Sepia}{\section{\sffamily Conclusions}\label{sec:conclusions}}
To summarize, in this work, we have addressed the following issues to study the OTOC in the context of supersymmetric quantum mechanics:\\
\begin{itemize}
	\item We apply the computational techniques of recently developing methodology of out of time ordered correlators (OTOCs) to study the phenomenon of time disorder averaging for a given quantum statistical ensemble or quantum randomness for two very well known integrable one dimensional quantum mechanical models viz. Harmonic Oscillator and 1D potential well in the context of one dimensional supersymmetric quantum mechanics. We show that to develop a complete understanding of the underlying randomness in the quantum system, not only the correlators constructed from different operators at different times scales are important, but also the correlators constructed from similar quantum mechanical operators at different time scales, play a pivotal role.
	
	\item We have constructed all the temperature independent micro-canonical and temperature dependent canonical un-normalized and normalized version of these OTOCs in the eigenstate representation of the supersymmetric time independent Hamiltonian of the quantum system and represent all of them in a general model independent way. From the previous study it is expected that the OTOCs in the eigenstate representation one should not expect any chaotic behaviour i.e. the exponential growth with respect to the time scale in the correlators. However, one can get a broader knowledge of some other aspects of quantum randomness, which might capture the random behaviour in the correlators in terms of non-chaoticity. From our analysis it is expected that a large class of quantum mechanical models will be covered which provide the signature of quantum mechanical randomness in general.
	
	\item The explicit calculation of the 4-pt correlator $C^{(1)}(t_1,t_2)$ for the Supersymmetric Harmonic Oscillator, shows the significance of the introduction of supersymmetry within the context of quantum mechanics compared to the framework of quantum mechanics without having any supersymmetry. Supersymmetry introduces an energy state dependence on the temperature independent micro-canonical correlators which is usually not appear in the framework without having supersymmetry in the quantum mechanics description. Apart from the dependence on the energy states the canonical correlators have an additional dependence on temperature, which is also a different and notable feature compared to the results obtained from quantum mechanical set up without having supersymmetry. This energy state and temperature dependence of the correlators differentiates a Supersymmetric and a Non-Supersymmetric Harmonic oscillator\cite{Hashimoto:2017oit} though the time dependence in the OTOCs remains same. Also, particularly for the Supersymmetric Harmonic Oscillator we have found that the normalized four point OTOCs that are made up of same operators at different time scales show exactly same behaviour, which implies they are not independent of each other. However, this statement might not be true for other integrable models. On the other hand, in the un-normalized version of these two correlators we have found exact same time dependence, but the overall frequency dependent normalization factors will be different, which implies that they are proportional to each other in this case.
	
	\item The classical limit however matches with the non supersymmetric case apart from a factor of $2$ which can be inferred from the fact that in a supersymmetric quantum mechanical model there are two degrees of freedom, one from the original potential and the other associated with the partner potential. The time dependence of the classical and the quantum version of the correlators are exactly identical. However the temperature dependence observed in the quantum case vanishes for its classical counterpart, which is obviously a important finding from our computation.
	
	\item We observe a similar temperature and state dependence on the correlators for the Supersymmetric 1D Infinite potential well. However, it is interesting to note that \cite{Hashimoto:2017oit} also observed this state and temperature dependence for Non-Supersymmetric Infinite potential well. The behaviour of the only correlator studied in \cite{Hashimoto:2017oit} is exactly identical to what we observe for the Supersymmetric case. The correlator showed increase in the amplitude with increasing state number and higher temperatures which is exactly what we observe here. Hence we conclude that introduction of supersymmetry does not introduce new features in the case of 1D Infinite potential well.
	
	\item The significance of supersymmetry in 1D potential well however can be realised from its classical counterpart which is markedly different from what is obtained in the non-supersymmetric case. The classical limit of OTOC for the 1D non supersymmetric infinite potential well is independent of time and is merely a constant, whereas for the supersymmetric case there is a non-trivial time dependence, which is obviously a new finding from our computations. 
\end{itemize}

The future prospects of this work are as follows: 
\begin{itemize}
	\item In this paper we have restricted ourselves in considering only 2 and 4-pt correlation functions, to study quantum randomness in various Supersymmetric QM models. However a more generalised approach would be to consider the multipoint correlation functions to have a better understanding about the randomness underlying the system. We have an immediate plan to carry forward the work along this direction very soon.
	
	\item The study of OTOCs can be used to understand quantum randomness in various quantum mechanical models which are of prime significance in various condensed matter, nuclear and atomic physics models. Particularly the time dependent Hamiltonians we have not studied in this paper where this eigenstate formalism and related simplification to determine OTOC will not work. In that case one needs to use the general definition and representation of OTOC in terms of the well known {\it Schwinger Keldysh} formalism. We have some future plan on that direction as well.
	
	\item Use of the other types of correlators defined in this paper can be used to study some of the well understood QM models, to have an insight on the lost information of quantum randomness. We are very hopeful that incorporating the study of these additional correlators which we have defined and evaluated in this paper can able to capture more broader perspective on time disorder averaging phenomena through quantum randomness, rather only give insight about quantum mechanical chaos from the temporal growth of the correlators.
	
	\item Very recently in \cite{Choudhury:2020dpf, Banerjee:2020ljo, Akhtar:2019qdn, Bhattacherjee:2019eml, Jana:2020vyx} the authors have studied various relevant measures for an entangled open quantum system. The study of OTOCs for such type of entangled systems \cite{Choudhury:2017qyl, Choudhury:2017bou, Choudhury:2016pfr, Choudhury:2020ivj} will be quite relevant for understanding the rtime disorder averaging phenomena and chaos for such an entangled OQS. 
	
	\item Last but not the least, very recent we have proposed a detailed mechanism and framework using which one can able to compute the OTOC within the framework of primordial cosmological perturbation theory by making use of the gauge invariant scalar perturbations and its associated canonically conjugate momenta \cite{Choudhury:2020yaa} and finally found out the chaotic like behaviour in the representative cosmological version of OTOC. However, in that paper we have not reported anything about the other possible two operators which we have introduced in this paper. At present we are working on that direction and very hopeful to get the certain non-trivial features of time disorder averaging which might have application to explain various cosmologically relevant phenomena within the evolution history of our universe, which was not explored earlier at all.  
\end{itemize}

\clearpage
\newpage

\newpage
	\subsection*{Acknowledgements}
	SC is thankful to Latham Boyle, Robert Myers, Andrew R. Liddle, Douglas Stanford, Alexi Y. Kitaev, Paul Joseph Steinhardt, Martin Bojowald, Eugenio Bianchi, Sudhakar Panda, Soumitra SenGupta, Sumit Ranjan Das, Igor R. Klebanov, Eva Silverstein, Leonardo Senatore, Subhashish Banerjee, Anupam Mazumdar, Savan Kharel for enormous helpful discussions, suggestions and support. The Post-Doctoral research fellowship of SC is supported by the ERC Consolidator grant 772295 ``Qosmology" of Professor Jean-Luc Lehners. SC take this opportunity to thank sincerely to
Jean-Luc Lehners for his constant support and inspiration. SC thank 
 Latham Boyle for inviting at Perimeter Institute for Theoretical Physics (PITP), Zohar Komargodski for inviting at Simons Center for Geometry and Physics (SCGP), Stony Brook University, Leonardo Senatore  for inviting at  Institute for Theoretical Physics, Stanford University, Juan Martín Maldacena for inviting at  Workshop on Qubits and Spacetime, Institute for Advanced Studies (IAS), Princeton, Paul Joseph Steinhardt  for inviting at  Department of Physics, Princeton University, Martin Bojowald  for inviting at 
The Institute for Gravitation and the Cosmos (IGC),  Department of Physics, Eberly College of Science, Pennsylvania State University (University Park campus), Sudhakar Panda  for inviting at School of Physical Sciences, National Institute of Science Education and Research (NISER), Bhubaneswar, Abhishek Chowdhury  for inviting at Department of Physics, Indian Institute of Technology (IIT), Bhubaneswar, Anjan Sarkar  for inviting at Department of Astrophysics, Raman Research Institute, Bengaluru, Aninda Sinha and Banibrata Mukhopadhyay for inviting at Center for High Energy Physics (CHEP) and Department of Astronomy and Astrophysics, Indian Institute of Science, Bengaluru, Uma Shankar for inviting at Department of Physics, Indian Institute of Technology (IIT), Bombay, Shiraz Minwalla for inviting at Department of Theoretical Physics, Tata Institute of Fundamental Research, Mumbai, Abhishek Mahapatra for inviting at National Institute of Technology (NIT), Rourkela, for official academic visit where the work was done partially. Part of this work was presented as a talk, titled "Cosmology from Condensed Matter Physics: A study of out-of-equilibrium physics" and " Cosmology Meets Condensed Matter Physics" at Perimeter Institute for Theoretical Physics (PITP) (See the link: \textcolor{red}{http://pirsa.org/19110117/}), Simons Center for Geometry and Physics (SCGP), Stony Brook University (See the link: \textcolor{red}{http://scgp.stonybrook.edu/video portal/video.php?id=4358}), Department of Physics, Princeton University, The Institute for Gravitation and the Cosmos (IGC),  Department of Physics, Eberly College of Science, Pennsylvania State University (University Park campus), workshop on "Advances in Astroparticle Physics and Cosmology, AAPCOS-2020" at Saha Institute of Nuclear Physics, Kolkata on the occasion of the 100 years of Saha Ionisation Equation by Prof. Meghnad Saha, Department of Physics, Scottish Church College, Kolkata, School of Physical Sciences, National Insitute of Science Education and Research (NISER), Bhubaneswar, Department of Physics, Indian Institute of Technology (IIT), Bhubaneswar, Department of Astrophysics, Raman Research Institute, Bengaluru, Center for High Energy Physics (CHEP), Indian Institute of Science, Bengaluru, Department of Physics, Indian Institute of Technology (IIT), Bombay, National Institute of Technology (NIT), Rourkela. SC would like to thank Quantum Gravity and Unified Theory and Theoretical Cosmology
Group, Max Planck Institute for Gravitational Physics, Albert Einstein Institute (AEI), Perimeter Institute for Theoretical Physics (PITP), Simons Center for Geometry and Physics (SCGP), Stony Brook University, Institute for Theoretical Physics, Stanford University, 
The Institute for Gravitation and the Cosmos (IGC),  Department of Physics, Eberly College of Science, Pennsylvania State University (University Park campus), School of Physical Sciences, National Insitute of Science Education and Research (NISER), Bhubaneswar, Department of Astrophysics, Raman Research Institute, Bengaluru, Department of Physics, Indian Institute of Technology (IIT), Bombay, Quantum Space-time Group (Earlier known as String Theory and Mathematical Physics Group), Department of Theoretical Physics, Tata Institute of Fundamental Research, Mumbai and National Institute of Technology (NIT), Rourkela for providing financial support for the academic visits at Canada, U.S.A. and India. SC would like to thank the natural beauty of Prague, Dresden, Hamburg, Leipzig, Potsdam, Berlin, Mumbai, Bangalore, Kolkata, Bhubaneswar which inspires to do work very hard during the weekend trips and various academic visits. Particularly SC want to give a separate credit to all the members of the EINSTEIN KAFFEE Berlin Alexanderplatz for providing work friendly environment, good espresso shots and cookies, which helped to write the most of the part of the paper in that coffee shop. SC also thank all the members of our newly formed virtual international non-profit consortium ``Quantum Structures of the Space-Time \& Matter" (QASTM) for elaborative discussions. SC also would like to thank all the speakers of QASTM zoominar series from different parts of the world  (For the uploaded YouTube link look at: \textcolor{red}{https://www.youtube.com/playlist?list=PLzW8AJcryManrTsG-4U4z9ip1J1dWoNgd}) for supporting my research forum by giving outstanding lectures and their valuable time during this COVID pandemic time. KYB, SC (Satyaki), NG, RND, AM, GDP, SP would like to thank IISC Bangalore, NISER Bhubaneswar, IISER Mohali, IIT Bombay, University of Waterloo, IIT Indore respectively for providing fellowships. \textit{\textcolor{blue}{We would like to dedicate this work for the people those who are helping us to fight against
COVID-19 pandemic across the globe}}. Last but not the least, we would like to acknowledge our debt to 
the people belonging to the various part of the world for their generous and steady support for research in natural sciences. 
	\clearpage

\phantomsection
\addcontentsline{toc}{section}{References}
\appendix


\textcolor{Sepia}{\section{\sffamily  Derivation of the normalization factors for the Supersymmetric HO}}

To normalize the obtained OTOC we divide it by the thermal average of the dynamical variables considered in calculating the OTOC.

\begin{align}
\braket{x(t_1)x(t_1)}_{\beta}&= \frac{1}{Z} \text{Tr}(e^{-\beta H} x(t_1) x(t_1)) 
= \frac{1}{Z}\sum_{n}e^{-\beta n\omega} \langle \Psi_n| p(t_1) p(t_1)|\Psi_n \rangle .
\end{align}
Now keeping in mind that the ground state is only bosonic whereas all the other eigenstates has a fermionic part associated with it. Hence, separating the ground state from the other states, the above equation can be written as
\begin{align}
\braket{ x(t_1)x(t_1)}_{\beta}&= \frac{1}{Z}\biggl[\braket{\psi_0|x(t_1)x(t_1)|\psi_0}+\sum_{n>0}^{} e^{-\beta n \omega}\braket{\psi_n|x(t_1)x(t_1)|\psi_n}\biggr]\\ \nonumber
&= \frac{1}{Z}\sum_{k}\biggl[\underbrace{\braket{\psi_0|x(t_1)\ket{\psi_k}\bra{\psi_k}x(t_1)|\psi_0}}_\text{\textcolor{red}{Term A}}+\underbrace{\sum_{n>0}^{} e^{-\beta n \omega}\braket{\psi_n|x(t_1)\ket{\psi_k}\bra{\psi_k}x(t_1)|\psi_n}}_\text{\textcolor{red}{Term B}}\biggr].
\end{align}
Now we are going to explicitly show the calculation of term A and term B.
Expanding the position operators using the Heisenberg picture for the evolution of operators
\begin{align}
\nonumber
\text{Term A} &= \sum_{k}\braket{\psi_0|x(t_1)\ket{\psi_k}\bra{\psi_k}x(t_1)|\psi_0} \\ 
&= \frac{1}{2}\sum_{k} e^{-ik\omega t_1}x_{0k}^B ~e^{ik\omega t_1}x_{k0}^B 
= \frac{1}{2}\sum_{k}x_{0k}^B~x_{k0}^B = \frac{1}{4\omega} .\\ \nonumber
\text{Term B} &= \sum_{n>0}^{} e^{-\beta n \omega}\sum_{k}\braket{\psi_n|x(t_1)\ket{\psi_k}\bra{\psi_k}x(t_1)|\psi_n} \\ \nonumber
&= \sum_{n>0} e^{-\beta n \omega} \braket{\psi_n|x(t_1)\ket{\psi_0}\bra{\psi_0}x(t_1)|\psi_n} + \sum_{n>0} e^{-\beta n \omega} \sum_{k>0}\braket{\psi_n|x(t_1)\ket{\psi_k}\bra{\psi_k}x(t_1)|\psi_n} \\ \nonumber
&= \frac{1}{2}\sum_{n>0} e^{-\beta n\omega} e^{in\omega t_1}x_{n0}~e^{-in\omega t_1}x_{0n} + \sum_{n>0}\sum_{k>0} x_{nk}(t_1)x_{kn}(t_1) \\ 
&= \frac{e^{-\beta \omega}}{4\omega}+ \frac{1}{4\omega}\sum_{n>0} e^{-\beta n\omega}\biggl(2n+\sqrt{n(n+1)}+\sqrt{n(n-1)}\biggr).
\end{align}

\begin{tcolorbox}[colframe=black,arc=0mm]
	\begin{align} \label{eq:normxxapp} 
	\braket{x(t_1)x(t_1)}_{\beta}=\frac{1}{4Z\omega}\biggl(1+ e^{-\beta \omega}+\sum_{n>0} e^{-\beta E_n} [2n + \sqrt{n(n+1)}+ \sqrt{n(n-1)} ] \biggr).
	\end{align}
\end{tcolorbox}
\vspace{0.5cm}
A similar calculation is carried out for the thermal average of the product of the momentum operators.
\begin{align}
\label{normspppapp}
\braket{p(t_2)p(t_2)}_{\beta}&= \frac{1}{Z} \text{Tr}(e^{-\beta H} p(t_2) p(t_2)) 
= \frac{1}{Z}\sum_{n}e^{-\beta n \omega} \langle \Psi_n|p(t_2) p(t_2)|\Psi_n \rangle .
\end{align}
Now keeping in mind that the ground state is only bosonic whereas all the other eigenstates has a fermionic part associated with it. Hence, separating the ground state from the other states, the above equation can be written as
\begin{align}
\braket{ p(t_2)p(t_2)}_{\beta}&= \frac{1}{Z}\biggl[\braket{\psi_0|p(t_2)p(t_2)|\psi_0}+\sum_{n>0}^{}\braket{\psi_n|p(t_2)p(t_2)|\psi_n}\biggr]\\ \nonumber
&= \frac{1}{Z}\sum_{k}\biggl[\underbrace{\braket{\psi_0|p(t_2)\ket{\psi_k}\bra{\psi_k}p(t_2)|\psi_0}}_\text{\textcolor{red}{Term A}}+\underbrace{\sum_{n>0}^{}\braket{\psi_n|p(t_2)\ket{\psi_k}\bra{\psi_k}p(t_2)|\psi_n}}_\text{\textcolor{red}{Term B}}\biggr].
\end{align}
Now we are going to explicitly show the calculation of term A and term B.
Expanding the position operators using the Heisenberg picture for the evolution of operators
\begin{align}
\nonumber
\text{Term A} &= \sum_{k}\braket{\psi_0|p(t_2)\ket{\psi_k}\bra{\psi_k}p(t_2)|\psi_0} \\ 
&= \frac{1}{2}\sum_{k} e^{iE_{0k}t_2}p_{0k}^B ~e^{iE_{k0}t_2}p_{k0}^B 
= \frac{1}{2}\sum_{k}p_{0k}^B~p_{k0}^B = \frac{\omega}{4}. \\ \nonumber
\text{Term B} &= \sum_{n>0}^{} e^{-\beta E_n}\sum_{k}\braket{\psi_n|p(t_2)\ket{\psi_k}\bra{\psi_k}p(t_2)|\psi_n} \\ \nonumber
&= \sum_{n>0} e^{-\beta E_n} \braket{\psi_n|p(t_2)\ket{\psi_0}\bra{\psi_0}p(t_2)|\psi_n} + \sum_{n>0}\sum_{k>0}\braket{\psi_n|p(t_2)\ket{\psi_k}\bra{\psi_k}p(t_2)|\psi_n} \\ \nonumber
&= \frac{1}{2}\sum_{n>0} e^{-\beta E_n} e^{iE_{n0}t_2}p_{n0}~e^{iE_{0n}t_2}p_{0n} + \sum_{n>0}\sum_{k>0} p_{nk}(t_2)p_{kn}(t_2) \\ 
&= \frac{\omega e^{-\beta \omega}}{4}+ \frac{\omega}{4}\sum_{n>0} e^{-\beta E_n}\biggl(2n+\sqrt{n(n+1)}+\sqrt{n(n-1)}\biggr).
\end{align}

\begin{tcolorbox}[colframe=black,arc=0mm]
	\begin{align}
	\braket{p(t_2)p(t_2)}_{\beta}=\frac{\omega}{4Z}\biggl(1+e^{-\beta \omega}+ \sum_{n>0} e^{-\beta E_n} \biggl[2n +\sqrt{n(n+1)}+\sqrt{n(n-1)}\biggr]\biggr).
	\end{align}
\end{tcolorbox}
\vspace{0.5cm}


\textcolor{Sepia}{\section{\sffamily Poisson bracket relation for the Supersymmetric partner potential associated with the 1D infinite well potential}\label{app:PoissonbrackSUSY1Dpot}}
For calculating the Poisson Bracket  $\{x_2(t_1),p_2(t_2)\}$ we need the partial derivaties of the dynamical variables characterising the partner hamiltonian with respect to their initial values. For the partner potential associated with the 1D box it can be seen from \Cref{classolxpar} and \Cref{classolppar} that the partial derivatives of the classical variables with respect to its initial value will not be trivial and will depend on the initial value chosen. For the sake of convenience we introduce some symbols in this section. 
We denote the partial derivative of position with respect to its initial value with the symbol $\frac{\mathcal{O}_1}{\mathcal{O}_2}$ i.e
\begin{align}
\frac{\partial x_2(t)}{\partial x_2(0)}=(-1)^n\frac{\mathcal{O}_1}{\mathcal{O}_2}.
\end{align}
where the symbols $\mathcal{O}_1$ and $\mathcal{O}_2$ have the following expressions:
\begin{align}
\nonumber
\mathcal{O}_1 &= \sqrt{1+4\pi^4}\cos(\sin^{-1}(\alpha)-t\beta)\sin(\pi x_2(0)) \biggr(p_2(0)^2 +4 p_2(0)^2 \pi^4 + \\ \nonumber &~~~
16 \pi^8 \cot^4 (\pi x_2(0))+ 8 \pi^4 \cot^2 (\pi x_2(0))(p_2(0)^2 2\pi^4 -2\pi^4 \text{cosec}^2(\pi x_2(0))) \\ \nonumber&~~~~~~~~~-16 \pi^8 \cot^3 (\pi x_2(0))~t~ c~\text{cosec}^3(\pi x_2(0)) \sqrt{\frac{p_2(0)^2 \sin^2(\pi x_2(0))}{c}} \\ &~~~~~~~~~~~~~~~~~~ 4\pi^4 t \cot(\pi x_2(0))c^{3} \text{cosec}^5(\pi x_2(0)) \biggr[\frac{p_2(0)^2 \sin^2(\pi x_2(0))}{c}\biggr]^{3/2}\biggr). \\
\mathcal{O}_2 &= c^{2} \sqrt{\frac{p_2(0)^2 \sin{(\pi x_2(0))}}{c}\eta(t)}.
\end{align}

Similarly the partial derivative of $x_2(t)$ with $p_2(0)$ we get
\begin{align}
\frac{\partial x_2(t)}{\partial p_2(0)} &= \frac{p_2(0)\sqrt{1+4\pi^4}\cos\biggr[\sin^{-1}(\alpha)-t\beta\biggr]\biggr(4\pi^4 cos(\pi x_2(0))+c^{3}t\biggr)\sqrt{\displaystyle \frac{p_2(0)^2 \sin^2(\pi x_2(0))}{c}}}{\displaystyle \pi c^{3}\beta \sqrt{\frac{p_2(0)^2 \sin^2(\pi x_2(0))}{c}\eta(t)} },~~~~~~~~~
\end{align}

where the symbols $\alpha$, $\beta$ and $\eta(t)$ used in the above equations has the following expressions.
\begin{align*}
\alpha &= \frac{\cos (\pi x_2(0))}{\displaystyle \sqrt{\frac{p_2(0)^2+4\pi^4 \cot^2(\pi x_2(0))}{p_2(0)^2+4\pi^4+4\pi^4 \cot^2(\pi x_2(0))}}},~~~~~~~~
\beta = \sqrt{p_2(0)^2+4\pi^4+4\pi^4 \cot^2(\pi x_2(0))}, \\
\eta(t)&=\biggr(1-(1+4\pi^4)\sin^2[\sin^{-1}(\alpha)-t\beta]\biggr).
\end{align*} 
The partial derivative of the momentum associated with the partner potential wrt its initial position and the momentum  are now explicitly evaluated. The partial derivative of the momentum wrt the initial position is given by
\begin{align}
\frac{\partial p_2(t)}{\partial x_2(0)} = \frac{4\pi^5 \biggr(-\cot(\pi x_2(0))\text{cosec}^2(\pi x_2(0)) + \frac{\mathcal{Y}_1}{\mathcal{Y}_2}-\frac{\mathcal{Y}_3}{\mathcal{Y}_4}\biggr)}{\displaystyle \sqrt{c^2- \frac{4\pi^4 (1+4\pi^4)\sin^2[\sin^{-1}(\alpha)-\beta t]}{\eta(t)}}}.
\end{align}

where the symbols $\mathcal{Y}_1$, $\mathcal{Y}_2$, $\mathcal{Y}_3$, $\mathcal{Y}_4$ used in the above equations refers to the following
\begin{align}
\nonumber
\mathcal{Y}_1 &= (1+4\pi^4)^2 \cos[\sin^{-1}(\alpha)-\beta t] \sin(\pi x_2(0)) \biggr(p_2(0)^4 +4 p_2(0)^2 \pi^4 +16 \pi^8 \cot^4(\pi x_2(0))  \\ \nonumber &~~+8 \pi^4 \cot^2(\pi x_2(0))~(p_2(0)^2+2\pi^4 -2\pi^4 \text{cosec}^2(\pi x_2(0)))-16 \pi^8 t \cot^2(\pi x_2(0))~ c~ \\ \nonumber&~~~~~ \text{cosec}^3(\pi x_2(0)) \sqrt{\frac{p_2(0)^2 \sin^2(\pi x_2(0))}{c^2}}-4\pi^4 t \cot(\pi x_2(0)) c^3 \text{cosec}^5(\pi x_2(0)) \\ &~~~~~~~~~ \biggr(\frac{p_2(0)^2 \sin^2(\pi x_2(0))}{c^2}\biggr)^{3/2}\biggr)\sin^3 [\sin^{-1}(\alpha)-t\beta], \\
\mathcal{Y}_2 &= c^3 \beta \sqrt{\frac{p_2(0)^2 \sin^2(\pi x_2(0))}{c^2}}\biggr(-1+(1+4\pi^4)\sin^2[\sin^{-1}(\alpha)-\beta t]\biggr)^2 ,\\
\mathcal{Y}_3 &= (1+4\pi^4) \sin(\pi x_2(0)) \biggr(p_2(0)^4 + 4p_2(0)^2 \pi^4 + 16 \pi^8 \cot^4(\pi x_2(0))  \\ \nonumber &~~+8 \pi^4 \cot^2(\pi x_2(0))~(p_2(0)^2+2\pi^4 -2\pi^4 \text{cosec}^2(\pi x_2(0)))-16 \pi^8 t \cot^2(\pi x_2(0))~ c~ \\ \nonumber&~~~~~ \text{cosec}^3(\pi x(0)) \sqrt{\frac{p(0)^2 \sin^2(\pi x_2(0))}{c^2}}-4\pi^4 t \cot(\pi x_2(0)) c^3 \text{cosec}^5(\pi x_2(0)) \\ &~~~~~~~~~ \biggr(\frac{p_2(0)^2 \sin^2(\pi x_2(0))}{c^2}\biggr)^{3/2} \sin[2\sin^{-1}(\alpha)-2\beta t]\biggr), \\
\mathcal{Y}_4 &= 2 c^3 \beta \sqrt{\frac{p_2(0)^2 \sin^2(\pi x_2(0))}{c^2}}\biggr(-1+(1+4\pi^4) \sin^2[\sin^{-1}(\alpha)-\beta t]\biggr).
\end{align}

The partial derivative of the momentum wrt the initial momentum is given by
\begin{align}
\frac{\partial p_2(t)}{\partial p_2(0)} =\frac{\biggr(2 p_2(0) + \frac{\mathcal{Z}_1}{\mathcal{Z}_2}-\frac{\mathcal{Z}_3}{\mathcal{Z}_4}\biggr)}{\displaystyle 2\sqrt{c^2+ \frac{4\pi^4 (1+4\pi^4)\sin^2[\sin^{-1}(\alpha)-\beta t]}{-1+(1+4\pi^4)\sin^2[\sin^{-1}(\alpha)-\beta t]}}}.
\end{align}
where the symbols $\mathcal{Z}_1$, $\mathcal{Z}_2$, $\mathcal{Z}_3$, $\mathcal{Z}_4$ represents the following expressions
\begin{align}
\nonumber
\mathcal{Z}_1 &= 8p_2(0)\pi^4 (1+4\pi^4)^2 \cos[\sin^{-1}(\alpha)-\beta t]\biggr(4 \pi^4 \cos(\pi x_2(0)) \\  &~~~~~~~~~~~~~~~~~~~~~~~~~~~~~~~~+c^3 t\sqrt{\frac{p_2(0)^2 \sin^2(\pi x_2(0))}{c^2}}\biggr)\sin^3[\sin^{-1}(\alpha)-\beta t] ,\\ 
\mathcal{Z}_2 &= c^3 \beta \sqrt{\frac{p_2(0)^2 \sin^2(\pi x_2(0))}{c^2}} \biggr(-1+(1+4\pi^4)\sin^2[\sin^{-1}(\alpha)-\beta t]\biggr),\\
\mathcal{Z}_3 &= 4p_2(0) \pi^4 (1+4\pi^4)\biggr(4\pi^4 \cos(\pi x_2(0))+\sqrt{2}t c^3 \\ &~~~~~~~~~~~~~~~~~~~~~~~~~~~~~~~~` \sqrt{\frac{p_2(0)^2 \sin^2(\pi x_2(0))}{p_2(0)^2+4\pi^4 +(-p_2(0)^2 +4\pi^4)\cos(2\pi x_2(0))}}\biggr) ,~~~~~~~ \\
\mathcal{Z}_4 &= c^3 \beta \sqrt{\frac{p_2(0)^2 \sin^2(\pi x_2(0))}{c}}\biggr(-1+(1+4\pi^4)\sin^2[\sin^{-1}(\alpha)-\beta t]\biggr).
\end{align}

In a similar way the Poisson Bracket involving the position and momentum variables at different times, we denote it by $\frac{\mathcal{X}_2}{\mathcal{P}_2}$ for the sake of convenience i.e 
\begin{align}
\{x_2(t_1),p_2(t_2)\} =(-1)^n \frac{\mathcal{X}_2}{\mathcal{Q}_2},
\end{align} 
where $\mathcal{X}_2$ and $\mathcal{Q}_2$ can be explicitly evaluated to have the following expressions
\begin{align}
\nonumber
\mathcal{X}_2 &= p_2(0)\sqrt{1+4\pi^4}~\biggl[3p_2(0)^4+20 p_2(0)^2 \pi^4 + 64 \pi^8- 4(p_2(0)^4+4p(0)^2\pi^4-16\pi^8) ~~~\\ \nonumber &~~~~\cos(2\pi x(0))+(p_2(0)^2-4p_2(0)^2\pi^4)\cos(4\pi x(0))\biggr]\cos\biggl[\sin^{-1}\alpha-\frac{t_1 \gamma}{\sqrt{2}}\biggl]\text{cosec}^3(\pi x_2(0))~~ \\ \nonumber
&~~~~~\biggl(\frac{3\gamma}{\sqrt{2}}-4\sqrt{2}\pi^4 \gamma +24\sqrt{2}\pi^8 \gamma + \frac{(1+4\pi^4)^2 \cos[4\sin^{-1}(\alpha)-2\sqrt{2}t_2\gamma]~\gamma}{\sqrt{2}} \\ \nonumber &~~~~~-2\sqrt{2}(-1+16\pi^8)\cos[2\sin^{-1}(\alpha)-\sqrt{2}t_2\gamma]-16\pi^4 t_1 \sin[2\sin^{-1}(\alpha)-\sqrt{2}t_2\gamma] \\ \nonumber &~~~~~~-64\pi^8 t_1 \sin[2\sin^{-1}(\alpha)-\sqrt{2}t_2\gamma] +16\pi^4 t_2 \sin[2\sin^{-1}(\alpha)-\sqrt{2}t_2\gamma]\\  &~~~~~~+64\pi^8 t_2 \sin[2\sin^{-1}(\alpha)-\sqrt{2}t_2\gamma]\biggr) .\\
\nonumber
\mathcal{Q}_2 &=64 c^2 \beta^2 p_2(0) \sin(\pi x_2(0))\sqrt{\eta(t_1)}~(\eta(t_2))^2 \sqrt{c^2-\frac{4\pi^4 (1+4\pi^4)\sin^2[\sin^{-1}(\alpha)-t_2\beta]}{\eta(t_2)}}.
\end{align}
The symbol $\gamma$ used in the above equations denotes the following expression
\begin{align}
\gamma = \sqrt{\biggl(p_2(0)^2 + 8\pi^4-p_2(0)^2\cos(2\pi x_2(0))\biggr)}~\text{cosec}(\pi x_2(0)).
\end{align}

The Poisson Bracket of the momentum  at different times is symbolically denoted by 
$\{p_2(t_1),p_2(t_2)\} =\mathcal{P}_1/\mathcal{P}_2$ ,
where $\mathcal{P}_1$ and $\mathcal{P}_2$ represents the following expressions:
\begin{align} 
\nonumber
\mathcal{P}_1 &= 2p(0)\pi^4 (1+4\pi^4) \sin(\pi x(0))\biggr(-\frac{1}{4}\biggr( 1-4\pi^4 + (1+4\pi^4)\cos[2\sin^{-1}(\alpha)-2\beta t_1]\biggr)\biggr)^2 \\ \nonumber &~~~~~~~~
\beta \sin[2\sin^{-1}(\alpha)-2t_2 \beta]+\sin[2\sin^{-1}(\alpha)-2t_1 \beta] \biggr(\frac{1}{4}\biggr(1-4\pi^4 + (1+4\pi^4) \\  &~~~~~~~~~~~\cos[2\sin^{-1}(\alpha)-2t_2 \beta]\biggr)^2 \beta -2\pi^4 (1+4\pi^4)(t_1-t_2)\sin[2\sin^{-1}(\alpha)-2t_2 \beta]\biggr),~~~~~~~~~ \\ \nonumber
\mathcal{P}_2 &= \biggr\{p(0)\sin(\pi x(0))\biggr((1+4\pi^4)\sin^2[\sin^{-1}(\alpha)-t_1 \beta]-1\biggr)^2 \\ \nonumber &~~~~~\times \sqrt{c^2+\frac{4\pi^4(1+4\pi^4)\sin^2[\sin^{-1}(\alpha)-t_1\beta]}{(1+4\pi^4)\sin^2[\sin^{-1}(\alpha)-t_1\beta]-1}} \biggr\} \times \biggr((1+4\pi^4)\sin^2[\sin^{-1}(\alpha)-t_2 \beta]-1\biggr)^2 \\  &~~~~~~~~~~~~~~~~~~~~~~~~~~~~~~~~~~~~~~~~~~~~ \times \sqrt{c^2+\frac{4\pi^4(1+4\pi^4)\sin^2[\sin^{-1}(\alpha)-t_2\beta]}{(1+4\pi^4)\sin^2[\sin^{-1}(\alpha)-t_2\beta]-1}}.
\end{align}


\end{document}